\documentclass[runningheads,openany]{svmult}

\usepackage{palatino}

\usepackage{euler}

\usepackage{helvet}

\usepackage{graphicx}  

\usepackage{physprbb}  

\usepackage{mathrsfs}

\usepackage{proc}  

\usepackage{epsfig}

\usepackage{amssymb}

\makeindex             

\usepackage{amsmath}   
\usepackage{multicol}
\usepackage{bm}

\let\tocauthor\authorrunning

\begin{document}

\frontmatter



\thispagestyle{empty}
\parindent=0pt

{\Large\sc Blejske delavnice iz fizike \hfill Letnik~7, \v{s}t. 2}

\smallskip

{\large\sc Bled Workshops in Physics \hfill Vol.~7, No.~2}

\smallskip

\hrule

\hrule

\hrule

\vspace{0.5mm}

\hrule

\medskip
{\sc ISSN 1580--4992}

\vfill

\bigskip\bigskip
\begin{center}

{\bfseries 
{\Large  Proceedings to the $9^\textrm{th}$ Workshop}\\
{\Huge What Comes Beyond the Standard Models\\}
\bigskip
{\Large Bled, September 16--26, 2006}\\
\bigskip
}

\vspace{5mm}

\vfill

{\bfseries\large
Edited by

\vspace{5mm}
Norma Manko\v c Bor\v stnik

\smallskip

Holger Bech Nielsen

\smallskip

Colin D. Froggatt

\smallskip

Dragan Lukman

\bigskip


\vspace{12pt}

\vspace{3mm}

\vrule height 1pt depth 0pt width 54 mm}

\vspace*{3cm}

{\large {\sc  DMFA -- zalo\v{z}ni\v{s}tvo} \\[6pt]
{\sc Ljubljana, december 2006}}
\end{center}
\newpage

\thispagestyle{empty}
\parindent=0pt
\begin{flushright}
{\parskip 6pt
{\bfseries\large
                  The 9th Workshop \textit{What Comes Beyond  
                  the Standard Models}, 16.-- 26. September 2006, Bled}

\bigskip\bigskip

{\bfseries\large was organized by}

{\parindent8pt
\textit{Department of Physics, Faculty of Mathematics and Physics,
University of Ljubljana}

}

\bigskip

{\bfseries\large and sponsored by}

{\parindent8pt
\textit{Slovenian Research Agency}

\textit{Department of Physics, Faculty of Mathematics and Physics,
University of Ljubljana}


\textit{Society of Mathematicians, Physicists and Astronomers
of Slovenia}}}
\bigskip
\medskip

{\bfseries\large Organizing Committee}

\medskip

{\parindent9pt
\textit{Norma Manko\v c Bor\v stnik}

\textit{Colin D. Froggatt}

\textit{Holger Bech Nielsen}}

\end{flushright}

\setcounter{tocdepth}{0}

\tableofcontents

\cleardoublepage

\chapter*{Preface}
\addcontentsline{toc}{chapter}{Preface}

The series of workshops on "What Comes Beyond the Standard Model?" started 
in 1998 with the idea of organizing a real workshop, in which participants 
would spend most of the time in discussions, confronting different 
approaches and ideas. The picturesque town of Bled by the lake of the 
same name, surrounded by beautiful mountains and offering pleasant walks, 
was chosen to stimulate the discussions.

The idea was successful and has developed into an annual workshop, which is
taking place every year since 1998.  Very open-minded and fruitful discussions
have become the trade-mark of our workshop, producing several published works.
It takes place in the house of Plemelj, which belongs to the Society of
Mathematicians, Physicists and Astronomers of Slovenia.

In this ninth workshop, which took place from 16 to 26 of September 2006 at
Bled, Slovenia, there were some changes: the date for this year's workshop was
moved due to participants' other obligations from a customary mid July to
September, and several members of prof. Sannino's group were present thanks to
bilateral Slovene-Danish collaboration project. They delivered talks,
write-ups of some of which you can read in this volume and enriched our
discussions. This ninth workshop differs from the previous ones in the fact
that it is a very short period between the workshop and the deadline for
sending the contributions for the proceedings. Because of this many a
participant has not succeeded to send the contribution in time. We promise to
include those, which were received too late to be included in this
proceedings, in the next year proceedings.  Also the discussion section, which
usually is quite the rich one, is in this time missing several contributions,
from the same reason - too short time.

We have tried to answer some of the open questions which the
Standard models leave unanswered, like:
 
\begin{itemize}
\item  Why has Nature made a choice of four (noticeable) dimensions? While all 
the others, if existing, are hidden?  And what are the properties of 
space-time in the hidden dimensions? 
\item  How could Nature make the decision about the breaking of symmetries 
down to the noticeable ones, if coming from some higher dimension $d$?
\item  Why is the metric of space-time Minkowskian and how is the choice 
of metric connected with the evolution of our universe(s)? 
\item  Why do massless fields exist at all? Where does the weak scale 
come from?
\item  Why do only left-handed fermions carry the weak charge? Why does 
the weak charge break parity?
\item  What is the origin of Higgs fields? Where does the Higgs mass come from?
\item  Where does the small hierarchy come from? (Or why are some Yukawa 
couplings so small and where do they come from?) 
\item  Where do the generations come from? 
\item  Can all known elementary particles be understood as different states of 
only one particle, with a unique internal space of spins and charges?
\item  How can all gauge fields (including gravity) be unified and quantized?
\item  How can different geometries and boundary conditions influence conservation laws?
\item  Does noncommutativity of coordinate manifest in Nature?
\item  Can one make the Dirac see working for fermions and bosons?
\item  What is our universe made out of (besides the baryonic matter)?
\item  What is the role of symmetries in Nature?
\end{itemize}

We have discussed these and other questions for ten days. Some results of this
efforts appear in these Proceedings. Some of the ideas are treated in a very
preliminary way.  Some ideas still wait to be discussed (maybe in the next
workshop) and understood better before appearing in the next proceedings of
the Bled workshops.  The discussion will certainly continue next year, again
at Bled, again in the house of Josip Plemelj.

Physics and mathematics are to our understanding both a part of Nature.  To
have ideas how to try to understand Nature, physicists need besides the
knowledge also the intuition, inspiration, imagination and much more.  These
fundamental questions also receive quite an attention by the general
public\,---\,see for example articles on these topics in several science
magazines. Among them we should perhaps mention the article '\textit{Create
  your own universe}' by Zeeya Merali in the New Scientist magazine (Issue
2559, 10 July 2006), mentioning the work of our contributor Eduardo Guendelman
(see the article by Ansoldi and Guendelman on this very topic in this issue).

The organizers are grateful to all the participants for the lively discussions
and the good working atmosphere. Support for the bilateral Slovene-Danish
collaboration project by the Research Agency of Slovenia is gratefully 
acknowledged.\\[2 cm]

\parbox[b]{40mm}{%
                 \textit{Norma Manko\v c Bor\v stnik}\\
                 \textit{Holger Bech Nielsen}\\
                 \textit{Colin Froggatt}\\
                 \textit{Dragan Lukman} }
\qquad\qquad\qquad\qquad\qquad\qquad\quad
\textit{Ljubljana, December 2006}

\newpage

\cleardoublepage


\mainmatter

\parindent=20pt

\setcounter{page}{1}


\title{Child Universes in the Laboratory}
\author{S. Ansoldi${}^1$ and E.I. Guendelman${}^2$}
\institute{%
${}^1$ International Center for Relativistic Astrophysics (ICRA)\\
and Universit\`{a} degli Studi di Udine, Udine, Italy\\
\textrm{email:} \texttt{ansoldi@trieste.infn.it}\thanks{%
Mailing address: Dipartimento di Matematica e Informatica,
Universit\`{a} degli Studi di Udine, via delle Scienze 206,
I-33100 Udine (UD), Italy}\\
${}^2$ Eduardo I. Guendelman\\
Ben Gurion Univeristy, Beer Sheva, Israel\\
\textrm{email:} \texttt{guendel@bgu.ac.il}
}

\titlerunning{Child Universes in the Laboratory}
\authorrunning{S. Ansoldi and E.I. Guendelman}
\maketitle

\begin{abstract}
Although cosmology is usually considered an
observational science, where there is little
or no space for experimentation, other approaches
can (and have been) also considered. In particular,
we can change rather drastically the above, more
passive, observational perspective and
ask the following question: could it be possible,
and how, to create a universe in a laboratory?
As a matter of fact, this seems to be possible,
according to at least two different paradigms;
both of them help to evade the consequences
of singularity theorems.
In this contribution we will review some of these models
and we will also discuss possible extensions
and generalizations, by
paying a critical attention to the still open issues as,
for instance, the detectability of child universes and the
properties of quantum tunnelling processes.

\end{abstract}

\section{The studies so far \dots{}}

The world Cosmology stems from the Greek word \textit{cosmos},
which meant \textit{beauty}, \textit{harmony}, and is the name
of that branch of science which studies the origin and evolution
of the universe. Thus, considering its name and the object of its study,
it is, perhaps, natural to take a ``passive'' point of
view when dealing with cosmological problems, where we use
the word \emph{passive} to emphasize that our experience of cosmology
is mainly observational in nature. This may undoubtedly
be a condition that seems hard to change \emph{in practice}: after
all we are dealing with problems, as the birth
of our universe and its evolution in the present
state, which do not appear suitable for a direct experimental approach.
On the other hand, we do not see any reason why this should prevent
us from changing our \emph{attitude} toward the problem, switching
from a contemplative to a more active one.
In our opinion, a stimulation in this sense is coming already
from the theory which first gave
us the opportunity to address cosmological
problems quantitatively, i.e.
General Relativity. General Relativity
raises for the first time the concept of
causality as a central one in physics.
This means that, taking a very pragmatic
point of view, we have to
admit that only a subset of what exists in
our universe can be experienced/observed
by us. This is not because of our limited
capabilities as humans, but, more
fundamentally, because of the restrictions
imposed by the spacetime structure
on the causal relations among objects.
At the same time causality also brings a challenge
to cosmologists in connection with the \emph{large scale} spacetime structure;
this is because the simplest models
of the universe which are built according to General Relativity
and whose late time predictions have a reasonable degree of consistency with what we observe,
seem doomed to have an initial singularity in their past so that field equations
break down exactly where we would like to set up the initial conditions.
This undesirable situation looks even more disappointing after the observation
that many parameters describing the state of the early universe
are quite far from the
domain of ``very large scales'' which characterizes
the present observable universe. Let us, for instance,
consider
a Grand Unified Theory scale of
$10 ^{14}$ Gev: the universe could then emerge
from a classical bubble which starts
from a very small size and has a mass of
of the order of about $10$ Kg (by using quantum tunnelling
the mass of the bubble could be
arbitrarily small, but the probability of
production of a new universe out
of it would be reduced).
The density of the universe would, admittedly,
have been quite higher than
what we could realize with present technologies,
but the orders of magnitude of the
other parameters make not unreasonable to
ask the question: might we have the possibility
of building a child universe in the laboratory?

As a matter of fact,
a positive answer to this question was already
envisaged some years ago (for a popular level
discussion see \cite{bib:NewSc20062559....32M}).
In particular Farhi \emph{et al.} suggested an interesting model able
to describe universe creation starting from a
non-singular configuration and involving
semiclassical effects. This proposal, actually,
leaves some open issues, for instance about
the semiclassical part of the process and the
global (Euclidean) structure of the solution.
Although since then, a few more proposal
have appeared, addressing in more detail
qualitative issues, it is interesting to observe
that most of the problems which emerged
in the earliest formulation are,
somehow, still open. It is our hope that
the present review of the different approaches
which have been developed along this interesting
research line, will stimulate to study
in more detail and with systematic
rigor these problems as well as other
realistic answers to the above question.
In our opinion, this question is not a purely
academic one, and might help not only to change
our perspective (passing from an observational
to an experimental one) in addressing cosmological
problems, but also to shed some light on the importance
of the interplay of gravitational and quantum phenomena.

We would also like to remark how, this complementary
perspective can be considered much more promising nowadays
than some years ago, thanks to the results of recent observations.
These observations are helping us in focusing our field of view
back in time, closer and closer to the earliest stages of life
of our universe and are providing us with a large amount of data
and information that will, hopefully, help us in sharpen our
theoretical models. This has already
allowed tighter constrains on the parameters of
models of the early universe, giving us the chance
for a more decisive attack of most of the still open
problems. This will be a great help also for a
``child universe formation in the laboratory''
program; it can make easier to identify the
fundamental elements (building blocks)
required to model the \emph{creation} of a
universe that will evolve in something similar to
the present one. At the same time, it will help
us to narrow our selection of the fundamental
principles that forged the earliest evolution of
the universe, and, as we said already before,
strengthen our hope to enlighten
a crucial one, which is the interplay between
General Relativity and Quantum Theory.

This said, in the rest of this section, keeping in mind the
above preliminary discussion, we are going
to give a concise review of the state of the
art in the field and to to make a closer contact
with some of the models for child universe formation;
in particular we are going to review some of the existing
works on the dynamics of vacuum bubbles and on
topological inflation
(both also considered
in a semiclassical framework).

Callan and Coleman initiated the study of vacuum decay more than
30 years ago \cite{bib:PhReD1977..15..2929C,bib:PhReD1977..16..1762C};
after their seminal papers the interest in the subject rapidly increased.
The possible interplay of true vacuum bubbles with gravitation was then
considered \cite{bib:PhReD1980..21..3305L,bib:PhReD1987..36..1088W}.
More or less at the same time and as opposed with the true vacuum
bubbles of Coleman \textit{et al.}, false vacuum bubbles were also considered.
The classical behavior of
regions of false vacuum coupled to gravity
was studied by Sato et \textit{al.}
\cite{bib:PrThP1981..65..1443M,%
bib:PrThP1981..66..2052M,bib:PrThP1981..66..2287S,%
bib:PhLeB1982.108....98K,bib:PhLeB1982.108...103M,%
bib:PrThP1982..68..1979S} and followed by the works
of Blau \textit{et al.} \cite{bib:PhReD1987..35..1747G}
and Berezin \textit{et al.} \cite{bib:PhReD1987..36..2919T,%
bib:PhReD1991..43.R3112T}. The analysis in \cite{bib:PhReD1987..35..1747G}
clarified some aspects in the study of false vacuum
dynamics coupled to gravity; in it, for the first time, the problem
was formulated using geodesically complete coordinate systems:
this made more clear the issue of wormhole formation,
with all its rich sequel of stimulating properties and consequences.

The presence of wormholes makes possible a feature of false vacuum
bubbles that is otherwise counterintuitive, which is
that these objects can undergo an exponential inflation
without displacing the space outside of the bubble itself;
this could seem strange at a first look and is due to the fact
that they have an energy density which is higher
than that of the surrounding spacetime and which is
responsible for keeping the required pressure difference.
Because of this, \emph{child universe} solutions appear
as expanding bubbles of false vacuum
which disconnect from the exterior region.
Apart from the already mentioned wormhole, they are also
characterized by the presence
of a \emph{white-hole like} initial singularity;
the simplest example can be obtained by modelling
the region inside the bubble with a
domain of de Sitter spacetime and the
region outside the bubble with a domain of Schwarzschild spacetime.
These two regions are then joined across the bubble
surface, using the well known Israel junction conditions
\cite{bib:NuCim1966.B44.....1I,%
bib:PhReD1991..43..1129I}; Einstein equations, which hold
independently in the two domains separated by the bubble, are
also satisfied on the bubble surface if interpreted in a distributional
sense; they determine the motion (embedding)
of the bubble in the two domains of spacetime.
Although there are various simple configurations of this system
(as well as more elaborate generalizations)
that are appropriate to describe the evolution
of a newly formed universe (i.e. they are such that
the expanding bubble can become very large),
these \emph{classical} models present also some
undesirable features.
    In particular it turns out that
    only bubbles with masses above
    some critical value can expand from very small
    size to infinity.
    But then these solutions necessarily have a (white-hole) singularity
     in their past; in fact, for all of them the
    hypotheses of singularity theorems are satisfied.

In connection with the restriction on the values of the total mass,
the situation could be improved in theories containing an appropriate multiplet
of additional scalars\cite{bib:PhReD1991..44..3152R}\footnote{{The
subject of inflation assisted by topological
defects was also studied later in
\cite{bib:PhReL1994..72..3137V} and
\cite{bib:PhLeB1994..327...208L}.}}:
then all bubbles that start evolving from zero radius can inflate
to infinity if the scalars are in
a ``hedgehog'' configuration, or global monopole
of big enough strength. This effect also
holds in the gauged case for magnetic monopoles
with large enough magnetic charge: in this way the
mass requirement is traded for requirements about the properties
of magnetic monopoles.

A possible connection of this approach with the problem of the
initial singularity appears, then, from the work
of Borde \textit{et al.}
\cite{bib:PhReD1999..59043513V}: they
proposed a mechanism which,
by means of the coalescence of two regular magnetic
monopoles (with \textit{below critical}
magnetic charge), is able to produce
a \textit{supercritical} one, which then
inflates giving rise to a child universe.
This idea might help addressing the
singularity problem and in this context it is very
interesting the work of Sakai \textit{et al.}
\cite{bib:gr-qc2006..02...084K}: in it
the interaction of a magnetic monopole
with a collapsing surrounding membrane
is considered; also in this case a
new universe can be created and the presence
of an initial singularity \emph{in the causal past
of the newly formed universe} can be avoided.

To solve the problem of initial singularity, there
are also other approaches which make a good use of
quantum effects. Needless to say, these ideas are
very suggestive because they require a proper interplay
of quantum and gravitational physics, for which a consistent
general framework is still missing. This is the main reason
why most of these investigations try to obtain a simplified
description of the system by requiring a high degree of
symmetry from the very beginning. In particular, if
we describe the bubble separating the inflating
spacetime domain from the surrounding spacetime
in terms of Israel junction conditions
\cite{bib:NuCim1966.B44.....1I,%
bib:PhReD1991..43..1129I}, under the additional assumption
of spherical symmetry, the dynamics of the system is
determined by the dynamics of an effective system with only one degree
of freedom: this is called the \textit{minisuperspace
approximation}; in this framework the problem
of the semiclassical quantization of the system, even
in the absence of an underlying
quantum gravity theory, can be undertaken with less (but
still formidable) technical problems using as a direct guideline
the semiclassical procedure with which we are familiar in ordinary
Quantum Mechanics. This has been the seminal idea of
Farhi \textit{et al.} \cite{bib:NuPhy1990B339...417G} and of Fishler \textit{et al.}
\cite{bib:PhReD1990..41..2638P,bib:PhReD1990..42..4042P}.
One additional difficulty in these approaches was in connection with the
stability of the classical initial state. Interestingly enough,
this could be solved by the introduction of massless scalars or gauge fields
that live on the shell and produce a classical stabilization effect
of false vacuum bubbles.
By quantum tunnelling, these bubbles can then become child universes
\cite{bib:ClQuG1999..16..3315P} and, at least in a $2+1$-dimensional example
\cite{bib:MPhLA2001..16..1079P}, it has been shown that
the tunnelling can be arbitrarily small.

\section{\dots{} and their future perspectives}

From the above discussion, we think it is already clear that there
are many interesting aspects in the study of models for child universe
creation in the laboratory. We would also like to remember how
most of these models are based on a very well-known
and studied classical system,
usually known as a \textit{general relativistic shell}
\cite{bib:NuCim1966.B44.....1I,%
bib:PhReD1991..43..1129I}. The classical dynamics
of this system is thus ``under control'',
many analytical results can be found in the literature
and numerical methods
have also been employed (see the introduction of
\cite{bib:ClQuG2002..19..6321A} for
additional references). On the other hand there has been
little progress in the development of the quantized
theory, which still remains a
non-systematized research field. We stress how a progress in
this direction would be decisive
for a more detailed analysis of the
semiclassical process of universe creation.

Before coming back to the quantum side of the problem,
let us first consider what could be done on
the classical one. We will concentrate mainly
on the works
of Borde \textit{et al.} \cite{bib:PhReD1999..59043513V}
and of Sakai \textit{et al.} \cite{bib:gr-qc2006..02...084K},
which suggest many interesting ideas for further
developments.
    For instance, it is certainly important to extend
    the analysis in \cite{bib:PhReD1999..59043513V}, which is
    mainly qualitative in nature, to take fully into account
    the highly non-linear details of the collision process
    by means of which a supercritical monopole is created
    (this is certainly instrumental for a quantitatively
    meaningful use of the idea of topological inflation).
    Also the study performed in \cite{bib:gr-qc2006..02...084K}
    should be extended; to obtain some definitive conclusion about
    the stability of the initial configuration, it is, in fact,
    necessary to study the spacetime structure of the model for
    all possible values of the parameters; it could then be possible
    to determine if stability is a general feature of
    monopole models or an \emph{accident} of some particular
    configurations.
From the classical point of view, in both the above models another
central point is the study of their causal structure; it can be
obtained by well-known techniques, but, again, a full classification
of all the possibilities that can arise
is certainly required to gain support for the proposed mechanisms.
Known subtleties which require closer scrutiny
(as for example, the presence of singularities
in the causal past of the created universe
\emph{but} not in the past of the experimenter
creating the universe in the laboratory or, sometimes,
the presence of timelike \emph{naked} singularities)
make a discussion of the
problem of initial conditions not only interesting
but necessary, especially in this context\footnote{The proper
analysis of the Cauchy problem will, in fact,
involve resolution or proper handling of these
singularities.}.

A suggestive complement to the classical aspects discussed
above, is represented, of course, by the quantum
(more precisely semiclassical) ones, where
quantum effects are advocated to realize
the tunnelling between classical solutions. If
(i) the classical solution used to describe the initial state can
be formed without an initial
singularity and is stable, (ii) the classical solution which represents the final state
can describe an inflating universe and (iii) we can master
properly the tunnelling process, then we could
use the quantum creation of an inflating
universe \emph{via} quantum tunnelling
to evade the consequences of singularity
theorems. The construction of proper initial and final states
has already been successfully accomplished. The
stability of the initial classical configuration has been,
instead, only partly analyzed \cite{bib:ClQuG1999..16..3315P}
and it would be certainly interesting to
consider the tunnelling process in more
general situations, where, for example,
the stabilization can be still classical
in origin. Although there is some evidence
\cite{bib:gr-qc2006..02...084K}
of a general way to solve this issue
in the context of monopole
configurations, as we mentioned
above, the analysis should be extended
to the whole of the parameter space.
At the same time a complementary possibility is
that \emph{semiclassical}
effects might stabilize the initial
configuration. In particular,
closely related to the problem of
instabilities present in many models,
is the fact that the spacetime surrounding the vacuum
bubble has itself an instability due to presence of a
white hole region (see, for instance,
\cite{bib:PhReD1977..16..3359T}). Also in this
context quantum effects might stabilize the system
and help solving the issue. This approach could
require the determination of the
stationary states of the system in the
WKB approximation, a problem for which
a generalization of the procedure presented in
\cite{bib:ClQuG2002..19..6321A} (where
this analysis was performed for the first
time in a simplified model) could be useful.

Another equally (if not more) important point for future
investigations is certainly related with the still
open issues in the semiclassical tunnelling procedure.
We will shortly discuss this
by following, for definiteness,
the clear, but non-conclusive,
analysis developed by Farhi
\textit{et al.} \cite{bib:NuPhy1990B339...417G}:
it is shown in their paper that, when considering
the tunnelling process, it is not possible
to devise a clear procedure to build the manifold
interpolating between the initial and final
classical configurations; this
manifold would describe the instanton that is
assumed to mediate the process. According to
the discussion of Farhi \textit{et al.}
it seems possible to build
only what they call a \emph{pseudo-manifold},
i.e. a manifold in which various points
have multiple covering. To make sense of
this, they are forced to introduce
a `covering space' different from the
standard spacetime manifold, in which
they allow for a change of sign of the
volume of integration required for
the calculation of the tunnelling action
and thus of tunnelling probabilities.
It would be important to put on a more solid
basis this interesting proposal, comparing it
with other approaches
which might help to give a more precise
definition of this
\emph{pseudo-manifold}. In particular
we would like to mention two possibilities.
A first one uses the \textit{two measures
    theory} \cite{bib:PhReD1999..60065004K}; considering
    four scalar fields it is possible to define
    an integration measure in the action
    from the determinant of the mapping between
    these scalar fields and the four
    spacetime coordinates; there can, of course, be
    configurations where this mapping is not of
    maximal rank and if we then interpret the scalar
    fields as coordinates
    in the \emph{pseudo-manifold}
    of \cite{bib:NuPhy1990B339...417G}, then the
    non-Riemannian volume
    element of the two measures theory would be
    related to the non-Riemannian
    structure that could be associated to the
    \emph{pseudo-manifold}. In this perspective,
    non-Riemannian
    volume elements could be essential to make
    sense of the quantum creation of a universe in
    the laboratory and it could be important to develop
    the theory of shell dynamics in the framework described
    by the two measures theory.

A second one, likely complementary,
    can come from a closer study of the Hamiltonian
    dynamics of the system.
    Let us preliminarily remember
    that the Hamiltonian for a general relativistic
    shell, which we are using as a model for the
    universe creation process, is a non quadratic
    function of the momentum (this comes from the
    non-linearities intrinsic to General Relativity);
    this makes the quantization procedure non-standard
    and quite subtle too. Moreover, although it is possible
    to determine an expression for the Euclidean
    momentum and use it to reproduce \cite{bib:ClQuG1997..14..2727A}
    standard results for the decay of vacuum bubbles (as for instance
    the results of Coleman \emph{et al.} \cite{bib:PhReD1980..21..3305L})
    this momentum can have unusual properties
    along the tunnelling trajectory; some of these inconsistencies
    disappear if we consider the momentum as
    a function valued on the circle instead than on
    the real line \cite{bib:gr-qc2006..xx...zzzS} but
    further investigations in this direction
    are required; they will likely help us to obtain
    a better understanding of the semiclassical tunnelling creation
    of this general relativistic system and, perhaps, show us
    some interesting properties of the interplay between the quantum
    and the gravitational realms. In this context, it should be also explored how the
    Euclidean baby universes \cite{bib:Proc.1989Pakistan...N} could be matched
    continuously to the real time universes and in this way provide new
    ways to achieve spontaneous creation of real time baby universes

To complement the above discussion, we would now like to provide some
additional contact points between theoretical
ideas and experimental evidence. We start
considering if all \emph{creation efforts} might end
in a child universe totally disconnected from its
\emph{creator} or not. Of course, there is not a definitive
answer also to this problem yet, since this is tightly bound
to the child universe creation model. Nevertheless, it
it is certainly stimulating to address
the question if, in some way, the new
universe might be detectable. There is an
indication in this direction from the
analysis performed in
\cite{bib:PhReD1991..44...333M}: here a
junction with a Vaidya radiating metric is employed,
so that the child universe could be detectable because
of modifications to the Hawking radiation.
Generalizations that apply
to solitonic inspired universe creation%
\footnote{It is, for instance, certainly possible to extend
the metric describing
the monopole, i.e.
the Rei\ss{}ner-Nordstr\"{o}m spacetime, to the
Rei\ss{}ner-Nordstr\"{o}m-Vaidya case.}
can be important, especially from the point of view
of a quantum-gravitational scenario in which the
exact and definite character of classical
causal relations might be \emph{waved}
by quantum effects.

Other issues that could be tackled after having
a more detailed model of child universe
creation, are certainly phenomenological ones.
They would also help to better understand the
differences between purely classical and partly quantum
processes, which is also a motivation to consider
them explicitly and separately. Also
the physical consequences of different values of
the initial parameters characterizing the child
universe formation process (initial conditions)
should be analyzed\footnote{In particular different
ways of creating a universe
in the laboratory could lead to different
coupling constants, gauge groups, etc..}
and in this context we would also like to recall the
idea of Zee \textit{et. al}
\cite{bib:phys.2005..10...102Z}, i.e. that a creator
of a universe could pass a message to the future
inhabitants of the created universe. From our
point of view this is can be a suggestive
way to represent the problem
of both initial conditions and
causal structure; this could be of relevance
also for the problem of defining probabilities
in the context of the multiverse theory and
of eternal inflation.

A final point of phenomenological relevance
would be in connection with observations that suggest
the universe as super-accelerating. This seems to
support the idea that some very unusual
physics could be governing the universe,
in the sense that standard energy conditions
might not be satisfied. In the context of
child universes creation in the laboratory
in the absence of an initial singularity,
it might very well be
that a generalized behavior of the universe to
try to raise its vacuum energy would manifest
itself locally with the
creation of bubbles of false vacuum (as seen
by the
surrounding spacetime), which would then led
to child universes. In \cite{bib:gr-qc2006..07...111G}
a proposal, based on the two
measures theory, to avoid initial
singularities in a homogeneous cosmology has already
been put forward. It would then be
desirable to apply it to the
non-singular child universe creation also.

To conclude we cannot miss to point out how
all the above discussion about the possibility
of producing child universes in the laboratory
could take a completely new and concrete perspective
in connection with the possible existence of new physics
at the TeV scale in theories with large
compact extra-dimensions, physics that might become
available to our experimental testing at the
colliders which will shortly start to operate.

\section*{Acknowledgements}

We would like to thank H. Ishihara and
J. Portnoy for conversations.

\newcommand{\cen}{\centerline}
\title{Relation between Finestructure Constants at the Planck Scale from 
Multiple Point Principle \thanks{Invited talk by D.L. Bennett}}
\author{D.L. Bennett${}^1$, L.V.~Laperashvili ${}^{2}$ and %
H.B.~Nielsen ${}^{3}$}
\institute{%
${}^{1}$ Brookes Institute for Advanced Studies, Copenhagen, Denmark\\
\textrm{email:}\texttt{bennett@nbi.dk}\\
${}^{2}$ The Institute of Theoretical and Experimental Physics, Moscow,
Russia\\
\textrm{email:}\texttt{laper@itep.ru}\\
${}^{3}$ The Niels Bohr Institute, Copenhagen, Denmark\\
\textrm{email:}\texttt{hbech@nbi.dk}}

\titlerunning{Relation between Finestructure Constants at the Planck Scale}
\authorrunning{D.L. Bennett, L.V.~Laperashvili and H.B.~Nielsen}
\maketitle

\begin{abstract}

 We derive a relation between the three finestructure constants in the
 Standard Model from the assumptions of what we call ``multiple point
  principle'' (MPP) and ``AntiGUT''. By the first assumption we mean that we
  require coupling constants and mass parameters to be adjusted - by our
  multiple point principle - to be just so as to make several vacua have the
  same cosmological constants (from our point of view, basically zero).  By
  AntiGUT we refer to our assumption of a more fundamental precursor to the
  usual Standard Model Group (SMG) consisting of the $N_{gen}$-fold cartesian
product of the usual SMG such that each of the three families of quarks and
leptons has its own set of gauge fields. The usual SMG comes about when
SMG$^3$ breaks down to the diagonal subgroup at roughly a factor 10 below the
Planck scale. Up to this scale we assume the absence of new physics.  Relative
to earlier work where the multiple point principle was used to get predictions
for the gauge couplings independently of one another, the point here is to
increase accuracy by considering a relation between all the gauge couplings
(i.e., for U(1), and SU(N) with N=2 or 3)as a function of a N-dependent
parameter $d_N$ that is a characteristic of SU(N) groups. In doing this, the
parameter $d_N$ that initially only takes discrete values corresponding to the
``N´´ in SU(N) is promoted to being a continuous variable (corresponding to
fantasy groups for $N \notin {\bf Z}$).  By an approprite extrapolation in the
variable $d_N$ to a fantasy group for which the $\beta$-function for the
magnetic coupling ${\tilde g}^2$ vanishes we avoid the problem of our
ignorance of the ratio of the monopole mass scale to the fundamental scale.
In addition to increasing the accuracy
of our predictions for the gauge couplings by circumventing the uncertainty in our knowledge of this ratio, 
we interpret our results as being very supportive of the multiple point principle and
AntiGut.


\end{abstract}

\section{Introduction}

In earlier work \cite{db1} we invented our Multiple Point Principle / AntiGUT 
(MPP / AntiGUT)  gauge group model for the purpose of predicting
the Planck scale values of the three Standard Model Group gauge couplings. These predictions were made independently for
the three gauge couplings. In this work we test an alternative method of treatment of MPP/AntiGUT in which we seek a relation that would
put a rather severe constraint on the values of the SMG couplings.

An important ingredient for the calculational technique in this paper is the Higgs monopole model description in which magnetic 
monopoles are thought of as particles described by a scalar field $\phi$ with an effective potential $V_{eff}$
of the Weinberg-Coleman type \cite{db2},\cite{db3}. The MPP is implemented by requiring that the two minima of $V_{eff}$ are degenerate. This requirement results
in a relation between the square of the monopole charge ${\tilde g}^2$ and the self-coupling $\lambda$ that defines a phase transition 
boundary between a Coulomb-like phase (with $ < \phi > =0$) and a phase with a monopole condensate (with $ < \phi > \neq 0$).
For Abelian monopoles this phase boundary has been presented in earlier work \cite{db4},\cite{db5}. This phase boundary condition which has a term  quadratic in
${\tilde g}^2$ (as well as terms linear and quadratic in $\lambda$) consists of tuples ($\lambda, {\tilde g}^2 $) of critical values of $\lambda$
and ${\tilde g}^2$. A characteristic feature of this boundary is that it has negative curvature as a function of $\lambda$ and hence a
maximum value  ${\tilde g}^2_{U(1)\; crit\, max}$ of ${\tilde g}^2$. It is important to keep in mind that this phase transition that we find by applying 
MPP to $V_{eff}$ is assumed to be at the scale of the monopole mass.
In the present work, the $V_{eff}$ is generalized in such a way that it also embodies nonAbelian $SU(N)$ monopoles using the technique  of Chan \& Tsou
in which a $SU(N)$ monopole is described as a \underline{\bf N} under a dual Yang-Mills vector potential ${\tilde A}_{\mu}$ \cite{db6}. The corresponding phase
boundary now has coefficients that depend on $N$ (i.e., "$N$" as in $SU(N)$). But otherwise the phase boundary for a $SU(N)$
monopole is qualitatively the same as that for the Abelian monopole the important difference being that  the maximum value of ${\tilde g}^2$ in the Abelian theory
is different than in the nonAbelian $SU(N)$ theory: $ {\tilde g}^2_{U(1)\; crit \; max} \neq {\tilde g}^2_{SU(N)\; crit \; max} $.

In fact a major trick in the present article is to consider a correspondance between
Abelian and non-Abelian monopoles (the latter in the understanding of 
Chan-Tsou to be explained in section 2 below). This correspondence is defined using a $N$-dependent parameter $C=C(N)$ defined by
\begin{equation}
C \hat{=} \frac{{\tilde g}^2_{SU(N)\; crit \; max}}{{\tilde g}^2_{U(1)\; crit \; max}}.
\end{equation}
We can think of this definition of parameter $C$ as a definition of the Abelian theory with ${\tilde g}^2_{U(1)\; crit \; max} $ that corresponds to the nonAbelian
$SU(N)$ theory with $ {\tilde g}^2_{SU(N)\; crit \; max}$. 

As shall be seen soon we need an unambiguous way to define the Abelian magnetic charge ${\tilde g}^2_{U(1)\;corresp. \: SU(N)}$ corresponding
to {\it any} nonAbelian $SU(N)$ magnetic charge ${\tilde g}^2_{SU(N)}$.
Our definition of such a correspondence is simply:
\begin{equation}
C \hat{=} \frac{{\tilde g}^2_{SU(N)}}{{\tilde g}^2_{U(1)\; corresp \; SU(N)}},
\end{equation}
where ${\tilde g}^2_{SU(N)}$ can have any value (not necessarily critical or critical maximum).
We can thereby think of say 
the fundamental scale nonAbelian dual - i.e. mono\-pole - coupling
also as an Abelian one. 

We now go to the Planck scale where the MPP/AntiGUT model was originally invented as a simple way of relating the 
experimental values of the SMG gauge couplings $g_{U(1)},\:g_{SU(2)}$and $g_{SU(3)}$ (extrapolated to the Planck scale 
in the absence of new physics underway) to the critical values of these three coupling as determined using lattice gauge theory.
This MPP/AntiGUT relation in terms of the (critical values of the) gauge couplings $g_i$ ($i\in \{U(1),SU(2),SU(3)\})$ is in this 
work reformulated in terms of the dual (critical values of the) of the magnetic charges using the Dirac relations 
$g{\tilde g}= 2\pi$ and  $g{\tilde g}= 4\pi$ for respectively Abelian and nonAbelian monopole theories. Recall that we already have a convention
for calculating the (squared) Abelian magnetic charge that corresponds to a (squared) nonAbelian magnetic charge ${\tilde g}^2_{SU(N)}$ 
using the parameter $C$:
\begin{equation}
{\tilde g}^2_{U(1)\;corresp. \ SU(N)}= \frac{{\tilde  g}^2_{SU(N)}}{C}.
\end{equation}
Now assuming for the moment that our MMP/AntiGUT model is in fact a law of Nature
it would not be unreasonable to 
discuss whether the Abelian correspondent couplings ${\tilde g}^2_{U(1) \; corresp. SU(2)}$
coupling is smooth or not as a function of a gauge group characteristica
such as $N$ for SU(N) groups.  Also  the U(1) gauge group can be taken
into consideration using our nonAbelian to Abelian correspondence relation in the special case 
\begin{equation}
{\tilde g}^2_{U(1) \; corresp. U(1)} \equiv \frac{{\tilde g}^2_{U(1)}}{C=1}
\end{equation}
which  will be seen below to correspond to $d_N=0$

Actually for later convenience we shall instead of $N$ use an $N$-dependent parameter $d_N$ as our independent variable and the quantitiy
$\frac{3{\tilde g}^2_{U(1) \; corresp. SU(2)}}{\pi}$ instead of ${\tilde g}^2_{U(1) \; corresp. SU(2)})$ as our on $d_N $ analytically dependent
variable.
We have now three points $(d_N, \frac{3{\tilde g}^2_{U(1) \, corresp. \; SU(N)}}{\pi})$ belonging to our hypothesized in $d_N$ 
analytic function 
namely the points
\begin{equation}
(0,\frac{3{\tilde g}^2_{U(1)}}{\pi}), \;\; (d_2, \frac{3{\tilde g}^2_{U(1) \; corresp. SU(2)}}{\pi}), \;\; (d_3, \frac{3{\tilde g}^2_{U(1) \; corresp. SU(3)}}{\pi}).
\end{equation}
Now we make the guess that the analytic function in $d_N$ that we seek is the parabolic fit obtained using these three points. 
It must be emphasized that our hypothesized function analytic in $d_N$ lives at the Planck scale 
while the phase transition boundary discussed above lives at the (unknown) scale of the monopole mass and consists of critical $\lambda$ and ${\tilde g}^2$ values  that both run with scale. So the problem is how to connect hypothesized Planck scale physics  in the form of our in $d_N$ analytic function $\frac{3{\tilde g}^2_{U(1) \; corresp. SU(2)}}{\pi}$ with say critical values
of the function ${\tilde g}^2_{U(1) \: corresp. \;SU(N) \; crit}$ at the unknown scale of the monopole mass.

There is one value of ${\tilde g}^2$ for which this connection would be trivial namely the special point $(d_{N_{spec}}, \frac{3{\tilde g}^2_{spec}}{\pi})$ lying at the 
intersection of our in $d_N$ analytically continued function with the function defined by requiring that the $\beta$-function for ${\tilde g}^2$ vanishes. I.e.,
$\beta_{{\tilde g}^2}=0$. Just this value of $\frac{3{\tilde g}^2_{spec}}{\pi}$ is the same at the Planck scale and at the (unknown) scale of the monopole mass.
We describe now briefly how we find the ``fantasy`` (and nonexistent!) gauge group "`$SU(N_{spec})$"' for which the corresponding ${\tilde g}^2$ does not run with scale. But first a little digression
on how $\beta_{{\tilde g}^2}$ depends on $d_N$

Starting from the definition of the Abelian magnetic charge correspondent to a nonAbelian magnetic charge we use the nonAbelian Dirac relation to obtain
\begin{equation}
{\tilde g}^2_{U(1) \; corresp. SU(N)}\hat{=}\frac{{\tilde g}^2_{SU(N)}}{C}=\frac{1}{C}\frac{16\pi^2}{g^2_{SU(N)}}
\end{equation}

We now require that the Dirac relation remain intact under scaling; i.e.,
\begin{equation}
 \frac{d}{dt}{\tilde g}^2_{U(1) \; corresp. SU(N)}=\frac{16\pi^2}{C}\frac{d}{dt}(\frac{1}{g^2_{SU(N)}})=\frac{16\pi^2}{C}\frac{1}{4\pi}\frac{d}{dt}(\alpha^{-1}_{SU(N)}).
\end{equation}

Using that  $\frac{d}{dt}(\alpha^{-1}_{SU(N)})=\frac{11N}{12\pi}$ (which is just the usual Yang-Mills contribution to the $\beta$-function for $\alpha^{-1}_{SU(N)}$
using t=ln$\mu^2$)
we get the Yang-Mills contribution to the running of ${\tilde g}^2_{U(1) \; corresp. \: SU(N)}$:
\begin{equation}
\beta_{{\tilde g}^2_{U(1) \; corresp. \; SU(N)}}|_{Y.-M. \;\; contrib}=\frac{16\pi^2}{C}\frac{1}{4\pi}\frac{11N}{12\pi}=\frac{11N}{3C}\hat{=}d_N.
\end{equation}  

\noindent So even though there is of course no Yang-Mills contribution to $\beta_{U(1)}$ we see that $\beta_{U(1)\;corresp. \; SU(N)}$ inherits a dependence on 
$\beta_{g^2_{SU(N)}}$ through the requirement that the Dirac relation remain intact under scale changes.
Using the known $\beta$-function for $\tilde g^2_{U(1)}$ (to 2-loops):
\begin{equation}
\beta_{\tilde g^2_{U(1)}}= \frac{{\tilde g}^4}{48\pi^2}+\frac{{\tilde g}^6}{(16\pi^2)^2}
\end{equation}
which leads to the $\beta$-function for ${\tilde g}^2_{U(1) \; corresp, \; SU(N)}$:

\begin{equation}
\beta_{{\tilde g}^2_{U(1) \; corresp, \; SU(N)}}=\beta_{{\tilde g}^2_{U(1)}}+d_N=\frac{{\tilde g}^4}{48\pi^2}+\frac{{\tilde g}^6}{(16\pi)^2}+d_N.
\end{equation}

\noindent The  intersection point (with no running of ${\tilde g}^2$)is readily found as the 
intersection of 
\begin{equation}
\beta_{{\tilde g}^2_{U(1) \; corresp, \; SU(N)}}=0 
\end{equation}
\noindent and our in $d_N$ analytically extrapolated function
$\displaystyle
\frac{3{\tilde g}^2_{U(1) \; corresp. \; SU(N)}}{\pi}$. At the intersection 
point we have
\begin{equation}
(d_N, \frac{3{\tilde g}^2}{\pi})=(-0.57,14.54).
\end{equation}

\begin{figure}
\centerline{\epsfxsize=0.9\textwidth \epsfbox{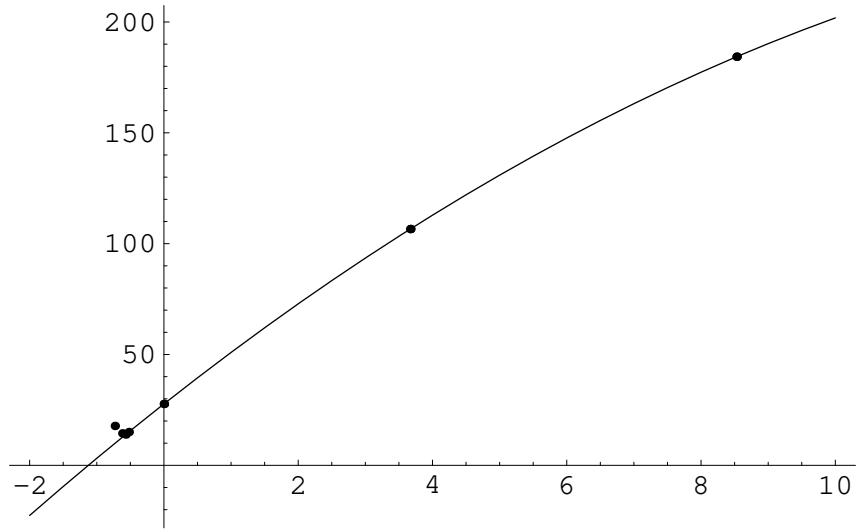}}
\caption[]{\label{db-fig1} Magnetic coupling $\frac{3{\tilde g}^2}{\pi}$ (ordinate) as a function of $d_N$ (abcissa). The curve is determined to be the parabola that
passes through the values of $\frac{3{\tilde g}^2}{\pi}$ corresponding to at ``experimental´´ values of gauge couplings (at positive $d_N$-values). 
The points at negative value of $d_N$ lying off the parabolic fit (solid line) correspond to maximum values of $3{\tilde g}^2/\pi $ on phase transition curves calculated to one and two loops (see closeup in Fig.~\ref{db-fig2}).} 
\end{figure}

\begin{figure} 
\centerline{\epsfxsize=0.9\textwidth \epsfbox{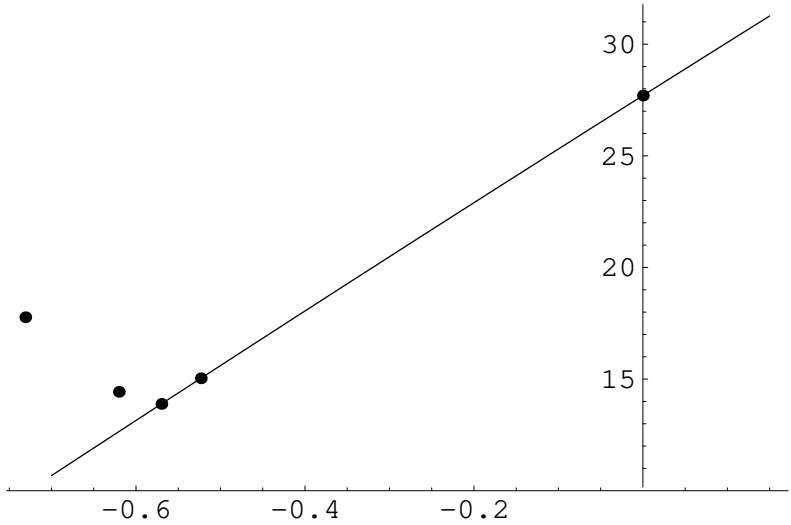}}
\caption[]{\label{db-fig2}  
Closeup of $\frac{3{\tilde g}^2}{\pi}$ vs. $d_N$ for slightly negative values of $d_N$. The two points (at negative $d$) lying on the parabolic fit (solid line) correspond to to the intersection of the parabolic fit with the curves $\beta_{g^2}=0$ calculated for one- and two loops the latter corresponding to the point at $d_N\approx -0.57$. The two points (also at negative values of $d_N$) not on the parabolic fit correspond to the maximum values of $3{\tilde g}^2/\pi$  on the phase transition curves calculated to one and two loop. The latter corresponds to the point at $d_N =\approx -0.62$. The points for highest order (i.e., at $d_N \approx -0.57$ on curve  and at $d_N \approx -0.62$ off the curve)
are seen to converge towards one another relative to the remaining two points calculated to one loop.}  
\end{figure}

\noindent So any intersection point would have a ${\tilde g}^2$ that necessarily 
has a RG trajectory parallel to the $\lambda$ axis in the space spanned by $(\lambda, {\tilde g}^2)$ where the phase transition boundary lives. But only one of the horizontal (i.e., parallel to the $\lambda$ axis)
RG trajectories can be tangent to the phase transition boundary and the point of tangency must neccessarily be at the top point of the phase transition curve.  The interesting result of this work  is that our intersection point (which depends of course on the experimental values of the SMG finestructure constants) to high accuracy 
is the value of ${\tilde g}^2$ with the RG trajectory that is tangent to the phase boundary at its top point where ${\tilde g}^2={\tilde g}^2_{U(1) \: crit \; max}=15.11$. Had our intersection point singled out any other RG trajectory our MPP would have been falsified (see Figures~\ref{db-fig1} and~\ref{db-fig2}). 



\section{The Chan-Tsou duality and monopole critical coupling calculation} 


Investigating nonAbelian theories, we have used the quantum Yang-Mills theory
by Chan-Tsou \cite{db6} for a system of fields with chromo-electric 
charge $g$  and chromo-magnetic charge $\tilde g$ (monopoles). This theory
describes symmetrically the non-dual and the dual sectors of theory with nonAbelian
vector potentials $A_{\mu}$ and ${\tilde A}_{\mu}$ covariantly interacting
with chromo-electric $j_{\mu}$ and chromo-magnetic ${\tilde j}_{\mu}$
currents respectively. As a result, the Chan-Tsou nonAbelian theory has a doubling
of symmetry from $SU(N)$ to
$$
       SU(N)\times \widetilde {SU(N)}
$$
and reveals the generalized dual symmetry which reduces to the well-known 
electromagnetic (Hodge star) duality in the Abelian case.

We want in principle to consider  three phase transitions connected with a single nonAbelian 
monopole which in the philosophy of the Chan-Tsou-theory to be described below
is an $\underline{N}$-plet under the by Chan-Tsou introduced dual Yang Mills 
four vector field $\tilde{A}_{\mu}$, namely 1) a confining phase, 2) a Coulomb phase, and 
3) a phase with monopole condensate. According to the Multiple Point Principle 
the coupling constants and mass parameters should then be adjusted in Nature to
just make these phase degenerate (i.e. same cosmological constant). Earlier we have  
used two loop calculations to obtain the Abelian gauge group a phase transition
between the monopole-condensate phase and the Coulomb phase using the Coleman-Weinberg 
effective potential technique, which led to a relation between the self coupling 
$\lambda$ for the monopole Higgs field and the monopole charge $\tilde{g}$. To determine 
the monopole mass we should, however, in principle involve one more phase, i.e. 
the monopole confining one, but that would need a string description in the 
language used for the two other phases and doing that 
sufficiently accurately for the fit towards which we aim in this article is not undertaken here. 
Hence we shall assume only that the ratio of the mass scale of the 
monopole condensate or approximately equivalently the monopole mass to fundamental
scale, taken here to be the Planck scale, is an analytical function of some group 
characteristic, which we shall take to be the quantity $d_N$ that we shall return to shortly.
The basic point is that we derive by the Coleman-Weinberg effective potential 
an a priori scale independent phase transition curve in as far as the monopole 
mass drops out of the relation describing the phase border between the Coulomb
phase and the monopole condensate one, so that the only scale dependence of this relation 
comes in via the renormalization group. The lack of a good technology for calculating 
the mass scale of the monopole therefore means that we have troubles in calculating the 
renormalization group correction of the by the Coleman-Weinberg-technique calculated relation
between $\lambda$ and $\tilde g^2$ to run it from the monopole mass scale to the 
fundamental scale. The major idea of the present article now is that this would be
no problem if the beta-function for the monopole coupling $\tilde g$ had been zero.
The trick now is to effectively achieve that zero beta function by extrapolating in the
gauge group so to speak to a ``fantasy'' group having zero beta-function. 

A priori magnetic monopole couplings for different gauge groups cannot be compared,
and so to make the statement that the phase transition coupling is analytic as a 
function of some group characteristic, call it $d_N$ say, is a priori not meaningfull.
This is so because in principle one could vary notation from group to group, and such 
a choice of notation would not a priori be analytical. We shall primarily be interested
in the phase transitions from a Coulomb phase to the monopole condensed phase 
as obtained from studying the effective potential as a function
of the norm of the vacuum expectation value of the monopole scalar field in the manner of Coleman-Weinberg.
We decide to take the a priori arbitrary ratio between the ratio of a gauge coupling for an $SU(N)$ gauge group and 
the coupling for the corresponding Abelian $U(1)$ theory to be the same as that for the critical values of these couplings.
We take the Lagrangian densities for a $U(1)$ theory and an $SU(N)$ Higgs Yang Mills theory
respectively as 
\begin{equation}
{\cal L} = -\frac{1}{4\tilde g^2} \tilde F_{\mu\nu}^2 + |\tilde D_{\mu}\phi|^2 - \frac{1}{2} \mu^2 \phi^2 - \frac{\lambda}{4}|\phi|^4
\label{L1}
\end{equation}       
and 
\begin{equation}
{\cal L} = -\frac{1}{4\tilde g^2} {\tilde F_{\mu\nu}}^{j2} +  |\tilde D_{\mu}\phi^a|^2 - \frac{1}{2} \mu^2 |\phi^a|^2 - 
\frac{\lambda}{4}(|\phi^a|^2)^2
\label{L2}
\end{equation}       
where $\phi^a$ is a monopole  $\underline{N}$-plet,  $\tilde D_{\mu} $ is the covariant  derivative for dual gauge field $\tilde A_{\mu}$  
and $\tilde g $ is magnetic charge,  such that the meaning of the mass $\mu$ and the 
self coupling $\lambda$ becomes related in as far as we decide to identify as having corresponding meaning of the length 
squares of the fields; i.e. we identify
\begin{equation}
|\phi|^2 = \sum_{a=1}^N |\phi^a|^2 
\label{L3} 
\end{equation}
as is natural, since from the derivative part in the kinetic term respectively 
$\frac{1}{2} |\partial_{\mu}\phi|^2$ and  $\frac{1}{2} |\partial_{\mu}\phi^a|^2$ we can claim that a given size 
of $|\phi|^2$ and  $|\phi^a|^2$ corresponds to a given density of Higgs particles, a number of particles per unit volume
being the same in both theories. Accepting (\ref{L3}) as a physically meaningfull identification we can also 
claim that the $\lambda$ and the $\mu$ in (\ref{L1}) and (\ref{L2}) are naturally identified i an $N$-independent way (i.e., $N$ as in SU(N)). 

Denoting (\ref{L3}) by just $|\phi|^2$ one can write - as is seen by a significant amount of calculation or by 
using Coleman-Weinberg \cite{db2} and Sher \cite{db3} - the one-loop effective potential for $U(1)$ and $SU(N)$ gauge groups as
$$
V_{eff} = -\frac{1}{2} \mu^2 |\phi|^2 +\frac{\lambda}{4}|\phi|^4+ 
\frac{|\phi|^4}{64\pi^2} $$ $$ [ 3B\tilde g^4 \ln \frac{|\phi|^2}{M^2}
 +(-\mu^2 +3 \lambda |\phi|^2 )^2 \ln\frac{-\mu^2 +3\lambda |\phi|^2}
{M^2}$$ \begin{equation}
 + A (-\mu^2 +\lambda |\phi|^2)^2 \ln\frac{-\mu^2 +\lambda |\phi|^2}{M^2} ],
\label{L4}
\end{equation}     
where
\begin{equation} 
A=B=1 \qquad \mbox{for Abelian case,} \label{L5}
\end{equation} 
and
\begin{eqnarray}
A&=&2N-1,\\
B&=& \frac{(N-1)(N^2 +2N-2)}{8N^2} \qquad \mbox{for SU(N) gauge group.}  \label{L6}
\end{eqnarray}
The $SU(N)$ formula we used here were derived using an $\underline{N}$-plet 
monopole and with a convention for  the covariant derivative 
\begin{equation}
\tilde D_{\mu} = \partial_{\mu} - i\tilde A^j_{\mu}\lambda^j/2  \label{L8}
\end{equation}
in the convention with absorbed coupling where the generators $\lambda^j/2$
were normalized to
\begin{equation}
Tr(\frac{\lambda^j}{2}\frac{\lambda^k}{2}) = \frac{1}{2}\delta_{jk},   \label{L9}   
\end{equation}
while for the Abelian theory we used the convention
\begin{equation}
\tilde D_{\mu} = \partial_{\mu} -i\tilde A_{\mu}. \label{L10}
\end{equation}
 
We have  stable or meta-stable vacua when we have  minima in the 
effective potential (\ref{L4}) which of course then means that the derivatives 
of it are zero there:
\begin{equation}
\frac{\partial V_{eff}(|\phi|^2)}{\partial |\phi|^2}|_{min \,\, i} =0  \label{L11}
\end{equation}
where $i$ enumerates the various minima.

\noindent Now our multiple point principle asserts that there should be as many degenerate vacua as possible - i.e., the more degenerate vacua the more intensive
parameters that become finetuned by the requirement of being at the multiple point (in parameter space). In the case we 
consider here there are just two degenerate vacua at say $ \phi=\phi_{min1}$ and $\phi=\phi_{min2}$. This means that if we take the 
degenerate minima to have zero energy density (cosmological constant) that
\begin{equation}
V_{eff}(|\phi|^2_{min1})=V_{eff}(|\phi|^2_{min2}) =0.     \label{L12}
\end{equation}

\noindent The joint solution of equations (\ref{L12}) and (\ref{L11})
for the effective potential (\ref{L4}) gives  
the phase transition border curve between a Coulomb phase and monopole condesced phase:
\begin{equation}
       3B{\tilde g}_{p.t.}^4 + (5+A)\lambda_{p.t.}^2 + 16\pi^2\lambda_{p.t.} = 0. \label{L13}
\end{equation}
All of the combinations $(\lambda,{\tilde g}^2)$ satisfying (\ref{L13}) are critical in the sense of separating phases. The maximum value of ${\tilde g}^2_{U(1)\;\; crit}$ - we have
called it ${\tilde g}^2_{U(1)\;\; crit \;\; max}$ turns out to be interesting for us.
Let us now find this top point of the phase boundary curve (\ref{L13}) (see also \cite{db4} and \cite{db5}).
 
\begin{equation}
      \frac{d{\tilde g}^4}{d\lambda}|_{crit} = 0,
\end{equation} which gives
\begin{equation}
            {\tilde g}^4_{crit} = \frac{{(16\pi^2)^2}}{12(5+A)B},      \label{L14}
\end{equation} and
\begin{equation}
              \lambda_{crit} = - \frac{16\pi^2}{2(5+A)}.           \label{L15}
\end{equation}
From Eq.~(\ref{L14}) we obtain:\\

for U(1) group:\\

A=1, B=1,

\begin{equation}
      {\tilde g}^2_{crit,U(1)} = \frac{8\pi^2}{3\sqrt 2} \approx 18.61,
                                                                \label{L16a} 
\end{equation} 

for N=2:\\

A=3, $B=\frac{3}{16},$

\begin{equation}
 {\tilde g}^2_{crit,SU(2)} = \frac{8\pi^2}{3\sqrt 2} \approx 37.22, \label{L16b}
\end{equation}

for N=3:\\

A=5, $B=\frac{13}{36},$

\begin{equation}
          {\tilde g}^2_{crit,SU(3)}= \sqrt \frac{108}{65}\cdot \frac{8\pi^2}{3\sqrt 2} \approx 11.99.
                                                                \label{L16c} 
\end{equation}
In general we define the parameter $C$ such that
\begin{equation}
 {\tilde g}^2_{crit,SU(N)}= C{\tilde g}^2_{crit,U(1)}, \quad {\mbox {where}}
\quad C=\sqrt \frac{6}{(5+A)B}.         \label{L16d}
\end{equation}
We shall assume that this relationship between the Abelian and nonAbelian couplings is also valid when the couplings are not critical. 
These results are given  at the scale of monopole mass or VEV.

\section{Compilation and correction of finestructure constants in AntiGut model}

Recall that MPP is to be applied to the $N_{gen}$-fold replication of SMG.
For $N_{gen}=3$ we have
\begin{equation}
      (SMG)^3 = U(1)^3\times SU(2)^3\times SU(3)^3 \label{L17} 
\end{equation}
that breaks down to the diagonal subgroup at roughly the Planck scale.
It is the couplings for the diagonal subgroup that are predicted to coincide 
with ¨experimental¨ gauge group couplings at the Planck scale \cite{db1}:
$$
      \alpha_{1,exp}^{-1} = 6\alpha_{1,crit}^{-1}, $$
$$
    \alpha_{2,exp}^{-1} = 3\alpha_{2,crit}^{-1}, $$
\begin{equation}
     \alpha_{3,exp}^{-1} = 3\alpha_{3,crit}^{-1}. \label{L18}
\end{equation}
According to the Particle Data Group results \cite{db7}, we have:
\begin{equation}
     \alpha_{1,exp}^{-1}(\mu_{Pl})\approx 55.4;\qquad
 \alpha_{2,exp}^{-1}(\mu_{Pl})\approx 49.0_3; \qquad 
\alpha_{3,exp}^{-1}(\mu_{Pl})\approx 53.00.                \label{L19}
\end{equation}
In the Abelian theory the Dirac relation is
\begin{equation}
g\tilde g =2\pi n \mbox{ where } n \in  {\bf Z}
\end{equation}
which leads to
\begin{equation}
\alpha \tilde \alpha = \frac 14      \label{L20}  
\end{equation} 
\noindent or
\begin{equation}     
\alpha^{-1} = 4 \tilde \alpha = \frac {{\tilde g}^2}{\pi}.
                                            \label{L21}
\end{equation} 
In the nonAbelian case the Dirac relation is
\begin{equation}
g \tilde g= 4\pi n \mbox{    where  } n \in {\bf Z}.
\end{equation} 
This leads to
\begin{equation}
\alpha \tilde \alpha = 1
\end{equation}
or
\begin{equation}
\alpha^{-1} =\tilde \alpha = \frac{\tilde {g}^2}{4\pi}
\end{equation}     
From Eqs.~(\ref{L18}) and (\ref{L19}) we have at the Planck scale:
$$
       \frac{3{\tilde g}^2_{U(1)}}{\pi}\approx 27.7,
$$ $$
    \frac{3{\tilde g}^2_{SU(2)/Z_2}}{\pi}\approx 196,
$$ \begin{equation}
 \frac{3{\tilde g}^2_{SU(3)/Z_3}}{\pi}\approx 212.0. \label{L22}
\end{equation}
But the correction of the running of finesructure constants 
from  AntiGUT group (\ref{L17}) in the interval $\Delta t = \sqrt {40}$
gives:
\begin{equation}
\alpha_{2,exp}^{-1}(\mu_{Pl})\approx 53.3; \qquad 
\alpha_{3,exp}^{-1}(\mu_{Pl})\approx 59.4.                \label{L23}
\end{equation}
The corresponding Abelian values of $\frac{3{\tilde g}^2}{\pi}$ are:

for U(1): 

\begin{equation}
\frac{3{\tilde g}^2_{U(1)}}{\pi}\approx 27.7, \label{L24}
\end{equation}

for N=2:

\begin{equation}
\frac{3{\tilde g}^2_{U(1)}}{\pi}\approx 106.6, \label{L25}
\end{equation}

for N=3:

\begin{equation}
\frac{3{\tilde g}^2_{U(1)}}{\pi}\approx 59.4\cdot \sqrt\frac{65}{27}
\approx 184.32. \label{L26}
\end{equation}

\section{The $d_N$-parameter.}

As we suggested in the introduction we shall consider a correspondance 
between the nonAbelian and  Abelian Chan Tsou monopoles 
and in this connection have a correspondance of couplings so that 
the ratio of corresponding monopole couplings is like that of the 
critical couplings for the same two theories; i.e., as in equation (\ref{L16d}). 
Since we want to avoid having to try to use our 
bad knowledge of the ratio of the monopole mass scale at which the phase 
transition couplings are rather easily estimated and the Planck or fundamental 
scale, we are interested in beta-functions. 
So the most important feature 
of the gauge group for the purpose here is how the monopole coupling will 
run as a function of the scale. For this purpose we use $\beta_{{\tilde g}^2}$. 
Now if we consider - as is the simplest -
an Abelian monopole we strictly speaking would have no group dependence 
of the beta function $\beta_{{\tilde g}^2}$, but now we imagine there 
there actually {\em is} an effect of the selfcouplings of the Yang Mills 
fields in the "electric" sector and include that. We shall do that by the 
assumption that there will be a term corresponding to it in such a way as to
make the Dirac relation (or its replacement for a non Abelian monopole) (\ref{L20})
be valid for the running couplings at all scales. In (\ref{L16d}) we have the 
relation between the Chan-Tsou ${\widetilde {SU(N)}}$ coupling ${\tilde g}^2_{SU(N)}$
for the critical coupling, which we here by definition of the relation 
between corresponding theories extend also to non-critical couplings:
\begin{equation}
 {\tilde g}^2_{SU(N)}= C{\tilde g}^2_{U(1)\;\; corresp \;\; SU(N)}, \quad {\mbox {where}}
\quad C=\sqrt \frac{6}{(5+A)B}. \label{L27}
\end{equation}          
In the Chan-Tsou formalism the replacement for the Dirac relation 
is that 
\begin{equation}
{\tilde g} g = 4\pi n, \quad n \in {\bf Z}. 
\label{L28}
\end{equation}
Combining (\ref{L27}) and (\ref{L28}) and taking $n$ to be unity we get for
the Abelian magnetic charge corresponding to that of
a nonAbelian $SU(N)$ 
\begin{equation}
{\tilde g}^2_{U(1) \;\;corr\:\; SU(N)} = \frac{(16\pi^2)}{Cg^2_{SU(N)}}.
\label{L29}
\end{equation}
With the postulate - but that is really true - that the running of the 
couplings shall be consistent with the Dirac relation we can take the 
scale dependence of this equation (\ref{L29}) on both sides to obtain
\begin{equation}
\beta_{{\tilde g}^2_{U(1) \;\; corresp \;\; SU(N)}} = - (16\pi^2/C)\cdot \beta_{g^2_{SU(N)}}/g_{SU(N)}^4.
\label{L30}
\end{equation}
Now there is the group dependent contribution to the well known 
beta function 
\begin{equation}
\beta_{g_{SU(N)}^2}|_{Y.M. contribution} =- g^4_{SU(N)} \cdot\frac{11N}{48\pi^2}
\label{L31}
\end{equation}
which to keep the Dirac relation valid at all scales must be tranfered to 
also exist in the beta function for the square of the monopole charge
\begin{equation}
\beta_{{\tilde g}^2}|_{Y.M. contribution} =  (16\pi^2/C)\cdot \frac{11N}{48\pi^2}= 
\frac{11N}{3C} \stackrel{\wedge}{=} d_N
\label{L32}
\end{equation}



\section{Our monopole coupling versus $d$ curve and successful agreement}

Let us make it quite clear what we mean by our {\it intersection point} and our 
so called {\it fourth point}. 

Our intersection point is the point at which
two functions that live in the space spanned by the variables 
$(d_N, {\tilde g}^2_{U(1)})$ intersect one another. One of these functions is
our in $d_N$ extrapolated curve of "experimental" values of function
${\tilde g}^2_{U(1)\;\; corresp \;\; SU(N)}$. The other function consists of
values $(d_N, {\tilde g}^2_{U(1)\;\; corresp \;\; SU(N)})$
that satisfy the condition

\begin{equation}
\beta_{{\tilde g}^2_{U(1) \; corresp, \; SU(N)}}=\beta_{{\tilde g}^2_{U(1)}}+d_N=\frac{{\tilde g}^4}{48\pi^2}+\frac{{\tilde g}^6}{(16\pi)^2}+d_N=0.
\label{betagtilde}\end{equation}

The value of ${\tilde g}^2_{U(1)\;\; corresp \;\; SU(N)}$ at the intersection point
remains the same under scale changes of course since its $\beta$-function vanishes.

Our fourth point - we could denote it as $(d_{N_{4th}}, {\tilde g}^2_{U(1)\;\; crit \;\; max}$) -lives in the space of variables $(d_N, {\tilde g}^2)$. By definition the second coordinate is the maximum value value of ${\tilde g}^2_{U(1) \;\; crit}$
on the phase transition bounday which lives in the space of the variables 
$(\lambda, {\tilde g}^2_{U(1)})$. We have above denoted this maximum alias top point
by the symbol ${\tilde g}^2_{U(1) \;\; crit \;\; max}$. The first coordinate of the fourth point - i.e., $d_{N_{4th}}$ - is the value of $d_N$ obtained when 
${\tilde g}^2_{U(1) \;\; crit \;\; max}$ is substituted into Eqn (\ref{betagtilde}) above.
We get for the fourth point

\begin{equation}
 (d_N, {\tilde g}^2_{U(1)\;\; crit \;\; max})= (-0.62, 15.11).
\end{equation}

The success that we have in this paper is that the intersection point coincides with the fourth point to very high accuracy.

Maybe it's instructive to think of choosing a bunch of ${\tilde g}^2_{U(1)}$ values 
satisfying Eqn (\ref{betagtilde})
(corresponding of course to a bunch of different
$d_N$ values). With this bunch of ${\tilde g}^2_{U(1)}$ values we know how to RG run them back and forth between
Planck scale and monopole mass scale - also  in the space spanned by $(\lambda, {\tilde g}^2)$
where the phase transition curve is located with its top point 
$$(\lambda_{crit}, {\tilde g}^2_{U(1) \, crit \; max})= (-7.13,15.11).$$
 This bunch of
${\tilde g}^2_{U(1)}$ values don't RG run at all by definition so they must be parallel to the $\lambda$ axis in the space in which the phase transition boundary lives. What happens to these parallel to $\lambda$ RG tracks of the bunch of
${\tilde g}^2_{U(1)}$ values depends only on the height (i.e., ${\tilde g}^2_{U(1)\;\; crit \;\; max}$) and not on the details (including the unknown scale) of the phase
transition curve. Our intersection point follows the one horizontal RG track that can become a tangent
to the phase transition boundary and the point of tangency is necessarily the top point. RG tracks below the one corresponding to our (fortuitous for MPP) intersection point would hit the phase boundary below the top point and therefore corresspond to having only a monopole condensate phase.  And being within the condensate phase and removed from the phase transition boundary so that the Coulomb phase is energetically inaccessible  would violate MPP. The (horizontal) RG tracks of ${\tilde g}^2_{U(1)}$ values above the
top point would miss hitting the phase transition boundary and hence correspond to 
being in the Coulomb phase more or less energetically prohibited from being in the monopole condensate phase depending on how far above the top point, that the RG track is. This is also in violation of MPP. The only RG trajectory allowed by MPP is the one
that goes through the fourth point. And our intersection point singles out just
this RG trajectory.

Actually, MPP is put to a very stringent test here. Had our intersection point picked out any other RG trajectory then the one that hits the fourth point our MPP would have been falsified.

\section{Conclusion and outlook}
In the figures we see the plot of the values of $3\tilde{g}^2 /\pi$ -- for the
``corresponding '' $U(1)$ monopole coupling -- versus
our group characterising quantity $d_N$. The extrapolation to the point where it
crosses the curve of $(d_N, 3\tilde{g}^2/\pi)$ combinations for which the
$\beta$-function for $\tilde{g}$ is  zero  (the crossing point is drawn both for the one
loop and two loop calculation and lies of course exactly on the extrapolated
curve but the intersecting curve $ \beta_{\tilde{g}^2}=0$ is not drawn)
it is remarkably close to being critical in the sense that its ordinate is very
close to the critical value $3\tilde{g}^2/\pi = 3\cdot 15.11/\pi=
14.46$ (the two loop critical value for the case of zero beta function of
$\tilde{g}^2$ is 15.11 corresponding to the top point of the phase transition curve). We have drawn on the plot the point with this ordinate
calculated using the relation $\beta_{{\tilde g}^2} =0$.

It is this coincidence which is either accidental or because our theory of
Anti-GUT conbined with MPP is working so well that it 
only deviates in say the inverse $U(1)$ fine structure constant at
the Planck scale by about 2$\cdot$1.5 = 3 units.

If the Planck scale had not been taken to be $\mu_{Pl} = 1.2 \cdot 10^{19}$ GeV, as was the case for the figure
but rather to the value obtained if we had used $8\pi G $
instead of $G$ as the quantity from which to determine by
dimensional arguments the Planck scale, the latter would have been a
factor $\sqrt{8\pi}$ smaller.   This would correspond to subtracting 1.61 from
the $\ln \mu_{Pl}$ or 3.22 from the $\ln \mu_{Pl}^2$. With this 
lower value for the fundamental energy scale the $U(1)$ running coupling
would be shifted up in value from the 55.4 of equation (23) by 1.84 to 57.2
which in turn would shift the 55.4/2=27.7 value plotted on the curve up by
0.92. This is seen by eye to shift the curve even closer to going through the
fourth point than was the case for the Planck scale calculated simply using $G$. Have in
mind that the $SU(2)$ point with almost compensating $\beta$-function
contributions from the Yang Mills and the fermion contributions would only be
moved very little by a slight change in the Planck scale so that this point
would be almost unchanged by the replacement of $G$ with  $8\pi G$, while
the $SU(3)$ point would go the opposite way, meaning down in $3\tilde{g}^2
/\pi$-value, so that the curve would essentially be tilted so as to rise more slowly with increasing $d_N$.

Assuming that the correct fundamental scale should be obtained from  $8\pi G$
rather than from $G$, the deviation would in terms of the inverse $U(1)$ fine
structure constant be only about one out of 27.7 meaning about 1 in 55.4, which is
only about 2\%. We can not meaningfully expect better coincidence unless we were to
calculate a three loop critical value since going  from one to two loop
in the critical monopole coupling
squared gave  a 20\% correction so that 4\% would be expected from three loops
But 4\% in the critical monopole coupling square would mean about 2\% in the
$U(1)$ finestructure constant (inverted) at the Planck scale.

We may also be concerned  that we have not yet made two loop calculations of our
correction factors $C$ giving the correspondance between the
nonAbelian couplings and the corresponding $U(1)$ coupling. But it is our experience numerically that the
remarkable result of the intersection point having critical coupling is rather insensitive to the exact $C$-system used.
We therefore believe
that even with the $C$'s only computed to one loop the accuracy of the calculation is already
effectively a two loop calculation.

The question of the $8\pi G$ versus the $G$ in determining the Planck scale or
fundamental  scale must be considered basically an uncertainty at least until after a very
detailed  discussion of this point.

Our result brings to mind the proverbial story of trying to find a needle in a large haystack. Actually we could even tell a better story.
Start by standing at a distance from the haystack with an electric torch with a very narrow beam. Initially the torch is turned off. Now imagine that our task is to aim the 
torch in the dark {it before } we turn the torch on - we are not allowed to move the torch after turning it on. Now ask about the likelyhood of capturing the needle in our fixed narrow beam
of light. We could improve the story by claiming that we stand with our torch
at the Planck scale and must aim it at a haystack at the scale of the monopole mass before turning it on. There is even the added complication that we a priori don't  know exatly what we're looking for. But we miraculously find the needle in our narrow beam of light. But maybe with the guiding light of our model this is not such a miracle after all. Nature may be telling us that we're on the right track with our MPP/AntiGUT model.

In this work we have only concerned ourselves with two of the three intersting phases for monopoles, namely the Coulomb phase and the phase with a monopole condensate. 
These phases have been treatable using the Higgs mono\-pole description of monopoles as particles. We would also like to have the mono\-pole confining phase at our multiple
point. For every new phase we bring to the multiple point there is one more intensive parameter that becomes finetunned by MPP. This could be ratio of the monopole mass scale to that of the Planck mass if we had the confining phase. However taking the confining phase into account  would require appending a string scenario to our approach here and doing this with the
the high accuracy we otherwise have in this work is not yet possible.

\title{On the Origin of Families of Fermions and Their Mass Matrices\,---\,%
Approximate Analyses of Properties of Four Families Within Approach 
Unifying Spins and Charges}
\author{M. Breskvar, D. Lukman and N.S. Manko\v c Bor\v stnik}
\institute{%
Department of Physics, University of Ljubljana,\\
Jadranska 19, 1000 Ljubljana}

\titlerunning{On the Origin of Families of Fermions and Their Mass Matrices}
\authorrunning{M. Breskvar, D. Lukman and N.S. Manko\v c Bor\v stnik}
\maketitle

\begin{abstract} 
The approach unifying all the internal degrees of freedom - proposed by one of 
us\cite{norma92,MDNnorma93,normasuper94,MDNnorma95,norma97,pikanormaproceedings1,holgernorma00,norma01,%
pikanormaproceedings2,MDNPortoroz03} - is offering a new way of understanding 
families of quarks and leptons. 
Spinors, namely, living in $d \;(=1+13)-$dimensional space, manifest  
in the observed $d(=1+3)$-dimensional space (at ''physical energies'')  
all the known charges of quarks and leptons (with the mass protection property of the Standard model 
that only the left handed quarks 
and leptons carry the  weak charge while the right handed ones are weak chargeless included), 
while a part of the starting Lagrange density  in $d \;(=1+13)$ transforms the right handed 
quarks and leptons into the left handed ones, manifesting a mass term in $d=1+3$. Since a spinor  
carries two kinds of spins and interacts accordingly with two kinds of the 
spin connection fields, the approach predicts families and the corresponding Yukawa couplings.
In the paper\cite{pikanorma05} the appearance of families of quarks 
and leptons within this approach was investigated and the explicit expressions for the 
corresponding Yukawa couplings, following from the approach after some approximations and 
simplifications, presented. 
In this paper we continue investigations of this new way of presenting families of quarks and leptons 
by further analyzing properties  of mass matrices, treating quarks and leptons in an equivalent way.  
Since it is a long way from the starting simple action for a Weyl spinor in $d=1+13$ to the 
observable phenomena at low energies, and we yet want to make  (at least rough) 
predictions of the approach, we connect free parameters of the approach with the known experimental 
data and investigate a  possibility that the fourth family of quarks and leptons appears at 
low enough energies to be observable with the new generation of accelerators.
\end{abstract}

\section{Introduction}
\label{s:MDNintroduction}

The Standard model of the electroweak and strong interactions (extended by assuming nonzero masses 
of the neutrinos) fits with around 25 parameters and constraints all the existing experimental 
data. However, it leaves  unanswered many open questions, among which are also the questions   
about the origin of the families, the Yukawa couplings of quarks and leptons and the 
corresponding Higgs mechanism.  
Understanding the mechanism for generating families, their masses and mixing matrices 
might be one of the most promising ways to 
physics beyond the Standard model. 

The approach, unifying spins and charges\cite{norma92,MDNnorma93,normasuper94,MDNnorma95,norma97,%
pikanormaproceedings1,holgernorma00,norma01,pikanormaproceedings2,MDNPortoroz03,pikanorma05}, might by   
offering a new way of describing families, give an explanation 
about the origin of the Yukawa couplings. 

It was demonstrated in the references\cite{pikanormaproceedings1,%
norma01,pikanormaproceedings2,MDNPortoroz03} that a left handed $SO(1,13)$ 
Weyl spinor multiplet includes, if the representation is analyzed 
in terms of the subgroups $SO(1,3)$, $SU(2)$, $SU(3)$ and the sum of the two $U(1)$'s,  all 
the spinors of the Standard model - that is the left handed $SU(2)$ doublets and the right 
handed  $SU(2)$ singlets of (with the group  $SU(3)$ charged) quarks and  (chargeless) leptons.
There are the (two kinds of)  spin connection  fields and  the vielbein fields in 
$d=(1+13)-$dimensional space, 
which might  manifest - after some  
appropriate compactifications (or some other kind of making the rest of $d-4$ space 
unobservable at low energies) - in the four dimensional 
space as all the gauge fields of the known charges, as well as the Yukawa couplings.

The paper\cite{pikanorma05}  analyzes, how do terms, which lead to masses of quarks 
and leptons, appear in the approach unifying spins and charges as a part of the spin 
connection and vielbein fields. No Higgs is needed in this approach to ''dress'' 
right handed spinors with the weak charge, since the terms of the 
starting Lagrangean, which include $\gamma^0\gamma^s,$ with $s=7,8,$ do the job of a Higgs 
field. 

Since we have done no analyses (yet) about the way of breaking symmetries  
of the starting group $SO(1,13)$ to $SO(1,7) \times U(1) \times SU(3)$ and 
further within our approach (except some very rough 
estimations in ref.\cite{hnrunBled02}), we do not know how might 
symmetry breaking in  the ordinary space  influence the fields (spin connections and vielbeins), 
which in the starting action determine the Yukawa couplings. We also do not know how do 
nonperturbative and other effects 
(like boundary conditions) influence after the break of symmetries the couplings of spinors 
to the part of the starting gauge fields which in $d=1+3$ manifest at low energy as 
the Yukawa couplings. It is namely a long way from 
a simple starting action for a spinor with only one parameter and only the gravity as a gauge field  
to the observable quarks and leptons interacting with all the known gauge fields.

We can accordingly in this investigation, by connecting 
Yukawa couplings with the experimental data, only discuss about the appearance of the 
''vacuum expectation values'' of the spin connection fields which enter into the Yukawa 
couplings, trying to guess how all the complicated breaks of the symmetries have lead from
the starting action to the observable massess and mixing angles. 
We also have no explanation yet why the second kind of the Clifford algebra objects do 
not manifest in $d=1+3$ any charges, which could appear in addition to the known ones. 

Since the generators of the Lorentz transformations and the generators of families commute, 
and since only the generators of families contribute to nondiagonal elements of mass matrices  
(which means, that the off diagonal matrix elements of quarks and leptons are strongly correlated),  
the question arises, what makes leptons so different from quarks in the 
proposed approach. Can it be that at 
some energy level they are very alike and that there are some kinds of boundary conditions 
together with the nonperturbative effects which 
lead to observable properties, or might it be that Majorana like objects, 
not taken into account in these investigations up to now, are responsible for the 
observed differences?

To evaluate whether this way of going beyond the Standard model of the electroweak and 
colour interaction is in agreement with the nature, we must first make a rough estimate 
of what the approach predicts, before going to more sophisticated and therefore also more 
trustable predictions. Starting  with only the gauge gravity in $d=1+13$ (or in any $d$) and 
then coming down to the ''the physical world'', is a huge project, which needs to be made 
in several successive steps. 

What turns out in our approach to be exciting  
is that one Weyl spinor in d=1+13 offers all the quarks and the leptons postulated by 
the Standard model and with just right quantum numbers answering the open question of 
the Standard model, how it is at all possible that the weak charge ''knows'' for the 
handedness (left handed weak charged quarks and leptons and right handed weak chargeless 
quarks and leptons), since in the Standard model the handedness and the weak charge 
belong to totally separated degrees of freedom.

Next exciting thing in the approach is, that it is  a part of the starting Lagrangean 
which does, what the Standard model requires from the Higgs: makes the nonzero matrix 
elements between the weak chargeless right handed quarks and leptons and weak charged 
quarks and leptons, correspondingly, which then manifest as the mass matrices. 

The third exciting thing of our approach is that it offers a mechanism for the appearance 
of families and accordingly the possibility to calculate the Yukawa couplings ''from 
the first principle''.

There are also severe problems, which the approach is confronting in the way down to 
the $d=1+3$ world. 

First severe problem is, how can at all exist at low energies any ''non Planck scale'' 
mass, if one starts from a very high scale? We are working hard\cite{holgernorma05}
to overcome this 
''Witten's no go'' theorem\cite{MDNwitten}, which concerns all the Kaluza-Klein-like theories.  
We found for a toy model ($d=1+5$) a particular boundary condition, which by requiring that 
spinors of only one handedness exist on the boundary makes that no Dirac mass can occur and 
therefore also the corresponding Yukawas do not appear after a particular break of a symmetry.  
We also found \cite{hn06maj} 
that a Weyl spinor with no charge has no Majorana mass only in some dimensions, 
and d=1+13 is one of those.    
In this paper we assume that the extension of the toy model can work also in our 
more generalized case. The justification is under consideration. 

One further severe problem is how to treat the ''history'' of spinors after the 
breaks of symmetries with all the perturbative but mostly nonperturbative effects, 
which are ''dressing'' spinors (quarks and leptons), since even in the hadron 
physics a similar problem is not yet solved.
In this paper we assume that the ''dressing'' manifests, after going beyond 
''the tree level'' in different ''vacuum expectation values'' of the omega fields. 

Even proceeding in this way, this paper  (making a first step towards more 
justified results by  allowing several approximations and assumptions in order 
to find out, whether something very essential and unexpected can go wrong with 
our approach) was quite a work. We indeed come in this paper and in the previous one, 
which this one is following to promising results, that the approach might have a 
chance to go successfully beyond the Standard model.

In this paper we try to understand properties of quarks and leptons within  the approach 
unifying spins and charges treating quarks and leptons equivalently. 
Within this approach we discuss also a possibility, that 
the fourth family of quarks and leptons appears at low enough energies to be observable 
with new accelerators.

In Sect.\ref{lagrangesec} of this paper we present the action  for a Weyl spinor in 
$(1+13)$-dimensional space and the part of the  Lagrangean, which manifests at ''physical energies''   
as an effective Lagrangean, with the Yukawa mass term included. This section is a brief 
repetition of the derivations presented 
in the ref.\cite{pikanorma05}.

Also Sect.\ref{Lwithassumptions} is a short summary of  the 
paper\cite{pikanorma05} in the part  in which  
the explicit expression for the four mass matrices of 
the four families of quarks and leptons is presented, 
derived under several assumptions  and simplifications from the starting action of the approach 
unifying spins and charges. 
In Subsect.\ref{simplified} we study properties of the mass matrices in the approximation, that
the "vacuum expectation values" of the gauge fields of the second kind of the 
Clifford algebra objects entering into mass matrices are the same 
for all the quarks and the leptons.
In Subsect.\ref{improvedmatrices} we study properties of the mass matrices  relaxing this 
requirement. 

In Subsect.\ref{negativemasses} we discuss the problem of the appearance of negative masses 
in connection with the internal parity, defined within the presented approach.

In Sect.\ref{numerical} we present the numerical results after fitting the free parameters of the 
mass matrices  with the experimental data, predicting masses and mixing matrices of 
the four families of quarks and leptons.

In Sect.\ref{discussions} we comment on the success of our approximate prediction of our approach.


\section{Weyl spinors in $d= (1+13)$ manifesting at ''physical energies'' families of 
quarks and leptons} 
\label{lagrangesec}

We assume a  left handed Weyl spinor  in $(1+13)$-dimensional space. A spinor carries 
only the spin (no charges) and interacts accordingly with only the gauge gravitational fields 
- with spin connections and vielbeins. We assume two kinds of the Clifford algebra objects 
and allow accordingly two kinds of gauge fields\cite{MDNnorma92,MDNnorma93,normasuper94,MDNnorma95,norma97,%
pikanormaproceedings1,holgernorma00,norma01,pikanormaproceedings2,MDNPortoroz03,pikanorma05}. 
One kind is the ordinary gauge field (gauging the Poincar\' e symmetry in $d=1+13$). 
The corresponding spin connection field appears for spinors as a gauge field of 
$S^{ab}= \frac{1}{4} (\gamma^a \gamma^b - \gamma^b \gamma^a)$, 
where $\gamma^a$ are the ordinary Dirac operators. 
The contribution of these fields to the mass matrices manifests in only the diagonal terms -  
connecting the right handed weak chargeless quarks or leptons to the left handed weak charged 
partners within one family of spinors. 

The second kind of gauge fields is in our approach responsible for families of spinors and 
couplings among families of spinors - contributing to diagonal matrix elements as well - and    
might explain the appearance of families of quarks and leptons and the Yukawa couplings of the 
Standard model of the electroweak and colour interactions. 
The corresponding spin connection fields appear for spinors as  gauge fields  
of $\tilde{S}^{ab}$ ($\tilde{S}^{ab} = \frac{1}{2} (\tilde{\gamma}^a \tilde{\gamma}^b-
\tilde{\gamma}^b \tilde{\gamma}^a)$) with $\tilde{\gamma}^a$, which are 
the Clifford algebra objects\cite{MDNnorma93,technique03}, like $\gamma^a$, but anticommute with $\gamma^a$.

Following the ref.\cite{pikanorma05} we write the action for a Weyl (massless) spinor  
in $d(=1+13)$ - dimensional space as follows\footnote{Latin indices  
$a,b,..,m,n,..,s,t,..$ denote a tangent space (a flat index),
while Greek indices $\alpha, \beta,..,\mu, \nu,.. \sigma,\tau ..$ denote an Einstein 
index (a curved index). Letters  from the beginning of both the alphabets
indicate a general index ($a,b,c,..$   and $\alpha, \beta, \gamma,.. $ ), 
from the middle of both the alphabets   
the observed dimensions $0,1,2,3$ ($m,n,..$ and $\mu,\nu,..$), indices from the bottom of 
the alphabets
indicate the compactified dimensions ($s,t,..$ and $\sigma,\tau,..$). We assume the signature 
$\eta^{ab} =
diag\{1,-1,-1,\cdots,-1\}$.
}
\begin{eqnarray}
S &=& \int \; d^dx \; {\mathcal L},  
\nonumber\\
{\mathcal L} &=& \frac{1}{2} (E\bar{\psi}\gamma^a p_{0a} \psi) + h.c. = \frac{1}{2} 
(E\bar{\psi} \gamma^a f^{\alpha}{}_a p_{0\alpha}\psi) + h.c.,
\nonumber\\
p_{0\alpha} &=& p_{\alpha} - \frac{1}{2}S^{ab} \omega_{ab\alpha} - \frac{1}{2}\tilde{S}^{ab} 
\tilde{\omega}_{ab\alpha}.
\label{lagrange}
\end{eqnarray}
 
Here $f^{\alpha}{}_a$ are  vielbeins (inverted to the gauge field of the generators of translations  
$e^{a}{}_{\alpha}$, $e^{a}{}_{\alpha} f^{\alpha}{}_{b} = \delta^{a}_{b}$,
$e^{a}{}_{\alpha} f^{\beta}{}_{a} = \delta_{\alpha}{}^{\beta}$),
with $E = \det(e^{a}{}_{\alpha})$, while  
$\omega_{ab\alpha}$ and $\tilde{\omega}_{ab\alpha} $ are the two kinds of the spin connection fields, 
the gauge 
fields of $S^{ab}$ and $\tilde{S}^{ab}$, respectively, corresponding to the two kinds of the Clifford 
algebra 
objects\cite{holgernorma02,MDNPortoroz03}, namely $\gamma^a$ and $\tilde{\gamma}^{a}$, with the 
properties
\begin{eqnarray}
\{\gamma^a,\gamma^b\}_{+} = 2\eta^{ab} =  \{\tilde{\gamma}^a,\tilde{\gamma}^b\}_{+},
\quad \{\gamma^a,\tilde{\gamma}^b\}_{+} = 0, 
\label{clifford}
\end{eqnarray}
leading to $\{ S^{ab}, \tilde{S}^{cd}\}_-=0$. 
We kindly ask the reader to learn about the properties 
of these two kinds of the Clifford algebra objects - $\gamma^a$ and $\tilde{\gamma}^a$  
and of the corresponding  $S^{ab}$ and $\tilde{S}^{ab}$ - and about our technique in the 
ref.\cite{pikanorma05} or the refs.\cite{holgernorma02,technique03}.

One Weyl spinor representation in $d=(1+13)$ with the spin as the only internal 
degree of freedom,   
manifests, if analyzed in terms of the subgroups $SO(1,3) \times
U(1) \times SU(2) \times SU(3)$ in 
four-dimensional physical space  as the ordinary ($SO(1,3)$) spinor with all the known charges 
of one family of  the left handed weak charged and the right handed weak chargeless 
quarks and leptons of the Standard model. The reader can see this analyses in the 
paper\cite{pikanorma05} (as well as in several references, like the one in the ref.\cite{MDNPortoroz03}).

We may rewrite the Lagrangean of Eq.(\ref{lagrange}) so that it manifests the usual 
$(1+3)-$dimensional spinor Lagrangean part and the term manifesting as a mass 
term\cite{pikanorma05} 
\begin{eqnarray}
{\mathcal L} &=& \bar{\psi}\gamma^{m} (p_{m}- \sum_{A,i}\; g^{A}\tau^{Ai} A^{Ai}_{m}) \psi 
+ \nonumber\\
& &  \sum_{s=7,8}\; 
\bar{\psi} \gamma^{s} p_{0s} \; \psi + {\rm the \;rest}.
\label{yukawa}
\end{eqnarray}
Index $A$ determines the charge groups ($SU(3), SU(2)$ and the two $U(1)$'s), index $i$ determines
the generators within one charge group. $\tau^{Ai}$ denote the generators of the charge groups 
\begin{eqnarray}
\tau^{Ai} = \sum_{s,t} \;c^{Ai}{ }_{st} \; S^{st},
\nonumber\\
\{\tau^{Ai}, \tau^{Bj}\}_- = i \delta^{AB} f^{Aijk} \tau^{Ak}, 
\label{tau}
\end{eqnarray}
 with $s,t \in 5,6,..,14$, while $A^{Ai}_{m}, m=0,1,2,3,$ 
denote the corresponding
gauge fields (expressible in terms of $\omega_{st m}$).

We have: $Y = \tau^{41} + \tau^{21}, \quad  Y' = \tau^{41} - \tau^{21}$, with 
$\tau^{11}: = \frac{1}{2} ( {\mathcal S}^{58} - {\mathcal S}^{67} )$,
$\tau^{12}: = \frac{1}{2} ( {\mathcal S}^{57} + {\mathcal S}^{68} )$,
$\tau^{13}: = \frac{1}{2} ( {\mathcal S}^{56} - {\mathcal S}^{78} )$,
$\tau^{21}: = \frac{1}{2} ( {\mathcal S}^{56} + {\mathcal S}^{78} )$,
$\tau^{31}: = \frac{1}{2} ( {\mathcal S}^{9\;12} - {\mathcal S}^{10\;11} )$,
$\tau^{32}: = \frac{1}{2} ( {\mathcal S}^{9\;11} + {\mathcal S}^{10\;12} )$,
$\tau^{33}: = \frac{1}{2} ( {\mathcal S}^{9\;10} - {\mathcal S}^{11\;12} )$,
$\tau^{34}: = \frac{1}{2} ( {\mathcal S}^{9\;14} - {\mathcal S}^{10\;13} )$,
$\tau^{35}: = \frac{1}{2} ( {\mathcal S}^{9\;13} + {\mathcal S}^{10\;14} )$,
$\tau^{36}: = \frac{1}{2} ( {\mathcal S}^{11\;14} - {\mathcal S}^{12\;13})$,
$\tau^{37}: = \frac{1}{2} ( {\mathcal S}^{11\;13} + {\mathcal S}^{12\;14} )$,
$\tau^{38}: = \frac{1}{2\sqrt{3}} ( {\mathcal S}^{9\;10} + {\mathcal S}^{11\;12} - 2{\mathcal S}^{13\;14})$,
$\tau^{41}: = -\frac{1}{3}( {\mathcal S}^{9\;10} + {\mathcal S}^{11\;12} + {\mathcal S}^{13\;14})$.

The subgroups are chosen so that the gauge fields in the physical region agree with the
known gauge fields. If the break of symmetries in the $\tilde{S}^{ab}$ sector demonstrates 
the same symmetry after the break as in the 
$S^{ab}$ sector, then also the corresponding operators with $\tilde{\tau}^{Ai}$ should be defined.

Making several assumptions, explained in details in the ref.\cite{pikanorma05} - 
we shall repeat them bellow - 
 needed to manifest the observable phenomena (and can not yet be derived, since we do not yet know 
how the break of symmetries influences the starting Lagrangean), we are able to rewrite the mass term 
of spinors (fermions) from Eq.(\ref{yukawa}) $(\sum_{s=7,8}\; \bar{\psi} \gamma^{s} p_{0s} \; \psi$,  
neglecting  ${\rm the \;rest}$) by assuming that they are small in 
comparison with what we keep  at "physical energies") 
as $L_Y$, demonstrating the  Yukawa couplings of  the Standard model 
\begin{eqnarray}
{\mathcal L}_{Y} = \psi^+ \gamma^0 \;  
\{ & &\stackrel{78}{(+)} ( \sum_{y=Y,Y'}\; y A^{y}_{+} + 
\sum_{\tilde{y}=\tilde{N}^{+}_{3},\tilde{N}^{-}_{3},\tilde{\tau}^{13},\tilde{Y},\tilde{Y'}} 
\tilde{y} \tilde{A}^{\tilde{y}}_{+}\;)\; + \nonumber\\
  & & \stackrel{78}{(-)} ( \sum_{y=Y,Y'}\;y  A^{y}_{-} +  
\sum_{\tilde{y}= \tilde{N}^{+}_{3},\tilde{N}^{-}_{3},\tilde{\tau}^{13},\tilde{Y},\tilde{Y'}} 
\tilde{y} \tilde{A}^{\tilde{y}}_{-}\;) + \nonumber\\
 & & \stackrel{78}{(+)} \sum_{\{(ac)(bd) \},k,l} \; \stackrel{ac}{\tilde{(k)}} \stackrel{bd}{\tilde{(l)}}
\tilde{{A}}^{kl}_{+}((ac),(bd)) \;\;+  \nonumber\\
 & & \stackrel{78}{(-)} \sum_{\{(ac)(bd) \},k,l} \; \stackrel{ac}{\tilde{(k)}}\stackrel{bd}{\tilde{(l)}}
\tilde{{A}}^{kl}_{-}((ac),(bd))\}\psi,
\label{yukawa4tilde0}
\end{eqnarray}
with
\begin{eqnarray}
\stackrel{ab}{(k)}: = \frac{1}{2} (\gamma^a + \frac{\eta^{aa}}{ik} \gamma^8 ),\quad 
\stackrel{ab}{\tilde{(k)}}= \frac{1}{2} (\tilde{\gamma}^a + 
\frac{\eta^{aa}}{ik} \tilde{\gamma}^b )  
\label{pm}
\end{eqnarray}
and with $k=\pm 1,$ if $\eta^{aa}\eta^{bb}=1$ and $ \pm i,$ if $\eta^{aa}\eta^{bb}=-1$.
While $\stackrel{ab}{(k)}$ are expressible in terms of ordinary $\gamma^a$ and $\gamma^b$,  
$\stackrel{ab}{\tilde{(k)}}$ are expressible in terms of the second kind of the Clifford algebra objects, 
namely in terms of $\tilde{\gamma}^a$ and $\tilde{\gamma}^b$.

The Yukawa part of the starting Lagrangean (Eq.(\ref{yukawa4tilde0})) has the diagonal terms, that is the 
terms manifesting the Yukawa couplings within each family, and the off diagonal terms, determining the 
Yukawa couplings among families.

The operators, which contribute to the non diagonal terms in mass matrices, 
are superposition of $\tilde{S}^{ab}$ (times the corresponding fields $\tilde{\omega}_{abc}$)  
and can be represented  as factors of nilpotents 
\begin{eqnarray}
\stackrel{ab}{\tilde{(k)}}\stackrel{cd}{\tilde{(l)}},  
\label{lowertilde}
\end{eqnarray}
with indices $(ab)$ and $(cd)$ which belong to the Cartan subalgebra indices 
and the superposition of the  fields $\tilde{\omega}_{abc}$.
We may write accordingly
\begin{eqnarray}
  \sum_{(a,b) } -\frac{1}{2} \stackrel{78}{(\pm)}\tilde{S}^{ab} \tilde{\omega}_{ab\pm} =
- \sum_{(ac),(bd), \;  k,l}\stackrel{78}{(\pm)}\stackrel{ac}{\tilde{(k)}}\stackrel{bd}{\tilde{(l)}} 
\; \tilde{A}^{kl}_{\pm} ((ac),(bd)),  
\label{lowertildeL}
\end{eqnarray}
where the pair $(a,b)$ in the first sum runs over all the  indices, which do not characterize  the Cartan 
subalgebra, with $ a,b = 0,\dots, 8$,  while the two pairs $(ac)$ and $(bd)$ denote only the Cartan 
subalgebra pairs
 (for SO(1,7) we only have the pairs $(03), (12)$; $(03), (56)$ ;$(03), (78)$;
$(12),(56)$; $(12), (78)$; $(56),(78)$) ; $k$ and $l$ run over four 
possible values so that $k=\pm i$, if $(ac) = (03)$ 
and $k=\pm 1$ in all other cases, while $l=\pm 1$.
The fields  $\tilde{A}^{kl}_{\pm} ((ac),(bd))$ can then 
be expressed by $\tilde{\omega}_{ab \pm}$ as follows 
\begin{eqnarray}
\tilde{A}^{++}_{\pm} ((ab),(cd)) &=& -\frac{i}{2} (\tilde{\omega}_{ac\pm} -\frac{i}{r} \tilde{\omega}_{bc\pm} 
-i \tilde{\omega}_{ad\pm} -\frac{1}{r} \tilde{\omega}_{bd\pm} ), \nonumber\\
\tilde{A}^{--}_{\pm} ((ab),(cd)) &=& -\frac{i}{2} (\tilde{\omega}_{ac\pm} +\frac{i}{r} \tilde{\omega}_{bc\pm} 
+i \tilde{\omega}_{ad\pm} -\frac{1}{r} \tilde{\omega}_{bd\pm} ),\nonumber\\
\tilde{A}^{-+}_{\pm} ((ab),(cd)) &=& -\frac{i}{2} (\tilde{\omega}_{ac\pm} + \frac{i}{r} \tilde{\omega}_{bc\pm} 
-i  \tilde{\omega}_{ad\pm} +\frac{1}{r} \tilde{\omega}_{bd\pm} ), \nonumber\\
\tilde{A}^{+-}_{\pm} ((ab),(cd)) &=& -\frac{i}{2} (\tilde{\omega}_{ac\pm} - \frac{i}{r} \tilde{\omega}_{bc\pm} 
+i  \tilde{\omega}_{ad\pm} +\frac{1}{r} \tilde{\omega}_{bd\pm} ),
\label{Awithomega}
\end{eqnarray}
with $r=i$, if $(ab) = (03)$ and $r=1$ otherwise.
 We simplify the index $kl$ in the exponent 
of the fields $\tilde{A}^{kl}{}_{\pm} ((ac),(bd))$ to $\pm $, omitting $i$. 

We must point out that a way of breaking  any of the two symmetries - the Poincar\' e one and 
the symmetry  determined by the generators $\tilde{S}^{ab}$ in $d=1+13$ - 
strongly influences the Yukawa couplings of 
Eq.(\ref{yukawa4tilde0}), relating  the parameters $\tilde{\omega}_{abc}$. 
Not necessarily any break of  the Poincar\' e   symmetry 
 influences  the break of the other symmetry and opposite. Although we expect that it does. 
 Accordingly 
the coefficients $c^{Ai}{}_{ab}$ determining 
the operators $\tau^{Ai}$ in Eq.(\ref{tau}) and the coefficients  
$\tilde{c}^{\tilde{A}i}{}_{ab}$ 
determining the operators $\tilde{\tau}^{\tilde{A}i}$ in the relations 
\begin{eqnarray}
\tilde{\tau}^{\tilde{A}i} = \sum_{a,b} \;\tilde{c}^{\tilde{A}i}{ }_{ab} \; \tilde{S}^{ab},
\nonumber\\
\{\tilde{\tau}^{\tilde{A}i}, \tilde{\tau}^{\tilde{B}j}\}_- = i \delta^{\tilde{A}\tilde{B}} 
\tilde{f}^{\tilde{A}ijk} 
\tilde{\tau}^{\tilde{A}k}
\label{tildetau}
\end{eqnarray}
might even  not be correlated. If correlated (through 
boundary conditions, for example) the break of symmetries might cause that off diagonal 
matrix elements of Yukawa couplings distinguish between quarks and leptons.

We made, when deriving the mass matrices of quarks and leptons from the approach 
unifying spins and charges, several assumptions, approximations and simplifications in order 
to be able to make at the end some rough  predictions about the properties of the 
families of quarks and leptons:

i.  The break of symmetries of the group $SO(1,13)$ (the  Poincar\' e group in 
$d=1+13$)  into $SO(1,7)\times SU(3)\times U(1)$ occurs 
in a way that only 
massless spinors in $d=1+7$ with the charge $ SU(3)\times U(1)$ survive, and yet the two $U(1)$ charges, 
following from $SO(6)$ and $SO(1,7)$, respectively, are related. (Our work on the compactification 
of a massless spinor in $d=1+5$ into   $d=1+3$ and a finite disk gives us some hope that such 
an assumption might be justified\cite{holgernorma05}.) 
The requirement that the terms with $S^{5a}\omega_{5ab}$ and $S^{6a}\omega_{6ab}$ 
do not contribute to the mass term, assures that the charge   
$Q= \tau^{41} + S^{56}$ is conserved at low energies. 

ii. The break of symmetries influences  both, the (Poincar\' e) symmetry 
described by $S^{ab}$ and the symmetries described by $\tilde{S}^{ab}$, and in a way that  
 there are no terms, which would  transform  
$\stackrel{{56}}{\tilde{(+)}}$ into $\stackrel{{56}}{\tilde{[+]}}$. This assumption  was made that 
at "low energies" only four families have to be treated and 
can be explained by a break of the symmetry $SO(1,7)$ into $SO(1,5)\times U(1)$ in the $\tilde{S}^{ab}$ 
sector so that all the contributions of the type $\tilde{S}^{5a}\tilde{\omega}_{5ab}$ and 
$\tilde{S}^{6a}\tilde{\omega}_{6ab}$ are equal to zero. We also assume that the terms which 
include components $p_s, s=5,..,14$, of the momentum $p^a$ do not 
contribute  to the mass matrices. We keep in mind that  any further break of symmetries 
strongly influences the relations among 
$\tilde{\omega}_{abc}$, appearing in the paper \cite{pikanorma05} as ''vacuum expectation values'' 
in mass matrices, so that predictions in Sect.\ref{numerical} strongly depend on the way of breaking. 
 
iii. We  make estimations on a ''tree level''.  

iv. We assume the mass matrices to  be real and symmetric (expecting that complexity and nonsymmetric 
properties will not influence considerably masses and 
mixing matrices of quarks and leptons).


\section{Four families of quarks and leptons}
\label{Lwithassumptions}

Taking into account the assumptions, presented in Sect.\ref{lagrangesec}, we end up with four 
families of quarks and leptons 
\begin{eqnarray}
I.\;& & \stackrel{03}{(+i)} \stackrel{12}{(+)} |\stackrel{56}{(+)} \stackrel{78}{(+)}||...\nonumber\\ 
II.\;& &\stackrel{03}{[+i]} \stackrel{12}{[+]} |\stackrel{56}{(+)} \stackrel{78}{(+)}||... \nonumber\\
III.& & \stackrel{03}{[+i]} \stackrel{12}{(+)} |\stackrel{56}{(+)} \stackrel{78}{[+]}||... \nonumber\\
IV. & & \stackrel{03}{(+i)} \stackrel{12}{[+]} |\stackrel{56}{(+)} \stackrel{78}{[+]}||... .
\label{fourfamilies}
\end{eqnarray}
The Yukawa couplings for these four families are for $u$-quarks and neutrinos  
presented on Table~\ref{TableI}, where 
$\alpha$ stays for $u$-quarks and neutrinos.
\begin{table}
\begin{center}
\begin{tabular}{|r||c|c|c|c|}
\hline
$\alpha$&$ I_{R} $&$ II_{R} $&$ III_{R} $&$ IV_{R}$\\
\hline\hline
&&&& \\
$I_{L}$   & $ A^I_{\alpha}  $ & $ \tilde{A}^{++}_{\alpha} ((03),(12))= $ & $  \tilde{A}^{++}_{\alpha} ((03),(78)) =$  &
$ -  \tilde{A}^{++}_{\alpha} ((12),(78)) = $ \\
$$&$$ &$ \frac{\tilde{\omega}_{327 \alpha} +\tilde{\omega}_{ 018 \alpha}}{2}$&$
\frac{\tilde{\omega}_{ 387 \alpha} +\tilde{\omega}_{078 \alpha}}{2} $&$
 \frac{\tilde{\omega}_{277 \alpha} +\tilde{\omega}_{187 \alpha}}{2}$ \\
&&&& \\
\hline
&&&&\\
$II_{L}$  & $ \tilde{A}^{--}_{\alpha} ((03),(12))= $ & $ A^{II}_{\alpha}= A^{I}_{\alpha} +$ & $ 
\tilde{A}^{-+}_{\alpha} ((12),(78)) = $ &
$ -  \tilde{A}^{-+}_{\alpha} ((03),(78)) = $ \\
$$&$ \frac{\tilde{\omega}_{327 \alpha} +\tilde{\omega}_{018 \alpha}}{2}$&$  
(\tilde{\omega}_{127 \alpha} - \tilde{\omega}_{038 \alpha})$
&$ -\frac{\tilde{\omega}_{277 \alpha} -\tilde{\omega}_{187 \alpha}}{2} $&$  
\frac{\tilde{\omega}_{387 \alpha} - \tilde{\omega}_{078 \alpha}}{2}$ \\
\hline 
&&&& \\
$III_{L}$ & $  \tilde{A}^{--}_{\alpha} ((03),(78)) =$ & $-
\tilde{A}^{+-}_{\alpha} ((12),(78))= $ & $ A^{III}_{\alpha}= A^{I}_{\alpha} + $ &
$   \tilde{A}^{-+}_{\alpha} ((03),(12)) =$ \\
$$&$  \frac{\tilde{\omega}_{387 \alpha} +\tilde{\omega}_{078 \alpha}}{2}$&$ -
\frac{\tilde{\omega}_{277 \alpha} -\tilde{\omega}_{187 \alpha}}{2}$
&$ (\tilde{\omega}_{787 \alpha} - \tilde{\omega}_{038 \alpha})$&$ -
\frac{\tilde{\omega}_{327 \alpha} -\tilde{\omega}_{018 \alpha}}{2} $\\
&&&& \\
\hline 
&&&& \\
$IV_{L}$  & $ \tilde{A}^{--}_{\alpha} ((12),(78)) =$ & $- \tilde{A}^{+-}_{\alpha} ((03),(78)) = $ & 
$ \tilde{A}^{+-}_{\alpha} ((03),(12))$ & $ A^{IV}_{\alpha} = A^{I}_{\alpha} + $ \\
$$&$ \frac{\tilde{\omega}_{277 \alpha} +\tilde{\omega}_{187 \alpha}}{2}$&$ 
\frac{\tilde{\omega}_{387 \alpha} -\tilde{\omega}_{078 \alpha}}{2}$
&$ -\frac{\tilde{\omega}_{327 \alpha} -\tilde{\omega}_{018 \alpha}}{2} 
$&$ (\tilde{\omega}_{127 \alpha} + \tilde{\omega}_{787 \alpha})$\\
&&&& \\
\hline\hline
\end{tabular}
\end{center}
\caption{\label{TableI}%
The mass matrix of four families of  $u$-quarks and neutrinos, obtained within the approach 
unifying spins and charges under the assumptions i.-iv. (see also the ref.\cite{pikanorma05}). 
According to Eq.(\ref{yukawadiagonal}) and Table~\ref{TableI} and~\ref{TableII}
there are 13 free parameters, expressed in terms
of the fields $A^{I}_{\alpha}$ and $\tilde{\omega}_{\alpha abc}$, if $\tilde{\omega}_{\alpha abc}$ 
are the same for different $\alpha $. They then accordingly determine (due to assumptions 
i.-iv.) all the properties of the  four families of the two types of quarks and the two types of 
leptons.} 
\end{table}

The corresponding mass matrix for the $d$-quarks and the electrons is presented on Table~\ref{TableII}, 
where $\beta$ stays for $d$-quarks and electrons.
\begin{table}
\begin{center}
\begin{tabular}{|r||c|c|c|c|}
\hline
$\beta$&$ I_{R} $&$ II_{R} $&$ III_{R} $&$ IV_{R}$\\
\hline\hline
&&&& \\
$I_{L}$   & $ A^I_{\beta}  $ & $ \tilde{A}^{++}_{\beta} ((03),(12))= $ & $  - 
\tilde{A}^{++}_{\beta} ((03),(78)) =$  &
$   \tilde{A}^{++}_{\beta} ((12),(78)) = $ \\
$$&$$ &$ \frac{\tilde{\omega}_{327 \beta} - \tilde{\omega}_{018 \beta}}{2}$&$ -
\frac{\tilde{\omega}_{387 \beta} - 
\tilde{\omega}_{078 \beta}}{2} $&$  -\frac{\tilde{\omega}_{277 \beta} +
\tilde{\omega}_{187 \beta}}{2}$ \\
&&&& \\
\hline
&&&&\\
$II_{L}$  & $ \tilde{A}^{--}_{\beta} ((03),(12))= $ & $ A^{II}_{\beta}= A^{I}_{\beta} + $ & $  -\tilde{A}^{-+}_{\beta} ((12),(78)) = $ &
$  \tilde{A}^{-+}_{\beta} ((03),(78)) = $ \\
$$&$ \frac{\tilde{\omega}_{327 \beta} -\tilde{\omega}_{018 \beta}}{2}$&$ 
(\tilde{\omega}_{127 \beta} + \tilde{\omega}_{038 \beta})$
&$ \frac{\tilde{\omega}_{277 \beta} -\tilde{\omega}_{187 \beta}}{2} $&$  - 
\frac{\tilde{\omega}_{387 \beta} +\tilde{\omega}_{078 \beta}}{2}$ \\
\hline 
&&&& \\
$III_{L}$ & $  - \tilde{A}^{--}_{\beta} ((03),(78)) =$ & $ \tilde{A}^{+-}_{\beta} ((12),(78))= 
$ & $ A^{III}_{\beta}=A^{I}_{\beta} + $ &
$   \tilde{A}^{-+}_{\beta} ((03),(12)) =$ \\
$$&$ -\frac{\tilde{\omega}_{387 \beta} -\tilde{\omega}_{078 \beta}}{2}$&$ 
\frac{\tilde{\omega}_{277 \beta} -\tilde{\omega}_{187 \beta}}{2}$
&$ (\tilde{\omega}_{787 \beta} + \tilde{\omega}_{038 \beta})$&$ -
\frac{\tilde{\omega}_{018 \beta} +\tilde{\omega}_{327 \beta}}{2} $\\
&&&& \\
\hline 
&&&& \\
$IV_{L}$  & $ -\tilde{A}^{--}_{\beta} ((12),(78)) =$ & $ \tilde{A}^{+-}_{\beta} ((03),(78)) = $ & 
$ \tilde{A}^{+-}_{\beta} ((03),(12))$ & $ A^{IV}_{\beta} A^{I}_{\beta} + $ \\
$$&$ -\frac{\tilde{\omega}_{277 \beta} +\tilde{\omega}_{187 \beta}}{2}$&$ -
\frac{\tilde{\omega}_{387 \beta} +\tilde{\omega}_{078 \beta}}{2}$
&$ -\frac{\tilde{\omega}_{018 \beta} +\tilde{\omega}_{327 \beta}}{2} $&$ 
(\tilde{\omega}_{127 \beta} + \tilde{\omega}_{787 \beta})$\\
&&&& \\
\hline\hline
\end{tabular}
\end{center}
\caption{\label{TableII}%
The mass matrix of four families of the $d$-quarks and electrons, $\beta$ stays for the $d$-quarks 
and the electrons. Comments are the same as on Table~\ref{TableI}.}
\end{table}

The explicit form of the diagonal matrix elements for the above choice of assumptions in terms of 
$\omega_{abc \delta}$'s, $\delta = \alpha, \beta$ and $A^{y}_{\pm}, y = Y$ and $Y'$, 
$\tilde{\omega}_{abc \delta}$  and $\tilde{A}^{41}_{\pm}$ is  as follows
\begin{eqnarray}
A^{I}_{u} &=& \frac{2}{3} A^{Y}_{- u} - \frac{1}{3} A^{Y'}_{- u}  + \tilde{\omega}^{I}_{- u},\quad \;\;\;
A^{I}_{\nu}=                                    - A^{Y'}_{- \nu}  + \tilde{\omega}^{I}_{- \nu},\nonumber\\ 
A^{I}_{d} &=&-\frac{1}{3} A^{Y}_{+ d} + \frac{2}{3} A^{Y'}_{+ d}  + \tilde{\omega}^{I}_{+ d},\;\;\; \;
A^{I}_{e}  =            - A^{Y}_{+ e} +                           \tilde{\omega}^{I}_{+ e},\nonumber\\
A^{II}_{\alpha}  &=&  A^{I}_{\alpha} + (i \tilde{\omega}_{03- \alpha} + \tilde{\omega}_{12 -\alpha }), 
\;\;\;
A^{II}_{\beta} =  A^{I}_{\beta} + (i \tilde{\omega}_{03+ \beta} + \tilde{\omega}_{12 + \beta}), \nonumber\\
A^{III}_{\alpha} &= & A^{I}_{\alpha} + (i \tilde{\omega}_{03- \alpha } + \tilde{\omega}_{78 - \alpha}), \;\;
A^{III}_{\beta} =  A^{I}_{\beta} + (i \tilde{\omega}_{03+ \beta} + \tilde{\omega}_{78 + \beta}), \nonumber\\
A^{IV}_{\alpha} &= & A^{I}_{\alpha} + ( \tilde{\omega}_{12- \alpha} + \tilde{\omega}_{78 - \alpha}), \quad
A^{IV}_{\beta} =  A^{I}_{\beta} + ( \tilde{\omega}_{12+ \beta} + \tilde{\omega}_{78 + \beta}), \nonumber\\
\label{yukawadiagonal}
\end{eqnarray}
with $\alpha= u,\nu$, $\beta=d,e $ and $-\tilde{\omega}^{I}{}_{\pm} = 
\frac{1}{2} (i \tilde{\omega}_{03 \pm} + \tilde{\omega}_{12 \pm} + 
\tilde{\omega}_{56 \pm} + \tilde{\omega}_{78 \pm}
+ \frac{1}{3} \tilde{A}^{41}_{\pm})$. We put the index $\alpha =u,\nu$ and $\beta = d,e$ to manifest that 
breaking of symmetries, boundary conditions and nonperturbative and other effects influence the 
"vacuum expectation values" gauge fields. If we assume that $\tilde{\omega}_{abc \delta}$ are 
equal if they 
belong to different $\delta$, the assumption that all the matrix elements  are real relates 
$\tilde{\omega}^{I}{}_{+}= \frac{1}{2} \tilde{\omega}_{03 8} + \tilde{\omega}, 
\tilde{\omega}^{I}{}_{-}= -\frac{1}{2} \tilde{\omega}_{03 8} + \tilde{\omega},$ where 
$\tilde{\omega}$ is (in case that breaking of symmetries does not influence quarks and leptons 
differently) one common parameter.

If the break of symmetries does not influence the quarks and the leptons in a different way, then 
under the assumptions i.-iv. the off diagonal matrix elements of mass matrices for quarks 
are the same as for the corresponding leptons (the off diagonal matrix elements  of the 
$u$-quarks and the neutrinos are the same,  and the off diagonal matrix elements for the
$d$-quarks and the electrons are the same) and since the diagonal matrix elements differ 
only in a constant times a unit matrix, the predicted mixing matrices of 
the quarks and the leptons  would be the same. 

We must ask ourselves at this point: Can we find a way of breaking symmetries 
 - allowing some special boundary conditions and taking into account all the 
 perturbative and nonperturbative effects - which would lead to 
 so different  properties of quarks and leptons 
as experimentally observed or must we take the 
Majorana like degrees of freedom into account additionally?

In this paper, we are not yet able to answer this question. We can only make some estimates trying 
to  learn from the approach unifying spins and charges 
about possible explanations 
for the properties of quarks and leptons.

We proceed by relating the experimental data and the mass matrices from the approach.  
Knowing  from the experimental data that 
the first two families of quarks and leptons 
are much lighter than the third one, while in the refs.\cite{okun,okunmaltoni,okunbulatov} 
 the authors, analyzing the  
experimental data, conclude that the experimental data do not forbid  masses of the fourth family of 
quarks to be between 
$200$ GeV and $300$ GeV,  of the fourth electron to be around $100$ GeV  and of the fourth neutrino
to be at around $50 $ GeV 
we make one more assumption, 
which seems quite reasonable also 
from the point of view of the measured matrix elements of the mixing matrix for quarks. 
Namely, {\em we assume  
that the mass matrices of the four families of quarks and leptons are diagonalizable in 
two steps, so that the first diagonalization transforms 
them into block-diagonal matrices with two  $2\times 2$ sub-matrices.}  
This assumption, which   means, 
that a real and symmetric 
$4 \times 4$ matrix is diagonalizable by only three rather than six angles,  
simplifies considerably further studies, making conclusions very transparent. Let us comment that 
such a property of  mass matrices could  be a consequence of an approximate break of symmetry 
in the $\tilde{S}^{ab}$ sector from  $SO(1,5)$ to  $SU(2)\times SU(2)\times U(1)$, which makes, 
for example, all the terms $\tilde{S}^{7a} \tilde{\omega}_{7ab \delta}$ and $\tilde{S}^{8a} 
\tilde{\omega}_{8ab \delta}$ contributing small terms to mass matrices.  
 The exact break of this type  makes that the lower two 
 families  completely decouple from the higher two.
(Similarly we have required, in order to end up with only four rather than eight families, 
 that $SO(1,7)$ breaks to $SO(1,5) \times U(1)$ so that  
 all $\tilde{S}^{5a}\;\tilde{\omega}_{5a \pm \delta}$ 
 and $\tilde{S}^{6a}\;\tilde{\omega}_{6a \pm \delta }$ contribute nothing to mass matrices.)

It is easy to prove that a $4\times 4$ matrix is diagonalizable in two steps only  
 if it has a  structure\cite{matjazdiploma}
\begin{equation}\label{deggen}
          \left(\begin{array}{cc}
                  A   &  B\nonumber\\
                  B   &  C=A+k B \nonumber\\
                    \end{array}
                \right).
\end{equation}
Since $A$ and $C$ are, as assumed on Table~\ref{TableI} and Table~\ref{TableII}, symmetric  
$2 \times 2$ matrices, 
so must then be also $B$. The parameter $k$ is assumed to be an unknown  number.

The assumption (\ref{deggen})  requires: i. $\tilde{\omega}_{277 \delta}= 0, \;
\tilde{\omega}_{327 \delta}= - \frac{k}{2}\tilde{\omega}_{187 \delta}, \; \tilde{\omega}_{787 \delta} = 
 \frac{k}{2}\tilde{\omega}_{387 \delta}, \;\tilde{\omega}_{038 \delta}=- 
 \frac{k}{2}\tilde{\omega}_{078 \delta}$, $\delta=u,\nu,d,e$,  
and, in the case that $\tilde{\omega}_{abc \delta}$ do not depend on $\delta$ but only on $abc$.  
ii. $k_u= -k_d$ and $k_{\nu}= - k_{e}$, where $k_u$ and $k_{\nu}$ 
are two independent parameters. (If $k=0$ in Eq. (\ref{deggen}), the angle of rotation is
$45^\circ$ - then, if also all the $2 \times 2$ matrices would have the same structure  
(namely equal diagonal and equal nondiagonal elements\cite{astridragannorma,bmnBled04}),  the 
corresponding mixing matrices for quarks and leptons would be the 
identity.
) 

\subsection{Evaluation of mass matrices under assumption that $\tilde{\omega}_{abc \delta}$ do not 
distinguish among quarks and leptons}
\label{simplified}

Let us first assume that neither boundary conditions nor nonperturbative or other effects influence 
the fields $\tilde{\omega}_{abc \delta}$ in a way that they would differ for different $\delta$.

We shall present in what follows some simple relations which  
demonstrate transparently  properties of mass matrices.
After the one step diagonalization determined by the angle of rotation
\begin{eqnarray}
\tan \varphi_{\alpha} &=&  \pm(\sqrt{1+ (\frac{k}{2})^2 } \pm \frac{k}{2}),\quad 
\tan \varphi_{\beta} =  \pm(\sqrt{1+ (\frac{k}{2})^2 } \mp \frac{k}{2}),\nonumber\\ 
{\rm with}\;&&
\tan (\varphi_{\alpha}-\varphi_{\beta}) = \pm \frac{k}{2},\; ({\rm or} \: \pm \frac{2}{k}) 
\label{firstangle}
\end{eqnarray}
we end up with two  by diagonal matrices, with $k=k_u$ for quarks and $k=k_{\nu}$ for leptons, 
while $\alpha$ concerns the $u$-quarks and 
$\nu$,  and $\beta$ the $d$-quarks  and electrons.

The first by diagonal mass matrix of the $u$-quarks ($\alpha = u$) and neutrinos ($\alpha = \nu$) is 
as follows 
\begin{equation}\label{aunuIIa}
 {\bf A^{a}} = \left(\begin{array}{cc}
 a_{\alpha} 
 ,& \frac{1}{2}(\tilde{\omega}_{018} - 
 \sqrt{1 + (\frac{k_{\alpha}}{2})^2}\;\tilde{\omega}_{187})\nonumber\\
  \frac{1}{2}(\tilde{\omega}_{018 } - 
 \sqrt{1 + (\frac{k_{\alpha}}{2})^2}\;\tilde{\omega}_{187 }), &  a_{\alpha} + 
 \tilde{\omega}_{127}  + 
   \sqrt{1 + (\frac{k_{\alpha}}{2})^2}\;\tilde{\omega}_{078}\nonumber\\
                    \end{array}
                \right),
\end{equation}
with $a_{u}= \frac{2}{3}A^Y - \frac{1}{3} A^{Y'} + \tilde{\omega} - 
\frac{1}{2} \tilde{\omega}_{038}
+ \frac{1}{2}( \frac{k_u}{2} - \sqrt{1 + (\frac{k_u}{2})^2}) 
 (\tilde{\omega}_{078} + \tilde{\omega}_{387})$ and $a_{\nu}= 
 - A^{Y'} + \tilde{\omega} - \frac{1}{2} \tilde{\omega}_{038}
+ \frac{1}{2}( \frac{k_{\nu}}{2} - \sqrt{1 + (\frac{k_{\nu}}{2})^2}) 
 (\tilde{\omega}_{078} + \tilde{\omega}_{387})$. 
 The mass matrix for the second two families of $u$-quarks ($\alpha = u$) and neutrinos 
 ($\alpha = \nu$) is equal to  
\begin{equation}\label{aunuIIb}
 {\bf A^{b}_{\alpha}} = \left(\begin{array}{cc}
 a_{\alpha} + \sqrt{1 + (\frac{k_{\alpha}}{2})^2} 
 \;(\tilde{\omega}_{078} + \tilde{\omega}_{387})
 ,& \frac{1}{2}(\tilde{\omega}_{018} + 
 \sqrt{1 + (\frac{k_{\alpha}}{2})^2}\;\tilde{\omega}_{187})\nonumber\\
  \frac{1}{2}(\tilde{\omega}_{018} +  
 \sqrt{1 + (\frac{k_{\alpha}}{2})^2}\;\tilde{\omega}_{187}), &  a_{\alpha} + 
 \tilde{\omega}_{127}  + 
   \sqrt{1 + (\frac{k_{\alpha}}{2})^2}\;\tilde{\omega}_{387}\nonumber\\
                    \end{array}
                \right).
\end{equation}

Accordingly we find for the first two families of $d$-quarks  ($\beta = d$) and 
electrons ($\beta = e $)
\begin{equation}\label{adeIIa}
 {\bf A^{a}_{\beta}} = \left(\begin{array}{cc}
 a_{\beta} 
 ,& -\frac{1}{2}(\tilde{\omega}_{018} - 
 \sqrt{1 + (\frac{k_{\alpha}}{2})^2}\;\tilde{\omega}_{187})\nonumber\\
  -\frac{1}{2}(\tilde{\omega}_{018} - 
 \sqrt{1 + (\frac{k_{\alpha}}{2})^2}\;\tilde{\omega}_{187}), &  a_{\beta} + 
 \tilde{\omega}_{127}  + 
   \sqrt{1 + (\frac{k_{\alpha}}{2})^2}\;\tilde{\omega}_{078}\nonumber\\
                    \end{array}
                \right),
\end{equation}
with $a_{d}= - \frac{1}{3}A^{Y} + \frac{2}{3} A^{Y'} + 
\tilde{\omega} + \frac{1}{2} \tilde{\omega}_{038}
- \frac{1}{2}( \frac{k_u}{2} + \sqrt{1 + (\frac{k_u}{2})^2}) 
 (\tilde{\omega}_{078} - \tilde{\omega}_{387})$ and $a_{e}= 
 - A^{Y} + \tilde{\omega} + \frac{1}{2} \tilde{\omega}_{038}
- \frac{1}{2}( \frac{k_{\nu}}{2} + \sqrt{1 + (\frac{k_{\nu}}{2})^2}) 
 (\tilde{\omega}_{078} - \tilde{\omega}_{387})$.  $k_{\alpha}$ in 
 (\ref{adeIIa}) is $k_u$ 
 for $d$-quarks and $k_{\nu}$ for electrons.
For the second two families of $d$-quarks ($\beta = d $) and electrons ($\beta = e$) it follows  
\begin{equation}\label{adeIIb}
 {\bf A^{b}_{\beta}} = \left(\begin{array}{cc}
 a_{\beta} + \sqrt{1 + (\frac{k_{\alpha}}{2})^2} \;
 (\tilde{\omega}_{078} - \tilde{\omega}_{387})
 ,& -\frac{1}{2}(\tilde{\omega}_{018} +  
 \sqrt{1 + (\frac{k_{\alpha}}{2})^2}\;\tilde{\omega}_{187})\nonumber\\
  -\frac{1}{2}(\tilde{\omega}_{018} +  
 \sqrt{1 + (\frac{k_{\alpha}}{2})^2}\;\tilde{\omega}_{187}), &  a_{\beta} + 
 \tilde{\omega}_{127}  - 
   \sqrt{1 + (\frac{k_{\alpha}}{2})^2}\;\tilde{\omega}_{387}\nonumber\\
                    \end{array}
                \right).
\end{equation}
Again, $k_{\alpha}$ in (\ref{adeIIb}) is $k_u$ for $d$-quarks and
 $k_{\nu}$ for electrons.

There are three angles, which in the two step orthogonal transformations rotate each mass matrix 
into a diagonal one. The angles of rotations for $u-$quarks  and $d-$quarks, and accordingly 
for neutrinos and electrons, 
 are related as seen from Eq.(\ref{firstangle}) for the first step rotation. It follows namely 
 that $\tan \varphi_{\alpha} = \tan^{-1} \varphi_{\beta}$ 
 and accordingly 
 \begin{eqnarray}
 \label{phirelated}
  \varphi_{\alpha} &=& \frac{\pi}{2}  -\varphi_{\beta}, \quad \varphi_{\alpha} = \frac{\pi}{4}-
  \frac{\varphi}{2}, \quad \varphi_{\beta} = \frac{\pi}{4}+   \frac{\varphi}{2},\nonumber\\
  \quad {\rm with}\;&& \varphi = \varphi_{\alpha} -\varphi_{\beta}.
  \end{eqnarray}
Similarly also the two angles of rotations of the two by two diagonal matrices are related.
Reminding the reader that in the unitary transformations ($S^{\dagger} S=I$)  
the  trace and the determinant are among the invariants, while the angle 
of rotation, which diagonalizes $2$ by $2$ matrices  (of the type (\ref{deggen})), and the 
values of the diagonal matrices are related as follows
\begin{eqnarray}
\tan \Phi &=& (\sqrt{1+(\frac{C-A}{2B})^2}\mp \frac{C-A}{2B}),\nonumber\\
\lambda_{1,2} &=& \frac{1}{2}((C+A) \pm \sqrt{(C-A)^2 + (2B)^2}),
\label{solutions}
\end{eqnarray}
where for $A,B,C$ the corresponding matrix elements from Eqs.(\ref{aunuIIa},%
\ref{aunuIIb},\ref{adeIIa},\ref{adeIIb})  must be taken, one   
  easily finds that  
\begin{eqnarray}
\label{angle}
{}^a \eta_{\alpha}= -{}^a \eta_{\beta},\quad {}^b \eta_{\alpha}= -{}^b \eta_{\beta},
 \quad \alpha = u,\nu,\; \beta = d, e,
\end{eqnarray}
where index $a$ denotes the first two families and $b$ the second two families of 
either quarks ($\alpha =u,\; \beta = d$) and leptons ($\alpha =\nu,\; \beta = e$) and 
$\eta = \frac{C-A}{2B}$. 
One then finds the relations, equivalent to those of  Eq.(\ref{phirelated}) 
 \begin{eqnarray}
 \label{abphirelated}
  {}^{a,b}\varphi_{\alpha} &=& \frac{\pi}{2}  - {}^{a,b}\varphi_{\beta}, 
  \quad {}^{a,b}\varphi_{\alpha} = \frac{\pi}{4}-
  \frac{{}^{a,b}\varphi}{2}, \quad {}^{ab}\varphi_{\beta} = \frac{\pi}{4}+   
  \frac{{}^{a,b}\varphi}{2},\nonumber\\
  \quad {\rm with}\;&& {}^{a,b}\varphi = {}^{a,b}\varphi_{\alpha} -{}^{a,b}\varphi_{\beta}.
  \end{eqnarray}
We accordingly find for the case, that $\tilde{\omega}_{abc \delta}$ do not depend on 
$\delta$ the following relations among the masses of the quarks and the leptons  
$ |m_{u_2} - m_{u_1}| = |m_{d_2} - m_{d_1}|,\; $ $ |m_{u_4} - m_{u_3}|  =  |m_{d_4} - m_{d_3}|,\;$ $
|m_{\nu_{2}} - m_{\nu_{1}}| = |m_{e_2} -m_{e_1}|, \; $ $ 
|m_{\nu_4} - m_{\nu_{3}}| = |m_{e_4} -m_{e_3}|,\;$ $
|(m_{u_4} + m_{u_3}) - (m_{u_2} + m_{u_1})| = |(m_{d_4} + m_{d_3}) - (m_{d_2} + m_{d_1})|, \;$ $ 
|(m_{\nu_4} + m_{\nu_{3}}) - (m_{\nu_{2}} + m_{\nu_1})| =   
|(m_{e_4} + m_{e_3}) - (m_{e_2} + m_{e_1})|, \;$ $
|m_{u_4} + m_{u_3}|  \approx  2\;\sqrt{1+(\frac{k_u}{2})^2} \;\tilde{\omega}_{387} 
 \approx |m_{d_4} + m_{d_3}|, \; $ $ 
 |m_{\nu_4} + m_{\nu_{3}}| \approx  2\; \sqrt{1+(\frac{k_{\nu}}{2})^2} \;\tilde{\omega}_{387} 
 \approx |m_{e_4} + m_{e_3}|.$  
We take the absolute values of the sums and the differences, since 
whenever an eigenvalue $\lambda_{1,2} $ (Eq.\ref{solutions}) appears to be negative,   
an appropriate change of a phase of the corresponding  state transforms the negative value into the 
positive one by changing simultaneously the internal parity of the particular state, 
as it will be discussed in Sect.\ref{negativemasses}. 

The above relations among masses of quarks and leptons do not agree with the  
experimental data, as expected. 
We can take them as a very rough estimation  in the limit when masses of the  fourth family 
are much higher than the mass $m_{t}$, knowing   
 that $m_{t} $ is more than $30$ times larger than the mass $m_{b}$.

\subsection{Predictions with relaxed assumptions}
\label{improvedmatrices}

We shall make in this subsection the evaluation of the mass matrices and their 
properties by allowing that $\tilde{\omega}_{abc \delta}$ depend on the  type of quarks and 
leptons, assuming that  boundary conditions connected with breaking of symmetries, 
perturbative, nonperturbative and other effects, appearing after each break of symmetries influence 
the "vacuum expectation values" of the gauge fields, entering into mass matrices so that 
$\tilde{\omega}_{abc \delta}$ are not the same for all the quarks and the leptons. To find out, how 
do they differ, one should make very sophisticated calculations, which even under the assumption 
that one can treat gravity in the region far away from the Planck scale as all the other gauge fields, 
is a very tough work, not only because  we do not yet know how to treat all the breaks 
of symmetries to end up with massless spinors before the Yukawa couplings take care of their mass, 
but also because the nonperturbative effects are not solved yet even in hadron physics.

Learning from Subsect.\ref{simplified} that diagonalization of 
$4 \times 4 $ matrix in two steps enables a transparent view on properties of the mass matrices 
and recognizing that such an assumption is at least not in disagreement with the 
known experimental data, we shall make calculations under the assumption that the 
{\em diagonalization in two steps}  is a meaningful simplification. We shall also keep 
the  relation from Subsect.\ref{simplified}, which requires that the   
{\em  angles of rotations for the $u$-quarks  and  the $d$-quarks mass 
matrices, as well as for the  neutrinos and the electrons mass matrices,   
which determine the first and the second step of diagonalization, are simply related, just 
as presented in}  Eqs.(\ref{phirelated},\ref{abphirelated}). These assumptions reduce 
considerably the number 
of free parameters (and might also help us to recognize the way of breaking symmetries in 
more sophisticated calculations).

 It follows then that in Eqs.(\ref{aunuIIa},\ref{aunuIIb},%
\ref{adeIIa},\ref{adeIIb}) all the fields $\tilde{\omega}_{abc \delta}$ carry an additional  
index $\delta = \alpha, \beta,$ 
 while we keep  the relations $k_{\alpha} = - k_{\beta}$, 
$\alpha = u, \nu$ and $\beta = d, e$, where $k_{\alpha,\beta}$ define the first step 
orthogonal transformations leading to $2$ by $2$ by diagonal mass matrices and  
the relations among the angles of rotations in the second step of 
orthogonal transformations determined by ${}^{a,b}\eta_{\alpha} = {}^{a,b}(\frac{2B}{C-A})_{\alpha}$, 
requiring that (Eq.(\ref{angle})) 
${}^{a,b}\eta_{\alpha}=  -{}^{a,b}\eta_{\beta}$ 
(where $A,B,C$ are to be replaced 
by the corresponding matrix elements for the first two families determined by the matrix 
$A^a_{\alpha,\beta}$ (Eqs.(\ref{aunuIIa}, \ref{adeIIa})) 
and the second two families determined by the matrix $A^b_{\alpha,\beta}$ (%
Eqs.(\ref{aunuIIb}, \ref{adeIIb}))).

Then it must be   
\begin{eqnarray}
\label{angleab}
{}^a\varepsilon_{\alpha}\;(\tilde{\omega}_{018 \beta}- 
\sqrt{1+ (\frac{k_{\alpha}}{2})^2 }\; \tilde{\omega}_{187 \beta}) &=&
 ( \tilde{\omega}_{018 \alpha}- \sqrt{1+ (\frac{k_{\alpha}}{2})^2 }\; 
\tilde{\omega}_{187 \alpha}),\nonumber\\
{}^b\varepsilon_{\alpha}\; (\tilde{\omega}_{018 \beta}+ \sqrt{1+ (\frac{k_{\alpha}}{2})^2 } \;
\tilde{\omega}_{187 \beta}) &=&
( \tilde{\omega}_{018 \alpha} + \sqrt{1+ (\frac{k_{\alpha}}{2})^2 } \;
\tilde{\omega}_{187 \alpha}),\nonumber\\
{}^a\varepsilon_{\alpha}\;(\tilde{\omega}_{127 \beta}+ \sqrt{1+ (\frac{k_{\alpha}}{2})^2 } \;
\tilde{\omega}_{078 \beta}) &=&
 ( \tilde{\omega}_{127 \alpha}+ \sqrt{1+ (\frac{k_{\alpha}}{2})^2 } \;
\tilde{\omega}_{078 \alpha}),\nonumber\\
{}^b\varepsilon_{\alpha}\;(\tilde{\omega}_{127 \beta}- \sqrt{1+ (\frac{k_{\alpha}}{2})^2 } \;
\tilde{\omega}_{078 \beta}) &=&
 ( \tilde{\omega}_{127 \alpha}- \sqrt{1+ (\frac{k_{\alpha}}{2})^2 } \;
\tilde{\omega}_{078 \alpha}),
\end{eqnarray}
where index $a$ and $b$ distinguish between the two by two matrices for the first two and 
the second two families, correspondingly, while $\alpha =u,\nu$ and $\beta=d,e$.
The two mixing matrices for the quarks and the leptons have  the form 
\begin{eqnarray}
\label{abcwithm}
{\bf V}_{\alpha \beta} = \left(\begin{array}{cccc}
c(\varphi)c({}^a\varphi)&-c(\varphi)s({}^a\varphi)&
-s(\varphi)c({}^a\varphi^b)& s(\varphi)s({}^a\varphi^b)\\
c(\varphi)s({}^a\varphi)&c(\varphi)c({}^a\varphi)&
-s(\varphi)s({}^a\varphi^b)& -s(\varphi)c({}^a\varphi^b)\\
s(\varphi)c({}^a\varphi^b)&-s(\varphi)s({}^a\varphi^b)&
c(\varphi)c({}^b\varphi)& -c(\varphi)s({}^b\varphi)\\
s(\varphi)s({}^a\varphi^b)&s(\varphi)c({}^a\varphi^b)&
c(\varphi)s({}^b\varphi)& c(\varphi)c({}^a\varphi)\\
 \end{array}
                \right),
\end{eqnarray}
with the angles (Eq.\ref{phirelated},\ref{abphirelated}) described  by the three parameters 
$k_{\alpha},{}^{a}\eta_{\alpha},{}^{b}\eta_{\alpha}$ 
as follows
\begin{eqnarray}
\label{anglesandmixmatr3ngles}
\varphi = \varphi_{\alpha}-\varphi_{\beta},
\quad {}^a\varphi = {}^a\varphi_{\alpha}-{}^a\varphi_{\beta},\quad 
{}^a\varphi^b = - \frac{{}^a\varphi + {}^b \varphi}{2}.
\end{eqnarray}
We recognize that with the mixing matrix for either quarks or leptons  describable 
by the three parameters each 
$k_{\alpha}, {}^a\eta_{\alpha}, {}^b\eta_{\alpha}; \alpha=u,\nu,$ we have just 
enough free parameters  to make any choice for the masses of the fourth family of quarks and leptons.   
To see this we just express $A,B,C$ in any of the two $2$ by $2$ 
matrices in terms of the corresponding diagonal values that is in terms of the masses $m_{\alpha i}, 
m_{\beta i};
i=1,2,3,4; \alpha= u,\nu,\beta=d,e,$ and the parameters 
$k_{\alpha}=-k_{\beta}, {}^a\eta_{\alpha} = - {}^a\eta_{\beta}, {}^b\eta_{\alpha} = - {}^b\eta_{\beta};
\alpha=u,\nu,$.
The matrix elements of ${\bf A}^{a}_{\alpha}$ (${}^aa_{\alpha},{}^ab_{\alpha},{}^ac_{\alpha}
$) for $u$-quarks and neutrinos are expressible with 
the masses $m_{\alpha 1}, m_{\alpha 2}$ of the first two families 
of $u$-quarks or neutrinos and the corresponding angles of 
rotations as follows
\begin{eqnarray}
\label{abcwithm1}
{\bf A}^{a}_{\alpha} = \left(\begin{array}{cc}
\frac{1}{2}( m_{\alpha 1} + m_{\alpha 2} - \frac{{}^a\eta_{\alpha} (m_{\alpha 2} - 
m_{\alpha 1})}{\sqrt{1+({}^a\eta_{\alpha})^2}}),& 
\frac{m_{\alpha 2} - m_{\alpha 1}}{2\sqrt{1+({}^a\eta_{\alpha})^2}})\nonumber\\
\frac{m_{\alpha 2} - m_{\alpha 1}}{2\sqrt{1+({}^a\eta_{\alpha})^2}}),&
\frac{1}{2}( m_{\alpha 1} + m_{\alpha 2} + 
\frac{{}^a\eta_{\alpha}(m_{\alpha 2} - m_{\alpha 1})}{\sqrt{1+({}^a\eta_{\alpha})^2}})
 \end{array}
                \right),
\end{eqnarray}
while the 
expressions for the matrix ${\bf A}^{b}_{\alpha}$ with matrix elements 
${}^ba_{\alpha}, {}^bc_{\alpha}, {}^bb_{\alpha}$ follow, if
we replace $m_{\alpha 1}$ with $m_{\alpha 3}$ and $m_{\alpha 2}$ with $m_{\alpha 4}$.
Equivalently we obtain the mass matrices for the $d$-quarks and the 
electrons by replacing $\alpha$ by $\beta$ in all expressions. 
Eq.(\ref{abcwithm}) below demonstrates that if once  the three parameters 
$k_{\alpha},{}^{a}\eta_{\alpha},{}^{b}\eta_{\alpha}$ are chosen to fit the
experimental data, any four masses for the fourth family of quarks and leptons 
agree with the proposed requirements.

The starting mass matrices ${\bf M}_{\alpha,\beta} $ - the Yukawa couplings - for 
quarks and leptons, 
which are $4$ by $4$ matrices of 
Table~\ref{TableI} and Table~\ref{TableII}, are expressible 
with the matrices  ${\bf A}^{a,b}_{\alpha,\beta}$ of Eq.(\ref{abcwithm}) as follows
\begin{eqnarray}
\label{MDNyukamainm}
 \left(\begin{array}{cc}
\frac{1}{2}[({\bf A}^a_{\alpha,\beta } + {\bf A}^b_{\alpha,\beta }) 
- \frac{({\bf A}^b_{\alpha,\beta } - {\bf A}^a_{\alpha,\beta
  })k_{\alpha,\beta}}%
       {2\sqrt{1+(\frac{k_{\alpha,\beta}}{2})^2}}],& 
\frac{{\bf A}^b_{\alpha,\beta} - {\bf A}^a_{\alpha,\beta}}{2\sqrt{1+
 (\frac{k_{\alpha,\beta}}{2})^2}} \nonumber\\
\frac{{\bf A}^b_{\alpha,\beta } - {\bf A}^a_{\alpha,\beta }}%
{2\sqrt{1+(\frac{k_{\alpha,\beta}}{2})^2}},&
\frac{1}{2}[({\bf A}^a_{\alpha,\beta} + {\bf A}^b_{\alpha,\beta}) 
+ \frac{({\bf A}^b_{\alpha,\beta} - {\bf A}^a_{\alpha,\beta})k_{\alpha,\beta}}%
{2\sqrt{1+(\frac{k_{\alpha,\beta}}{2})^2}}]
 \end{array}
                \right).
\end{eqnarray}
(One easily sees that the matrix 
${\bf M_{\alpha,\beta}}$ is 
equal to a democratic matrix with all the elements equal 
to $m_{\alpha_4}/4, $ with $\alpha = u, \nu, d, e,$ if all the angles of rotations are equal to $\pi/4$   
(that is for $k_{\alpha}=0$, ${}^{a,b}\eta_{\alpha} = 0$), while 
$m_{\alpha_i,\beta_i}=0, i=1,3$,  and that the mixing matrices are 
then the identity.)

Once knowing the matrices ${\bf M_{\alpha,\beta}} $ one easily finds for the parameters 
 $\tilde{\omega}_{abc\alpha,\beta},$ with $(abc)$ equal to $(018),
(078), (127),(187),(387)$, entering in Table~\ref{TableI} 
and Table~\ref{TableII} the expressions 
\begin{eqnarray}
\label{omegatilde}
\tilde{\omega}_{018\alpha} &=&\frac{1}{2}[\frac{m_{\alpha 2} - 
m_{\alpha 1}}{\sqrt{1+ ({}^a\eta_{\alpha})^2}} +
\frac{m_{\alpha 4 } - m_{\alpha 3}}{ \sqrt{1+ ({}^b\eta_{\alpha})^2}}], \nonumber\\
\tilde{\omega}_{078\alpha} &=&\frac{1}{2 \;\sqrt{1+ (\frac{k_{\alpha}}{2})^2}}
[\frac{{}^a\eta_{\alpha}\;(m_{\alpha 2} - m_{\alpha 1})}{\sqrt{1+ ({}^a\eta_{\alpha})^2}} - 
\frac{{}^b\eta_{\alpha}\;(m_{\alpha 4 } - m_{\alpha 3})}{\sqrt{1+ ({}^b\eta_{\alpha})^2}}], \nonumber\\
 \tilde{\omega}_{127\alpha} &=&\frac{1}{2}[\frac{{}^a\eta_{\alpha}\;(m_{\alpha 2} - 
m_{\alpha 1})}{\sqrt{1+ ({}^a\eta_{\alpha})^2}} +
\frac{{}^b\eta_{\alpha}\;(m_{\alpha 4 } - m_{\alpha 3})}{\sqrt{1+ ({}^b\eta_{\alpha})^2}}], \nonumber\\
\tilde{\omega}_{187\alpha} &=&\frac{1}{2 \sqrt{1+(\frac{k_{\alpha}}{2})^2}}[-\frac{
m_{\alpha 2} - m_{\alpha 1}}{\sqrt{1+ ({}^a\eta_{\alpha})^2}} +
\frac{m_{\alpha 4 } - m_{\alpha 3}}{ \sqrt{1+ 
({}^b\eta_{\alpha})^2}}], \nonumber\\
\tilde{\omega}_{387\alpha} &=& \frac{1}{2 \sqrt{1+(\frac{k_{\alpha}}{2})^2}}
[(m_{\alpha 4 } + m_{\alpha 3}) - (m_{\alpha 2} + m_{\alpha 1})],\nonumber\\
a^a_{\alpha}&=& \frac{1}{2}( m_{\alpha 1} + m_{\alpha 2} - \frac{{}^a\eta_{\alpha}\;(m_{\alpha 2} - 
m_{\alpha 1})}{\sqrt{1+({}^a\eta_{\alpha})^2}}).
\end{eqnarray}
%

 
\subsection{Negative masses and parity of states}
\label{negativemasses}

We have mentioned in the previous  section that after the diagonalization of mass matrices 
of quarks and leptons, masses of either positive or negative sign can appear and that by changing 
the phase of the basis and accordingly also the internal parity of states we change the sign of 
the mass. Let us prove this. 

First we recognize that while the starting Lagrange density for spinors (Eq.\ref{lagrange}) 
commutes with the operator of handedness in $d(=1+13)$-dimensional space $\Gamma^{(1,13)} $ 
($\Gamma^{(1,13)} \;= i 2^{7} 
\; S^{03} S^{12} S^{56} \cdots S^{13 \; 14} $), it does not commute with the operator of handedness 
in $d(=1+3)$-dimensional space $\Gamma^{(1,3)} $ ($\Gamma^{(1,3)}\;=  - i 2^2 S^{03} S^{12}$). 
Accordingly also the term, which 
manifests at "physical energies" as the mass term $ \hat{m}$ 
\begin{eqnarray}
\gamma^0 \hat{m} &=&
\gamma^0 \;  
\biggl\{ \stackrel{78}{(+)} ( \sum_{y=Y,Y'}\; y A^{y}_{+} + 
\sum_{\tilde{y}=\tilde{N}^{+}_{3},\tilde{N}^{-}_{3},\tilde{\tau}^{13},\tilde{Y},\tilde{Y'}} 
\tilde{y} \tilde{A}^{\tilde{y}}_{+}\;)\; \nonumber \\
  &+& 
   \stackrel{78}{(-)} ( \sum_{y=Y,Y'}\;y  A^{y}_{-} +  
\sum_{\tilde{y}= \tilde{N}^{+}_{3},\tilde{N}^{-}_{3},\tilde{\tau}^{13},\tilde{Y},\tilde{Y'}} 
\tilde{y} \tilde{A}^{\tilde{y}}_{-}\;) \nonumber\\
 &+& 
 \stackrel{78}{(+)} \sum_{\{(ac)(bd) \},k,l} \; \stackrel{ac}{\tilde{(k)}} \stackrel{bd}{\tilde{(l)}}
\tilde{{A}}^{kl}_{+}((ab),(cd)) \nonumber\\
 &+&  
  \stackrel{78}{(-)} \sum_{\{(ac)(bd) \},k,l} \; \stackrel{ac}{\tilde{(k)}}\stackrel{bd}{\tilde{(l)}}
\tilde{{A}}^{kl}_{-}((ab),(cd))\biggr\}\nonumber,
\end{eqnarray} 
(Eq.\ref{yukawa4tilde0})), does not commute with the $\Gamma^{(1,3)}$,  
they instead anticommute 
$$\{\Gamma^{(1,3)}, \gamma^0 \hat{m}\}_+ =0.$$
But the rest of the "effective" Lagrangean (Eq.\ref{yukawa}) commutes with the operator of handedness 
in $d=(1+3)$-dimensional space: 
$$\{\gamma^0\gamma^{m} (p_{m}- \sum_{A,i}\; 
g^{A}\tau^{Ai} A^{Ai}_{m}),\Gamma^{(1,3)}\}_-=0.  $$

It then follows that the Lagrange density 
\begin{eqnarray}
{\mathcal L} = 
(\Gamma^{(1,3)}\psi)^{\dagger} \; \left[ \gamma^0\gamma^{m} (p_{m}- \sum_{A,i}\; 
g^{A}\tau^{Ai} A^{Ai}_{m}) - \Gamma^{(1,3)}\gamma^0 \hat{m}\Gamma^{(1,3)}\right]\; (\Gamma^{(1,3)}\psi)\nonumber\\
{  }
\label{mto-m}
\end{eqnarray}
for the Dirac spinor $\Gamma^{(1,3)}\psi$ differs from the one from Eq.(\ref{yukawa}) 
in the sign of the 
mass term, while the function $\Gamma^{(1,3)}\psi$ differs from $\psi$ in the internal parity, if 
$\psi$ is the solution for the  Dirac equation. 
Since the internal parity is just the convention, the negative mass changes sign if the 
internal parity of the spinor changes. The same argument was used in the ref.(\cite{fritsch}),
while the ref.(\cite{chengli}) uses the equivalent argument, namely, that the choice of the phase 
of either the right or the 
left handed spinors can always be changed and that accordingly also the signs of particular 
mass terms change. 

Let us demonstrate now on Table~\ref{TableIII} how does the operator of parity ${\cal P}$, if 
postulated as 
\begin{eqnarray}
\label{parity}
{\cal P} = \gamma^0 \gamma^8 I_x, {\rm with}\;\; I_x x^m (I_x)^{(-1)} =x_m,   
\end{eqnarray}
 transform a
right handed $u$-quark into the left handed $u$-quark: ${\cal P} u_{R} = \alpha u_{L}$, where 
$\alpha $ is the proportionality factor.

\begin{table}
\begin{center}
\begin{tabular}{|r|c||c||c|c||c|c|c||c|c|c||r|r|}
\hline
i&$$&$|^a\psi_i>$&$\Gamma^{(1,3)}$&$ S^{12}$&$\Gamma^{(4)}$&
$\tau^{13}$&$\tau^{21}$&$\tau^{33}$&$\tau^{38}$&$\tau^{41}$&$Y$&$Y'$\\
\hline\hline
&& ${\rm Octet},\;\Gamma^{(1,7)} =1,\;\Gamma^{(6)} = -1,$&&&&&&&&&& \\
&& ${\rm of \; quarks}$&&&&&&&&&&\\
\hline\hline
1&$u_{R}^{c1}$&$\stackrel{03}{(+i)}\stackrel{12}{(+)}|\stackrel{56}{(+)}\stackrel{78}{(+)}
||\stackrel{9 \;10}{(+)}\stackrel{11\;12}{(-)}\stackrel{13\;14}{(-)}$
&1&$\frac{1}{2}$&1&0&$\frac{1}{2}$&$\frac{1}{2}$&$\frac{1}{2\sqrt{3}}$&$\frac{1}{6}$&$\frac{2}{3}$&$-\frac{1}{3}$\\
\hline 
2&$u_{R}^{c1}$&$\stackrel{03}{[-i]}\stackrel{12}{[-]}|\stackrel{56}{(+)}\stackrel{78}{(+)}
||\stackrel{9 \;10}{(+)}\stackrel{11\;12}{(-)}\stackrel{13\;14}{(-)}$
&1&$-\frac{1}{2}$&1&0&$\frac{1}{2}$&$\frac{1}{2}$&$\frac{1}{2\sqrt{3}}$&$\frac{1}{6}$&$\frac{2}{3}$&$-\frac{1}{3}$\\
\hline
3&$d_{R}^{c1}$&$\stackrel{03}{(+i)}\stackrel{12}{(+)}|\stackrel{56}{[-]}\stackrel{78}{[-]}
||\stackrel{9 \;10}{(+)}\stackrel{11\;12}{(-)}\stackrel{13\;14}{(-)}$
&1&$\frac{1}{2}$&1&0&$-\frac{1}{2}$&$\frac{1}{2}$&$\frac{1}{2\sqrt{3}}$&$\frac{1}{6}$&$-\frac{1}{3}$&$\frac{2}{3}$\\
\hline 
4&$d_{R}^{c1}$&$\stackrel{03}{[-i]}\stackrel{12}{[-]}|\stackrel{56}{[-]}\stackrel{78}{[-]}
||\stackrel{9 \;10}{(+)}\stackrel{11\;12}{(-)}\stackrel{13\;14}{(-)}$
&1&$-\frac{1}{2}$&1&0&$-\frac{1}{2}$&$\frac{1}{2}$&$\frac{1}{2\sqrt{3}}$&$\frac{1}{6}$&$-\frac{1}{3}$&$\frac{2}{3}$\\
\hline
5&$d_{L}^{c1}$&$\stackrel{03}{[-i]}\stackrel{12}{(+)}|\stackrel{56}{[-]}\stackrel{78}{(+)}
||\stackrel{9 \;10}{(+)}\stackrel{11\;12}{(-)}\stackrel{13\;14}{(-)}$
&-1&$\frac{1}{2}$&-1&$-\frac{1}{2}$&0&$\frac{1}{2}$&$\frac{1}{2\sqrt{3}}$&$\frac{1}{6}$&$\frac{1}{6}$&$\frac{1}{6}$\\
\hline
6&$d_{L}^{c1}$&$\stackrel{03}{(+i)}\stackrel{12}{[-]}|\stackrel{56}{[-]}\stackrel{78}{(+)}
||\stackrel{9 \;10}{(+)}\stackrel{11\;12}{(-)}\stackrel{13\;14}{(-)}$
&-1&$-\frac{1}{2}$&-1&$-\frac{1}{2}$&0&$\frac{1}{2}$&$\frac{1}{2\sqrt{3}}$&$\frac{1}{6}$&$\frac{1}{6}$&$\frac{1}{6}$\\
\hline
7&$u_{L}^{c1}$&$\stackrel{03}{[-i]}\stackrel{12}{(+)}|\stackrel{56}{(+)}\stackrel{78}{[-]}
||\stackrel{9 \;10}{(+)}\stackrel{11\;12}{(-)}\stackrel{13\;14}{(-)}$
&-1&$\frac{1}{2}$&-1&$\frac{1}{2}$&0&$\frac{1}{2}$&$\frac{1}{2\sqrt{3}}$&$\frac{1}{6}$&$\frac{1}{6}$&$\frac{1}{6}$\\
\hline
8&$u_{L}^{c1}$&$\stackrel{03}{(+i)}\stackrel{12}{[-]}|\stackrel{56}{(+)}\stackrel{78}{[-]}
||\stackrel{9 \;10}{(+)}\stackrel{11\;12}{(-)}\stackrel{13\;14}{(-)}$
&-1&$-\frac{1}{2}$&-1&$\frac{1}{2}$&0&$\frac{1}{2}$&$\frac{1}{2\sqrt{3}}$&$\frac{1}{6}$&$\frac{1}{6}$&$\frac{1}{6}$\\
\hline\hline
\end{tabular}
\end{center}
\caption{\label{TableIII}%
The 8-plet of quarks - the members of $SO(1,7)$ subgroup, belonging to one Weyl left 
handed ($\Gamma^{(1,13)} = -1 = \Gamma^{(1,7)} \times \Gamma^{(6)}$) spinor representation of 
$SO(1,13)$. 
It contains the left handed weak charged quarks and the right handed weak chargeless quarks 
of a particular colour ($(1/2,1/(2\sqrt{3}))$). Here  $\Gamma^{(1,3)}$ defines the 
handedness in $(1+3)$ space, 
$ S^{12}$ defines the ordinary spin (which can also be read directly from the basic vector, 
since $S^{ab} \stackrel{ab}{(k)} = \frac{k}{2} \stackrel{ab}{(k)}$ and 
$S^{ab} \stackrel{ab}{[k]} = \frac{k}{2} \stackrel{ab}{[k]}$, if $S^{ab}$ belong to the Cartan 
subalgebra set), 
$\tau^{13}$ defines the weak charge, $\tau^{21}$ defines the $U(1)$ charge, $\tau^{33}$ and 
$\tau^{38}$ define the colour charge and $\tau^{41}$ another $U(1)$ charge, which together with the
first one defines $Y$ and $Y'$.}
\end{table}

\noindent
One notices that, if the operator ${\cal P}$ is applied on a state, which represents the  
right handed weak chargeless ($\tau^{13} =0$) $u$-quark of one of the three colours  
and is presented in terms of nilpotents in the first row of Table~\ref{TableIII}, it transforms this 
state into the state, which can be found in the seventh row of the same table and represents 
the left handed $u$-quark of the same colour and spin and it is weak charged. Taking into
account Eq.(\ref{parity}) and Eq.(12,16) from ref.\cite{pikanorma05} one finds 
${\cal P} u_{R} = i u_{L}$, while 
${\cal P}{\cal P}=I$. By changing appropriately the phases of this two basic states 
($u_{R} $ and $ u_{L}$) we can easily achieve that ${\cal P} u_{R} =  u_{L},
{\cal P} u_{L} =  u_{R}$. We should in addition keep in mind that ${\cal P}$ must 
take into account also the appearance of families. We shall study discrete symmetries 
of our approach in the "low energy region" in a separate paper.



\section{Numerical results}
\label{numerical}

In this section we connect parameters  $\tilde{\omega}_{abc \delta}$ of the Yukawa couplings 
following from our approach unifying spins and charges (after several assumptions, approximations 
and simplifications) with the experimental data. We investigate, how do the parameters of the 
approach reflect the known data. We also investigate a possibility of making some predictions.


\subsection{Experimental data for quarks and leptons}
\label{experimentaldata}

We present in this subsection those experimental data, which are relevant for our study:  
that is the measured values for the masses of the three 
families of quarks and leptons and the measured mixing matrices. 

We take in our calculations the experimental masses for the known three 
families from the ref.\cite{expckm,expmixleptons}.
\begin{eqnarray}
\label{masses}
m_{u_i}/GeV &=& (0.0015- 0.004, 1.15-1.35, 174.3-178.1
),\nonumber\\
m_{d_i}/GeV &=& (0.004-0.008, 0.08-0.13, 4.1-4.9 
),\nonumber\\
m_{\nu_i}/GeV &=& (1 \; 10^{-12}, 1 \; 10^{-11}, 5 \; 10^{-11} 
),\nonumber\\
m_{e_i}/GeV &=& (0.0005,0.105,1.8
).
\end{eqnarray}
Predicting four families of quarks and leptons at ''physical'' energies, we require 
the unitarity condition for the mixing matrices for four rather than three measured families 
of quarks\cite{expckm}  
\begin{eqnarray}
\label{expckm}
 \left(\begin{array}{ccc}
 0.9730-0.9746 & 0.2174-0.2241 & 0.0030-0.0044\\
 0.213-0.226   & 0.968-0.975  & 0.039-0.044\\
 0.0 - 0.08    & 0.0-0.11      & 0.07-0.9993\\
\end{array}
                \right).
\end{eqnarray}
The experimental data are for the mixing matrix for leptons known very 
weakly\cite{expmixleptons} 
\begin{eqnarray}
\label{expckmleptons}
 \left(\begin{array}{ccc}
 0.79-0.88 & 0.47-0.61 &  {} < 0.20\\
 0.19-0.52 & 0.42-0.73 & 0.58-0.82\\
 0.20-0.53 & 0.44-0.74 & 0.56-0.81\\
\end{array}
                \right).
\end{eqnarray}
We see that within the experimental accuracy both mixing matrices - for  
quarks and leptons - may be assumed to be symmetric up to a sign. We then fit with these 
two matrices the six parameters $k_{\alpha}, {}^a\eta_{\alpha}, {}^b\eta_{\alpha}$, $\alpha =u,\nu.$ 


\subsection{Results}
\label{numericalresults}

We started with the explicit expressions for the Yukawa couplings suggested by the approach 
unifying spins and charges after making several additional assumptions to the starting 
assumptions of the approach, also some approximations and simplifications, 
in order to be able to make some approximate predictions.  
We ended up with the four families of quarks and leptons, with the mass matrices 
symmetric and real and diagonalizable in two steps with accordingly three angles of rotation 
for each mass matrix and yet the angles of rotations for the $u$-quarks are related 
to those of the $d$-quarks and so are related also 
the angles of the two kinds of leptons. Accordingly the two 
mixing matrices are diagonalizable with three angles of rotations each.

Since to do the rigorous calculations is a huge project and the evaluations in this 
paper are only the first rough step towards more sophisticated very demanded calculations, 
we on this step only are parametrizing the influence of breaking symmetries of 
either the Poincar\' e group or the group defining families  and of nonperturbative 
and other effect by parametrizing 
the "vacuum expectation values" of gauge fields entering into mass matrices. 
The assumptions, which we made, take care of simplifying the evaluation as much as possible 
while making the properties of mass matrices and mixing matrices as transparent as possible. 

Let we repeat that according to  Subsect.\ref{improvedmatrices} (in particular 
Eq.(\ref{omegatilde})) any choice for the masses of the fourth family fits the experimental data, once 
twice the three angles of the orthogonal transformations, determining the two mixing
matrices are chosen. Assuming that some kind of symmetry (the charge $Y$ and $Y'$, for 
example) makes the "vacuum expectation values" of the gauge fields entering into the mass matrices 
further related, we are going to test how does the requirement that the ratios of the parameters 
$\tilde{\omega}_{abc \delta}$ are for a chosen set $[a,b,c]$ as close to rational numbers as possible 
influence the properties of the fourth family of quarks and leptons.

We also repeat the recognition made in Subsect.\ref{experimentaldata} (Eqs.(\ref{expckm}, 
\ref{expckmleptons})) that experimental data agree,  
within the experimental accuracy,  for either 
quarks or leptons, that mixing matrices are symmetric and  
determined with only three angles. (We do not pay attention on $CP$ non conservation.)

We shall now  connect the parameters of the approach, which are left free,  
 with the experimental data and  
try to find out what can we learn from the corresponding results.

{\em We fit} for quarks and leptons  
{\em the three angles} of Eqs.(\ref{phirelated},\ref{abphirelated}) 
{\em with the Monte-Carlo method under the requirement that the ratios of the parameters 
$\tilde{\omega}_{abc \delta}$ are for a chosen set $[a,b,c]$ so close to a rational number as possible.}

We allow the masses of the fourth family as follows: 
The two quark masses must lie in the range from $200$ GeV to $1$ TeV, the fourth neutrino mass 
must be within the interval $50 - 100$ GeV and of the fourth electron mass within $50 - 200$ GeV. 

\begin{figure}
\includegraphics[width=8cm]{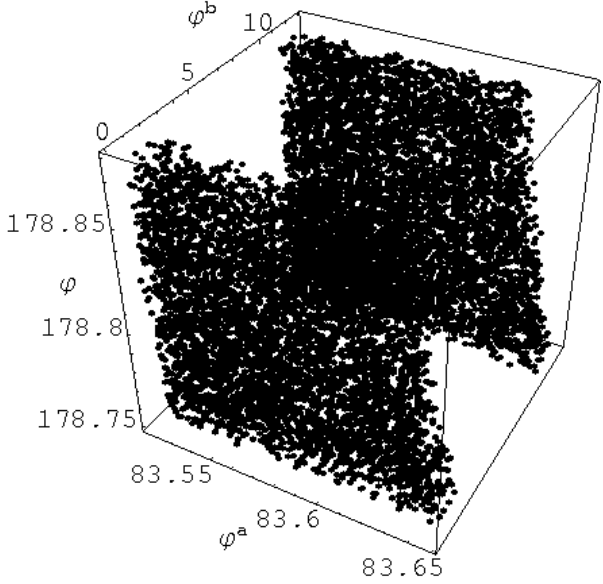}
\caption{\label{fig1}%
Figure shows the Monte-Carlo fit\cite{matjazdiploma} of the experimental mixing matrix 
for quarks (Eq.\ref{expckm}) with the three angles of Eq.(\ref{abcwithm}). The three angles  
define the three parameters $k_u, {}^a\eta_u$ and ${}^b\eta_u$ 
(Eqs.\ref{firstangle},\ref{solutions}). We make a choice among those values for the best  fit, 
 which makes the ratios $\tilde{\omega}_{abc u}/ \tilde{\omega}_{abc d}$ as close to 
 rational numbers as possible while assuring that the masses of the three known families stay 
 within the   acceptable values from Eq.(\ref{masses}), with no constraints on $a_{\alpha}$ and 
 the two quark masses of the fourth family lie in the range $200-1000$ MeV.}
\end{figure}

Fig.~\ref{fig1} shows the results of the Monte-Carlo simulation for the three angles determining the 
mixing matrix for quarks. There are the  experimental inaccuracies, which determine the allowed regions 
for the three angles. 

The results for the quarks are presented on Table~\ref{TableIV} and~\ref{TableV} (together with 
the corresponding values for leptons).

\begin{figure}
\includegraphics[width=8cm]{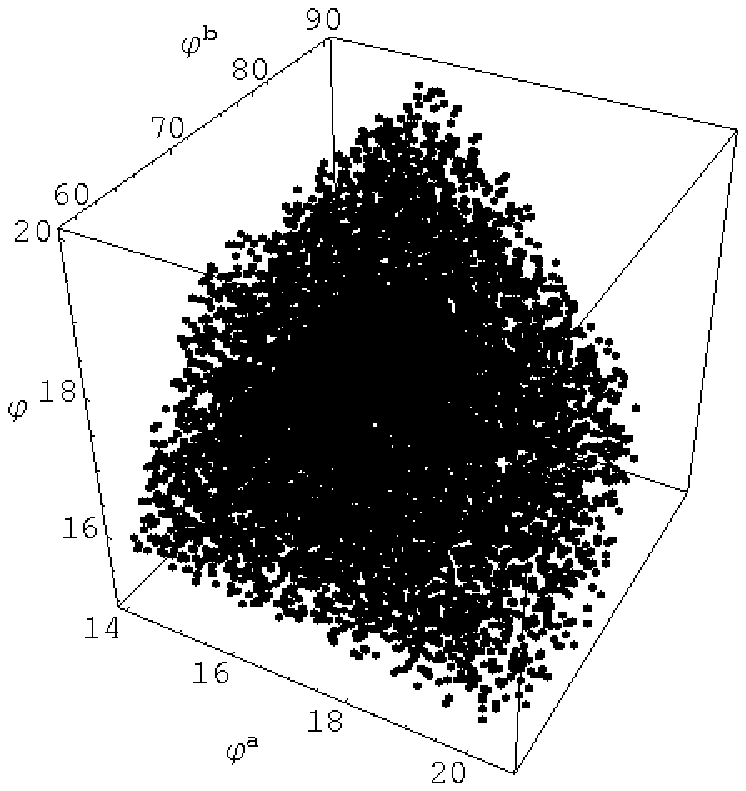}
\caption{\label{fig2}%
Figure shows  the Monte-Carlo fit of the experimental data for the mixing matrix for 
leptons(Eq.(\ref{expckmleptons})).  The three angles  
define the three parameters $k_{\nu}, {}^a\eta_{\nu}$ and ${}^b\eta_{\nu}$ 
(Eqs.\ref{firstangle},\ref{solutions}). Again we make a choice among those values for the best  fit, 
 which make the ratios $\tilde{\omega}_{abc u}/ \tilde{\omega}_{abc d}$ as close to 
 rational numbers as possible. }
\end{figure}

Fig.~\ref{fig2} shows the Monte-Carlo fit for the three angles determining the mixing matrix for
leptons. There are the experimental inaccuracies which limit the values of the three angles. 
Since in the lepton case the mixing matrix for the 
 three known families as well as the masses for the three  neutrinos  are weakly known, the 
 calculations for four families bring much less information than in the quark case.

The results for the leptons are presented together with the results for the quarks 
 on Table~\ref{TableIV} and~\ref{TableV}.

\begin{table}
\centering
\begin{tabular}{|c||c|c|c|c|}
\hline 
&$u$&$d$&$\nu$&$e$\tabularnewline
\hline
\hline 
$k$       & -0.085 &  0.085&-1.254&  1.254\tabularnewline
\hline 
${}^a\eta$& -0.229 &  0.229& 1.584& -1.584\tabularnewline
\hline
${}^b\eta$&  0.420 & -0.440& -0.162& 0.162\tabularnewline
\hline
\end{tabular}\\
\caption{\label{TableIV}%
The Monte-Carlo fit to the experimental data\cite{expckm,expmixleptons} 
for the three parameters $k$, ${}^a\eta$ and   ${}^b\eta$ determining the mixing matrices 
for  the four families of quarks and leptons are presented.}
\end{table}

\begin{table}
\centering
\begin{tabular}{|c||c|c|c||c|c|c|}
\hline 
&$u$&$d$&$u/d$&$\nu$&$e$&$\nu/e$\tabularnewline
\hline
\hline 
$|\tilde{\omega}_{018}|$& 21205& 42547& 0.498&    10729& 21343 &0.503\tabularnewline
\hline 
$|\tilde{\omega}_{078}|$& 49536& 101042& 0.490&   31846& 63201 &0.504\tabularnewline
\hline 
$|\tilde{\omega}_{127}|$& 50700& 101239& 0.501&   37489& 74461 &0.503\tabularnewline
\hline 
$|\tilde{\omega}_{187}|$& 20930& 42485& 0.493&    9113&  18075 &0.505\tabularnewline
\hline 
$|\tilde{\omega}_{387}|$& 230055& 114042& 2.017&  33124& 67229 &0.493\tabularnewline
\hline
$a^{a}$&94174& 6237& &   1149&1142 &\tabularnewline
\hline
\end{tabular}\\
\caption{\label{TableV}%
Values for the parameters $\tilde{\omega}_{abc \delta}$ (entering into 
the mass matrices for the $u-$quarks, the $d-$quarks, the neutrinos and the 
electrons, as suggested by the approach unifying spins and charges after making 
the additional assumptions and simplifications as decribed in this paper) 
as following after the Monte-Carlo fit, 
relating the parameters and the experimental data.}
\end{table}

In Eq.(\ref{resultmasses}) we present masses for the four families of quarks and leptons as 
obtained after the Monte-Carlo fit
\begin{eqnarray}
\label{resultmasses}
m_{u_i}/GeV &=& (0.0034, 1.15, 176.5, 285.2),\nonumber\\
m_{d_i}/GeV &=& (0.0046, 0.11, 4.4, 224.0), \nonumber\\
m_{\nu_i}/GeV &=& ( 1\; 10^{-12}, 1 \; 10^{-11}, 5 \; 10^{-11},  84.0 ),\nonumber\\
m_{e_i}/GeV &=& (0.0005,0.106,1.8, 169.2).
\end{eqnarray}

The results of the Monte-Carlo fit show\cite{matjazdiploma} that the requirement, that the ratios of the 
corresponding parameters of $\tilde{\omega}_{abc \delta}$, for a chosen set $[a,b,c]$, 
for the quarks and the leptons  should 
be as close to the rational numbers as possible, forces the masses of the fourth 
family to lie pretty low. We recognize that the values agree with the experimentally 
allowed values as evaluated by the 
refs.\cite{okun,okunmaltoni,okunbulatov}. 
Eq.(\ref{omegatilde}), however, tells us, that it is mainly the top mass and 
the masses of the fourth family which mostly influence these ratios. 
But since only the integer $2$ and the half integer ($1/2$) are involved 
(the ratios go to one 
only when all the masses of the fourth family are equal and 
are high in comparison with the top mass), we could use 
this result as a guide when looking for the way of breaking symmetries on the way down 
to $d=1+3$ in (much) more sophisticated calculations.

The Monte-Carlo fit  leads to the following mixing matrix for the quarks
\begin{eqnarray}
\label{resultckm}
 \left(\begin{array}{cccc}
 0.974 & 0.223 & 0.004 & 0.042\\
 0.223 & 0.974 & 0.042 & 0.004\\
 0.004 & 0.042 & 0.921 & 0.387\\
 0.042 & 0.004 & 0.387 & 0.921\\
 \end{array}
                \right)
\end{eqnarray}
and for the leptons
\begin{eqnarray}
\label{resultckmleptons}
 \left(\begin{array}{cccc}
 0.697 & 0.486 & 0.177 & 0.497\\
 0.486 & 0.697 & 0.497 & 0.177\\
 0.177 & 0.497 & 0.817 & 0.234\\
0.497  & 0.177 & 0.234 & 0.817\\ 
\end{array}
                \right).
\end{eqnarray}
The estimated mixing matrix for the four families of quarks predicts quite a strong couplings  
between the fourth and the other three families, limiting  (due to the assumptions and 
approximations we made, which manifest in the symmetric mixing matrices) some of the matrix elements 
of the three families as well.  

The estimated mixing matrix for the four families of leptons predicts very probably far 
too strong couplings 
between the known three and the fourth family (although they are not in contradiction  with  
the report in\cite{expckm}).

Let us now repeat the number of input data and the number of predictions:

\noindent
i. We take as the input data the experimental masses (the three $u$-quark masses, the 
three $d$-quark masses, the three electron masses and the three (very weakly known) neutrino 
masses),\\
ii. The quark mixing matrix and the (weakly known) lepton mixing matrix.

By taking into account relations among the mass matrix elements for the four 
types of spinors ($u$-quarks, $d$-quarks, neutrinos and electrons) as suggested by the 
approach unifying spins and charges (after some additional assumptions, approximations 
and simplifications, which all seam reasonable from the point of view of the experimental 
data and the fact that we want to obtain at least some rough estimations 
and come to at least some rough 
predictions for the approach to see whether it is not in a very severe contradiction with the 
experimental data) for the "low energy region", we were able to fit within the 
experimental accuracy and using the Monte-Carlo procedure all the {\em known experimental data 
and predict four masses} 
(the masses of the quarks and leptons of the fourth family) and {\em the corresponding 
mixing matrices for quarks and for leptons}.

Let us end  up this section by repeating that all the predictions must be taken as  a rough 
estimate, since they follow from the approach unifying spins and charges after several  
approximations and assumptions, which
we made to be able to come in quite a short way to simple and also (very) 
transparent predictions, used as a first step to much more sophisticated calculations.


\section{Discussions and conclusions}
\label{s:MDNdiscussions}

In this paper and in the previous one\cite{pikanorma05} we study a possibility 
that the approach of one of 
us\cite{MDNnorma92,MDNnorma93,normasuper94,MDNnorma95,norma97,pikanormaproceedings1,holgernorma00,norma01,%
pikanormaproceedings2,MDNPortoroz03}, unifying spins and charges, might be a new right way 
for answering those of 
the open questions of the Standard model of the electroweak and colour interaction, 
which are connected with the appearance of  families of fermions, of the Yukawa couplings and of  
the weak scale: Why do only the left handed spinors carry the weak 
charge, while the right handed are weak chargeless? 
Where do the families of the quarks and the leptons come from? 
What does determine the strenghts of the Yukawa 
couplings and the weak scale?

We can conclude that {\em within the approach unifying spins and charges 
the answer to  the question, why do 
only the left handed spinors carry the weak charge, while the right handed ones are 
weak chargeless, does exist}: The representation of one Weyl spinor of the group SO(1,13), 
analyzed with respect to the properties of the subgroups $SO(1,7) \times SU(3) \times U(1)$ of this 
group and further with respect to SU(2) and the second U(1), manifests the left 
handed weak charged quarks and leptons and the right handed weak chargeless quarks and 
leptons. 

{\em The approach answers as well the question about a possible origin of }(by the Higgs 
weak charge) {\em ''dressing'' of the right 
handed quarks and leptons in the  Standard model}: The approach proposes 
the Lagrange density for fermions in $d(=1+13)$-dimen\-sional space (a simple one) 
in which fermions interact with only the  gravity (the gauge fields of the momentum - vielbeins -
and  the two kinds of the Clifford algebra objects - spin 
connections).  
It is  a part of the spin connection field of the Poincar\' e group, which connects 
the right handed weak 
chargeless spinors with the left handed weak charged ones, playing the role of the Higgs field 
(and the Yukawa couplings within a family) of the Standard model. 

{\em The approach is answering also the question about the origin of the families of 
quarks and leptons}: Two kinds of the Clifford algebra objects gauging 
two kinds of spin connection fields, are assumed. One kind 
takes care of the spin and the charges and of connecting right handed weak chargeless 
fermions with left handed weak charged fermions. The other kind takes care 
of  the families of fermions and consequently 
of the Yukawa couplings among the families contributing also to the diagonal elements. 

In the previous paper\cite{pikanorma05} we derived from the approach 
the expressions for the Yukawa couplings 
for four families of quarks and leptons. 

It is a long way from the starting simple Lagrange density for spinors carrying only the spins 
and interacting with only the vielbeins and spin connections to 
the observable quarks and leptons. To treat breaking of symmetries properly, taking into account 
all perturbative and nonperturbative effects, boundary connditions and other effects  
(by treating gauge gravitational fields in the same way as ordinary gauge fields, 
since the scale of breaking $SO(1,13)$ is supposed to be far from the Planck scale) 
is a huge project. 

The purpose of this paper is to estimate whether has the approach unifying spins and 
charges at all a chance to be the right way beyond the Standard model. Accordingly we tried to 
estimate in a rough way what does the approach predict for a low energy physics in $d=1+3$. To be able 
to make any predictions (in a simple enough way) we made  
several approximations, assumptions and simplifications, which look acceptble from the point 
of view of the known experimental data and the fact that only a very preliminary 
prediction is looked for. 

Approximations, assumptions and simplifications we made enable   
simple and transparent view on the mass matrices for the 
four families of quarks and leptons 
in terms of the spin connection fields of the two kinds and allow to  
predict masses of the fourth 
family of quarks and leptons and of the corresponding matrix elements of the mixing matrices. 
We treat quarks and leptons equivalently, no Majorana leptons are 
taken into account in this study.

The assumptions, approximations and simplifications we made  (which are 
not the starting assumtion of our approach unifying spins and charges) are presented in the 
ref.\cite{pikanorma05} and in Sect.\ref{lagrangesec}: 

i.  The break of symmetries of the group $SO(1,13)$ (the  Poincar\' e group in 
$d=1+13$)  into $SO(1,7)\times SU(3)\times U(1)$ occurs 
in a way that only 
massless spinors in $d=1+7$ with the charge $ SU(3)\times U(1)$ survive. (Our work on 
the compactification 
of a massless spinor in $d=1+5$ into   $d=1+3$ and a finite disk gives us some hope that such 
an assumption might be justified\cite{holgernorma05}.) 
The requirement that the terms with $S^{5a}\omega_{5ab}$ and $S^{6a}\omega_{6ab}$ 
do not contribute to the mass term, assures that the charge   
$Q= \tau^{41} + S^{56}$ is conserved at low energies. 

ii. The break of symmetries influences  both, the (Poincar\' e) symmetry 
described by $S^{ab}$ and the symmetries described by $\tilde{S}^{ab}$, and in a way that  
 there are no terms, which would  transform  
$\stackrel{{56}}{\tilde{(+)}}$ into $\stackrel{{56}}{\tilde{[+]}}$. This assumption  
can be explained by a break of the symmetry $SO(1,7)$ into $SO(1+5)\times U(1)$ in 
the $\tilde{S}^{ab}$  sector. We also assume that the terms which 
include components $p_s, s=5,..,14$, of the momentum $p^a$ do not 
contribute  to the mass matrices. 
 
iii. We  make estimates on a ''tree level'', taking effects bellow the tree level 
into account by allowing the matrix elements to depend on the type of fermions in a 
way, that the corresponding ratios are   
(very close to) rational numbers -   which seems to be acceptable by the approach.

iv. We assume the mass matrices to  be real and symmetric and diagonalizable in two steps,
first in two by two by diagonal matrices and then further to diagonal ones (which 
is suggested by the fact that this would happen if $SO(1,5)$ breaks (approximately) to $SU(2)\times
SU(2)\times U(1)$ in the $\tilde{S}$ sector, while the break of 
 $SO(1,5)$ to $SU(3)\times U(1)$ instead, would make the fourth family (approximately) 
decoupled  from the first three and would accordingly strongly change our - very preliminary - 
results. (While such a break seems to  be even acceptable when describing properties of leptons, 
it would  predict for quarks much too strong couplings between the third and the first two families  
with respect to the measured values.)

Taking all the above assumptions into account we then relate the free parameters 
of the mass matrices with the measured experimental data within the experimental accuracy 
treating quarks and leptons equivalently (and not taking into account a possible 
existence of  Majorana neutrinos).  
Making a rough prediction of the  properties of the fourth family of the two kinds of 
quarks and the two kinds of leptons: of their masses and the corresponding matrix 
elements of the mixing matrices, we found that our results are in agreement with the analyses of 
refs.\cite{okun,okunmaltoni,okunbulatov}. We predict the fourth family masses $m_{u_4}=285$ GeV,
$m_{d_4}=224$ GeV, $m_{\nu_4}=65$ GeV, $m_{e_4}=129$ GeV. 
Predictions for the couplings between the fourth and the other three families seem reasonable for quarks, 
while for leptons the corresponding mixing matrix elements might suggest  that either different break of 
symmetries in the 
$\tilde{S}^{ab}$ sector from the assumed one, or the Majorana neutrinos, or both effects should at least be 
further studied.

To try to answer within the approach unifying spins and charges the open question of the 
Standard model: Why the weak scale appears as it does? a more detailed study of the breaks 
of symmetries in both sectors  ($S^{ab}$ and $\tilde{S}^{ab}$) is needed.

What we can conclude, after making in this paper a first rough step towards more 
justified results (by  allowing several approximations and assumptions) in order 
to find out, whether something very essential and unexpected can go wrong with 
our approach is, that the approach of one of us unifying spins and charges might have a 
real chance to go successfully beyond the Standard model. 

\section*{Acknowledgments} We would like to express many thanks to ARRS for the
grant.
It is a pleasure to thank all the participants of the   workshops entitled 
"What comes beyond the Standard model", 
taking place  at Bled annually in  July, starting at 1998,  for fruitful discussions, 
in particular to H.B. Nielsen.


\title{Cosmoparticle Physics: Cross-disciplinary Study of Physics Beyond %
the Standard Model}
\author{M.Yu. Khlopov}
\institute{%
Center for Cosmoparticle Physics "Cosmion"\\
Miusskaya Pl. 4\\
 125047 Moscow\\
Russia\\
\textrm{e-mail:} \texttt{Maxim.Khlopov@roma1.infn.it}}

\titlerunning{Cosmoparticle Physics}
\authorrunning{M.Yu. Khlopov}
\maketitle

\begin{abstract}
Cosmoparticle physics is the natural result of development of
mutual relationship between cosmology and particle physics. Its
prospects offer the way to study physics beyond the Standard Model
and the true history of the Universe, based on it, in the proper
combination of their indirect physical, astrophysical and
cosmological signatures. We may be near the first positive results
in this direction. The basic ideas of cosmoparticle physics are
briefly reviewed.
\end{abstract}

\section{Cosmoparticle physics as the solution of Uhroboros puzzle}
 \label{Uhroboros}
Cosmoparticle physics originates from the well established
relationship between microscopic and macroscopic descriptions in
theoretical physics. Remind the links between statistical physics
and thermodynamics, or between electrodynamics and theory of
electron. To the end of the XX Century the new level of this
relationship was realized. It followed both from the cosmological
necessity to go beyond the world of known elementary particles in
the physical grounds for inflationary cosmology with
baryosynthesis and dark matter as well as from the necessity for
particle theory to use cosmological tests as the important and in
many cases unique way to probe its predictions.

The convergence of the frontiers of our knowledge in micro- and
macro worlds leads to the wrong circle of problems, illustrated by
the mystical Uhroboros (self-eating-snake). The Uhroboros puzzle
may be formulated as follows: {\it The theory of the Universe is
based on the predictions of particle theory, that need cosmology
for their test}. Cosmoparticle physics \cite{ADS}, \cite{MKH},
\cite{book} offers the way out of this wrong circle. It studies
the fundamental basis and mutual relationship between micro-and
macro-worlds in the proper combination of physical, astrophysical
and cosmological signatures.
\section{Cosmological pattern of particle physics}
 \label{pattern}
Let's specify in more details the set of links between fundamental
particle properties and their cosmological effects.

The role of particle content in the Einstein equations is reduced
to its contribution into energy-momentum tensor. So, the set of
relativistic species, dominating in the Universe, realizes the
equation of state $p= \varepsilon/3$ and the relativistic stage of
expansion. The difference between relativistic bosons and fermions
or various bosonic (or fermionic) species is accounted by the
statistic weight of respective degree of freedom. The very
treatment of different species of particles as equivalent degrees
of freedom physically assumes strict symmetry between them.

Such strict symmetry is not realized in Nature. There is no exact
symmetry between bosons and fermions (e.g. supersymmetry). There
is no exact symmetry between various quarks and leptons. The
symmetry breaking implies the difference in particle masses. The
particle mass pattern reflects the hierarchy of symmetry breaking.

Noether's theorem relates the exact symmetry to conservation of
respective charge. The lightest particle, bearing the strictly
conserved charge, is absolutely stable. So, electron is absolutely
stable, what reflects the conservation of electric charge. In the
same manner the stability of proton is conditioned by the
conservation of baryon charge. The stability of ordinary matter is
thus protected by the conservation of electric and baryon charges,
and its properties reflect the fundamental physical scales of
electroweak and strong interactions. Indeed, the mass of electron
is related to the scale of the electroweak symmetry breaking,
whereas the mass of proton reflects the scale of QCD confinement.

Extensions of the standard model imply new symmetries and new
particle states. The respective symmetry breaking induces new
fundamental physical scales in particle theory. If the symmetry is
strict, its existence implies new conserved charge. The lightest
particle, bearing this charge, is stable. The set of new
fundamental particles, corresponding to the new strict symmetry,
is then reflected in the existence of new stable particles, which
should be present in the Universe and taken into account in the
total energy-momentum tensor.

Most of the known particles are unstable. For a particle with the
mass $m$ the particle physics time scale is $t \sim 1/m$, so in
particle world we refer to particles with lifetime $\tau \gg 1/m$
as to metastable. To be of cosmological significance metastable
particle should survive after the temperature of the Universe $T$
fell down below $T \sim m$, what means that the particle lifetime
should exceed $t \sim (m_{Pl}/m) \cdot (1/m)$. Such a long
lifetime should find reason in the existence of an (approximate)
symmetry. From this viewpoint, cosmology is sensitive to the most
fundamental properties of microworld, to the conservation laws
reflecting strict or nearly strict symmetries of particle theory.

However, the mechanism of particle symmetry breaking can also have
the cosmological impact. Heating of condensed matter leads to
restoration of its symmetry. When the heated matter cools down,
phase transition to the phase of broken symmetry takes place. In
the course of the phase transitions, corresponding to given type
of symmetry breaking, topological defects can form. One can
directly observe formation of such defects in liquid crystals or
in superfluid He. In the same manner the mechanism of spontaneous
breaking of particle symmetry  implies restoration of the
underlying symmetry. When temperature decreases in the course of
cosmological expansion, transitions to the phase of broken
symmetry  can lead, depending on the symmetry breaking pattern, to
formation of topological defects in very early Universe. The
defects can represent the new form of stable particles (as it is
in the case of magnetic monopoles), or the form of extended
structures, such as cosmic strings or cosmic walls.

In the old Big bang scenario the cosmological expansion and its
initial conditions was given {\it a priori}. In the modern
cosmology the expansion of the Universe and its initial conditions
is related to the process of inflation. The global properties of
the Universe as well as the origin of its large scale structure
are the result of this process. The matter content of the modern
Universe is also originated from the physical processes: the
baryon density is the result of baryosynthesis and the nonbaryonic
dark matter represents the relic species of physics of the hidden
sector of particle theory. Physics, underlying inflation,
baryosynthesis and dark matter, is referred to the extensions of
the standard model, and the variety of such extensions makes the
whole picture in general ambiguous. However, in the framework of
each particular physical realization of inflationary model with
baryosynthesis and dark matter the corresponding model dependent
cosmological scenario can be specified in all the details. In such
scenario the main stages of cosmological evolution, the structure
and the physical content of the Universe reflect the structure of
the underlying physical model. The latter should include with
necessity the standard model, describing the properties of
baryonic matter, and its extensions, responsible for inflation,
baryosynthesis and dark matter. In no case the cosmological impact
of such extensions is reduced to reproduction of these three
phenomena only. The nontrivial path of cosmological evolution,
specific for each particular realization of inflational model with
baryosynthesis and nonbaryonic dark matter, always contains some
additional model dependent cosmologically viable predictions,
which can be confronted with astrophysical data. The part of
cosmoparticle physics, called cosmoarcheology, offers the set of
methods and tools probing such predictions.

\section{Cosmoarcheology of new physics}
 \label{Cosmoarcheology}

Cosmoarcheology considers the results of observational cosmology
as the sample of the experimental data on the possible existence
and features of hypothetical phenomena predicted by particle
theory. To undertake the {\it Gedanken Experiment} with these
phenomena some theoretical framework to treat their origin and
evolution in the Universe should be assumed. As it was pointed out
in \cite{Cosmoarcheology} the choice of such framework is a
nontrivial problem in the modern cosmology.

Indeed, in the old Big bang scenario any new phenomenon, predicted
by particle theory was considered in the course of the thermal
history of the Universe, starting from Planck times. The problem
is that the bedrock of the modern cosmology, namely, inflation,
baryosynthesis and dark matter, is also based on experimentally
unproven part of particle theory, so that the test for possible
effects of new physics is accomplished by the necessity to choose
the physical basis for such test. There are two possible solutions
for this problem: a) a crude model independent comparison of the
predicted effect with the observational data and b) the model
dependent treatment of considered effect, provided that the model,
predicting it, contains physical mechanism of inflation,
baryosynthesis and dark matter.

The basis for the approach (a) is that whatever happened in the
early Universe its results should not contradict the observed
properties of the modern Universe. The set of observational data
(especially, the light element abundance and spectrum of black
body radiation) put severe constraint on the deviation from
thermal evolution after 1 s of expansion, what strengthens the
model independent conjectures of approach (a).

One can specify the new phenomena by their net contribution into
the cosmological density and by forms of their possible influence
on parameters of matter and radiation. In the first aspect we can
consider strong and weak phenomena. Strong phenomena can put
dominant contribution into the density of the Universe, thus
defining the dynamics of expansion in that period, whereas the
contribution of weak phenomena into the total density is always
subdominant. The phenomena are time dependent, being characterized
by their time-scale, so that permanent (stable) and temporary
(unstable) phenomena can take place. They can have homogeneous and
inhomogeneous distribution in space. The amplitude of density
fluctuations $\delta \equiv \delta \varrho/\varrho$ measures the
level of inhomogeneity relative to the total density, $\varrho$.
The partial amplitude $\delta_{i} \equiv \delta
\varrho_{i}/\varrho_{i}$ measures the level of fluctuations within
a particular component with density $\varrho_{i}$, contributing
into the total density $\varrho = \sum_{i} \varrho_{i}$. The case
$\delta_{i} \ge 1$ within the considered $i$-th component
corresponds to its strong inhomogeneity. Strong inhomogeneity is
compatible with the smallness of total density fluctuations, if
the contribution of inhomogeneous component into the total density
is small: $\varrho_{i} \ll \varrho$, so that $\delta \ll 1$.

The phenomena can influence the properties of matter and radiation
either indirectly, say, changing of the cosmological equation of
state, or via direct interaction with matter and radiation. In the
first case only strong phenomena are relevant, in the second case
even weak phenomena are accessible to observational data. The
detailed analysis of sensitivity of cosmological data to various
phenomena of new physics are presented in \cite{book}. This set of
astrophysical constraints confronts phenomena, predicted by
cosmophenomenology as cosmological consequences of particle
theory.

\section{Cosmophenomenology of new physics}
 \label{Cosmophenomenology}

To study the imprints of new physics in astrophysical data
cosmoarcheology implies the forms and means in which new physics
leaves such imprints. So, the important tool of cosmoarcheology in
linking the cosmological predictions of particle theory to
observational data is the {\it Cosmophenomenology} of new physics.
It studies the possible hypothetical forms of new physics, which
may appear as cosmological consequences of particle theory, and
their properties, which can result in observable effects.
\subsection{Primordial particles} The simplest
primordial form of new physics is the gas of new stable massive
particles, originated from early Universe. For particles with the
mass $m$, at high temperature $T>m$ the equilibrium condition, $n
\cdot \sigma v \cdot t > 1$ is valid, if their annihilation cross
section $\sigma > 1/(m m_{Pl})$ is sufficiently large to establish
the equilibrium. At $T<m$ such particles go out of equilibrium and
their relative concentration freezes out. More weakly interacting
species decouple from plasma and radiation at $T>m$, when $n \cdot
\sigma v \cdot t \sim 1$, i.e. at $T_{dec} \sim (\sigma
m_{Pl})^{-1}$. The maximal temperature, which is reached in
inflationary Universe, is the reheating temperature, $T_{r}$,
after inflation. So, the very weakly interacting particles with
the annihilation cross section $\sigma < 1/(T_{r} m_{Pl})$, as
well as very heavy particles with the mass $m \gg T_{r}$ can not
be in thermal equilibrium, and the detailed mechanism of their
production should be considered to calculate their primordial
abundance.

Decaying particles with the lifetime $\tau$, exceeding the age of
the Universe, $t_{U}$, $\tau > t_{U}$, can be treated as stable.
By definition, primordial stable particles survive to the present
time and should be present in the modern Universe. The net effect
of their existence is given by their contribution into the total
cosmological density. They can dominate in the total density being
the dominant form of cosmological dark matter, or they can
represent its subdominant fraction. In the latter case more
detailed analysis of their distribution in space, of their
condensation in galaxies, of their capture by stars, Sun and
Earth, as well as of the effects of their interaction with matter
and of their annihilation provides more sensitive probes for their
existence. In particular, hypothetical stable neutrinos of the 4th
generation with the mass about 50 GeV are predicted to form the
subdominant form of the modern dark matter, contributing less than
0,1 \% to the total density. However, direct experimental search
for cosmic fluxes of weakly interacting massive particles (WIMPs)
may be sensitive to the existence of such component \cite{rpp}. It
was shown in \cite{Fargion99,Grossi,Belotsky} that annihilation of
4th neutrinos and their antineutrinos in the Galaxy can explain
the galactic gamma-background, measured by EGRET in the range
above 1 GeV, and that it can give some clue to explanation of
cosmic positron anomaly, claimed to be found by HEAT. 4th neutrino
annihilation inside the Earth should lead to the flux of
underground monochromatic neutrinos of known types, which can be
traced in the analysis of the already existing and future data of
underground neutrino detectors \cite{Belotsky}.

New particles with electric charge and/or strong interaction can
form anomalous atoms and contain in the ordinary matter as
anomalous isotopes. For example, if the lightest quark of 4th
generation is stable, it can form stable charged hadrons, serving
as nuclei of anomalous atoms of e.g. anomalous helium
\cite{BKS,4h}. Massive negatively charged particles bind with
$^4He$ in atom-like systems, as soon as helium is formed in Big
Bang Nucleosynthesis \cite{Tera}. Since particles $Q^-$ with
charge -1 form positively charged "ion" $[^4He Q^-]^+$, which
behave as anomalous hydrogen, the grave problem of anomalous
hydrogen overproduction is inevitable for particle models
predicting such particles, as it is the case for the model of
"Sinister" Universe \cite{Glashow}. Particles $Q^{-2}$ with charge
-2 form neutral O-helium "atom" $[^4He Q^{-2}]$, in which the
charge of $\alpha$ particle is shielded \cite{anutium,AC}. It
removes Coulomb barrier between $\alpha$ particle and nuclei,
giving rise to new paths of nuclear transformations, catalyzed by
O-helium.

Primordial unstable particles with the lifetime, less than the age
of the Universe, $\tau < t_{U}$, can not survive to the present
time. But, if their lifetime is sufficiently large to satisfy the
condition $\tau \gg (m_{Pl}/m) \cdot (1/m)$, their existence in
early Universe can lead to direct or indirect traces. Cosmological
flux of decay products contributing into the cosmic and gamma ray
backgrounds represents the direct trace of unstable particles. If
the decay products do not survive to the present time their
interaction with matter and radiation can cause indirect trace in
the light element abundance or in the fluctuations of thermal
radiation. If the particle lifetime is much less than $1$s the
multi-step indirect traces are possible, provided that particles
dominate in the Universe before their decay. On the dust-like
stage of their dominance black hole formation takes place, and the
spectrum of such primordial black holes traces the particle
properties (mass, frozen concentration, lifetime) \cite{Polnarev}.
The particle decay in the end of dust like stage influences the
baryon asymmetry of the Universe. In any way cosmophenomenoLOGICAL
chains link the predicted properties of even unstable new
particles to the effects accessible in astronomical observations.
Such effects may be important in the analysis of the observational
data.

So, the only direct evidence for the accelerated expansion of the
modern Universe comes from the distant SN I data. The data on the
cosmic microwave background (CMB) radiation and large scale
structure (LSS) evolution (see e.g. \cite{rpp}) prove in fact the
existence of homogeneously distributed dark energy and the slowing
down of LSS evolution at $z \leq 3$. Homogeneous negative pressure
medium ($\Lambda$-term or quintessence) leads to {\it relative}
slowing down of LSS evolution due to acceleration of cosmological
expansion. However, both homogeneous component of dark matter and
slowing down of LSS evolution naturally follow from the models of
Unstable Dark Matter (UDM) (see \cite{book} for review), in which
the structure is formed by unstable weakly interacting particles.
The weakly interacting decay products are distributed
homogeneously. The loss of the most part of dark matter after
decay slows down the LSS evolution. The dominantly invisible decay
products can contain small ionizing component \cite{Berezhiani2}.
Thus, UDM effects will deserve attention, even if the accelerated
expansion is proved.
\subsection{Archioles}
Parameters of new stable and metastable particles are also
determined by a pattern of particle symmetry breaking. This
pattern is reflected in the succession of phase transitions in the
early Universe. Phase transitions of the first order proceed
through bubble nucleation, which can result in black hole
formation (see e.g. \cite{kkrs} and \cite{book2} for review and
references). Phase transitions of the second order can lead to
formation of topological defects, such as walls, string or
monopoles. The observational data put severe constraints on
magnetic monopole and cosmic wall production, as well as on
parameters of cosmic strings. A succession of phase transitions
can change the structure of cosmological defects. More complicated
forms, such as walls-surrounded-by-strings can appear. Such
structures can be unstable, but their existence can lead the trace
in the nonhomogeneous distribution of dark matter and in large
scale correlations in the nonhomogeneous dark matter structures,
such as {\it archioles} \cite{Sakharov2}, which arise in the
result of cosmological evolution of pseudo-Nambu-Goldstone field.

A wide class of particle models possesses a symmetry breaking
pattern, which can be effectively described by
pseudo-Nambu--Goldstone (PNG) field and which corresponds to
formation of unstable topological defect structure in the early
Universe (see \cite{book2} for review and references). The
Nambu--Goldstone nature in such an effective description reflects
the spontaneous breaking of global U(1) symmetry, resulting in
continuous degeneracy of vacua. The explicit symmetry breaking at
smaller energy scale changes this continuous degeneracy by
discrete vacuum degeneracy.

At high temperatures such a symmetry breaking pattern implies the
succession of second order phase transitions. In the first
transition, continuous degeneracy of vacua leads, at scales
exceeding the correlation length, to the formation of topological
defects in the form of a string network; in the second phase
transition, continuous transitions in space between degenerated
vacua form surfaces: domain walls surrounded by strings. This last
structure is unstable, but, as was shown in the example of the
invisible axion \cite{Sakharov2,kss,kss2}, it is reflected in the
large scale inhomogeneity of distribution of energy density of
coherent PNG (axion) field oscillations. This energy density is
proportional to the initial value of phase, which acquires
dynamical meaning of amplitude of axion field, when axion mass is
switched on in the result of the second phase transition.

The value of phase changes by $2 \pi$ around string. This strong
nonhomogeneity of phase leads to corresponding nonhomogeneity of
energy density of coherent PNG (axion) field oscillations. Usual
argument (see e.g. \cite{sikivie} and references therein) is that
this nonhomogeneity is essential only on scales, corresponding to
mean distance between strings. This distance is small, being of
the order of the scale of cosmological horizon in the period, when
PNG field oscillations start. However, since the nonhomogeneity of
phase follows the pattern of axion string network this argument
misses large scale correlations in the distribution of energy
density of field oscillations.

Indeed, numerical analysis of string network (see review in
\cite{vs}) indicates that large string loops are strongly
suppressed and the fraction of about 80\% of string length,
corresponding to "infinite" strings, remains virtually the same in
all large scales. This property is the other side of the well
known scale invariant character of string network. Therefore the
correlations of energy density should persist on large scales, as
it was revealed in \cite{Sakharov2,kss,kss2}.

The large scale correlations in topological defects and their
imprints in primordial inhomogeneities is the indirect effect of
inflation, if phase transitions take place after reheating of the
Universe. Inflation provides in this case the equal conditions of
phase transition, taking place in causally disconnected regions.
\subsection{Primordial clouds of massive PBH}
If the phase transitions take place on inflational stage new forms
of primordial large scale correlations appear. The example of
global U(1) symmetry, broken spontaneously in the period of
inflation and successively broken explicitly after reheating, was
recently considered in \cite{KRS}. In this model, spontaneous U(1)
symmetry breaking at inflational stage is induced by the vacuum
expectation value $\langle \psi \rangle = f$ of a complex scalar
field $\Psi = \psi \exp{(i \theta)}$, having also explicit
symmetry breaking term in its potential $V_{eb} = \Lambda^{4} (1 -
\cos{\theta})$. The latter is negligible in the period of
inflation, if $f \gg \Lambda$, so that there appears a valley
relative to values of phase in the field potential in this period.
Fluctuations of the phase $\theta$ along this valley, being of the
order of $\Delta \theta \sim H/(2\pi f)$ (here $H$ is the Hubble
parameter at inflational stage) change in the course of inflation
its initial value within the regions of smaller size. Owing to
such fluctuations, for the fixed value of $\theta_{60}$ in the
period of inflation with {\it e-folding} $N=60$ corresponding to
the part of the Universe within the modern cosmological horizon,
strong deviations from this value appear at smaller scales,
corresponding to later periods of inflation with $N < 60$. If
$\theta_{60} < \pi$, the fluctuations can move the value of
$\theta_{N}$ to $\theta_{N} > \pi$ in some regions of the
Universe. After reheating, when the Universe cools down to
temperature $T = \Lambda$ the phase transition to the true vacuum
states, corresponding to the minima of $V_{eb}$ takes place. For
$\theta_{N} < \pi$ the minimum of $V_{eb}$ is reached at
$\theta_{vac} = 0$, whereas in the regions with $\theta_{N} > \pi$
the true vacuum state corresponds to $\theta_{vac} = 2\pi$. For
$\theta_{60} < \pi$ in the bulk of the volume within the modern
cosmological horizon $\theta_{vac} = 0$. However, within this
volume there appear regions with $\theta_{vac} = 2\pi$. These
regions are surrounded by massive domain walls, formed at the
border between the two vacua. Since regions with $\theta_{vac} =
2\pi$ are confined, the domain walls are closed. After their size
equals the horizon, closed walls can collapse into black holes.
The minimal mass of such black hole is determined by the condition
that it's Schwarzschild radius, $r_{g} = 2 G M/c^{2}$ exceeds the
width of the wall, $l \sim f/\Lambda^{2}$, and it is given by
$M_{min} \sim f (m_{Pl}/\Lambda)^{2}$. The maximal mass is
determined by the mass of the wall, corresponding to the earliest
region $\theta_{N} > \pi$, appeared at inflational stage.  This
mechanism can lead to formation of primordial black holes of a
whatever large mass (up to the mass of AGNs \cite{AGN}). Such
black holes appear in a form of primordial black hole clusters,
exhibiting fractal distribution in space \cite{KRS}. It can shed
new light on the problem of galaxy formation.
\subsection{Antimatter stars in Galaxy}
Primordial strong inhomogeneities can also appear in the baryon
charge distribution. The appearance of antibaryon domains in the
baryon asymmetrical Universe, reflecting the inhomogeneity of
baryosynthesis, is the profound signature of such strong
inhomogeneity \cite{CKSZ}. On the example of the model of
spontaneous baryosynthesis (see \cite{Dolgov} for review) the
possibility for existence of antimatter domains, surviving to the
present time in inflationary Universe with inhomogeneous
baryosynthesis was revealed in \cite{KRS2}. Evolution of
sufficiently dense antimatter domains can lead to formation of
antimatter globular clusters \cite{GC}. The existence of such
cluster in the halo of our Galaxy should lead to the pollution of
the galactic halo by antiprotons. Their annihilation can reproduce
\cite{Golubkov} the observed galactic gamma background in the
range tens-hundreds MeV. This observed background puts upper limit
on the total mass of antimatter stars in Galaxy ($M \le
10^5M_{\odot}$). The prediction of antihelium component of cosmic
rays \cite{ANTIHE}, accessible to future searches for cosmic ray
antinuclei in PAMELA and AMS02 experiments, as well as of
antimatter meteorites \cite{ANTIME} provides the direct
experimental test for this hypothesis. In this test planned
sensitivity of AMS02 experiment will reach the lower limit for the
mass of antimatter stars in Galaxy. This limit ($M \ge
10^3M_{\odot}$) follows from the condition that antimatter domain
should be sufficiently large to survive and sufficiently dense to
provide star formation.

So the primordial strong inhomogeneities in the distribution of
total, dark matter and baryon density in the Universe is the new
important phenomenon of cosmological models, based on particle
models with hierarchy of symmetry breaking.
\section{The encircled pyramid}
New physics follows from the necessity to extend the Standard
model. White spots in representations of symmetry groups,
considered in extensions of Standard model, correspond to new
unknown particles. Extension of gauge symmetry puts into
consideration new gauge fields, mediating new interactions. Global
symmetry breaking results in the existence of Goldstone boson
fields.

For a long time necessity to extend the Standard model had purely
theoretical reasons. Aesthetically, because full unification is
not achieved in the Standard model; practically, because it
contains some internal inconsistencies. It does not seem complete
for cosmology. One has to go beyond the Standard model to explain
inflation, baryosynthesis and nonbaryonic dark matter. Recently
there has appeared a set of experimental evidences for the
existence of neutrino oscillations, for effects of new physics in
anomalous magnetic moment of muon ($g-2$), for cosmic WIMPs  and
for double neutrinoless beta decay (see for recent review
\cite{rpp}). Whatever is the accepted status of some of these
evidences, they indicate that the experimental searches may have
already crossed the border of new physics.

In particle physics direct experimental probes for the predictions
of particle theory are most attractive. Predictions of new charged
particles, such as supersymmetric particles or quarks and leptons
of new generation, are accessible to experimental search at
accelerators of new generation, if their masses are in 100GeV-1TeV
range. However, predictions related to higher energy scale need
non-accelerator or indirect means for their test.

Search for rare processes, such as proton decay, neutrino
oscillations, neutrinoless beta decay, precise measurements of
parameters of known particles, experimental searches for dark
matter represent the widely known forms of such means.
Cosmoparticle physics offers the nontrivial extensions of indirect
and non-accelerator searches for new physics and its possible
properties. In experimental cosmoarcheology the data is to be
obtained, necessary to link the cosmophenomenology of new physics
with astrophysical observations (See \cite{Cosmoarcheology}). In
experimental cosmoparticle physics the parameters, fixed from the
consitency of cosmological models and observations, define the
level, at which the new types of particle processes should be
searched for (see \cite{expcpp}).
\subsection{New quarks and leptons}
The theories of everything should provide the complete physical
basis for cosmology. The problem is that the string theory
\cite{Green} is now in the form of "theoretical theory", for which
the experimental probes are widely doubted to exist. The
development of cosmoparticle physics can remove these doubts. In
its framework there are two directions to approach the test of
theories of everything.

One of them is related with the search for the experimentally
accessible effects of heterotic string phenomenology. The
mechanism of compactification and symmetry breaking leads to the
prediction of homotopically stable objects \cite{Kogan1} and
shadow matter \cite{Kogan2}, accessible to cosmoarcheological
means of cosmoparticle physics. The condition to reproduce the
Standard model naturally leads in the heterotic string
phenomenology to the prediction of fourth generation of quarks and
leptons \cite{Shibaev} with a stable massive 4th neutrino
\cite{Fargion99}, what can be the subject of complete experimental
test in the near future. The comparison between the rank of the
unifying group $E_{6}$ ($r=6$) and the rank of the Standard model
($r=4$) implies the existence of new conserved charges and new
(possibly strict) gauge symmetries. New strict gauge U(1) symmetry
(similar to U(1) symmetry of electrodynamics) is possible, if it
is ascribed to the fermions of 4th generation. This hypothesis
explains the difference between the three known types of neutrinos
and neutrino of 4th generation. The latter possesses new gauge
charge and, being Dirac particle, can not have small Majorana mass
due to sea saw mechanism. If the 4th neutrino is the lightest
particle of the 4th quark-lepton family, strict conservation of
the new charge makes massive 4th neutrino to be absolutely stable.
Following this hypothesis \cite{Shibaev} quarks and leptons of 4th
generation are the source of new long range interaction
($y$-electromagnetism), similar to the electromagnetic interaction
of ordinary charged particles. New strictly conserved local U(1)
gauge symmetries can also arise in the alternative approach to
extension of standard model \cite{AC} based on almost commutative
geometry \cite{Connes}.

It is interesting, that heterotic string phenomenology  embeds
even in its simplest realisation both supersymmetric particles and
the 4th family of quarks and leptons, in particular, the two types
of WIMP candidates: neutralinos and massive stable 4th neutrinos.
So in the framework of this phenomenology the multicomponent
analysis of WIMP effects is favorable.

The motivation for existence of new quarks and leptons also
follows from geometrical approach to particle unification
\cite{Mankoc} and from models of extended technicolor
\cite{Sannino,Kouvaris}.

New quarks and charged leptons can be metastable and have
lifetime, exceeding the age of the Universe. It gives rise to a
new form of stable matter around us - to composite dark matter,
whose massive charged constituents are hidden in atom-like
systems. Inevitable by-product of creation of such matter in
Universe is the existence of O-helium "atom", in which positive
charge of $\alpha$ particle is shielded by negative charge of
massive component. O-helium can be a fraction \cite{AC} or even
dominant form of dark matter \cite{anutium} and search for its
charged constituents at accelerators and cosmic rays \cite{KPS06}
acquires the significance of direct experimental test for this
form of dark matter.

In the above approach some particular phenomenological features of
simplest variants of string theory are studied. The other
direction is to elaborate the extensive phenomenology of theories
of everything by adding to the symmetry of the Standard model the
(broken) symmetries, which have serious reasons to exist. The
existence of (broken) symmetry between quark-lepton families, the
necessity in the solution of strong CP-violation problem with the
use of broken Peccei-Quinn symmetry, as well as the practical
necessity in supersymmetry to eliminate the quadratic divergence
of Higgs boson mass in electroweak theory is the example of
appealing additions to the symmetry of the Standard model. The
horizontal unification and its cosmology represent the first step
on this way, illustrating the approach of cosmoparticle physics to
the elaboration of the proper phenomenology for theories of
everything \cite{Sakharov1}.

We can conclude that from the very beginning to the modern stage,
the evolution of Universe is governed by the forms of matter,
different from those we are built of and observe around us. From
the very beginning to the present time, the evolution of the
Universe was governed by physical laws, which we still don't know.
Observational cosmology offers strong evidences favoring the
existence of processes, determined by new physics, and the
experimental physics approaches to their investigation.

Cosmoparticle physics \cite{ADS} \cite{MKH}, studying the
physical, astrophysical and cosmological impact of new laws of
Nature, explores the new forms of matter and their physical
properties, what opens the way to use the corresponding new
sources of energy and new means of energy transfer. It offers the
great challenge for the new Millennium. Its solution for the
Uhroboros puzzle is as paradoxical, as "encircled pyramid" -
cosmoparticle physics implies complex cross-dis\-ciplinary studies,
offering the multi-dimensional exit from the plane with wrong
circle of problems in the joint of cosmology and particle physics.
\section*{Acknowledgements}
 I am grateful to Organizers of 9th Workshop
"What Comes Beyond the Standard Models" for kind
 hospitality.

\title{Discussion Section on $4^\textrm{th}$ Generation}
\author{M.Yu. Khlopov}
\institute{%
Center for Cosmoparticle Physics "Cosmion"\\
Miusskaya Pl. 4\\
125047 Moscow\\
Russia\\
\textrm{e-mail:} \texttt{Maxim.Khlopov@roma1.infn.it}}

\titlerunning{Discussion Section on $4^\textrm{th}$ Generation}
\authorrunning{M.Yu. Khlopov}
\maketitle

\begin{abstract}
I briefly stipulate here some ideas, which were  considered in
the Discussion Section.
\end{abstract}

\section{Some phenomenological aspects of 4th generation in geometrical approach}
 \label{Norma}
In the presented realization of geometrical approach \cite{DS1Mankoc}
4th generation by construction is linked to 3d generation. In such
realization mixing and transitions between these generations are
inevitable. It makes all the particles of 4th generation unstable.
It would be interesting to estimate their lifetime. But, in any
case, this feature is important for accelerator search for quarks
and leptons of 4th generation. Processes of creation and decay of
these particles at accelerators have rather distinct experimental
signatures and can be clearly discriminated. On the other hand,
being unstable, hadrons and leptons of 4th generation should not
be present in cosmic rays.

The predicted values of 4th generation particles are in some cases
accessible to test even with the use of existing experimental data
\cite{DS1rpp}. For instant, unstable neutrino of 4th generation with
the mass $\sim 80$GeV should have been seen in LEP.

In the approach \cite{DS1Mankoc} another problem is also of interest:
if it is possible to have the lightest particles of the 4th
generation stable. In this case the 4th generation is decoupled
from three known families and their possible mass pattern can not
be directly deduced from mixing with known particles and from
their known properties.
\section{Composite dark matter from Technicolor}
 \label{technicolor}
Cosmological aspects \cite{DS1Kouvaris} of technicolor models
\cite{DS1Sannino} were concentrated on studies of possible WIMP-like
candidates for dark matter species. However, this approach also
provides the possibility of stable techniparticles $T^{--}$ with
charge -2.

If the model provides the possibility to generate in early
Universe excess of these particles, corresponding to cosmological
dark matter density, atomic bound states of these particles with
primordial $^4He$ can play the role of composite dark matter in
the form of techno-O-helium $^4HeT^{--}$. Formation and evolution
of this composite dark matter will follow the trend of O-helium
dark matter, studied in \cite{DS1anutium} for the case of stable
quarks of 4th generation. Experimental search for stable
techniparticles $T^{--}$ in cosmic rays and at accelerators is
possible similar to the case of new stable leptons
\cite{DS1AC,KPS06}.

\section{Mass self-adjustment for CLEP neutrinos}
 \label{CLEP}
In the model of CLEP states, offered in \cite{DS1Eduardo}, the
following mechanism of self-adjustment of neutrino mass can be
realized. Near a galaxy as the neutrinos expand to the outside
their density decreases and neutrino mass increases. Then as the
mass increases the neutrinos have now a tendency to clump; after
they clump, they become dense and therefore the neutrinos are not
anymore in the CLEP state, but rather in a higher density state
with a corresponding lower mass, which is now preventing neutrinos
to be clustered. Therefore neutrino gas expands again to reach low
density, to return in CLEP state.

If in the course of cosmological expansion neutrino mass in CLEP
state becomes too big some other interesting possibilities arise.
For instance a nontrivial realisation is possible for Unstable
Neutrino Cosmology and other effects of non-equilibrium particles
from CLEP state neutrino decays if this mass becomes too big some
other interesting possibilities arise. For instance a nontrivial
realisation is possible for Unstable Neutrino Cosmology and other
effects of non-equilibrium particles from CLEP state neutrino
decays\cite{DS1book}.

 \def\Tr{{\rm Tr}}
\title{Involution Requirement on a Boundary Makes Massless Fermions %
 Compactified on a Finite Flat Disk Mass Protected}
\author{N.S. Manko\v c Bor\v stnik${}^1$ and  H.B. Nielsen${}^2$}
\institute{%
${}^1$ Department of Physics, FMF,\\
University of Ljubljana,\\
 Jadranska 19, 1000 Ljubljana\\
${}^2$ Department of Physics, Niels Bohr Institute,\\
Blegdamsvej 17,\\
Copenhagen, DK-2100}
 
\titlerunning{Involution Requirement on a Boundary Makes \ldots}
\authorrunning{N.S. Manko\v c Bor\v stnik and  H.B. Nielsen} 
\maketitle 
 
\begin{abstract} 
The genuine Kaluza-Klein-like theories---with no fields in addition to gravity---have difficulties 
with the existence of massless spinors after the compactification
of some space dimensions \cite{witten}. We proposed in ref.\ \cite{hnkk06} a  
boundary condition which allows massless spinors compactified on a flat disk to be of only one handedness. 
Massless spinors then chirally couple to the corresponding background gauge gravitational field  
(which solves equations of motion for a free field linear in the Riemann curvature). 
In this paper we study the same toy model: $M^{(1+3)} \times M^{(2)}$, looking this time 
for an involution  which  
transforms a space of solutions of Weyl equations in $d=1+5$ 
from the outside of the flat disk in $x^5$ and $x^6$ into its inside  (or conversely). 
The natural boundary condition that on the wall 
an outside solution must coincide with the corresponding inside one leads  to 
massless spinors of only one handedness (and accordingly mass protected), chirally coupled to 
the corresponding background gauge gravitational field. We introduce the Hermitean  
operators  
of momenta and discuss the orthogonality of solutions, ensuring that to each mass only
one solution of equations of motion corresponds.

\end{abstract}

\section{Introduction}
\label{orbif:introduction}

The major problem of the compactification procedure in all Kaluza-Klein-like theories with 
only gravity and no additional gauge fields is how to ensure that massless spinors 
be mass protected after the compactification. Namely, even if we start with 
only one Weyl spinor in some even dimensional space of  $d=2$ 
modulo $4$ dimensions (i.e.\ in $d=2(2n+1),$ $n=0,1,2,\cdots$) so that there appear no Majorana 
mass if no conserved 
charges exist and families are allowed, as we have proven in 
ref.\ \cite{hnm06}, and accordingly with the mass protection from the very beginning, 
a compactification of $m$ dimensions gives rise to a spinor of one handedness in $d$  
with both handedness in $d-m$ and is accordingly  not mass protected any longer. 

And  in addition, since a spin (or the total conserved angular momentum) 
in the compactified part of space will    
in $d-m$ space appear as a charge and will manifest both values (positive and negative ones) 
and since in the second quantization procedure anti particles 
of opposite charges appear anyhow, doubling the number of massless 
spinors of both---positive and negative---charges when coming from 
$d(=2(2n+1))$-dimensional space down to $d=4$ and after 
a second quantized procedure is not in agreement  
with what we observe. Accordingly there must be some requirements, 
some boundary conditions, which 
ensure in a compactification procedure that only spinors of one handedness survive, if 
Kaluza-Klein-like theories have some meaning. However, the idea of Kaluza and Klein of 
having only gravity as a gauge field seems  too beautiful not to have the realization in Nature. 

One of us\cite{norma92,norma93,norma94,norma95,Portoroz03} has for long tried to 
unify the spin and all the charges to only the spin, so that spinors would in 
$d\ge 4$ carry nothing but 
a spin and interact accordingly with only the gauge fields of the Poincar\' e group, that is with 
vielbeins $f^{\alpha}{\!}_{a}$ \footnote{$f^{\alpha}{}_{a}$ are inverted vielbeins to 
$e^{a}{}_{\alpha}$ with the properties $e^a{}_{\alpha} f^{\alpha}{\!}_b = \delta^a{\!}_b,\; 
e^a{\!}_{\alpha} f^{\beta}{\!}_a = \delta^{\beta}_{\alpha} $. 
Latin indices  
$a,b,..,m,n,..,s,t,..$ denote a tangent space (a flat index),
while Greek indices $\alpha, \beta,..,\mu, \nu,.. \sigma,\tau ..$ denote an Einstein 
index (a curved index). Letters  from the beginning of both the alphabets
indicate a general index ($a,b,c,..$   and $\alpha, \beta, \gamma,.. $ ), 
from the middle of both the alphabets   
the observed dimensions $0,1,2,3$ ($m,n,..$ and $\mu,\nu,..$), indices from 
the bottom of the alphabets
indicate the compactified dimensions ($s,t,..$ and $\sigma,\tau,..$). 
We assume the signature $\eta^{ab} =
diag\{1,-1,-1,\cdots,-1\}$.
} and   
spin connections $\omega_{ab\alpha}$, which are the gauge fields of the Poincar\'e group. 

In this paper  we take (as we did in the ref.\ \cite{hnkk06}) the covariant momentum of a spinor, 
when applied on a spinor function $\psi$, to be
\begin{eqnarray}
p_{0 a} &=& f^{\alpha}{\!}_{a}p_{0 \alpha}, \quad p_{0 \alpha} \psi = p_{ \alpha} - \frac{1}{2} S^{cd} 
\omega_{cd \alpha}. 
\label{covp}
\end{eqnarray}
A kind of a total covariant derivative of $e^a{\!}_{\alpha}$ (a vector with both---Einstein and 
flat index) will be taken to be 
$p_{0 \alpha} e^{a}{\!}_{\beta} = i e^a{\!}_{\beta ; \alpha} = i (e^a{\!}_{\beta , \alpha} +
\omega^a{\!}_{d \alpha} e^d{\!}_{\beta} - 
\Gamma^{\gamma}{\!}_{\beta \alpha} e^a{\!}_{\gamma}),$ with the 
require that this derivative of a vielbein is zero: $e^a{\!}_{\beta ; \alpha} =0.$

The corresponding Lagrange density ${\cal L}$  for   a Weyl spinor has the form
${\cal L} = E \frac{1}{2} [(\psi^{\dagger}\gamma^0 \gamma^a p_{0a} \psi) + 
(\psi^{\dagger} \gamma^0\gamma^a p_{0 a}
\psi)^{\dagger}]$ and leads to
\begin{eqnarray}
{\cal L} &=& E\psi^{\dagger}\gamma^0 \gamma^a   ( p_{a} - \frac{1}{2} S^{cd}  \omega_{cda})\psi,
\label{weylL}
\end{eqnarray}
with $ E = \det(e^a{\!}_{\alpha}) $ \footnote{To generate more than one family, we actually observe 
up to now three families, a second kind of the Clifford algebra objects has also been 
introduced\cite{norma93,norma94,holgernorma2002,holgernorma2003}, 
which anti commute with the ordinary Dirac $\gamma^a$ matrices ($\{\gamma^a,\tilde{\gamma^b}\}_+=0$) 
and generate equivalent representations with respect to the generators 
$S^{ab}=\frac{i}{4}(\gamma^a \gamma^b - \gamma^b \gamma^a)$ and are used accordingly 
to generate families\cite{norma93,norma94,holgernorma2002,holgernorma2003}. 
In this work we shall not take families into account.}.

The authors of this paper have tried to find a way out of this ''Witten's no go theorem'' 
 for a toy model of 
$M^{(1+3)} \times$ a flat finite disk in $(1+5)$-dimensional space \cite{hnkk06} by postulating a 
particular boundary condition, which allows a spinor to carry  after the compactification 
only one handedness. 
Massless spinors then chirally couple to the corresponding background gauge gravitational field, 
which solves equations of motion for a free field, linear in the Riemann curvature, while 
the current through the wall is for a massless and massive solutions equal to zero. 

In the ref.\ \cite{hnkk06} the boundary condition was written in a covariant way  as 
\begin{eqnarray}
\hat{\cal{R}}\psi|_{\rm wall} &=& 0,\nonumber\\
\hat{\cal{R}} &=& \frac{1}{2}(1-i n^{(\rho)}{\!\!}_{a}\, n^{(\phi)}{\!\!}_{b}\, 
\gamma^a \gamma^b ),\quad \hat{\cal{R}}^2 = \hat{\cal{R}}
\label{diskboundary}
\nonumber
\end{eqnarray}
with $n^{(\rho)}=(0,0,0,0,\cos \phi, \sin \phi),\; n^{(\phi)}= 
(0,0,0,0,-\sin \phi, \cos \phi)$, which  
are the two unit vectors 
perpendicular and tangential to the boundary of the disk (at $\rho_0$), respectively. 
The projector $\hat{\cal{R}}$ can for the above choice of the two vectors 
$n^{(\rho)}$ and $n^{(\phi)}$ be written as  
\begin{eqnarray}
\hat{\cal{R}}   &=& {\stackrel{56}{[-]} }= \frac{1}{2} (1-i\gamma^5\gamma^6).
\label{prodiskboundary}
\end{eqnarray}
The reader can find more about the Clifford algebra objects 
$\stackrel{ab}{(\pm)}, \stackrel{ab}{[\pm]}$ in the Appendix (section \ref{appendixtechnique}). 

The boundary condition requires that only  massless states (determined by  
Eq.(\ref{weylL})) of one (let us say right) handedness 
with respect to the compactified disk degrees of freedom are allowed. Accordingly 
also massless states of only one handedness are allowed also in $d=1+3$.

In this paper we reformulate the above boundary condition as an {\em involution}, which 
transforms solutions of equations of motion from outside the boundary of the disk 
into its inside. We do this by  the intention that the limitation of 
$M2$ on a finite disk would have 
a natural explanation, originated in a symmetry relation.
We also define the Hermitean momentum $p^s$  and 
comment on the orthogonality relations of solutions of equations of motion, which 
fulfill the  boundary conditions. 

We  make use of the technique presented in ref.\ \cite{holgernorma2002,holgernorma2003}
when writing the equations of motion and their solutions. It turns out 
that all the  derivations and discussions appear to be very transparent when using this technique. 
We briefly repeat this technique in  
Appendix \ref{appendixtechnique}.

\section{Equations of motion and solutions}
\label{s:equations}

We assume that a two dimensional space, spanned by $x^5$ and $x^6$, is an 
Euclidean manifold $M^{(2)}$ (with no gravity)   
\begin{eqnarray}
f^{\sigma}{\!}_{s} = \delta^{\sigma}{\!}_{s},\; \omega_{56 s} =0.
\label{disk}
\end{eqnarray}
and accordingly with the rotational symmetry around an origin. 

Wave functions  describing spinors in $(1+5)$-dimensional space demonstrating  
$M^{(1+3)}$ $\times M^{(2)}$ symmetry  are 
required to obey the equations of motion
\begin{eqnarray}
\gamma^0 \gamma^a p_a \psi^{(6)} =0, \;a=m,s;\;m=0,1,2,3; \;s=5,6 .
\label{equations}
\end{eqnarray}
The most general solution for a free particle in $d=1+5$ should be written as a superposition
of all  four ($2^{6/2 -1}$) states of one Weyl representation. We ask the reader to see 
Appendix \ref{appendixtechnique} for the technical details how to write one Weyl representation 
in terms of the Clifford algebra objects after making a choice of the Cartan sub algebra 
set, for which we make a choice: $S^{03}, S^{12}, S^{56}$. 
In our technique \cite{holgernorma2002} the four 
states, which all are the eigenstates of the Cartan sub algebra set, are expressed with 
the following four products of projections ($\stackrel{ab}{[k]}$) and nilpotents 
($\stackrel{ab}{(k)}$):
\begin{eqnarray}
\varphi^{1}_{1} &=& \stackrel{56}{(+)} \stackrel{03}{(+i)} \stackrel{12}{(+)}\psi_0,\nonumber\\
\varphi^{1}_{2} &=&\stackrel{56}{(+)}  \stackrel{03}{[-i]} \stackrel{12}{[-]}\psi_0,\nonumber\\
\varphi^{2}_{1} &=&\stackrel{56}{[-]}  \stackrel{03}{[-i])} \stackrel{12}{(+)}\psi_0,\nonumber\\
\varphi^{2}_{2} &=&\stackrel{56}{[-]} \stackrel{03}{(+i)} \stackrel{12}{[-]}\psi_0,
\label{weylrep}
\end{eqnarray}
where  $\psi_0$ is a vacuum state.
If we write 
the operators of handedness in $d=1+5$ as $\Gamma^{(1+5)} = \gamma^0 \gamma^1 
\gamma^2 \gamma^3 \gamma^5 \gamma^6$ ($= 23 i S^{03} S^{12} S^{56}$), in $d=1+3$ 
as $\Gamma^{(1+3)}= -i\gamma^0\gamma^1\gamma^2\gamma^3 $ ($= 22 i S^{03} S^{12}$) 
and in the two dimensional space as $\Gamma^{(2)} = i\gamma^5 \gamma^6$ 
($= 2 S^{56}$), we find that all four states are left handed with respect to 
$\Gamma^{(1+5)}$, with the value $-1$, the first two are right handed and the second two 
left handed with respect to 
$\Gamma^{(2)}$, with  the values $1$ and $-1$, respectively, while the first two are left handed 
and the second two right handed with respect to $\Gamma^{(1+3)}$ with the values $-1$ and $1$, 
respectively. 

Taking into account Eq.(\ref{weylrep}) we may write a wave function  $\psi^{(6)}$ in $d=1+5$ as
\begin{eqnarray}
\psi^{(6)} = ({\cal A} \stackrel{56}{(+)} + {\cal B} \stackrel{56}{[-]})\psi^{(4)}, 
\label{psi6}
\end{eqnarray}
where ${\cal A}$ and ${\cal B}$ depend on $x^5$ and $x^6$, while $\psi^{(4)}$ determines the spin 
and the coordinate dependent part of the wave function $\psi^{(6)}$ in $d=1+3$. 

Spinors, which manifest masslessness in $d=1+3$, must obey the equation
\begin{eqnarray}
\gamma^0\gamma^s p_s \psi^{(6)}=0,\quad s=5,6,
\label{m}
\end{eqnarray}
since what will demonstrate as an effective action in $d=1+3$ is
\begin{eqnarray}
\int \prod_m dx^m  \Tr_{0123}(\int dx^5 dx^6 \Tr_{56} (\psi^{(6)\dagger}
\gamma^0 (\gamma^m p_m + \gamma^s p_s) \psi^{(6)})) = \nonumber\\
\int \prod_m dx^m \Tr_{0123}(\psi^{(4)\dagger} 
\gamma^0 \gamma^m p_m \psi^{(4)} ) - \int \prod_m dx^m \Tr_{0123}(\psi^{(4)\dagger} 
\gamma^0 m \psi^{(4)} ),
\label{m1}
\end{eqnarray}
where integrals go over all the space on which the solutions are defined. 
$\psi^{(6)}$ and $ \psi^{(4)} $ are the solutions in $d=1+5$ and $d=1+3$, respectively. 
$ \Tr_{0123} $ and $ \Tr_{56} $ mean the trace over the spin degrees 
of freedom in $x0, x1, x2, x3$ and in $x^5,x^6$, respectively. 
(One finds, for example, that $ \Tr (\stackrel{56}{[\pm]})=1.$)
For massless spinors it must be that  $\int dx^5 dx^6 \Tr_{56} 
(\psi^{(6)\dagger}\gamma^0\gamma^s p_s \psi^{(6)})=
\psi^{(4)\dagger}\gamma^0(-m) \psi^{(4)}$$ = 0$.

To find the effective action in $1+3$ for massive spinors, we recognize that for the mass term 
we have 
\begin{eqnarray}
\psi^{(4)\dagger} \gamma^0 (-{\cal A}^* \stackrel{56\;}{(+)^{\dagger}}
+ {\cal B}^* \stackrel{56\;}{[-]^{\dagger}})\gamma^s p_s 
({\cal A} \stackrel{56}{(+)}
+ {\cal B} \stackrel{56}{[-]})\psi^{(4)} = \nonumber\\
\psi^{(4)\dagger} \gamma^0 (-{\cal A}^* \stackrel{56\;}{(+)^{\dagger}}
+ {\cal B}^* \stackrel{56\;}{[-]^{\dagger}})(-m) 
(-{\cal A} \stackrel{56}{(+)}
+ {\cal B} \stackrel{56}{[-]})\psi^{(4)},
\label{m2}
\end{eqnarray}
with $s=5,6$, $\stackrel{56\;}{(\pm)^{\dagger}}=
- \stackrel{56}{(\mp)}$ and  $\stackrel{56\;}{[\pm]^{\dagger}}= \stackrel{56}{[\mp]}$, 
while $(^*)$ means complex conjugation. 
We took into account that $\gamma^0 \stackrel{56}{(+)} = - \stackrel{56}{(+)}\gamma^0$, while 
$\gamma^0\stackrel{56}{[-]}= \stackrel{56}{[-]}\gamma^0.$ 
We find that $ \Tr_{56}(\stackrel{56\;}{(+)^{\dagger}} \stackrel{56}{(+)}) = \Tr_{56} 
(\stackrel{56}{[-]})=1$ and $ \Tr_{56}(\stackrel{56\;}{[-]^{\dagger}} \stackrel{56}{[-]}) = \Tr_{56} 
(\stackrel{56}{[-]})=1$. In order that 
$\int dx^5 dx^6 \Tr_{56}(\psi^{(4)\dagger} \gamma^0 (-{\cal A}^* \stackrel{56\;}{(+)^{\dagger}}
+ {\cal B}^* \stackrel{56\;}{[-]^{\dagger}})\gamma^s p_s 
({\cal A} \stackrel{56}{(+)}
+ {\cal B} \stackrel{56}{[-]})\psi^{(4)}$ will appear in $d=1+3$ as a mass term $  
\psi^{(4)\dagger} \gamma^0 (-m) \psi^{(4)}$, we must solve the equation 
$ \gamma^s p_s ({\cal A} \stackrel{56}{(+)}
+ {\cal B} \stackrel{56}{[-]}) = 
(-m) (-{\cal A} \stackrel{56}{(+)}
+ {\cal B} \stackrel{56}{[-]})$.

We can rewrite equations of motion in terms 
of the two complex superposition of $x^5$ and $x^6$: $z: = x^5 + ix^6$ and  $\bar{z}: =  
x^5 - ix^6$
and their derivatives, defined as $\frac{\partial}{\partial z}: = \frac{1}{2}(
\frac{\partial}{\partial x^5} -i \frac{\partial}{\partial x^6}), $ 
$\frac{\partial}{\partial \bar{z}}: = \frac{1}{2}(
\frac{\partial}{\partial x^5} + i \frac{\partial}{\partial x^6}) $ and in terms of the 
two projectors $\stackrel{56}{[\pm]}: = \frac{1}{2}(1\pm i \gamma^5 \gamma^6)$
as follows
\begin{eqnarray}
2i\gamma^5 \{ \frac{\partial}{\partial z} \stackrel{56}{[-]} 
+ \frac{\partial}{\partial \bar{z}} \stackrel{56}{[+]} \}
({\cal A} \stackrel{56}{(+)} + {\cal B} \stackrel{56}{[-]})= 
- m (-{\cal A} \stackrel{56}{(+)} + {\cal B} \stackrel{56}{[-]}).  
\label{equationsin56}
\end{eqnarray}
Since in Eq.(\ref{equationsin56}) $\psi^{(4)}$ would be just  a spectator, we skipped it.

In the massless case the superposition of the first two states 
($\psi^{(6)m=0}_{+}=
\stackrel{56}{(+)} \psi^{(4)m=0}_{+}$, with $\psi^{(4)m=0}_{+} = (\alpha \stackrel{03}{(+i)} 
\stackrel{12}{(+)} + \beta \stackrel{03}{[-i]} \stackrel{12}{[-]})\psi_0$) 
or the second two states ( $\psi^{(6)m=0}_{-}=
\stackrel{56}{[-]} \psi^{(4)m=0}_{-}$, with $\psi^{(4)m=0}_{-} = (\alpha \stackrel{03}{[-i]} 
\stackrel{12}{(+)} + \beta \stackrel{03}{(+i)} \stackrel{12}{[-]})\psi_0$) 
of the left handed Weyl representation presented in Eq.(\ref{weylrep}) must be taken, with the ratio 
of the two parameters $\alpha$ and $\beta$ 
determined by the dynamics in $x^m$ space. In the  massive case  
$\psi^{(6)m}$ is the superposition of 
all the states to which $\gamma^5$ and $\gamma^0$ separately transform the starting state:  
$\psi^{(6)m} = ({\cal A} \stackrel{56}{(+)} + {\cal B} \stackrel{56}{[-]}) \psi^{(4)m}_{\pm}$, 
with $\psi^{(4)m}_{\pm} = \{\alpha [ \stackrel{03}{(+i)} 
\stackrel{12}{(+)} \pm  \stackrel{03}{[-i]} \stackrel{12}{(+)}] + \beta
[\stackrel{03}{[-i]} \stackrel{12}{[-]} \pm \stackrel{03}{(+i)} \stackrel{12}{[-]}]\}\psi_0. $
The sign $\pm$ denotes the eigenvalue of $\gamma^0$ on these states. 

We shall therefore simply write (as suggested in Eq.(\ref{psi6}))
$\psi^{(6)} = ({\cal A} \stackrel{56}{(+)} + {\cal B} \stackrel{56}{[-]}) \psi^{(4)}$
in the massless and the massive case, taking into account that in the massless case either 
${\cal A}$ or ${\cal B}$ is nonzero, while in the massive case both are nonzero. Accordingly 
also $\psi^{(4)}$ differs in the massless and the massive case.

We want our states to be eigenstates of the total angular momentum operator $M^{56}$ around 
a chosen origin in the flat 
two dimensional manifold ($M^{(2)}$)  
\begin{eqnarray}
M^{56} = z \frac{\partial}{\partial z} - \bar{z} \frac{\partial}{\partial \bar{z}} + 
S^{56}.
\label{mab}
\end{eqnarray}

Taking into account that $\gamma^5 \stackrel{56}{(+)} = - \stackrel{56}{[-]}$, 
$\gamma^0 \stackrel{56}{(+)} = - \stackrel{56}{(+)}\gamma^0$ and 
$\gamma^5 \stackrel{56}{[-]} =  \stackrel{56}{(+)}$,
$\gamma^0 \stackrel{56}{[-]} =  \stackrel{56}{[-]} \gamma^0 
$  (see Appendix\ref{appendixtechnique}) 
we end up with equations for ${\cal A}$ and ${\cal B}$
\begin{eqnarray}
\frac{\partial {\cal B}}{\partial z} + \frac{im}{2} {\cal A} =0,\nonumber\\
\frac{\partial {\cal A}}{\partial \bar{z}}  + \frac{im}{2} {\cal B}=0.  
\label{equationsin56red1}
\end{eqnarray}

For $m=0$ we get as solutions 
\begin{eqnarray}
\psi^{(6)m=0}_{n+1/2} &=& a_n z^n \stackrel{56}{(+)} \psi^{(4)}_{+},\nonumber\\
\psi^{(6)m=0}_{-(n+1/2)} &=& b_n \bar{z}^n \stackrel{56}{[-]} \psi^{(4)}_{-},\; n \ge 0.
\label{solmeq0}
\end{eqnarray}
We required $n\ge 0$ to ensure the integrability of solutions at the origin. The solutions
have  the eigenvalues of $M^{56}$ equal to $(n+1/2)$ and $-(n+1/2)$, respectively. 

Since in the massless case the contribution from $(p^5)2$ compensates the one from 
$(p^6)2$ for all the solutions from Eq. (\ref{solmeq0}) with 
$n \ge 1$ and has therefore obviously one of the two contributions to the zero $m2$ 
a negative real value unless $n=0$, it seems natural to  expect that the 
only massless solutions are the two solutions 
 with the eigenvalues $M^{(56)}$ equal to $1/2$ for the 
right handed spinor ($\psi^{(6)m=0}_{1/2} = a_0 \stackrel{56}{(+)} \psi^{(4)}_{+}$) and to 
$-1/2$ for the left handed spinor ($\psi^{(6)m=0}_{-1/2} = b_0 \stackrel{56}{[-]} \psi^{(4)}_{-}$),
and accordingly with the corresponding $\psi^{(4)m=0}_{+}$ and $\psi^{(4)m=0}_{-}$ of the left and 
right handedness in $d=1+3$, respectively. 
We shall reformulate the operator of momentum to be Hermitean
on the vector space of solutions fulfilling the involution boundary condition in sect.\ \ref{hermiticity}.
Having solutions of both handedness we must conclude that in such cases there is no mass protection.

For $m\ne0$ we get
\begin{eqnarray}
\psi^{(6)m}_{n+1/2} = a_n (J_n  \stackrel{56}{(+)}   -i J_{n+1}e^{i\phi} 
\stackrel{56}{[-]})e^{\pm in\phi} \psi^{(4)m},\; {\rm for \; n \ge 0},
\label{solmeqm}
\end{eqnarray}
where $J_n$ is the Bessel's functions of the first order. 
The easiest way to see that  $J_n$ and $J_{n+1}$ determine the massive solution  
is to  use Eq.(\ref{equationsin56red1}), take into account that $z=\rho e^{i\phi}$, 
define  $r=m \rho, \rho=\sqrt{(x^5)2 + (x^6)2}$,
recognize that $\frac{\partial}{\partial z} = \frac{1}{2} e^{-i\phi} (\frac{\partial}{\partial \rho} 
- \frac{i}{\rho} \frac{\partial}{\partial \phi})$ and we find 
 ${\cal B}= - \frac{2}{im} \frac{\partial {\cal A}}{\partial \bar{z}}$. 
 Then for the choice ${\cal A} = J_n 
e^{in \phi}$ it follows  that ${\cal B} = -i e^{i(n+1)\phi} (\frac{n}{r} J_{n} -  
 \frac{\partial J_n}{\partial r}) $, which tells that  
${\cal B} = -i J_{n+1} e^{i(n+1)\phi}.$

\section{Boundary conditions and involution}
\label{involution}

In the ref.\ \cite{hnkk06} we make a choice of particular solutions of the equations of motion 
by requiring that  $\hat{\cal{R}}\psi|_{\rm wall} = 0,$ where the wall were put on the circle 
of the radius $\rho_0$ of 
the finite disk (Eq.(\ref{prodiskboundary})). 

This boundary condition requires that 
in the massless case (since $\stackrel{56}{[-]} \stackrel{56}{(+)} =0$ while 
$\stackrel{56}{[-]} \stackrel{56}{[-]} = \stackrel{56}{[-]} $)  only the right handed 
solution (Eq.\ref{solmeq0}) $\psi^{(6)m=0}_{1/2} = a_0 \stackrel{56}{(+)} \psi^{(4)m=0}_{+}$ 
(that is the left handed with respect to $SO(1,3)$) is allowed, while the left 
handed solution must be zero ($b_n=0$) making the mass protection mechanism work in $d=1+3$.  

In the massive case the boundary condition determines masses of solutions, since only the 
solutions with $J_{n+1}|_{\rho = \rho_0} =0$ are allowed from the same reason as discussed 
for the massless case. This boundary condition determines masses of spinors through the relation 
$m_{n+1/2} \rho_0$ is equal to a zero of $J_{n+1}$:
 $$J_{n+1}(m_{n+1/2} \rho_0)=0.$$ 



This time we look for the {\em involution boundary conditions}.

First we recognize that for a flat $M2$ - $\{0\}$ manifold, with the origin $x^5=0=x^6$ 
excluded, the $Z_2$ or involution symmetry can be recognized: {\em The transformation 
$\rho/\rho_{0} \rightarrow \frac{\rho_0}{\rho}$} (which can be written also as $z/\rho_0 \rightarrow 
\frac{\rho_0}{\bar{z}}$) {\em transforms the exterior of the disk into the interior of the disk} 
and conversely. 

Then we extend the involution operator to operate also on the space of solutions
\begin{eqnarray}
\hat{{\cal O}} &=& (I - 2 \hat{{\cal R'}})|_{z/\rho_0 \rightarrow 
\rho_0/\bar{z}}, \nonumber\\
\hat{{\cal O}}^2 &=& I.
\label{involutionO}
\end{eqnarray}
The involution condition $\hat{{\cal O}}^2 =I$ requires, that $\hat{{\cal R'}}$ 
is a projector 
\begin{eqnarray}
(\hat{{\cal R'}})^2=  \hat{{\cal R'}}
\label{involutionR}
\end{eqnarray}
and can be written as $\hat{{\cal R'}}=\hat{{\cal R}} +  \hat{{\cal R}}_{add}$, 
where $\hat{{\cal R}}_{add}$ must be a nilpotent operator  fulfilling 
the conditions 
\begin{eqnarray}
(\hat{{\cal R}}_{add})^2= 0,\;\; \hat{{\cal R}}_{add} \hat{{\cal R}} =0,\;\;
\hat{{\cal R}} \hat{{\cal R}}_{add} = \hat{{\cal R}}_{add}, 
\label{involutionRadd}
\end{eqnarray}
 We had $\hat{{\cal R}} = \stackrel{56}{[-]},$ which is the projector. 
Since we find that $ \stackrel{56}{[-]} \stackrel{56}{(-)} = \stackrel{56}{(-)}$ 
(see Appendix \ref{appendixtechnique}), while 
$ \stackrel{56}{(-)} \stackrel{56}{[-]} = 0,$ we can choose $\hat{{\cal R}}_{add} = \alpha
 \stackrel{56}{(-)}$, where $\alpha$ is any function of  $z$ and $\frac{\partial}{\partial z}$. 
Let us point out that $\hat{{\cal R}}_{add}$ is not a Hermitean operator, since 
$\stackrel{56\;}{(-)^{\dagger}}= -\stackrel{56}{(+)}$ and $z^{\dagger} = \bar{z}, 
(\frac{\partial}{\partial z})^{\dagger}
=\frac{\partial}{\partial {\bar z}}$. Accordingly also neither $\hat{{\cal R'}}$ nor 
$\hat{{\cal O}}$ is a Hermitean operator.

We now make a choice of a natural boundary conditions on the wall $\rho = \rho_0$ 
\begin{eqnarray}
\{\hat{{\cal O}} \psi&=& \psi \}|_{\rm wall}, 
\label{involutionOwall}
\end{eqnarray}
saying that what ever the involution operator is, the state $\psi $ and its 
involution $\hat{{\cal O}}\psi$ must be the same on the wall, that is at $\rho=\rho_0$.

It is worthwhile to write the involution operator $\hat{{\cal O}}$ and correspondingly the 
projector $\hat{{\cal R'}}$ in a covariant way. 
Recognizing that $n^{(\rho)}{}_{\!\! a} \,\gamma^a n^{(\rho)}{}_{\!\! b} \, p^b = 
[e^{2i\phi} \frac{1}{2}(p^5 - i p^6) + 
\frac{1}{2}(p^5 + i p^6)]\stackrel{56}{(-)} +  [\frac{1}{2}(p^5 + i p^6) + 
e^{-2i\phi} \frac{1}{2}(p^5 + i p^6)] \stackrel{56}{(+)}$, we may write $\frac{1}{2}
 (1-i n^{(\rho)}{\!\!}_{a} \,n^{(\phi)}{\!\!}_{b} \,
\gamma^a \gamma^b ) (1 - \beta n^{(\rho)}{\!\!}_a \,\gamma^a \,n^{(\rho)}{\!\!}_b \,p^b ) = 
\stackrel{56}{[-]}( I + \beta i [e^{2i\phi} \frac{\partial}{\partial z} +  
\frac{\partial}{\partial \bar{z}}]\stackrel{56}{(-)})$. This is just 
our generalized projector $\hat{{\cal R'}}$, if we  make a choice for   
$\alpha$ from Eq.(\ref{involutionRadd}) as follows: $\alpha = \beta i 
[e^{2i\phi} \frac{\partial}{\partial z} +  
\frac{\partial}{\partial \bar{z}}]$ (since $\stackrel{56}{[-]}\stackrel{56}{(-)} =
\stackrel{56}{(-)})$.  We then have
\begin{eqnarray}
\hat{{\cal R'}} &=& \stackrel{56}{[-]}( I + \beta i [e^{2i\phi} \frac{\partial}{\partial z} +  
\frac{\partial}{\partial \bar{z}}]\stackrel{56}{(-)}),
\label{involutionR'gen}
\end{eqnarray}
where $\beta $ is any complex number.

\section{Current through the wall}
\label{s:current}

The current perpendicular to the wall can be written as
\begin{eqnarray}
n^{(\rho)s} j_s &=&\psi^{\dagger} \gamma^0 \gamma^s n^{(\rho)}_s \psi = 
\psi^{\dagger}\gamma^0 (-)\{ e^{-i\phi} \stackrel{56}{(+)} + 
e^{i\phi} \stackrel{56}{(-)}\} \psi =  \psi^{\dagger}\hat{j}_\perp \psi, \nonumber\\
\hat{j}_\perp &=& - \gamma^0 \{ e^{-i\phi} \stackrel{56}{(+)} + 
e^{i\phi} \stackrel{56}{(-)}\}. 
\label{current}
\end{eqnarray}
We need to know the current through the wall, which for physically acceptable cases when 
spinors are localized inside the disk (involution transforms outside the 
disk into its inside, or equivalently, it transforms inside the 
disk into its outside)  must be zero.
We find for the current
\begin{eqnarray}
\{\psi^{\dagger}\hat{j}_\perp\psi\}|_{\rm wall} =  
 \{\psi^{\dagger} {\hat{\cal O}}^{\dagger}\hat{j}_\perp{\hat{\cal O}} \psi\}|_{\rm wall}. 
\label{current1}
\end{eqnarray}
Since $\hat{{\cal O}}^{\dagger} = I - 2 (\hat{{\cal R}} + \hat{{\cal R}}^{\dagger}_{add})$
and $\hat{{\cal R}}^{\dagger}_{add} = 
(\alpha  \stackrel{56}{(-)})^{\dagger} = 
-\alpha^*  \stackrel{56}{(+)}$, it follows that
$\hat{{\cal O}}^{\dagger}\hat{j}_\perp\hat{{\cal O}} = - \hat{j}_\perp
- 2\alpha^* \gamma^0 e^{i\phi}  \stackrel{56}{[+]}
- 2\alpha \gamma^0 e^{-i\phi}  \stackrel{56}{[+]}$.

It must then be
\begin{eqnarray}
\{\psi^{\dagger}\hat{j}_\perp\psi\}|_{\rm wall} =  
 (- \psi^{\dagger} \{\hat{j}_\perp  + 2 \gamma^0 (\alpha^* 
 e^{i\phi}  + \alpha e^{-i\phi} 
 ) \stackrel{56}{[+]}\} \psi )|_{\rm wall}. 
\label{currentwall}
\end{eqnarray}

First we check  the current on the wall for the "old" case, when $\alpha =0$ 
and $\hat{\cal O} = I- 2 \hat{\cal R},  
\hat{\cal R} = \stackrel{56}{[-]} $.  
Not to be in contradiction with Eq.(\ref{currentwall}) the current on the wall must 
for either massless or massive case be zero.
In the case of massless solutions (Eq.(\ref{solmeq0})) only $\psi^{(6)m=0}_{n+1/2}$ can  
fulfill this boundary condition ($\psi^{(4)m=0 \dagger} {\bar z}^n (-) \stackrel{56}{(-)} 
\{-\gamma^0(e^{-i\phi} \stackrel{56}{(+)} + e^{i\phi} \stackrel{56}{(-)})\} z^n \stackrel{56}{(+)} 
\psi^{(4)m=0}_{+})|_{\rm wall} = 0$, for each nonnegative $n$. 
The chosen boundary condition accordingly 
allows only the right handed solutions.
We shall conclude when discussing Hermiticity of the operators that 
only $n=0$ is the physically acceptable solution.  

%
In the massive case the solutions of equations of motion (Eq.(\ref{solmeqm})) contribute 
no current through the wall, if $J_{n+1}|_{\rm wall}=0,$ which is exactly what the boundary 
condition (Eq.(\ref{involutionOwall}))  $\hat{\cal O}\psi|_{wall}= \psi|_{wall}$  required.  

Then we check the general case with $\hat{\cal O}= I - 2\hat{\cal R'}$, 
where $\hat{\cal R'} = \hat{\cal R} +
\hat{\cal R}_{add} = \stackrel{56}{[-]} + \beta i [e^{2i\phi} \frac{\partial}{\partial z} +  
\frac{\partial}{\partial \bar{z}}]\stackrel{56}{(-)}$.
For massless solutions 
it is not difficult to see that {\em for any nonzero choice of $\beta$ 
 only one handedness - 
the right handed one - survives and that only $n=0$ is allowed}.  
In the massive case we find
\begin{eqnarray}
\{-i\beta [ e^{2i\phi} \frac{\partial {\cal A}}{\partial z} +  
\frac{\partial {\cal A}}{\partial \bar{z}} ] + {\cal B} \}|_{\rm wall}=0.
\label{masscond2}
\end{eqnarray}
Since equations of motion require that ${\cal B} = - \frac{2}{im} 
\frac{\partial {\cal A}}{\partial \bar{z}}$ and since
$\frac{\partial}{\partial \bar{z}} = \frac{1}{2} 
e^{i\phi}(\frac{i}{\rho} \frac{\partial }{\partial \phi} + 
\frac{\partial }{\partial \rho})$, we fulfill the involution condition on the wall 
for ${\cal A} = J_n e^{in \phi}$ only if  ${\cal B} = -i J_{n+1} e^{i(n+1) \phi}$, with 
the requirement that $J_n|_{\rm wall} =0$ and $\beta = \frac{1}{m}$. 
While $J_n|_{\rm wall} =0$ can always be fulfilled, the second requirement $\beta = \frac{1}{m}$ 
means, since $\beta$ can not be an arbitrary number,  that our generalized 
condition is  not 
written in an covariant form, and is accordingly not the acceptable boundary condition.

\section{Hermiticity of operators and the orthogonality of solutions}
\label{hermiticity}


In this section we comment on the Hermiticity properties of the 
operators, in particular of $ p_s$ 
and on the orthogonality properties of those solutions of the equations of 
motion which fulfill the involution boundary conditions. 
 We  expect the solutions\\
 i) to be orthogonal 
 ($ \int d^2x 
 \psi_{i}{\!}^{\dagger}((p^5)2 + (p^6)2)  \psi_j = 
\int d^2x \psi_{i}{\!}^{\dagger}  \psi_j m2  \delta_{ij}$) and that \\
ii) on the space of these solutions   
the operators $ p_s$ are Hermitean and have accordingly expectation values of the 
operators $(p^s)2$ positive contribution  to  $ m2$ for each $s$.

Let us first check the orthogonality relations of the massive and massless solutions. 
We immediately see that the massive solutions $\psi^{(6)m}_{n+1/2}$ belonging to 
different $n$ are all orthogonal due to the orhogonality of the functions $e^{in\phi}$. 
We find $\int d^2x \Tr_{56}
(\psi^{(6)m\dagger}_{ n+1/2} 
\psi^{(6)m}_{\! k+1/2})= \delta_{n k} \; a^{*}_{\!n} a_{n} \frac{1}{2}
\psi^{(4)m\dagger} \psi^{(4)m}$$
\int^{\rho_0}_{0} ( J^{*}_{\!n}J_n + J^{*}_{\!n+1}J_{n+1}) \rho d\rho$.

It also  turns out that the massless solutions 
($\psi^{(6)m=0}_{n+1/2}$ (Eq.(\ref{solmeq0}))  
are orthogonal to all the massive states  (Eq.(\ref{solmeqm}))  
due to the properties of the $J_n$ Bessel's function. Namely, 
$$\int d^2x \Tr_{56}
(\psi^{(6)m=0\dagger}_{n+1/2} 
\psi^{(6)m}_{\! k+1/2}) = \delta_{nk} a^{0*}_{\!n+1/2} 
a_{k} \frac{1}{\sqrt{2}}
\psi^{(4)m=0\dagger} \psi^{(4)m}
\int^{\rho_0}_{0} \rho^n  J_n  \rho d\rho=0,$$ 
since $\int^{\rho_0}_{0} \rho^n J_ n \rho d\rho= 
\rho_0^{n+1} J_{n+1}(\rho_0)$, but $J_{n+1}(\rho_0)$ must be zero in order that 
the massive state with 
$n+1/2$ obeys the involution boundary condition. 
Massless solutions are again due to the $e^{in\phi} $ part orthogonal among themselves. 

{\em So we conclude that all the states, which obey the equations of motion and 
the involution boundary condition,  are orthogonal.}

Are $p_s$ Hermitean on the space of these solutions?

We know that $p_s = -i\frac{\partial}{\partial x^s} $  is Hermitean on the
 vectors space $\psi_i$ if for any two functions $\psi_i$  and $ \psi_j$ from the vector 
  space of solutions   $\Tr_{56}(\int d^2x p_s(\psi_{i}^{\dagger}  \psi_j)) =0$ (since then 
 $\int d^2x \psi_{i}{\!}^{\dagger} p_s \psi_{j} +  
 \int d^2x (-p_s \psi_{i})^{\dagger} \psi_{j}=0$). 

 We find that 
 \begin{eqnarray}
 p_s= i\frac{\partial}{\partial x^s} = i  
 \begin{pmatrix}\cos{\phi} \frac{\partial}{\partial \rho} - 
 \sin{\phi} \frac{1}{\rho} \frac{\partial}{\partial \phi} \\
 \sin{\phi} \frac{\partial}{\partial \rho} +  
 \cos{\phi} \frac{1}{\rho} \frac{\partial}{\partial \phi}\end{pmatrix},
 \label{ps}
 \end{eqnarray}
for $s=5$ (first row) and $6$(second row). Since 
 either messless ($\psi^{(6)m=0}_{n+1/2}$, Eq.(\ref{solmeq0})) or massive 
 ($\psi^{(6)m}_{n+1/2}$,  Eq.(\ref{solmeqm})) states can be written as a product of $e^{in\phi}$ 
 and the rest, say $\psi_n$, we see that $\int d^2x p_s(\psi_{n}^{\dagger} \psi_{k})$ 
is nonzero only if $|n-k| =1$. 


In this case we get that the integral 
$\Tr_{56}(\int d^2x p_s(\psi_{n}{\!}^{\dagger}  \psi_{n\pm 1})), s=x^5,x^6,$ proportional to 
$i\pi (^{1}_{i}) |\rho \psi^{\dagger}_n \psi_{n+1}|_{\rho_{0}}$, 
with $|\rho \psi^{\dagger}_n \psi_{n+1}|^{\rho_0}_0 $ 
equal to \\
i)$a^{m=0*}_{n}a^{m=0}_{n+1} (\rho_0)^{2(n+1)}$ in the case that two massless states are concerned, \\ 
ii) $a^{m=0*}_{n}a^{m}_{n+1} \rho_{0}^{n+1} J_{n+1}(\rho_0)$ in the case that one massless and one  
massive state are concerned, \\
iii) $a^{m*}_{n}a^{m}_{n+1} \rho_0(J_{n}J_{n+1} + J_{n+1}J_{n+2})|_{\rho_0}$ in 
 the case that two massive states are concerned. None of these integral is zero, since the two 
 $J_n$ and $J_{n+1}$ are not correlated ($J_n$ and $J_{n+1}$ are correlated, if both 
 belong to the solution with the same mass, determined by $J_{n+1}(m\rho =m\rho_0)=0$).
 We conclude that for none of the solutions $p_s$ are Hermitean operators. 
 
 {\em One can check}, however, that $\hat{p}_s$ 
  \begin{eqnarray}
  \hat{p}_s= i\{ \frac{\partial}{\partial x^s} -  \frac{1}{2} \frac{x^s}{\rho}
  \delta(\rho-\rho_0)\stackrel{56}{[+]}\},
  \label{ps1}
  \end{eqnarray}
{\em are Hermitean operators on the space of massive and massless solutions, fulfilling the
involution boundary conditions}. It contains the part with the $\delta$ function which corrects  
those parts of solutions, which are nonzero on the wall---the radial parts which appear 
with $\stackrel{56}{(+)}$.  
It can be shown that the integral over the part with the
 $\delta(\rho-\rho_0)$ function  contributes just the terms which compensate the nonzero 
 contribution in each of the three cases i)-iii).
 
 
 What we must check now is, what appears in this new definition of the operator of the momenta 
 (Eq.(\ref{ps})) as $\gamma^s p_s \gamma^t p_t$ and whether now the integral 
 $\Tr_{56}(\int d^2x \,\psi^{\dagger} \,\gamma^s p_s \gamma^t p_t \,\psi), s=x^5,x^6,$ 
 is still manifesting as just the mass term for those  
$\psi$ which we accept as solutions of equations of 
motion (Eq.(\ref{solmeq0},\ref{solmeqm})). 


One finds 
  \begin{eqnarray}
  \gamma^s \hat{p}_s \gamma^t \hat{p}_t &=& p_s p^s \nonumber\\
  &+&\frac{1}{2}\{ [\frac{\partial}{\partial \rho}\delta(\rho -\rho_0) + \frac{1}{\rho} 
  \delta(\rho-\rho_0) + \delta(\rho-\rho_0)
  ( \frac{\partial}{\partial \rho} - \frac{i}{\rho} \frac{\partial}{\partial \phi})]
  \stackrel{56}{[+]}\nonumber\\ 
  &+& \delta(\rho-\rho_0) (\frac{\partial}{\partial \rho} + 
 \frac{i}{\rho}  \frac{\partial}{\partial \phi})\stackrel{56}{[-]}\}.
  \label{pssquared}
  \end{eqnarray}
One notices that the first row of Eq.(\ref{pssquared}) represent the usual momentum squared. 
The last two terms are zero everywhere except on the wall. What we must check is the integral 
of the last two terms for all solutions fulfilling our involution boundary condition.

We find 
that the integral $\Tr_{56}(\int d^2x \,(z \stackrel{56}{(+)})^{\dagger} 
\,\gamma^s p_s \gamma^t p_t \,z^n \stackrel{56}{(+)})$ is for massless 
solutions (Eq.(\ref{solmeq0})) obeying the involution boundary condition proportional 
to $n$ and it is zero only for $n=0$. 

The integral $\Tr_{56}(\int d^2x \,(z \stackrel{56}{(+)})^{\dagger} 
\,\gamma^s p_s \gamma^t p_t \,z^n \stackrel{56}{(+)})$  demonstrates for massive solutions 
(Eq.(\ref{solmeqm})) the mass  term squared originating in the first row of Eq.(\ref{pssquared}), 
while the rest contributes zero.

{\em The requirement that the integral $\Tr_{56}(\int d^2x \,\psi^{\dagger} 
\,\gamma^s p_s \gamma^t p_t \,\psi), s=x^5,x^6,$  must be zero for massless solutions, makes a choice 
of only one among all possible massless solutions: the one with $n=0$.}


Our the only possible solution is in the massless case $\psi^{(6)m=0}_{1/2}$.
For the massive solutions we have   
$\psi^{(6)m}_{1/2}= a_{1/2} \frac{1}{\sqrt{2}} 
(J_0 \stackrel{56}{(+)} -i J_{1} e^{i\phi} \stackrel{56}{[-]})$, 
with $m_{1/2}\rho_0 $  as a zero of 
$J_{1}$, $\psi^{m}_{3/2}= a_{3/2} \frac{1}{\sqrt{2}} 
(J_1 -i J_{2} e^{i\phi}) e^{i\phi})$, with $m_{3/2}\rho_0 $ as a zero of 
$J_{2}$, 
$\psi^{(6)m}_{-1/2}= a_{-1/2} \frac{1}{\sqrt{2}} 
(J_1 \stackrel{56}{(+)} -i J_{0} \stackrel{56}{[-]}e^{-i\phi}) ) e^{-i\phi})$, 
with $m_{-1/2}\rho_0 $ equal to a zero of $J_{0} $
and so on.

\section{Properties of spinors in $d=1+3$}
\label{properties1+3}

To study how do spinors couple to the Kaluza-Klein gauge fields in the case of $M^{(1+5)}$, ''broken'' to 
$M^{(1+3)} \times $ a flat disk with $\rho_0$ and with the involution boundary condition, 
which allows only right handed spinors
at $\rho_0$,
we first look for (background) gauge gravitational fields, which preserve the rotational symmetry 
on the disk. Following ref.\ \cite{hnkk06} we find 
for the background vielbein field  
\begin{eqnarray}
e^a{}_{\alpha} = 
\begin{pmatrix}\delta^{m}{}_{\mu}  & e^{m}{}_{\sigma}=0 \\
 e^{s}{}_{\mu} & e^s{}_{\sigma} \end{pmatrix},
f^{\alpha}{}_{a} =
\begin{pmatrix}\delta^{\mu}{}_{m}  & f^{\sigma}{}_{m} \\
0= f^{\mu}{}_{s} & f^{\sigma}{}_{s} \end{pmatrix},
\label{f6}
\end{eqnarray}
with $f^{\sigma}{}_{m} = A_{\mu} \delta ^{\mu}{}_{m}
\varepsilon^{\sigma}{}_{\tau} x^{\tau}$
and  the spin connection field 
\begin{eqnarray}
\omega_{st \mu} = - \varepsilon_{st}  A_{\mu},\quad \omega_{sm \mu} = 
-\frac{1}{2} F_{\mu \nu} \delta^{\nu}{}_{m}
\varepsilon_{s \sigma} x^{\sigma}.
\label{omega6}
\end{eqnarray}
%
%
 The $U(1)$ gauge field $A_{\mu}$ depends only on $x^{\mu}$.
All the other components of the spin connection fields are zero, since for simplicity we allow no gravity in
$(1+3)$ dimensional space.

To determine the current, coupled to the Kaluza-Klein gauge fields $A_{\mu}$, we
analyze the spinor action
\begin{eqnarray}
{\cal S} &=& \int \; d^dx E \bar{\psi}^{(6)} \gamma^a p_{0a} \psi^{(6)} = \int \; 
d^dx  \bar{\psi}^{(6)} \gamma^m \delta^{\mu}{}_{m} p_{\mu} \psi^{(6)} + \nonumber\\
&& \int \; d^dx   \bar{\psi}^{(6)} \gamma^m (-)S^{sm} \omega_{sm \mu} \psi^{(6)}  + 
\int \; d^dx  \bar{\psi}^{(6)} \gamma^s \delta^{\sigma}{}_{s} p_{\sigma} \psi^{(6)} +\nonumber\\
&& \int \; d^dx   \bar{\psi}^{(6)} \gamma^m  \delta^{\mu}{}_{m} A_{\mu} 
(\varepsilon^{\sigma}{}_{\tau} x^{\tau}
 p_{\sigma} + S^{56}) \psi^{(6)}.
\label{spinoractioncurrent}
\end{eqnarray}
 $\psi^{(6)}$ are solutions of the Weyl equation in $d=1+3$ .
 $E$ is for $f^{\alpha}{}_{a}$ from (\ref{f6}) equal to 1. 
The first term on the right hand side  of Eq.(\ref{spinoractioncurrent}) is the kinetic term
(together with the last  term defines  
the  covariant derivative $p_{0 \mu}$ in $d=1+3$).  
The second term on the right hand side  contributes nothing when integration over 
the disk is performed, since it is proportional to $x^{\sigma}$ ($\omega_{sm \mu} = -\frac{1}{2}
F_{\mu \nu} \delta^{\nu}{}_{m} \varepsilon_{s \sigma} x^{\sigma}$).

We end up with 
\begin{eqnarray}
j^{\mu} = \int \; d^2x \bar{\psi}^{(6)} \gamma^m \delta^{\mu}{}_{m} M^{56}  \psi^{(6)}
\label{currentdisk}
\end{eqnarray}
as  the current in $d=1+3$.  The charge in $d=1+3$ is  proportional to the total 
angular momentum  $M^{56} =L^{56} + S^{56}$ on a disk, for either massless or massive spinors.


\section{Conclusions}
\label{discussions}

In this paper we were looking for what we call a "natural boundary condition"---a  
condition which would, due to some symmetry relations, make massless spinors which live in  
$M^{1+5}$ and carry nothing but the charge to live in $M^{(1+3)} \times $ a 
flat disk, manifesting in $M^{(1+3)}$, if massless,  
as a left handed spinor (with no right handed partner) and would accordingly be mass 
protected. The spin in $x^5$ and $x^6$ of the left handed massless spinor should  in $M^{(1+3)}$
manifest as the charge  and   
should chirally couple with  the Kaluza-Klein charge of only one 
value to the corresponding gauge field, in order that  after 
the second quantization procedure a particle and an antiparticle would not appear each of both 
charges.

We found the involution boundary condition
\begin{eqnarray}
\{\hat{{\cal O}} \psi&=& \psi \}|_{\rm wall}, \quad 
{\cal O} = (I - (I-i n^{(\rho)}{}_{a} n^{(\phi)}{}_{b} 
\gamma^a \gamma^b ))_{\frac{\rho}{\rho_0}\rightarrow \frac{\rho_0}{\rho}}, \quad {\cal O}^2 = I,
\label{involutionOwallc}
\end{eqnarray}
which transforms  solutions of the Weyl equations inside the flat disk 
into outside of it (or conversely) and allows in the massless case only the right  
handed spinor to live  
on the disk and accordingly manifests left handedness in $M^{(1+3)}$.  
The massless solution carries in the fifth and sixth dimension  (only) the spin $1/2$, 
which then manifests  as the charge in $d=1+3$. 

We defined a generalized momentum $p_s$
  \begin{eqnarray}
  \hat{p}_s= i\{ \frac{\partial}{\partial x^s} -  \frac{1}{2}\begin{pmatrix}\cos{\phi}\\
  \sin{\phi} \end{pmatrix} \delta(\rho-\rho_0) \stackrel{56}{[+]}\},
  \label{psc}
  \end{eqnarray}
which is the Hermitean operator in the case of our involution boundary condition.

The requirement that $\gamma^s \hat{p}_s \gamma^t \hat{p}_t$ manifests as the square of the mass
leads in the massless case to to the solution with the total angular momentum $1/2$ as the only solution, 
while the massive solutions carry all half integer angular momenta: $\pm 1/2, \pm 3/2, \cdots$.
The angular momenta in the fifth and sixth dimensions then manifests as the charge in 
the $1+3$ dimension. The massless solution with the spin $1/2$ is mass 
protected and chirally coupled to the corresponding Kaluza-Klein field. 

The negative charge of the  massless $1/2$ charge state appears only after the second quantization 
procedure in agreement with what we observe.  

All the solutions fulfilling the involution 
boundary conditions are orthogonal and in this vector space and the generalized operators are 
Hermitean.  

The involution boundary condition of Eq.(\ref{involutionOwallc}) are equivalent 
to the boundary condition, which we present in the ref.\ \cite{hnkk06}. Both take care 
that massless solutions of one handedness appear in $d=1+3$. 

We were looking for generalized boundary conditions with
\begin{eqnarray}
{\cal {\hat O}} &=& I-2 \hat{{\cal R'}}, \nonumber\\
\hat{{\cal R'}} &=& \stackrel{56}{[-]}( I + \beta i [e^{2i\phi} \frac{\partial}{\partial z} +  
\frac{\partial}{\partial \bar{z}}]\stackrel{56}{(-)}),
\label{involutionR'genc}
\end{eqnarray}
where $\beta $ is any complex number. This generalized boundary $\hat{{\cal R'}}$ can be 
written in a covariant way  as
\begin{eqnarray}
\hat{{\cal R'}} &=& \frac{1}{2}
 (1-i n^{(\rho)}{\!\!}_{a} \,n^{(\phi)}{\!\!}_{b} \,
\gamma^a \gamma^b ) (1 - \beta n^{(\rho)}{\!\!}_a \,\gamma^a
\,n^{(\rho)}{\!\!}_b \,p^b )\nonumber\\
 &=& 
\stackrel{56}{[-]}( I + \beta i [e^{2i\phi} \frac{\partial}{\partial z} +  
\frac{\partial}{\partial \bar{z}}]\stackrel{56}{(-)}).
\label{genc}
\end{eqnarray}
But while in the massless case the generalized boundary condition 
$\{\hat{{\cal O}} \psi= \psi \}|_{\rm wall}$
forbids all but $s=1/2$ solution, it fails  in the 
massive case  to demonstrate the covariance and is accordingly not an acceptable 
boundary condition.


\section{Appendix: Spinor representation technique in terms of Clifford algebra objects}
\label{appendixtechnique}

We define\cite{holgernorma2002} spinor representations as superposition of 
products of the Clifford algebra objects 
$\gamma^a$ so that they are  
eigen states of the chosen Cartan sub algebra of the Lorentz algebra $SO(d)$, 
determined by the generators 
$S^{ab} = i/4 (\gamma^a \gamma^b - \gamma^b \gamma^a)$.
By introducing the notation
\begin{eqnarray}
\stackrel{ab}{(\pm i)}: &=& \frac{1}{2}(\gamma^a \mp  \gamma^b),  \quad 
\stackrel{ab}{[\pm i]}: = \frac{1}{2}(1 \pm \gamma^a \gamma^b), \;{\rm  for} \; \eta^{aa} \eta^{bb} =-1, \nonumber\\
\stackrel{ab}{(\pm )}: &= &\frac{1}{2}(\gamma^a \pm i \gamma^b),  \quad 
\stackrel{ab}{[\pm ]}: = \frac{1}{2}(1 \pm i\gamma^a \gamma^b), \;{\rm for} \; \eta^{aa} \eta^{bb} =1,
\label{eigensab}
\end{eqnarray}
it can be checked that  the above binomials are really ''eigenvectors''  of  the generators 
$S^{ab}$
\begin{eqnarray}
S^{ab} \stackrel{ab}{(k)}: &=&  \frac{k}{2} \stackrel{ab}{(k)}, \quad 
S^{ab} \stackrel{ab}{[k]}:  =  \frac{k}{2} \stackrel{ab}{[k]}.
\label{eigensabev}
\end{eqnarray}
Accordingly we have
\begin{eqnarray}
\stackrel{03}{(\pm i)}: &=& \frac{1}{2}(\gamma^0 \mp  \gamma^3),  \quad 
\stackrel{03}{[\pm i]}: = \frac{1}{2}(1 \pm \gamma^0 \gamma^3), \nonumber\\
\stackrel{12}{(\pm )}: &= &\frac{1}{2}(\gamma^1 \pm i \gamma^2),  \quad 
\stackrel{12}{[\pm ]}: = \frac{1}{2}(1 \pm i\gamma^1 \gamma^2), \nonumber\\
\stackrel{56}{(\pm )}: &= &\frac{1}{2}(\gamma^5 \pm i \gamma^6),  \quad 
\stackrel{56}{[\pm ]}: = \frac{1}{2}(1 \pm i\gamma^5 \gamma^6), \nonumber\\
\label{eigensab031256}
\end{eqnarray}
with eigenvalues of $S^{03}$ equal to $\pm \frac{i}{2}$ for $\stackrel{03}{(\pm i)}$ and 
$\stackrel{03}{[\pm i]}$, and to $\pm \frac{1}{2}$ for  $\stackrel{12}{(\pm )}$ and 
$\stackrel{12}{[\pm ]}$, as well as for for $\stackrel{56}{(\pm )}$ and 
$\stackrel{56}{[\pm ]}$. 

We further find 
\begin{eqnarray}
\gamma^a \stackrel{ab}{(k)}&=&\eta^{aa}\stackrel{ab}{[-k]},\quad 
\gamma^b \stackrel{ab}{(k)}= -ik \stackrel{ab}{[-k]}, \nonumber\\
\gamma^a \stackrel{ab}{[k]}&=& \stackrel{ab}{(-k)},\quad \quad \quad
\gamma^b \stackrel{ab}{[k]}= -ik \eta^{aa} \stackrel{ab}{(-k)}.
\label{graphgammaaction}
\end{eqnarray}

We also find 
\begin{eqnarray}
\stackrel{ab}{(k)}\stackrel{ab}{(k)}= 0, & & \stackrel{ab}{(k)}\stackrel{ab}{(-k)}
= \eta^{aa}  \stackrel{ab}{[k]}, \quad 
\stackrel{ab}{[k]}\stackrel{ab}{[k]} =  \stackrel{ab}{[k]}, \;\;\quad \quad
\stackrel{ab}{[k]}\stackrel{ab}{[-k]}= 0, 
 \nonumber\\
\stackrel{ab}{(k)}\stackrel{ab}{[k]} = 0, & &  \stackrel{ab}{[k]}\stackrel{ab}{(k)}
=  \stackrel{ab}{(k)}, \quad \quad \quad
\stackrel{ab}{(k)}\stackrel{ab}{[-k]} =  \stackrel{ab}{(k)},
\quad \quad \stackrel{ab}{[k]}\stackrel{ab}{(-k)} =0.
\label{graphbinoms}
\end{eqnarray}

To represent one Weyl spinor in $d=1+5$, one must make a choice of the
operators belonging to the Cartan sub algebra of $3$ elements of the group $SO(1,5)$ 
\begin{eqnarray}
S^{03}, S^{12}, S^{56}.
\label{cartan}
\end{eqnarray}
Any eigenstate of the Cartan sub algebra (Eq.(\ref{cartan})) must be a product of 
three binomials, each of which is an eigenstate of one of the three elements. 
A left handed spinor ($\Gamma^{(1+5)} = -1$) representation with $2^{6/2-1}$ basic states 
is presented in Eq.(\ref{weylrep}). 
for example, the state  $\stackrel{03}{(+ i)}\stackrel{12}{(+)}
\stackrel{56}{(+)}\psi_0,$ where $\psi_0$ is a vacuum state (any, which is not annihilated 
by the operator in front of the state) has the eigenvalues  of $ S^{03}, S^{12} $ and $S^{56}$ equal 
to $\frac{i}{2}$, $ \frac{1}{2}$ and $\frac{1}{2}$, correspondingly. All the other states
of one representation of $SO(1,5)$ follow from this one by just the application of all possible 
$S^(ab)$, which do not belong to Cartan sub algebra.

\section{Acknowledgement } One of the authors (N.S.M.B.) would like to warmly thank Jo\v ze Vrabec 
for his very fruitful discussions, which  help a lot to clarify the concept of involution and 
 to use it in the right way.

\author{R. Mirman}
\title{How Can Group Theory be Generalized so Perhaps Providing
Further Information About Our Universe?}
\institute{%
14U\\
155 E 34 Street\\
New York, NY  10016\\
\textrm{e-mail:} \texttt{sssbbg@gmail.com}
}

\titlerunning{How Can Group Theory be Generalized \ldots}
\authorrunning{R. Mirman}
\maketitle

\begin{abstract}
Group theory is very familiar, perhaps too much so. We are thus
prejudiced about it, leading to views that are far too narrow. Yet it
is significantly richer than usually realized ~(\cite{imp}). Here we
wish to understand the restrictions giving the familiar forms and how
by changing these we can get added richness. Might these add to our
knowledge of nature? A purpose of this note is to stimulate thinking
about this.
\end{abstract}

\section{Geometry, through its transformations groups, is very
information, but so far not enough}\label{s1}

It is clear that much (all?) of physics is determined by geometry,
especially through its transformation groups~(\cite{imp}; \cite{gf};
\cite{ml}; \cite{qm}; \cite{cnfr}; \cite{bna}; \cite{bnb}; \cite{bnc};
\cite{op}; \cite{ia};  \cite{pt}). Yet it is necessary to go much
further. Can additional progress be made using group theory? This is a
very open question but worth exploring. One aspect to be explored is
whether group theory itself can be generalized. That could be of
interest for purely mathematical reasons. And it has many
applications. Generalizing it can thus be useful in various ways. This
we wish to explore here.

\section{What is the best way to try to understand fundamental
physics?}\label{s2}

What is the best approach to try to understand physics? The big fad
nowadays is to come up with the wildest, most unlikely ideas, ones
furthest from reality, ones totally unrelated to anything known, ones
having no rationale whatever. History and common sense show that this
approach is destined to lead nowhere except to even more wild ideas
(as it has).

Those who do that will find themselves badly cut by Occam's razor.
Unfortunately a large part of the physics community is doing just
that. Applying Occam's razor to the physics community will greatly
help physics to advance.

Another approach that is very likely to lead to failure, certainly if
there is no other rationale for it, is to base laws on how we measure,
on ourselves. We do not determine the laws of nature (something many
scientists, especially physicists, do not believe). How we measure is
limited by physical laws, but does not limit them. Studying
measurement can help us understand physics, but cannot determine it.

What then shall we do, what approach shall we use?
The best approach is the most conservative using requirements that are
certain, or at least likely, to be correct, or ones that deviate the
least from these.

\section{Reasonable requirements for developing theories}\label{s3}

What requirements can we impose?

First is consistency. Fundamental physical theories must be
consistent. (Phenomenological theories, classical physics is an
example, can be inconsistent mishmashes.) This is more difficult than
it might seem, so can be quite powerful.

Geometry imposes requirements, restrictions. Physics takes place in
geometry so must be in accord with the rules it leads to. This also is
powerful as we have seen (particularly) in the references.

What can we say about geometry? We always assume that it is a manifold
(locally flat) and that its coordinates are real numbers (rather say
than complex numbers or quaternions). It is very unlikely that physics
would be possible otherwise, but this can be investigated. A
fundamental property of geometry is its dimension. However it has long
been known that physics would be impossible unless the dimension is
3+1~(\cite{gf}; \cite{imp}). This is required by consistency,
illustrating its importance, for only with this dimension is a
consistent physics possible.

\section{What can we say about geometry?}\label{s4}

Is space curved? The curvature of space is given by a function over
it, the connection~(\cite{ml}). Can every space that is a manifold be
regarded as flat but with a function, the connection, so that all
curved spaces can be reduced to flat ones with different such
functions? This is an interesting question that we raise but do not
try to answer here. Also (many) curved spaces can be mapped (in
reasonable ways) into flat ones~(\cite{cnfr}). Thus we consider only
flat spaces, but these questions should be looked into.

What properties do flat spaces have? Beyond the reality of coordinates
and the dimension is a most fundamental property: the transformation
groups of the spaces (which are not symmetry groups~(\cite{imp}),
although it is quite interesting that they are that also). For our
space, apparently the only one in which physics is possible, the
largest symmetry group is the conformal group~(\cite{cnfr}), which has
subgroups the Poincar\'e group, its subgroup the Lorentz group and the
subgroup of that, the rotation group (SO(3)). The last gives that
angular momentum must be integral or half-odd-integral~(\cite{ia}),
illustrating how transformations limit physics. The massless
representations of the Poincar\'e group determine electromagnetism and
gravitation~(\cite{ml}). Clearly these are quite informative, but
clearly insufficient. It is possible that the conformal group can also
be quite informative but how is less clear.

\section{How might group theory be generalized?}\label{s5}

Can we go further? What we wish to do here is study whether what is
known about group theory can be generalized. We are all too familiar
with semisimple groups, like the rotation and Lorentz groups. But
group theory is far richer, even for these groups~(\cite{ml};
\cite{cnfr}). Perhaps it is richer than we realize. That is what we
consider here.

This may not have anything to do with fundamental laws. But it helps
to understand group theory, decreasing prejudice and broadening our
views, and produces interesting mathematical results. And they are
likely to lead to useful, even important, applications.

\section{Indexed groups}\label{s6}

Start by considering the curve, x = r$cos\theta$, y = r$sin\theta$,
which describes a circle. We move around a circle using the
two-dimensional rotation group O(2). By putting a constant into the
representation matrix we can generate an ellipse. But what about, say,
the curve x = r$(cos\theta)3$, y = r$(sin\theta)3$. What set of
transformations moves along this curve and why don't they form a
group? Clearly there is an identity, we do not have to move, and for
every transformation there is an inverse. Moreover the product of two
transformations is a transformation; if we move from A to B and the
from B to C, we can find a transformation from A to C. However the
transformations are not associative. It is for this reason that they
do not form a group. The transformation from B to C depends on where B
is (it is in a sense history dependent, depending on the previous
transformations). That is the operator going from B to C has a form
that depends on B, unlike rotations. This causes associativity to
fail.

Thus for a circle the product of the transformation matrix for
$\theta_1$ and for $\theta_2$ is that for $\theta_1$ + $\theta_2$,
which is not true for this curve.

While there are transformations along any (reasonable) n-dimensional
surface it is only in special cases that they form a group (of the
form usually considered). This emphasizes the relevance of
associativity and the restrictions it places. Many of the properties
of groups and their representations come from associativity. While
restricting, it also allows us to obtain properties that are so useful
in applications of groups.

How do we deal these more general transformations, say ones along
arbitrary surfaces? We introduce the concept of indexed groups. To do
this we assume that the transformations can be mapped (at least)
one-to-one onto a {\em group} of transformations. For each general
transformation we have a corresponding group element and this element
is the index of the general transformation. Thus for three dimensions
the index can be an element of the rotation group. For a group the
matrix representing the transformation that is the product of two is
the matrix product of the matrices of the two transformations. It is
here that indexed representations differ. For these the matrix
labeling the product is the matrix product of the two {\em labeling}
matrices. But the product matrix of the indexed transformation is not
the product of the matrices of the two transformations it is a product
of. It is the index that is given by the product, not the
transformation. With this definition of group product, differing from
the normal definition, these transformations form a group.
Associativity holds, since it follows from the associativity of the
indexing group, but only because of the revised definition of a
product.

To give a group we list its members and their products (thus the
spaces on which they act). But now with each set of members we have an
infinite number of product rules (determined by the mapping of the
transformations into the indexing groups, of which there may be
several) thus an infinite set of groups. Each n-dimensional surface
has its own group.

\section{Product rules determine groups}\label{s7}

This shows how properties of a group are dependent on the definition
of its product and how by revising this definition we can generalize
the type of structures that form groups. This adds to the richness of
group theory.

Groups which can be realized as matrices, thus whose products are
matrix products, we call standard groups. Ones whose elements are
indexed so whose products are given by the matrix products of their
indices we call indexed groups.

Indexed transformation groups exist for any surface that can be mapped
(properly, in a way that must be investigated) to the defining space
of a Lie group. For the three-dimensional rotation group SO(3) that is
a sphere, and there is a third parameter which can be considered as
giving the direction of a vector at each point of the sphere. Then
each point is mapped to a point on one of the generalized (overlying)
space, and each direction to one on that (which perhaps might be
considered as an internal symmetry). Likewise we can use SO(2,1) to
get another set of such surfaces. So we have associated with each
group an infinite set of groups, each given by different product rules
(from different mappings), or another way of saying this, an infinite
set of realizations or representations.

We assign to each group element a matrix, that of the regular
(adjoint) representation. That this is possible follows from the group
axioms. Then the group product is a matrix product, and this is the
usual group product. We call this the standard product.

The rotation group, besides its defining representation of 3 X 3
matrices has an infinite set of others. Consider the 5 X 5 one, say.
This is a subrepresentation of SO(5). We can map a surface to the
defining surface of SO(5) and the group defined over it (one for each
of the infinite number of such surfaces, ignoring aspects like
inversions), form representations of SO(3), but with additional
transformations which might be taken as internal ones. Since SO(3) has
an infinite number of representations (and it is simple so other
groups have infinite sets each of infinite numbers of representations)
this admits a huge number of transformation sets to be defined over
it.

But we can do more. The conformal group algebra is (isomorphic to)
that of SO(4,2) and SU(3,1). The group algebras are the same but
realized in terms of different variables. Instead of 4+2 real ones, or
3+1 complex ones, the conformal algebra is realized over 3+1 real
ones~(\cite{cnfr}). We might also realize it over more, rather than
fewer, variables. These again can be taken as internal ones.

We now map surfaces (choosing from an infinite number) to the
3+1-dimen\-sional real space on which the conformal group acts. The
product of the transformations on these is given by the product of
conformal transformations taking points and directions of the
3+1-space to others. These conformal transformations are the indices
of the transformations on the preimage space. Note that we have three
groups, SO(4,2), SU(3,1) and the conformal group, plus all their
representations, which can act as indices. This shows the great
richness introduced.

\section{Groups of 3+1 space as illustrations}\label{s8}

Thus there are two groups defined by a 3+1-dimensional real space,
one, ISO(3,1) is the Poincar\'e group, the other is the conformal
group. This itself shows richness known but not understood in group
theory. Take a surface (one of an infinite number) that is mapped to
real 3+1-dimensional space. The transformations on it can be indexed
by the transformations of the base space. However there are two sets,
the Poincar\'e group and the conformal group. So from ordinary space
we can find two infinite sets of groups (realizations,
representations). And the transformations of our space can be taken as
subsets of larger groups, giving further labels and products, of
infinite number. Some of this additional freedom might be relevant to
internal transformations.

Why should we consider these, aside from their showing the
assumptions? Groups are useful in many ways and these can extend their
usefulness. For example special functions are group representation
basis states and many properties can be derived from this.
Generalizing the concepts of representation can lead to other special
functions, perhaps with useful properties. However it is not clear
that properties can be found for these as they can for standard
representations, or indeed that they have simple properties. This must
be investigated. Associativity is important in determining these
properties, and allowing simple properties (that we can get rules
for). This procedure allows such great generalization that it is
likely that only a few cases, at most, can give simple rules. But
there might be some and these could be useful.

Also physical objects are statefunctions~(\cite{imp}) that are group
representation basis states. By expanding the set of these we may be
able to expand the set of objects that are such states. There are
clear limitations as the known states are those of standard
representations. It may be that the requirement that objects be
observers, and conversely~(\cite{qm}), provides strict limits. Yet
this is not known and these new representations allow study of this.
And some may have physical applications, perhaps to these fields.

\section{Why standard products are matrix product and why are these
usually relevant?}\label{s9}

While these indexed groups (groups with indexed products) may seem
unusual they raise the question why standard products are the relevant
ones, for those cases in which they are? This has to be considered for
each application. A general class is applications to geometry. The
transformations, but only for certain geometries, have standard
products. This is true for lines, circles, planes, spheres and
generalizations. For these, why are the products of transformations
the standard products?

Here symmetry enters. The action of a group transformation on the base
space is independent of the point in that space --- since all these
points are identical. Thus a group transformation taking a point to
another, acting on a second such transformation, gives a
transformation with identical action (as can be seen with a circle).
> From this associativity follows. These transformations thus form the
regular representation~(\cite{ia}). But this representation can be
given by matrices, and the group product is the same as a matrix
product, the standard product.

Other spaces do not have symmetry so their transformations cannot be
represented by matrices. Here we see how symmetry gives group
operations, and limits these. For spaces without symmetry we must use
other products.

\section{Conclusion}\label{sc}

Groups are determined very much by their products. Usually these give
matrix products. Here we have considered a set of different product
rules. These illustrate how such rules determine the properties of
groups, and the role of symmetry in the standard product rules.
Whether some of these generalized rules are useful has to be studied.
Lie groups have an extensive structure including their algebras. Do
these generalizations allow corresponding structures, including
algebras? That is another field of study. There is much that can be
done, some at least profitably.

\section*{Acknowledgements}

This discussion could not have existed without Norma Manko\v c Bor\v stnik.

\newcommand\CL{\mathcal{L}}
\newcommand\CD{\mathscr{D}}
\newcommand\CO{\mathcal{O}}
\newcommand\CS{\mathscr{C}}
\newcommand\bC{\mathbb{C}}
\newcommand\e{\mathrm{e}}
\newcommand\bZ{\mathbb{Z}}
\newcommand\bR{\mathbb{R}}
\newcommand\CU{\mathscr{U}}
\newcommand\Ds{D\hspace{-2.5mm}/} 
\title{Future Dependent Initial Conditions from Imaginary Part in Lagrangian}
\author{H.B. Nielsen${}^1$ and M. Ninomiya${}^{2,3}$}
\institute{%
${}^1$ Niels Bohr Institute,\\
Blegdamsvej 17,\\ 
DK2100 Copenhagen, Denmark\\
${}^2$ Yukawa Institute for Theoretical Physics,\\
Kyoto University, Kyoto 606-8502, Japan\\
${}^3$ Okayama Institute for Quantum Physics,\\
 Kyoyama-cho 1-9, Okayama City 700-0015, Japan}

\titlerunning{Future Dependent Initial Conditions from Imaginary Part %
in Lagrangian}
\authorrunning{H.B. Nielsen and M. Ninomiya}
\maketitle

\begin{abstract}
We want to unify usual equation of motion laws of nature with ``laws" 
about initial conditions, second law of thermodynamics, cosmology.
By introducing an imaginary part -- of a similar form but different 
parameters as the usual real part -- for the action to be used in the 
Feynman path way integral we obtain a model determining 
(not only equations of motion but) also the initial conditions, 
for say a quantum field theory.
We set up the formalism for e.g. expectation values, classical approximation
in such a model and show that provided the imaginary part gets unimportant
except in the Big Bang era the model can match the usual theory.
Speculatively requiring that there be place for Dirac strings and 
thus in principle monopoles in the model we can push away  the effects of the
imaginary part to be involved only with particles not yet found.
Most promising for seeing the initial condition determining effects from
the imaginary part is thus the Higgs particle.
We predict that the width of the Higgs particle shall likely turn out
to be (appreciably perhaps) broader than calculated by summing usual 
decay rates.
Higgs machines will be hit by bad luck.
\end{abstract}

\section{Introduction}\label{sec:1}
Usually when we talk about ``theory of everything" as 
superstring theory is hoped to be, 
it is not really meant that the initial state
of the universe is included in the model immediately.
Rather one needs to make extra assumptions -- cosmology, 
second law of thermodynamics\cite{1,2,3,7}, etc. --
about the initial conditions or one simply leaves it for the applicator 
of the theory to somehow himself manage to find out what the initial
conditions are for the experiment he wants to describe with the theory.
It is, however, the intention of the series of articles \cite{slaw,4,5,%
6,22} to which 
this article belongs to set up assumptions telling the initial conditions 
in a way that can be called that these initial condition assumptions are 
unified\cite{9} with the part of the theory describing the equations of motion
and the particle content (the usually T.O.E.).
Our unification may though be mainly a bit formal in as far as our main
point is to use in the Feynman path integral an action which has both
a real $S_R$ and an imaginary part $S_I$.
Usually of course the action is real and the imaginary part $S_I=0$ 
(Total $S=S_R + iS_I$).
We may quickly see that the imaginary part gives a typically hugely
different extra factor in the probability for different paths obeying
equations of motion.
Thus such an imaginary part essentially fix the path obeying equations
of motion which should almost certainly be the realized one.
In this way we can claim that to a good approximation an imaginary part
of the action will choose/settle the initial conditions.

In the present article it is not the point to settle on any choice of 
the in usual sense ``theory of everything".
Rather we shall present our idea of introducing an imaginary part in the
Lagrangian and thereby also in the action as a modification that can be
made on any theory as represented by the real action $S$.

We have already published a few articles on essentially a classical 
formulation of the present model.
We sought in these articles to be a little more general by simply defining
a probability weight called $P{(\mathrm{path})}$ defined for all possible
paths.
In classical theory it is really only the paths which obey the classical 
equations of motion for which we need to define $P$.
We already in the earlier articles suggested that this probability 
$P{(\mathrm{path})}$ for a certain track, path, to be the one realized in
nature should be given as the exponential of an expression depending on
the track, path, of the form of a space-time integral over a locally
defined quantity $\CL_I$ depending on the fields in the development, path.
Really this quantity $\CL_I$ (really $-\CL_I$) comes into determining the 
probability as if it were the imaginary part of the Lagrangian density.

A major point of the present article is to set up the quantum formulation
of our already published model, now really settling on taking the suggestive
idea of just making the action complex, but with a priori a different set
of coupling constants and $m^2$ for real and imaginary part separately.

A genuine problem with our kind of model is that very likely it predicts
that special simple configurations leading to big probability may be 
arranged at a priori any time.
That is to say, with our type of model it needs an explanation that one 
in particle almost never see any great arrangements being organized to occur
later on.
Really such arrangements might seem to us to be something like a hand of God,
but they seem very seldom.
Thus at first it looks that our type of model is already falsified by the
non-appearance of arrangements. Really such a problem is almost obviously 
expected to occur in a model that like ours {\em does not a priori make any 
time 
reversal asymmetric assumption at the fundamental level}. Unless in the 
Hartle-Hawking no boundary postulate \cite{9} we add some timereversal 
asymmetry spontaneously other otherwise that theory will be up to similar 
problems \cite{8,23}.

A model-language  describing how final states can be imposed by a density 
matrix $\rho_f$ is put forward by Hartle and Gell-Mann \cite{hg}.   

In the present work we hope for that a certain 
moment in the `middle of times' will turn out to become dominant w.r.t.
fixing the special solution selected as the realized one, and that this time 
can then be interpreted as a close to Big Bang time ( there may not really be 
a true big bang but just an inflation era coming out of a deflation era 
continuously). Then since we live in the time after this decisive Big Bang 
simulating era there is for us a time reversal asymmetry, nevertheless it
is a problem that like ours is even timetranslational invariant w.r.t. the 
law that finally settle the `initial conditions' to explain that there 
are not more prearranged events than one seemingly see.

However, we believe to have found some explanations able to 
suppress so many of these prearrangements that our model can be 
made compatible with present experience of essentially no prearrangements.

For really avoiding it we shall assume consistency of Dirac strings, but let
us postpone that discussion to section~\ref{sec:13} below.

Our model is really inspired from the considerations of time machines\cite{tm} 
and the 
troubles of needs for prearrangements in order to avoid the so called grand 
mother paradoxes, meaning the inconsistencies occurring when one seeks to go 
back in time and changes the events there.

We shall present the work by making {\em two} attempts to assumption about how 
to 
interpret the Feynman path integrals with the imaginary part of the action 
non-zero. In the first part of the paper we start out from letting the average 
of a dynamical variable ${\cal O}$  be given by equation (\ref{9}) below, but 
that this is a priori not so good is seen by it not being (safely) real 
even if the 
dynamical 
variable ${\cal O}$ is real. Therefore in section~\ref{sec:7} we restart the discussion
so now from the side of the interpretation of the Feynman path way integrals 
in our model.

{\bf First trial of interpretation}

In the next section~\ref{sec:2}, we shall put forward the basic formula for expectation
values with our complex action model and the philosophy that this model
even deliver the initial condition, or better the solution of equation of
motion to be the one realized.

In section~\ref{sec:3} we review our earlier reasons for that future should have only
little influence on what happens.

In section~\ref{sec:4} we then shall argue for some approximate treatment of the 
functional integral in the late times $t$, the future.

In section~\ref{sec:5} we shall make use of the approximation of the future to obtain
the usual quantum mechanics expressions at least in the case where our
imaginary part $S_I$ of the action can be ignored.
(It should be stressed that we actually have used already a philosophy 
based on this $S_I$ being non zero, so it is not fully zero.)

In section~\ref{sec:6} it turns out that we -- perhaps not completely convincing
though -- can make the effect be that we return to probability in practical
scattering experiments say get conserved.

{\bf Second trial of interpretation:}

In section~\ref{sec:7} we restart the discussion of making the interpretation 
formula for the Feynman path way integral, which after some talk takes 
the way of using the classical approximation weighted with the exponential 
of minus 2 times the imaginary part of the action. 
In a subsection~\ref{subsec:7.5} we formally connect
our model to our  earlier one based on the probability weight 
$P{(\mathrm{path})}$.

In section~\ref{sec:8} we develop a rather general formula for the correlated 
probability for a series of dynamical quantities or operators ${\cal O}_i$
at different moments of time take values inside small ranges specified.

In section~\ref{sec:9} we go a bit further in making the expressions like the ones 
one uses in practice in usual theories. Most importantly we again consider 
how to approximate the future when the effects of the imaginary part of the 
action is very small.

In section~\ref{sec:10} we put the  simplest example of the more general formula,
namely a formula for the probability of just one operator at one time 
being in a given range. ( This question would be impossible to predict 
even in principle in other theories, but we in principle can, but in practice 
not usually). But the resulting formula has what we call ``squared form'' in 
the sense that the projector comes in twice as a factor in it. The finding 
of a reduction to an unsquared form is left to section~\ref{sec:12}, while we 
in section~\ref{sec:11} then give an example of application of very interesting 
physical significance. In fact section~\ref{sec:11} predicts a broadening of the 
width of the Higgs particle due to the imaginary action.

In section~\ref{sec:12} we then bring about a connection between the postulated 
interpretation formulas for probabilities put forward in part I and part II.
In fact we find that they coincide under rather suggestive  assumptions.

In section~\ref{sec:13} we bring the promised argument for removing the effects
of the imaginary action $S_I$ from the domain of older accelerators, since
otherwise our model would have been falsified. The argument is based on
assuming monopoles.

In section~\ref{sec:14} we conclude and give a bit of outlook.
\vspace{.5 cm}

\noindent
{\huge \bf Part I, First Trial of Interpretation}

\section{Philosophy and formula}\label{sec:2}
Our basic modification of introducing an imaginary part in the actions leads 
to that integrand $\e^{iS}$ or $\e^{\frac{i}{\hbar} S}$ of the Feynman path 
way integrand
$\int \e^{iS} \CD \phi$ or $\int \e^{\frac{i}{\hbar}S} \CD \phi$
(if the Planck constant is written explicitly) varies a lot in magnitude,
and not only in phase as usual.
This effect is likely to make some regions in the space of paths 
-- or we could restrict to the space paths with $\delta S=0$, i.e. the space
classical solutions -- 
get a very much bigger weight in the integral than others.
Actually it can likely happen that only a very narrow range of paths or
better solutions (= paths obeying $\delta S=0$) will quite dominate the 
integral
\begin{eqnarray}\label{1}
\int \e^{iS} \CD \phi .
\end{eqnarray}
 
That should naturally be taken to mean that the presumed narrow range of
dominating paths represent the paths being actually realized in nature.
It is in this way that we hope our model to essentially predict the initial
state for the realized solutions.
It is important to have in mind that such an effect of the imaginary action
$S_I$ of selecting narrow bunches of solutions can make the boundary 
conditions at an initial and a final time for a period to be studied say,
superfluous.
A bit optimistically we might imagine that the imaginary part of action 
makes the functional integral converge even without boundary condition
specifications.
Note that being allowed to throw boundary conditions away
-- having them replaced by effects of $S_I$ --
is a great/nice simplification.
We consider this achievement as an aesthetically very nice feature of our
model!
Supposing that this works to deliver a meaningful Feynman-path integral
(\ref{1}) even without boundary conditions this way we must now decide how
one is supposed to extract information now in principle for the true
expectation value as it should occur even without further input.
Note here that we are -- but only in principle --
proposing an exceedingly ambitious model compared to usual quantum field
theories:

We want to predict expectation values without any further input than the
mere complex action!
This of course corresponds to that our level of ambition is to in addition
to the usual time-development laws of nature \underline{also} predict the
initial conditions, i.e. what really happens!

To write down the formula for some physical quantity let us first exercise
by a quantity $\CO (\varphi |_t)$ which is a function of the fields $\varphi |_t$
restricted to some time $t$, where $\varphi$ is a general symbol for all the
fields in the model.

If we for instance use the Standard Model as the starting model, then
providing it with an imaginary part of the Lagrangian density, then the 
symbol $\varphi (x) \,(x\in \grave{R}^4)$ is really a set
\begin{eqnarray}\label{2}
\varphi = (A^a_\mu, \psi^b, \,H)
\end{eqnarray}
where the indices on the Fermion fields runs through the combination of
flavor and color and/or $W$-spin components, while the index on the 
gauge fields run through the 12 gauge fields
-- 8 gluon color combination plus $(B_\mu)$ the $U (1)$-component
and 3 $W$'s --.
Finally $H$ is the two complex component Higgs field.

The quantity $\CO(\varphi |_t)$ can of course be considered a functional of
the whole field development $\CO(\varphi)$ also, i.e. it could be consider
a functional of  the path of one wants.

The simplest proposal for what the average quantity $\CO(\varphi)$ 
would be
\begin{eqnarray}\label{3}
\langle \CO(\varphi) \rangle = \frac{\int \e^{iS[\varphi]}\CO(\varphi)\CD\varphi}
{\int \e^{iS[\varphi]}\CD \varphi}.
\end{eqnarray}
This would mean that we have a ``sort of probability" given by
\begin{eqnarray}\label{4}
\mathrm{``Probability\, of}\, \CO \,\mathrm{being} \,\CO_0"
=\frac{\int \delta(  \CO(\varphi)-\CO_0) \e^{iS[\varphi]}\CD\varphi}
{\int \e^{iS[\varphi]}\CD\varphi}
\end{eqnarray}
Now, however, we must admit that conceiving of this expression as  a 
probability is upset by the severe \underline{problem} that it will 
typically be a complex number.
There is no guarantee that it is positive or zero.

Thus a priori one would say that this simple expression for the probability
density is quite untenable.

Nevertheless it is our intention to claim that we should 
-- and that is then part of our model -- 
use the simple expression (\ref{4}) and the corresponding (\ref{3}) and 
the expression to be given below for more general operators $\CO$ 
corresponding also to (\ref{3}) and (\ref{4}).

First let us again stress that it is our a priori philosophy that somehow
the imaginary part $S_I$ managed to fix both a state in future and 
in past.
Thereby asking the average of some quantity $\CO$ becomes much like 
in an already finished double slit experiment (Bohr-Einstein) in which
a particle already have been measured on the photographic plate
(presumable on an interference line) what \underline{were} the average
position of the particle when it past the double slit screen.
Really asking such a question concerning a quantity $\CO$ that were not 
measured and could not have been measured without having disturbed the
outcome of something later is one of the forbidden questions in quantum
mechanics.
Indeed it is by asking this sort of questions which are not answerable by
measurement that Einstein can find ammunition against quantum mechanics.
In other words our proposal (\ref{4}) for ``probability distribution" is 
a priori -- with our present philosophy of a future essentially determined
by $S_I$ -- an answer to a quantum mechanically forbidden question.
Niels Bohr would say we should not ask it.

In that light it may of course not be so serious that our formula gives
a rather stupid or crazy answer, a complex probability!

But now we have the problem of justifying that if we made a true measurement
the answer would turn out to give positive (or zero) probability.

Let us take as the important feature of a measurement of some quantity 
$\CO$  that there is an apparatus which makes a lot of degrees of freedom,
$\xi$ say (really macroscopic systems) develop in a way depending on value 
of $\CO$.
Such an amplification of the effect of the actual value of $\CO$ is
characteristic for a measurement.
Unless somehow there are special reasons for that $S_I$ be insensitive to $\xi$
(as we shall actually later seek to show but do not assume to be the case)
we expect that $S_I$ typically will depend on the macroscopically many
d.o.f. $\xi$ being influenced by $\CO$-value measured.
Now we argue like this:  Since there is a huge (macroscopical) number of
variables $\xi$
depending on the value of $\CO$ ``measured", the imaginary part $S_I$ of
the action is likely to depend very strongly on this measured value -- very
rapidly varying.

We here think of $S_I$ as the integral over the imaginary part of the
Lagrangian $L_I$ over all times $t \in ] -\infty, \infty [ $
\begin{eqnarray}
\left(
S_I = \int_{-\infty}^{\infty} L_I \mathrm{dt} 
\right)
\end{eqnarray}

Because of the great complications in an actual measuring apparatus, let
alone the further developments  depending the measured value, publications and 
so on, the imaginary action $S_I$ can easily be a very complicated 
function of the measured $\CO$ value.
Even if $S_I$ as function of the measured $\CO$ value should in principle
be continuous it may in practice vary so much up and down
-- caused by accidents influenced by the broadcasted measuring value --
that very likely the smallest value of  $S_I$  occurs for a seemingly
accidental value of the measured $\CO$.
If the $S_I$-variation with the ``measured $\CO$" is indeed very strong
so that the $S_I$ variations are big the exponential weight $\e^{-S_I}$
contained in (\ref{3}) and (\ref{4}) will have a completely dominant
value for only one measured $\CO$-value.

In this way our model has the integral in the numerator of (\ref{4}) be
much bigger for one single value of $\CO_0$.
If so, then the ratio (\ref{4}) is actually $\propto 1$ for this 
$\CO_0$-value and negligible for all other  $\CO_0$-values.
This means that our model much like usual measurement theory
(in Copenhagen interpretation) predicts that crudely only one value of 
a measured quantity is realized.
In principle it is even so that, bearing a very special situation,
the result of the measurement is calculable by essentially minimizing
the imaginary action $S_I$.
In practice, however, such calculation will only be doable in extremely 
rare cases.
(If we impress a special result by threatening with a Higgs-producing
machine).

We postpone the argumentation for that the probability distribution to be
obtained in practice shall be the one of usual quantum mechanical
measurement theory partly to the later sections and partly to a 
subsequent paper.

At the end of this section let us extend slightly our formula (\ref{3})
and thereby also (\ref{4}), to the case where the quantity $\CO$ 
corresponds in usual quantum mechanics to an operator that do not commute
with the fields $\varphi$.

An operator corresponding to a quantity measurable at a moment of time $t$
will in general in the quantum field theory considered be given by a 
matrix with a columns and rows in correspondence with field functions
$\varphi |_t$ restricted to the time $t$.
I.e. $\CO$ is given by a ``matrix"
\begin{eqnarray}\label{5}
\left(\varphi' |_t \big| \CO \big| \varphi |_t\right)
= 
\underbrace{\CO (\varphi' |_t, \,\varphi |_t)}
.
\end{eqnarray}

What should be the formulas replacing (\ref{3}) and (\ref{4}) in this
more general case?

Well, our main starting point were that we assumed our imaginary part $S_I$ 
to (essentially) fix both a further $| B \rangle$ and a past state 
$| A \rangle$.
A natural notation to introduce is in fact -- for the past --
\begin{eqnarray}\label{6}
\langle \varphi|_t | A \rangle
= A[\varphi|_t ]
= \int_{\mathrm{ending\, at}\, \varphi|_t
} \e^{iS_{-\infty \,\mathrm{to}\,t}}\CD \varphi
\end{eqnarray}
and analogously
\begin{eqnarray}\label{7}
\langle B|\varphi|_t  \rangle
= \langle \varphi|_t | B \rangle^{*}
= B[\varphi|_t]^{*}
= \int_{\mathrm{beginning\, at}\, \varphi|_t
} \e^{iS_{t\,\mathrm{to}\,+\infty }}\CD \varphi
\end{eqnarray}
 
In this notation our previous formulas (\ref{3}) and (\ref{4}) are for
the in $\varphi|_t$ diagonal operators $\CO(\varphi|_t)$ become
\begin{eqnarray}\label{8}
\langle \CO \rangle &=&
\frac{\int \e^{iS}\CO(\varphi|_t)\CD\varphi}
{\int \e^{iS}\CD\varphi}
\nonumber\\
&=& \frac{\oint _{\varphi|_t}\langle B|\varphi|_t  \rangle \CO(\varphi|_t)
\langle \varphi|_t | A \rangle}
{\oint _{\varphi|_t}\langle B|\varphi|_t  \rangle \langle \varphi|_t | A \rangle}
\nonumber\\
&=& \frac{\langle B|\CO|A \rangle}{\langle B|A \rangle}
\end{eqnarray}
and (\ref{4}) becomes
\begin{eqnarray}\label{9}
\mathrm{``Probability\, of}\, \CO \,\mathrm{being} \,\CO_0"
=\frac{\langle B|\delta (\CO-\CO_0)|A \rangle}
{\langle B|A \rangle}
\end{eqnarray}
Now really we want to suggest that formula (\ref{8}) and (\ref{9}) can
also be used for operators that are not simply functions of the 
fields $\varphi|_t$ at
time $t$,  used in the functional integral.

In order to justify that extension of our interpretation formulas 
we want to remark:
\begin{enumerate}
  \item Provided a Hermitean operator $\CO$ has either $|A\rangle$ or
$|B\rangle$ as eigenstate then the eigenvalue $\CO'$ in question can of course
be extracted as
\begin{eqnarray}\label{10}
\CO'=
\frac{\langle B|\CO|A \rangle}
{\langle B|A \rangle}
\end{eqnarray}
 \item One can quite generally 
-- by Fourier transformations at every step in a time lattice --
rewrite a functional integral of the Feynman path way integral form
from some set of variables $\varphi$ to a conjugate set:
\begin{eqnarray}\label{11}
\int \e^{iS}\CD \varphi &\stackrel{\mathrm{latticitation}}{=}&
\int \prod_{t\in \{ t \mathrm{-lattice}\}} \CD^{(3)} \varphi |_t
\e^{i\sum_{t\in \{\mathrm{lattice}\}} L_{\mathrm{discr}}
\left( \varphi|_t , \frac{\varphi|_{t+\Delta t}-\varphi|_t}{\Delta t}
\right) \Delta t}
\nonumber\\
&=& \int \prod_{t\in \{ t \mathrm{-lattice}\}} \CU
(\varphi|_{t+\Delta t}, \varphi|_t) \CD^{(3)}\varphi|_t
\end{eqnarray}
\end{enumerate}
where
\begin{eqnarray}\label{12}
\CU (\varphi|_{t+\Delta t}, \varphi|_t)
= \e^{i L \left( \varphi|_t, \frac{\varphi|_{t+\Delta t}- \varphi|_t}
{\Delta t}\right)},
\end{eqnarray}
can be rewritten into $\hat{\CU}(\Pi |_{t+\Delta t}, \Pi |_t )$
matrices obtained from the $\CU(\varphi |_{t+\Delta t}, \varphi |_t )$
by Fourier functional transformations
\begin{eqnarray}\label{13}
\hat{\CU}(\Pi |_{t+\Delta t}, \Pi |_t ) \stackrel{\mathrm{def}}{=}
\int \CD^{(3)} \varphi |_{t+\Delta t}
\,\e^{+i \varphi |_{t+\Delta t} \Pi |_{t+\Delta t}}
\CU(\varphi |_{t+\Delta t}, \varphi |_t)
\e^{-i \varphi |_t \Pi |_t} \CD^{(3)} \varphi |_t\nonumber\\
~~~
\end{eqnarray}
Now of course for long chains of $\hat{\CU}$-matrices 
(ignoring end problems) you have
\begin{eqnarray}\label{14}
\int \prod_{t\in \{ t \mathrm{-lattice}\}} \CU
(\varphi|_{t+\Delta t}, \varphi|_t) \CD^{(3)}\varphi|_t
\nonumber\\
\stackrel{\mathrm{except\, for\, end\, problems}}{=}
\int \prod_{t\in \{ t \mathrm{-lattice}\}} \hat{\CU}
(\Pi |_{t+\Delta t}, \Pi |_t) \CD^{(3)}\Pi |_t
\end{eqnarray}
Supposedly you can put the right hand side into a form
\begin{eqnarray}\label{15}
\int \e^{i S^{(\mathrm{in}\,\Pi)}[\Pi, \,\Delta \Pi \mathrm{-defferences}]}
\CD \Pi
\end{eqnarray}

Now you may argue with the same intuitive suggestion for getting
\begin{eqnarray}\label{16}
\CO (\Pi |_t)=
\frac{\int \e^{i S^{(\mathrm{in}\,\Pi)}}\CO(\Pi |_t) \CD \Pi
}
{\int \e^{i S^{(\mathrm{in}\,\Pi)}}\CD \Pi}
\end{eqnarray}
as we did for (\ref{3}).
By thinking of doing the just presented Fourier transformation partly we
might argue for a similar average formula for any operator
\begin{eqnarray}\label{17}
\langle \CO (\varphi |_t, \,\Pi\_t)\rangle &=&
\frac{\int \e^{i S} \CO(\varphi |_t, \Pi |_t) \CD \varphi}
{\int \e^{i S} \CD \varphi}
\nonumber\\
&=& 
\frac{\langle B_t | \CO(\varphi |_t, \Pi |_t)| A_t \rangle }
{\langle B_t|A_t \rangle}.
\end{eqnarray}
Really this proposal looks very bad because of several lacks of good
correspondence with usual quantum mechanics a priori:
\begin{enumerate}
  \item[a)] Obviously $|A_t\rangle$ is here (a sort of) wave function of the 
universe at time $t$, but our probability density (\ref{4}) or
\begin{eqnarray}\label{18}
\mathrm{``Probability\, for}\, \CO \,\mathrm{being} \,\CO_0"
=\frac{\langle B_t | \delta (\CO - \CO_0) | A_t \rangle }
{\langle B_t|A_t \rangle}
\end{eqnarray}
is not quadratic in $|A_t \rangle$ as we expect from the usual corresponding
formula 
\begin{eqnarray}\label{19}
\mathrm{``Probability\, for}\, \CO \,\mathrm{being} \,\CO_0 \,\mathrm{usual}"
=\frac{\langle A_t | \delta (\CO - \CO_0) | A_t \rangle }
{\langle A_t|A_t \rangle}.
\end{eqnarray}
  \item[b)] As already stated the ``probability density" (\ref{18}) is
even usual complex and  needs the above measurement special case to become
just positive. 
\end{enumerate}

We shall below argue for an approximate treatment of the future part 
$|B_t \rangle$ of the integral thereby achieving indeed a 
{\it rewriting into an expression which is of the form with $|A_t \rangle$
coming squared}.
Indeed we shall rewrite (\ref{18}) into (\ref{19}) below.

\subsection{
Justification of philosophy from semiclassical approximation}
In semiclassical approximation one simply evaluates different contributions 
to the functional integral $(1)$ by seeking the different extrema
for $\e^{iS}$ or equivalent $S=S_R + iS_I$.
Around such an extemum it is extremely well known that one can approximate
$S$ by the Taylor expansion up to second order
\begin{eqnarray}\label{triangle1}
S &=& S(\mathrm{extremum})
 + \frac{1}{2}\int \frac{\partial^2 S
}{\partial \varphi_{1} (x_1) \partial \varphi_{2}(x_2)}
\nonumber\\&&
\cdot
\left( \varphi_{1} (x_1) - \varphi_{1}^{\mathrm{extr}} (x_1) \right)
\left( \varphi_{2} (x_2) - \varphi_{2}^{\mathrm{extr}} (x_2) \right)
+ \cdots ~~~
\mathrm{d}^{4}x_1\mathrm{d}^{4}x_2
\end{eqnarray}
where then the linear terms
\begin{eqnarray}\label{triangle2}
\int\frac{\partial S}{\partial \varphi_1(x_1)}
\left(\varphi_1(x_1)- \varphi_{1}^{\mathrm{extr}}(x_1)
\right)\mathrm{d}^{4}(x_1)
\end{eqnarray}
vanish because of the extremiticity condition.
Here $ \varphi_{1}^{\mathrm{extr}} (x_1)$ and
$\varphi_{2}^{\mathrm{extr}} (x_2)$ denote the fields at the extremum
field configuration development.
Such an extremum as is well known corresponds to a solution to
\begin{eqnarray}\label{triangle3}
\delta S = 0
\end{eqnarray}
i.e. solving the variational principle leading to classical equations of
motion.

The main term in the exponent $iS$(extremum) is in the usual real action
case purely imaginary and thus only gives rise to a phase factor so that
in this approximation the contribution has the same size for
all the classical solutions, provided they can go on for real field
configurations.
With our $S_I$ included, however, we tend to get even to the approximation
of the first term in the Taylor expansion (\ref{triangle1}) a real term
$-S_I$ into the exponent and thus the order of magnitude for one classical
solution compared to another can easily become tremendous
\begin{eqnarray}
| \e^{iS (\mathrm{extremum})} | = \e^{-S_{I} (\mathrm{extremum})}.
\end{eqnarray}
It is our philosophy that only relatively very few classical solution
have terms $\e^{-S_{I} (\mathrm{extremum})}$ dominating violently the rest.
In this sense we expect and assumed that such one or a very few
classical solutions could be considered the only one realized.
With very big size of $S_I$
-- and that can easily come about for a couple of reasons --
it gets relatively only exceedingly few classical solutions that are
competitive in the sense that for most classical solutions
(of (\ref{triangle3})) you have exceedingly small $\e^{-S_I}$ compared
to the few dominant ones.
As the reasons for $S_I$ being big when it is not forbidden by gauge
invariance and the condition that Dirac strings shall be  unobservable we 
can give:
\begin{enumerate}
  \item There is in analogy to the $S_R$-term a $\frac{1}{\hbar}$-factor
in front of $S_I$. 
For practical purposes we know that we shall consider the Planck constant
$\hbar$ to be very small.
  \item We could easily get Avogadros number come in as a factor in
the $S_I$ because it would get such a factor a priori since there are 
typically in the world of macroscopic bodies of that order magnitude molecules.
\end{enumerate}

\section{Approximate treatment of future part of functional integral
(treatment of $|B_t \rangle$)}
\label{sec:3}
In our earlier works\cite{6}
-- in which we mainly worked in the classical approximation --
we presented some arguments that in the era which have been going on since
short time of after some effective (or real) Big Bang the imaginary
Lagrangian or action $L_I$ or $S_I$ effectively became very trivial.
That should mean that under the times starting after some early Big Bang
and extending into the future we could approximately take $L_I$ and
the part of $S_I$ coming from this era as independent of what are the
practical possibilities for what can go on.
Thus we should in this present era supposed to extend into even the 
infinite future be allowed to ignore in first approximation the imaginary 
parts $L_I$ or $S_I$.

The reasons, which we presented for that were that this present era
including supposedly all future is dominated by two types of particles:
\begin{enumerate}
  \item Massless particles (really the entropy of the universe is today
dominated by the massless microwave back ground radiation of photons).
  \item Non-relativistic particles carrying practically conserved
quantum numbers (the nucleons and the electrons are characterized by
their charges and baryon or lepton number so as to make their decays 
into lighter particles impossible).
\end{enumerate}

The argument then went that we could write the action
-- actually both real $S_R$ and imaginary $S_I$ --
for these particles, treated as particles, as a sum having each giving a 
contribution
proportional to the eigentimes for them: 
\begin{eqnarray}\label{3.2}
S_R, \,S_I = \sum_{\mathrm{particles}\, P} K_{P \{{ R\atop I} \} } \cdot \tau_P .
\end{eqnarray}

That is to say that each of the particles contribute to $S_I$ say a 
contribution proportional to the eigentime
\begin{eqnarray}
S_{I \,\mathrm{from}\, P} \propto \tau_P .
\end{eqnarray}

Now for massless particles any step in eigentime
\begin{eqnarray}
\Delta \tau_P =0~~ \mathrm{(for\, massless)}
\end{eqnarray}
and for nonrelativistic ($\simeq$ slow) particles, such a step is
\begin{eqnarray}
\Delta \tau_P = \Delta t
\end{eqnarray}
equal to the usual time.
Since the number of the conserved quantum numbers protected
particles are all the time the same the whole contribution to the $S_I$
from the present era becomes very trivial:

Zero from the massless, and just a constant integrated over coordinate
time for the conserved particles.

In addition there are terms  from interactions contributing a priori to
say $S_I$ also.
Since, however, in the era since a little after Big Bang the density of
particles were low in fundamental units presumably also the interaction
contributions would be much suppressed in this after Big Bang era.

So all together we estimate that it is only the very early Big Bang times
that will dominate $S_I$.
Thus the solution to the equations of motion being in a model with an
imaginary action $S_I$ selected to be the realized one will mainly depend 
on what happened in that solution in the early Big Bang era.
This means that it will be in our era as if it were the initial state that
were a rather special one determined by having an especially small
contribution to $S_I$ from Big Bang times.
This would mean a rather well determined starting state roughly which 
interpreted as a macrostate would be one with low entropy.
That is at least a good beginning for obtaining the second law of
thermodynamics, since then there are supposedly no strong effects of 
$S_I$ any more to enforce the universe to go to any special macrostate.
Rather it will go into bigger and bigger macrostates meaning that they
have higher and higher entropy.

Although we have now argued for approximately seeing no effects of $S_I$
in the era after Big Bang implying that our model should have no effects
in this era, this is however, presumably not
being quite sufficiently accurate.

We shall, however, below in section~\ref{sec:9} invent or find arguments that will 
allow us to get completely rid of the $L_I$ or $S_I$ from the in the
Standard Model already found particles.
Only for the Higgs involving processes our arguments in section 9
based on gauge symmetry and the assumption of unobservability of Dirac
strings associated with monopoles
will not quite function.
Thus we still expect that an $S_I$-contribution pops up with Higgs-particles.
But since Higgs-particles are so far not well studied such an effect of
$S_I$ might well have been overlooked so far.

\section{Treatment of $|B_t \rangle$ or
Treatment of the future factor in the functional integral}\label{sec:4}

In equation (\ref{7}) above we defined what one could
call ``the future part" of the functional integral relative to the time
$t$.
It should however be kept in mind that it is a part in the sense that the 
full integral is a contraction (a sort of product) of the past part and 
this future part,
\begin{eqnarray}\label{3.1}
\int \e^{iS}\CD \varphi = \langle B_t|A_t \rangle .
\end{eqnarray}
Now we must remember that according to the second law of thermodynamics
the state of the universe if at all obtainable (calculable) should be so
by considering the development in the past having lead to it.
The future, however, should be rather shaped after what happened earlier.
This suggests that we should mainly have the possibility to guess or know
$|A_t \rangle$ but determined from the fundamental Lagrangian as our model
suggests.
Really in order not to disagree drastically with the second law of 
thermodynamics the future should be shaped from the past and reflect the
latter.
However, there should not be -- at least not much --
adjustment of the happenings at say time $t$ in order to arrange something
special simple happening in future.
This means in or formalism that the by the $S_I$ future contributions
determined $|B_t \rangle$ should according to second law better disappear
quite from our formula for predicting probabilities for operator values,
i.e. from (\ref{4}) or more generally (\ref{18}).

Now, however, as we argued in foregoing section -- section~\ref{sec:3} --
reviewing previous articles working in the classical approximation
it should be the state of a solution to the equations of motion in the
early Big Bang time that dominates the selection of such a solution to
be the realized one.
The future on the other hand has only a small effect, if any, on
choosing the true or realized solution.
With the arguments to be given in section~\ref{sec:9} we argue for the effects of
$S_I$ being even smaller in the future.
Nevertheless we have if we talk exactly also effects of $S_I$ even in
the future.
Otherwise the hypothesis that the integral (
\ref{7}) defining $| B_t \rangle$
would be senseless since the $\e^{- S_I}$-weighting is needed to suppress
the integrand $\e^{-S_I}$ enough to make hope of a sensible practical
convergence.

However, we have in section~\ref{sec:3} and will in section~\ref{sec:9} argue for that 
$S_I$ varies much less in the future than in Big Bang era.

It is now the purpose of the present section to use this only weak $S_I$
variation with the fields in the future to argue for an approximation
in density matrix terminology for the future part $| B_t \rangle$ of 
the functional integral.

Let us indeed perform the following considerations for estimating the
crude treatment of $| B_t \rangle$  which we shall use:
\begin{enumerate}
  \item[a)] Since $S_I$ has in practice only small non-trivial 
contributions in the future it is needed to involve contributions in 
the integral
\begin{eqnarray}\label{4.1}
S_{It' \,\mathrm{to}\,+\infty}
=\int_{t'}^{\infty} \mathrm{dt} \, \int \mathrm{d}\vec{x} \, L_I
\end{eqnarray}
from very large $t\ge t'$.
  \item[b)] At these enormous $t$ regions then at the end we get finally
a rather restricted range of solutions.
-- we can think of classical solutions here, if we like --
  \item[c)] Now the solutions from the enormously late times under a) 
have to be developed backward in time to the time $t'$ say to deliver
the state $| B_{t'} \rangle$ 
(really we first get $\langle  B_{t'}|\phi \rangle$ from equation 
(\ref{7})).
  \item[d)] Now we make the assumption that the system/world is 
sufficiently ``ergodic" and the large times so large and so smeared
out (also because of the smallness of the $L_I$-effects) that we can
take it that there is almost the same probability for finding the system
in state $| B_{t'} \rangle$ at any place in phase space allowed by
the conserved quantum numbers of the theory practically valid in the 
future era.
  \item[e)] Ignoring for simplicity the conserved quantities we
thus argued that with equal probability; equally distributed in phase
space, we have that $| B_{t'} \rangle$ will be any state.
  \item[f)] We can especially imagine that we have chosen a basis of
wave packet states $| w \rangle$ in the field configuration space so
that they fill smoothly the phase space
-- accessible without violating the conservation laws relevant --.
Taking these to be -- approximately -- orthonormal 
$\langle w| w' \rangle \approx \delta ww'$ we clearly get for the 
average expectation of the projection operator
\begin{eqnarray}\label{4.2}
P_{B_{t'}} = | B_{t'} \rangle \langle B_{t'} |
\end{eqnarray}
the estimate
\begin{eqnarray}\label{4.3}
\mathrm{av} (| B_{t'} \rangle \langle B_{t'} |)=
\frac{1}{N} \sum_{w} |w \rangle \langle w| \simeq \frac{1}{N} \underline{1}
\end{eqnarray}
where $N$ is the number of states in the basis
\begin{eqnarray}\label{4.4}
| w \rangle , ~w=1, \,2,\,\cdots , \,N.
\end{eqnarray}

That is to say we have argued for that our weak $S_I$-influence in future
combined with an assumed approximate ergodicity leads to that we can
approximate 
\begin{eqnarray}\label{4.5}
| B_{t'} \rangle \langle B_{t'} | \approx \frac{1}{N} \underline{1}
\end{eqnarray}
in practice for all $t'$ at least a bit later than the earliest
Big Bang.
\end{enumerate}

The crude estimate that we could replace $|B_t \rangle \langle B_t |$
by $\frac{1}{N}\underline{1}$ derived as formula
$(|B_t \rangle \langle B_t | \approx \frac{1}{N}\underline{1})$
were based on that $L_I$ were in practice small.

\section{Deriving a more usual probability formula}
\label{sec:5}
We shall now make use of approximation (\ref{4.5}) for the ``future
factor" in the functional integral in order to obtain an expression
rewriting the formulas like (\ref{3}), (\ref{4}) and
(\ref{17}) and (\ref{18}) into expressions analogous to (\ref{19}).

The calculation is in fact rather trivial, starting say from the most
general of our postulated expressions (\ref{17}):
\begin{eqnarray}
\langle \CO (\varphi|_t , \,\Pi|_t )\rangle
&=& \frac{\int \e^{iS} \CO (\varphi|_t , \,\Pi|_t ) \CD \varphi
}{\int \e^{iS}\CD \varphi}
\nonumber\\
&=& \frac{\langle B_t | \CO (\varphi|_t , \,\Pi|_t ) | A_t \rangle
}{\langle B_t | A_t \rangle}
\nonumber\\
&\stackrel{\mathrm{trivial\, step}}{=}&
\frac{\langle A_t | B_t \rangle 
\langle B_t |\CO (\varphi|_t , \,\Pi|_t ) |A_t \rangle
}{\langle A_t | B_t \rangle \langle B_t | A_t \rangle}
\nonumber\\
&\stackrel{\mathrm{using\, (\ref{4.5})}}{=}&
\frac{\langle A_t |\frac{1}{N} \underline{1} \CO 
(\varphi|_t , \,\Pi|_t ) |A_t \rangle
}{\langle A_t |\frac{1}{N} \underline{1} |A_t \rangle}
\nonumber\\
&=& \frac{\langle A_t | \CO (\varphi|_t , \,\Pi|_t ) |A_t \rangle
}{\langle A_t | A_t \rangle}
\end{eqnarray}
which is the completely usual quantum mechanical expression
for the expectation value of the operator $\CO (\varphi|_t , \,\Pi|_t )$
in the wave functional state $|A_t \rangle$.

With this expression we see that we should be allowed, as we anyway would
expect, to use $|A_t \rangle$ as the quantum state of the universe.

It should be noted though that our $|A_t \rangle$ is
{\it in principle} calculable from the ``theory" when as we shall of course,
count also the $S_I$-expression as part of the theory.
In this way our model is widely more ambitious than usual quantum mechanics:

We have -- much like the Hartle-Hawking no boundary proposal --
a functional integral (\ref{6}) delivering in principle the wave functional
$|A_t \rangle$.
In usual quantum mechanics the wave function is left for the experimental
physicist to find out from his somewhat difficult job of preparing the state.
In practice we would presumably have to let him be so helped by observation
and arrangements under the preparation that we almost leave to him the usual
job.
We should, however, have in mind that in preparing a state one will usually
need to trust that some material is a rather pure chemical substance or
that no disturbing cosmic radiation spoils the preparation.
These kinds of trusts are usually based on some empirical experience which
in turn makes use of that big assembles of pure substances are 
easily/likely  available and that generally cosmic ray has low intensity.
Such trusts however, are at the very root connected with the starting
state -- the cosmology -- of our world.
But this starting state for practical purposes is in our model based on
the activity of our $L_I$ in early Big Bang times of the initial state
of the universe.

Thus it is even in the practical way of preparing a quantum state a lot of
reference to our $S_I$.

If, however, somehow the universe develops into states where $L_I$ is
no longer negligible we should expect corrections to such an 
approximation $(|B_t \rangle \langle B_t | \approx \frac{1}{N}\underline{1})$.

\section{Time development and $S_I$ corrections to $|B_t \rangle$}
\label{sec:6}
{}From the definitions (\ref{6}) and (\ref{7}) of $|A_t \rangle$ and 
$|B_t \rangle$ it is trivial to derive the time development formulas
for these Hilbert space vectors (say for $t' > t$)
\begin{eqnarray}\label{6.1}
|A_{t'} \rangle &=& \int_{\mathrm{over\, time}-\infty \, \mathrm{to}\, t'}
 \e^{iS_{-\infty \,\mathrm{to}\, t'}} \CD \varphi
\nonumber\\
&=& \int_{\mathrm{over\,}t \, \mathrm{to}\, t'}
 \e^{iS_{t \,\mathrm{to}\, t'}} A_t [\varphi|_t] \CD \varphi
\nonumber\\
&=& \CU (t', \, t) |A_t \rangle
\end{eqnarray}
where $\CU (t', \, t)$ is the operator corresponding to the matrix
(with columns and rows marked by $\varphi|_t$ configurations)
\begin{eqnarray}\label{6.2}
\CU (\hat{\varphi}|^{'}_{t'},\, \hat{\varphi}|_t)
= \int_{\mathrm{over\,} t \, \mathrm{to}\, t'
\,\mathrm{with}\, \varphi|_{t'}=\hat{\varphi}|_{t'}
\,\mathrm{and}\, \varphi|_{t}=\hat{\varphi}|_{t}
}
\e^{iS_{t \,\mathrm{to}\, t'}}  \CD \varphi.
\end{eqnarray}
Similarly we have from (\ref{7}) for $t' > t$ again, first taking the 
complex conjugate of (\ref{7})
\begin{eqnarray}\label{6.3}
\langle \varphi |_t \,| B_t \rangle 
= \int_{\mathrm{beginning\, at}\, \varphi |_{t}}
\e^{-iS^{*}_{t \, \mathrm{to}\, +\infty}}\CD \varphi
\end{eqnarray}
and thus
\begin{eqnarray}\label{6.4}
\langle \varphi |_t \,| B_t \rangle 
= \int_{\mathrm{over}\, t \, \mathrm{to}\, t'}
\e^{-iS^{*}_{t \, \mathrm{to}\, +\infty}}
\langle \varphi |_{t'} | B_{t'} \rangle \CD \varphi
\end{eqnarray}
which can be written
\begin{eqnarray}\label{6.5}
|B_t \rangle = \CU_{\mathrm{with}\, L_I \to -L_I} (t', \,t)^+
\,|B_{t'} \rangle.
\end{eqnarray}
Here we used that e.g.
\begin{eqnarray}
S_{t \, \mathrm{to}\, +\infty} =
\int_{t}^{\infty} \mathrm{dt}
\int \mathrm{d}^3\vec{X} (\CL_R + i\CL_I)
\end{eqnarray}
where $\CL_R$ and $\CL_I$ are respectively the real and the imaginary
parts of the Lagrangian densities.
So
\begin{eqnarray}\label{6.6}
S^{*}_{t \, \mathrm{to}\, +\infty} =
\int_{t}^{\infty} \mathrm{dt}
\int \mathrm{d}^3\vec{x} (\CL_R - i\CL_I),
\end{eqnarray}
and now restricting ourselves for \{pedagogics/simplicity\} 
at first to boson fields we have (usually) that for them 
$\CL_R$ and $\CL_I$ are even order in the time derivatives which are
under latticification
\begin{eqnarray}\label{6.7}
\partial_t \varphi_{(t,\, \vec{x})}
\approx \frac{\varphi(t+\Delta t,\, \vec{x})- \varphi(t,\, \vec{x})
}{\Delta t} .
\end{eqnarray}
Thus conceived as operators between the configuration at the two close
by times $t$ and $t+ \Delta t$, i.e. with columns and rows marked by
$\varphi |_{t+\Delta t}$ and $\varphi |_t$ we have e.g.
\begin{eqnarray}\label{6.8}
(\CL_R + i\CL_I)^+ = \CL_R -i\CL_I
\end{eqnarray}
because
\begin{eqnarray}\label{6.9}
\CL_R^T = \CL_R ~~\mathrm{and}~~ \CL_I^T = \CL_I
\end{eqnarray}
and
\begin{eqnarray}\label{6.10}
\CL_R^* = \CL_R ~~\mathrm{and}~~ \CL_I^* = \CL_I.
\end{eqnarray}
In formula (\ref{6.5}) of course the meaning of the under symbol text
in the expression $\CU_{\mathrm{with}\,L_I \to -L_I}(t',\, t)^+$ is that
in addition to taking the Hermitian conjugation of $\CU(t',\, t)$ 
as defined by the matrix representation (\ref{6.2}) one shall shift the 
sign for all occurrences of the $L_I$-part of the Lagrangian or of the 
$L_I$-part of the Lagrangian density.
One should have in mind that it is easily seen that 
\begin{eqnarray}\label{6.11}
\CU(t',\, t)^{-1} =\CU_{\mathrm{with}\,L_I \to -L_I}(t',\, t)^+ .
\end{eqnarray}
Especially the ``usual" case of $L_I=0$ means that $\CU(t',\, t)$ becomes
unitary.
This relation (\ref{6.11}) together with (\ref{6.5}) and (\ref{6.1})
ensures that
\begin{eqnarray}\label{6.12}
\langle B_t | A_t \rangle =
\int \e^{iS_{-\infty \,\mathrm{to}\, +\infty}}\CD \varphi
\end{eqnarray}
can be true independent of the time $t$ chosen on the left hand side.

Since (6.1) represents a completely usual time development of the 
`wave function' $|A_t\rangle$ we have of course analogously to the usual 
theory
\begin{equation}
i\frac{d|A_t\rangle}{dt} = H|A_t\rangle 
\end {equation} 
where then H is the to the action   
\begin{equation}
S=S_R + S_I 
\end{equation}
corresponding Hamiltonian.
As we saw under point a) in section~\ref{sec:2} formula (\ref{19}) we can consider 
\begin{equation}
|A_t\rangle
\end{equation}
the wave function for the universe essentially. But really because of the 
normalizing denominator in (\ref{19}) it is rather the normalized  $|A_t\rangle$, 
namely
\begin{equation}
|A_t\rangle_{norm} = |A_t\rangle/\sqrt{\langle A_t|A_t\rangle}
\end{equation}
which is the true wave function. 

It is important to remeark that precisely because we now find that we 
shall use the normalized wave function rather than $|A_t\rangle$ itself we do not 
get as could be feared a lack of conservation of probability due to the 
non-unitarity of the time development. Have in mind that the to a non-real 
action corresponding Hamiltonian H will not be Hermitean! But with the 
normalizaion comming from the $\langle A_t|A_t\rangle$ in the denominator in (\ref{19}) the 
total probability will anyway remain unity.This result matches nicely with the 
from the slightly different start evaluated (9.22) below.      

\newpage
{\huge \bf Part II, Second Trial of Interpretation}

\section{Second Interpretation of the functional integral}
\label{sec:7}
Usually one only uses the functional integral over a time interval to
evaluate a transition matrix element from an initial time 
$t_i$ to a final time $t_f$
\begin{eqnarray}
U\left(\psi_f (\phi|_f),\psi_i (\phi|_i)   \right)
=\int  \CD^{\mathrm{ fixed\, time}} \phi|_f \int  \CD^{\mathrm{fixed\, time}} \phi|_i
\CD\phi\, \e^{iS_{t_{i}\,\mathrm{to}\,t_{f}}\left[\phi\right]}
\end{eqnarray}
where
\begin{eqnarray}
S_{t_{i}\,\mathrm{to}\,t_{f}}=
\int_{t_{i}}^{t_{f}} \int \CL(x)d^3\vec{x}dt
\end{eqnarray}
and the functional integral over $\CD\phi$ is restricted to $\phi$-functions
(field developments, or paths) which at times $t_i$ and $t_f$ respectively
coincides with $\phi|_i$ and $\phi|_f$ respectively.

In the present article we, however, have the ambition of having the
functional integral determine a priori not only the development with time 
but also say something about the initial conditions so that we a priori might 
ask for the probability of some dynamical variable $\CO$  say having certain
value $\CO$ at a certain time without imposing any initial conditions.
In order to obtain a formula or proceedure or how to obtain such
probabilities for what shall happen we have to assume such a formula.

We therefore need some intuitive and phenomenological guess leading to such
a formula/prescription.

In order to propose such a formula in a sensible way we shall first
consider a semiclassical approximation for our functional integral supposed
to be connected with and describing the development of the Universe
\begin{eqnarray}
\int \CD \phi \,\e^{i S[\phi]}
\end{eqnarray}
where we remember that in our model the $S[\phi]$ is not as usual real but 
is allowed to be complex.

\subsection{Semiclassical approach}
For first orientation let us imagine that the imaginary part of the action
$S[\phi]$ is effectively small in the sense that we can obtain the most
significant contributions to the functional integral by asking for saddle
points for the real part $S_R$.
That is we ask for field development solutions to the variational principle
\begin{eqnarray}
\delta S_R =0.
\end{eqnarray}
Without specifying the boundary conditions at $t \to \pm \infty$ in our
functional integral there should be (essentially) one solution for any
point in the (classical) phase space of the field theory described.
For the enumeration of the various development solutions $\phi$ we could
use the field and conjugate field configuration at any chosen moment of
time, to say.
However, now our hope and speculation is that the imaginary part should give
a probability weight distribution over the set ($\simeq$ phase space) of
these classical solutions.

\subsection{A first but wrong thinking}
It is clear that we must make a definition of an expectation value for 
function(al) $\CO$ say of the field development $\phi$ so that if a single
(semi) classical solution $\phi_{\mathrm{sol}}$ comes to be highly weighted then
this expectation value should be $\CO [\phi_{\mathrm{sol}}]$.

We might therefore at first think of 
\begin{eqnarray}\label{1.5}
\langle\CO\rangle = \frac{\int \CD \phi \CO [\phi] \,\e^{i S[\phi]}}
{\int \CD \phi \,\e^{i S[\phi]} }.
\end{eqnarray}

If really a single classical path contributed completely dominantly to both
numerator and denominator, then indeed we would obtain that this proposal
would obey
\begin{eqnarray}
\langle\CO\rangle = \CO[\phi_{\mathrm{sol\, dom}} ]
\end{eqnarray}
where $\phi_{\mathrm{sol\, dom}}$ is this single dominant solution.

It is however likely that if will be more realistic to imagine that 
there is a huge number of significant classical solutions $\phi_{\mathrm{sol}}$.
But then appears the ``problem" that in the expansion of the numerator 
functional integral into contributions from the various (semi) classical
solutions $\phi_{\mathrm{sol}\,i}$:
\begin{eqnarray}
\int \CD \phi \CO [\phi] \e^{i S[\phi]}
= \sum_{\phi_{\mathrm{sol}\,i\, \mathrm{all\, the\, classical\, solutions}}}
 \e^{i S[\phi_{\mathrm{sol}\,i}]} \CO[\phi_{\mathrm{sol}\,i}]
 \sqrt{\mathrm{det}_i}^{-1}
\end{eqnarray}
the various contributions contribute with quite different signs or rather
phases due to the appearance of the phase factor
$\e^{i {S_R} [\phi_{\mathrm{sol}\,i}]}$.
The proposal just put forward thus is not as it stands a usual average, 
it lacks the usual requirement of an average of being performed with 
a positive weight.
Rather the summation over the contribution becomes a summation with random
phases to a good approximation.
That means that if we classify in some ways the different solutions
$\phi_{\mathrm{sol}\,i}$ into classes, then what would sum up when such 
classes are combined would be the squared contributions rather than the 
contributions themselves.
In other words, if we define a contribution to
\begin{eqnarray}
\int \CD \phi \,\e^{i S[\phi]} \CO[\phi]
= \sum_{\phi_{\mathrm{sol}\,i}} \sqrt{\mathrm{det}_i}^{-1}
 \e^{i S[\phi_{\mathrm{sol}\,i}]} \CO[\phi_{\mathrm{sol}\,i}]
\end{eqnarray}

where

\begin{equation}
{\mathrm{det}_i} = \det \left ( 
\frac{\delta^2}{\delta \phi_1(x_1) \delta \phi_2(x_2)}
\right )
\end{equation}

from a certain class of semi classical solution $\CS_k$
then the quantities such as
\begin{eqnarray}
\int \CO \,\e^{i S} \CD\phi
\Big|_{\mathrm{from\, class\, \CS_k}}
\equiv 
 \sum_{\phi_{\mathrm{sol}\,i\in \CS_k}}
 \sqrt{\mathrm{det}_i}^{-1}
 \e^{i S[\phi_{\mathrm{sol}\,i}]} \CO[\phi_{\mathrm{sol}\,i}]
\end{eqnarray}
obey approximately
\begin{eqnarray}\label{asq}
&&\Bigg| \int \CO \,\e^{i S} \CD\phi
\Big|_{\mathrm{from\, class\, \CS_1}} \Bigg|^2 +
\Bigg| \int \CO \,\e^{i S} \CD\phi
\Big|_{\mathrm{from\, class\, \CS_2}} \Bigg|^2
\nonumber\\&&
\approx 
\Bigg| \int \CO \,\e^{i S} \CD\phi
\Big|_{\mathrm{from\, class\, \CS_{1}\cup \CS_{2}}} \Bigg|^2.
\end{eqnarray}
However we do not have a similar addition formula for numerical values as
\begin{eqnarray}
\Bigg| \int \CO \,\e^{i S} \CD\phi \Big|_{\mathrm{from\, class\, \CS_1}} \Bigg|^2,
\end{eqnarray} 
when they are not squared.
However, of course, we do have
\begin{eqnarray}
&& \int \CO \,\e^{i S} \CD\phi
\Big|_{\mathrm{from\, class\, \CS_1}}  +
 \int \CO \,\e^{i S} \CD\phi
\Big|_{\mathrm{from\, class\, \CS_2}} 
\nonumber\\&&
\approx 
 \int \CO \,\e^{i S} \CD\phi
\Big|_{\mathrm{from\, class\, \CS_{1}\cup \CS_{2}}} 
\end{eqnarray}
but this relation has terms of typically rather random phases.

\subsection{Approaching a probability assumption}
If we take $\CO$ to be a ``projection operator" in the sense of being a 
functional of $\phi$ only taking the values 0 and 1 then
$\int \CO \,\e^{i S} \CD\phi \Big|_{\mathrm{from\, class\, \CS_k}}$
should give the chance for solutions in the class $\CS_k$ to pass through the
configuration-class for which 
$\CO[\phi]=1$.
Because of the (random) phase and the lack of simple numerical additivity
mentioned if the foregoing subsection we are driven to assume that the probability
for $\phi$ being in the $\CO[\phi]=1$ region must be given by the squared 
contributions
\begin{eqnarray}
\Bigg| \int \CO \,\e^{i S} \CD\phi
\Big|_{\mathrm{from\, class\, \CS_k}} \Bigg|^2.
\end{eqnarray}

Calling the region in the space of $\phi$'s consisting of the $\phi$'s obeying
$\CO[\phi]=1$ with our ``project $\CO$", for region $M$, we get
\begin{eqnarray}
\mathrm{Prob}(M) \propto 
\Bigg| \int M \,\e^{i S} \CD\phi
\Big|_{\mathrm{from\, class\, \CS_k}} \Bigg|^2
\end{eqnarray}
for restriction to the class $\CS_k$.

This means using the probability of the complementary set $\bC M$ of $M$
\begin{eqnarray}
\mathrm{Prob}(\bC M) \propto 
\Bigg| \int \bC M \,\e^{i S} \CD\phi
\Big|_{\mathrm{from\, class\, \CS_k}} \Bigg|^2
\end{eqnarray}
and the additivity (\ref{asq})
\begin{eqnarray}
\Bigg| \int \,\e^{i S} \CD\phi
\Big|_{\mathrm{from\, class\, \CS_k}} \Bigg|^2 =
\Bigg| \int_M \,\e^{i} \CD\phi
\Big|_{\mathrm{from\, class\, \CS_k}} \Bigg|^2 +
\Bigg| \int_{\bC M} \,\e^{i S} \CD\phi
\Big|_{\mathrm{from\, class\, \CS_k}} \Bigg|^2\nonumber\\
{   } 
\end{eqnarray}
we derive
\begin{eqnarray}
\displaystyle
\mathrm{Prob}(M)  &=&
\frac{\Bigg| \displaystyle\int_{M} \,\e^{i S} \CD\phi 
\Big|_{\mathrm{from\, class\, \CS_k}} \Bigg|^2}
{\Bigg|\displaystyle \int \,\e^{i S} \CD\phi
\Big|_{\mathrm{from\, class\, \CS_k}} \Bigg|^2
}.\nonumber\\
&=&
\frac{\Bigg|\displaystyle \sum_{\phi_{\mathrm{sol}\,i}\,\mathrm{in}\,M} \,\e^{i S} 
\sqrt{\mathrm{det}_i}^{-1}
\Big|_{\mathrm{from\, class\, \CS_k} }\Bigg|^2}
{\Bigg|\displaystyle \sum_{\phi_{\mathrm{sol}\,i}} \,\e^{i S} 
\sqrt{\mathrm{det}_i}^{-1}
\Big|_{\mathrm{from\, class\, \CS_k} }\Bigg|^2
}
\nonumber\\
&\stackrel{\mathrm{using\, random\, phases}}{\cong} &\displaystyle
\frac
{\displaystyle\sum_{\phi_{\mathrm{sol}\,i}\,\mathrm{in}\,M\cap \CS_k} \,\e^{-2 S_I} 
\mathrm{det}^{-1}
}
{\displaystyle
\sum_{\phi_{\mathrm{sol}\,i}\,\mathrm{in}\,\CS_k} \,\e^{-2 S_I} 
\mathrm{det}^{-1}
}.
\end{eqnarray}
Here in principle of a classical approximation the $\e^{-2 S_I}$ factor
is much more important than the ``quantum correction" $\mathrm{det}^{-1}$.
Thus we would ignore the determinant $\mathrm{det}^{-1}$ factor in first
approximation.

Then we arrived to the picture here that the probability distribution over
phase space - at some chosen time, that due to Liouville's theorem does not
matter - is given by $\e^{-2 S_I [\phi_\mathrm{sol}]}$ where $\phi_\mathrm{sol}$
is the classical field solution associated with the point in phase space for 
which $\e^{-2 S_I [\phi_\mathrm{sol}]}$ shall be the probability density.

\subsection{About the effect of $S_I$ in the classical approximation}
To appreciate the just given probability density 
$\e^{-2 S_I [\phi_\mathrm{sol}]}$ over phase space
\begin{eqnarray}
P\left(\phi |_{t_0}, \,\Pi |_{t_0} \right)
\CD \phi |_{t_0} \CD \Pi |_{t_0} 
\propto 
\e^{-2 S_I [\phi_\mathrm{sol}]} 
\CD \phi |_{t_0}, \CD \Pi |_{t_0}
\end{eqnarray}
one should have in mind that in the classical approximation of the universe
developing along a solution $\phi_\mathrm{sol}$ to the equations of motion
\begin{eqnarray}
\delta S_R = 0,
\end{eqnarray}
the development is given quite uniquely by the equations of motion.
The only place in which the imaginary part then comes in is in weighting with
various probability densities the various ``initial state data"
$\left(\phi |_{t_0}, \,\Pi |_{t_0} \right)$
-- i.e. the phase space point --.
Once you know the initial state of the (sub)system considered the equation
of motion determines everything in the classical approximation determined by
$S_R$ just described, the $S_I$ gets totally irrelevant.
In other words it is \underline{only} to know something about the
``initial state" that $S_I$ has relevance.
Here the usual terminology of ``initial state" shall especially in our model 
not be taken too seriously in as far as it with the word ``initial" refers
to a beginning moment, the Big Bang start say. No,
as we just mentioned one can use any moment of time $t_0$ for the 
description of the phase space describing the set of classical solutions
$\phi_{\mathrm{sol}}$.
This $t_0$ time does not have to be the first moment
-- even if such one should exist --.
Rather we can use any moment of time as $t_0$.
In the usual theory we would tend to use  $t_0$ being the initial moment and 
the state at this moment should then be one of very low entropy describing
our start of universe state.
However, in our model there is the rather unusual feature that the probability
weight $\e^{-2 S_I [\phi_{\mathrm{sol}}] }$ 
is given via a functional $S_I [\phi_{\mathrm{sol}}]$ 
depending on how the solution $\phi_{\mathrm{sol}}$ behaves at all
different times $t$ and not only at $t_0$.
Since we by the classical equations of motion can calculate the whole time
development $\phi_{\mathrm{sol}}$ from the fields and their conjugate
$\left(\phi |_{t_0}, \,\Pi |_{t_0} \right)$ at some chosen time $t_0$, 
we can of course also consider $\e^{-2 S_I [\phi_{\mathrm{sol}}] }$ 
as a function of only the data at $t_0$, 
$\left(\phi |_{t_0}, \,\Pi |_{t_0} \right)$.

So it is only by the fact that in our model
$\e^{-2 S_R [\phi_{\mathrm{sol}}] }$ is a rather \underline{simple} function
of $\phi_{\mathrm{sol}}$ and thus because of the often chaotic development 
of the fields by the classical equations of motion typically a complicated
function(al) of the time $t_0$ data.
With a more usual model one might think of the initial state in a 
``first moment" $t_0$ would be specified by some sort of cosmological model
or no boundary condition.
In this case the probability density should be rather simple in terms of the
first moment data.
The \underline{simplicity} of $\e^{-2 S_R [\phi_{\mathrm{sol}}] }$ as 
functional of the $\phi_{\mathrm{sol}}$-behavior even at late times to
some extend is extremely dangerous for our model showing observable effects
not observed experimentally.
Indeed an especially high probability for initial states leading to a 
special sort of happenings today say would look as a hand of God effect
seeking to arrange just this type of happenings to occur.
In practise we never know the state of the universe totally at a moment of
time.
So there would usually be possibilities to adjust a bit the initial 
conditions. That ciould then in our model have happened in such a way 
that events or things to happen in the future gets arranged, if it can be done 
so as to organize especially big $\e^{-2 S_R [\phi_{\mathrm{sol}}] }$ i.e.
an especially low $S_I$.
So a priori there would in our model be such ``hand of God effects".
In a later section we shall, however, invent or propose a possible
explanation that could make the era of today be of very little significance
for the value of $S_I$ so that in the first approximation it mainly the early
time features of a solution that counts for its probability density
\begin{eqnarray}
\e^{-2 S_I [\phi_{\mathrm{sol}}]}
\sim f(\phi_{\mathrm{sol}} |_{\hbox{``early times''}} )
= f(\mathrm{early\, time\, part\, of\, \phi_{\mathrm{sol}}})
\end{eqnarray}

\subsection{Relation to earlier publications}\label{subsec:7.5}
We have earlier published articles working in the classical approximation
seeking to produce a model behind the second law of thermodynamics by
assigning a probability $P$ over the phase space of the Universe.
It were also there the point that this probability density $P$ in our
model should depend in the same way on the state at all times.
We already proposed that this $P$ were obtained by imposing an imaginary
part for the action $S_I$.
According to the above we clearly have
\begin{eqnarray}
P \propto \e^{-2 S_I}.
\end{eqnarray}

\section{Suggestion of the quantum formula}
\label{sec:8}
We already suggested above that if $M$ denotes a subset of paths, 
e.g. those taking values in certain subset of $\phi |_t$-configuration
space in a moment of time $t$, then the probability for the true path being
in $M$ would be 
\begin{eqnarray}
\mathrm{Prob}(M)=
\frac{|\int_{M} \e^{i S} \CD \phi |^2}
{|\int \e^{i S} \CD \phi |^2}.
\end{eqnarray}
We imagine the paths to be described by the field $\phi$ as function over
$\bR^4$, the Minkowski space.
Thus we could use such an $M$ to describe e.g. the project of the possible
development $\phi$ to some subspace of configuration space $M_i$ for a 
series of moments $t_i, \,i=1, 2, \cdots, n$.
In fact then we would have
\begin{eqnarray}
M = \left\{ \phi \in \{\mathrm{paths} \}\Big| 
\phi |_{t_i} \in M_i \,\mathrm{for\, all}\, i \right\}.
\end{eqnarray}
It would in this case be natural to think of the functional integral
\begin{eqnarray}
\int_M \e^{i S} \CD \phi 
\end{eqnarray}
as a product of a series of functional integral associated with the various
time intervals in the series of times
$-\infty < t_1 < t_2 < \cdots < t_n < \infty$.
In fact let us define
\begin{eqnarray}
\CU_{t_i \,\mathrm{to}\, t_{i+1}}(\phi |_{t_{i+1}}, \phi |_{t_{i}})\equiv 
\int_{\mathrm{BOUNDARY}~\atop \phi|_{t_{i}} \,\mathrm{and}\,
\phi |_{t_{i+1}} \,\mathrm{kept}}
\e^{i S_{t_i \,\mathrm{to}\, t_{i+1} [\phi]}} \CD \phi
. 
\end{eqnarray}
Here 
\begin{eqnarray}
S_{t_i\,\mathrm{to}\,t_{i+1}}[\phi]
=\int^{t_{i+1}}_{t_i}\int\CL(x)
d^3\vec{x} dt
\end{eqnarray}
and remember that we here have the complex $d(x)$, 
\begin{eqnarray}
\CL(x)=\CL_R(x)+i\CL_I(x).
\end{eqnarray}
We then can write
\begin{eqnarray}
&&
\int_M\e^{iS}\CD\phi
\nonumber\\&&
=\int_i\CD^{(3)}\phi|_{t_i}
\CU_{t_n\,\mathrm{to}\,\infty}(\phi|_\infty,\phi|_{t_n})
\theta_{M_n}[\phi|_{t_n}]
\CU_{t_{n-1}\,\mathrm{to}\,t_n}
(\phi|_{t_n},\phi|_{t_{n-1}})\nonumber\\&&
\qquad\qquad\theta_{M_{n-1}}[\phi|_{t_{n-1}}]
\cdots
\theta_{M_1}[\phi|_{t_1}]
\CU
(\phi|_{t_1},\phi|_{t_{-\infty}})
\end{eqnarray}
where
$\theta_i[\phi|_{t_i}]$
is the function
\begin{eqnarray}
\theta_i[\phi|_{t_i}]=
\left\{
  \begin{array}{ll}
  1     &\mbox{for}~\phi|_i\in\CU_i    \\
  0     &\mbox{for}~\phi|_i\notin\CU_i      \\
  \end{array}
\right.
\end{eqnarray}
We can also write this expression in language of a genuine operator
product
\begin{eqnarray}
\int_M\e^{iS}\CD\phi=
\CU_{t_n\,\mathrm{to}\,\infty}\theta_n
\CU_{t_{n-1}\,
\mathrm{to}\,t_n}\theta_{n-1}\cdots
\theta_1\CU_{-\infty\,\mathrm{to}\,t_1}~.
\end{eqnarray}
where the $\theta_i$'s are now conceived of as projection operators on
the space of wave functionals characterized by being zero outside $M_i$,
\begin{eqnarray}
\theta_i\psi(\phi|_{t_i})=
\theta_i(\phi|_{t_i})\psi(\phi|_{t_i})
=\left\{
  \begin{array}{ll}
 0      & \mbox{for}~\phi|_{t_i}\notin M_i   \\
 \psi(\phi|_{t_i})      & \mbox{for}~\phi|_{t_i}\in M_i    \\
  \end{array}
\right.
\end{eqnarray}
Here $\psi$ is a possible/general wave functional for the state of the universe,
in the formula presented at the moment $t_i$.

In this operator formalism our probability formula takes the form
\begin{eqnarray}\label{fpmp64}
\mathrm{Prob}(M)&=&
\mathrm{Prob}(M_1,M_2,\cdots,M_n)
\nonumber\\
&=&
\frac{\left|
\CU_{t_n\,\mathrm{to}\,\infty}\theta_n
\CU_{t_{n-1}\,\mathrm{to}\,t_n}\theta_{n-1}
\cdots\theta_1\CU_{-\infty\,\mathrm{to}\,t_1}
\right|^2}{\left|
\CU_{t_n\,\mathrm{to}\,\infty}
\CU_{t_{n-1}\,\mathrm{to}\,t_n}
\cdots
\CU_{-\infty\,\mathrm{to}\,t_1}
\right|^2}
\end{eqnarray}
Since we are anyway in the process of arguing along to
just make an assumption about how to interpret in terms of 
probabilities for physical quantities of our complex action
functional integral, we might immediately see that it would be
suggestive to extend the validity of this formula for probabilities for 
field variables to also be valid for distributions in the conjugate
fields $\Pi|_{t_i}$ or in combinations,
\begin{eqnarray}\label{fpmp67}
&&
\mathrm{Prob}(\CO_1\in\tilde M_1,
\CO_2\in\tilde M_2,\cdots,
\CO_n\in\tilde M_n
)
\nonumber\\&&
=
\frac{
\left|
\CU_{t_n\,\mathrm{to}\,\infty}
P_{\CO_n\in\tilde M_n}
\CU_{t_{n-1}\,\mathrm{to}\,t_n}
P_{\CO_{n-1}\in\tilde M_{n-1}}
\cdots
P_{\CO_{1}\in\tilde M_{1}}
\CU_{-\infty\,\mathrm{to}\,t_1}
\right|^2}{
\left|
\CU_{t_n\,\mathrm{to}\,\infty}
\cdots
\CU_{-\infty\,\mathrm{to}\,t_1}
\right|^2}
\end{eqnarray}

Provided this proposal is not inconsistent to assume, we will assume
it because it would be quite reasonable to assume that the analogous 
formula to (\ref{fpmp64}) should be valid for any change for variables
between $\phi$ and $\Pi$ in the formulation of our functional integral.

\subsection{Consistency and no need for boundary conditions}
It should be kept in mind that we expect that due to the presence of the 
imaginary part $S_I$ in the action $S$ it is not needed to require any
boundary conditions at $t \to \pm \infty$ so that we basically can
remove as not relevant the $\phi |_\infty$ and $\phi |_{- \infty}$
boundaries which one would at first have considered to be needed in the 
expressions
$\CU_{-\infty\,\mathrm{to}\,t_1}(\phi|_{t_1},\phi|_{-\infty})$
and
$\CU_{t_n\,\mathrm{to}\,+\infty}(\phi|_{\infty},\phi|_{t_n})$.
The imaginary part $S_I$ is in fact expected to weight various contributions
so strongly different that whenever the by this weighting flavored
component in $\phi |_\infty$ say is at all allowed by a potential choice
of boundary condition then that contribution will dominate so much that all 
over contributions will be relatively negligible.
So after taking the ratio for normalization such as (\ref{fpmp67})
the choice of the boundary conditions for $\phi |_\infty$ and 
$\phi |_{- \infty}$ becomes irrelevant.
This irrelevance of the boundary conditions would indeed allow us to formally
put in according to our convenience of calculation whatever boundaries we
might like provided it does not precisely kill the boundary wave function
component flavored by the $S_I$.
For instance we could put in at the infinity density matrices taken to be
unity, since it does not matter anyway what we put and a unit matrix
$\rho_1 =1$ and $\rho_k =1$ would not suppress severely any state such as
the flavored one(s).

By this trick we could write our formula for probability
\begin{eqnarray}\label{2.13}
&&
\mathrm{Prob}(\CO_1\in\tilde M_1,
\CO_2\in\tilde M_2,\cdots,
\CO_n\in\tilde M_n
)
\nonumber\\&&
=\mathrm{Tr}(
\CU_{t_n\,\mathrm{to}\,\infty}
P_{\CO_n\in\tilde M_n}
\CU_{t_{n-1}\,\mathrm{to}\,t_n}
P_{\CO_{n-1}\in\tilde M_{n-1}}
\cdots
P_{\CO_{1}\in\tilde M_{1}}
\CU_{-\infty\,\mathrm{to}\,t_1}
\nonumber\\&&~~~~~~~~
\CU^\dag_{-\infty\,\mathrm{to}\,t_1}
P_{\CO_1\in\tilde M_1}
\cdots
P_{\CO_{n-1}\in\tilde M_{n-1}}
\CU^\dag_{t_{n-1}\,\mathrm{to}\,t_n}
P_{\CO_{n}\in\tilde M_{n}}
\CU^\dag_{t_{n}\,\mathrm{to}\,\infty}
)
\nonumber\\&&~~~
/\mathrm{Tr}
(
\CU_{t_n\,\mathrm{to}\,\infty}
\CU_{t_{n-1}\,\mathrm{to}\,t_n}
\cdots
\CU_{-\infty\,\mathrm{to}\,t_1}
\CU^\dag_{-\infty\,\mathrm{to}\,t_1}
\cdots
\CU^\dag_{t_{n-1}\,\mathrm{to}\,t_n}
\CU^\dag_{t_{n}\,\mathrm{to}\,\infty}
)~.~~~
\end{eqnarray}
Here the reader should have in mind that because of the imaginary part
in the action $S = S_R + i S_I$ the different development operators
\begin{eqnarray}
\CU_{t_i\,\mathrm{to}\, t_{i+1}}(\phi|_{t_{i+1}},\phi|_{t_{i}})
=\int_{\mathrm{WITH\,BOUNDARIES}~\atop
\phi|_{t_{i+1}}\,\mathrm{and}\,\phi|_{t_{i}}
\,\mathrm{at}\,t_{i+1}\,\mathrm{and}\,t_i\,
\mathrm{respectively} 
}\e^{iS[\phi]}\CD\phi
\end{eqnarray}
are in general not as usual unitary.
Therefore it is quite important to distinguish,
\begin{eqnarray}
\CU^\dag_{t_i\,\mathrm{to}\,t_{i+1}}
\stackrel{\mathrm{in\,general}}\neq
\CU^{-1}_{t_i\,\mathrm{to}\,t_{i+1}}.
\end{eqnarray}

\section{Practical Application Formulas}\label{sec:9}
\subsection{Practical application philosophy}\label{practical application}

Although in principle our theory is so much a theory of everything that
it should even tell what really happens and not only what is allowed by
the equations of motion, we must of course admit that even
we knew the parameters of both $S_I$ and $S_R$ it would be so exceedingly
hard to calculate what really happens that can\underline{not} do that.

We are thus first of all interested in using some reasonable approximations
to derive (in a spirit of a correspondence principle) some rules 
coinciding under practical conditions with the quantum mechanics
(or quantum field theory rather) rules we usually use.

Now as is to be explained in this section 
we can by means of requirements of monopoles and using the Standard Model
gauge symmetries and homogeneity of the Lagrangian in the fermion fields
argue that there will in the present era where Higgs particles are seldom
and Standard model applicable be only very small effects of $S_I$.
We also seem to have justified to make the same assumption for very huge time
spans in the future so that also until very far out in future the 
influence of $S_I$ is small.
Even if we imagine that in the very long run $S_I$ selects almost
uniquely the state or rather development - as we used above to argue that
the boundary conditions  $\phi |_{- \infty}$ and $\phi |_{+ \infty}$
were unimportant - then in the practical (i.e. rather near) future we
would expect the far future determination to deliver under an ergodicity
assumption an effective density matrix proportional to the unit matrix.
 
\subsection{Insertion of practical future treatment  into interpretation 
formula}
The above suggestion for the treatment of the practical future to be equally
likely in ``all" (practical) states is implemented by replacing what is
basically taking the place of a future density matrix
$\rho_f$ in our interpretation formula (\ref{2.13}) namely
\begin{eqnarray}
\rho_f \approx \CU^+_{t_{n}\,\mathrm{to}\,\infty}
\CU_{t_{n}\,\mathrm{to}\,\infty}
\end{eqnarray}
by a normalized unit density matrix
\begin{eqnarray}
\rho_f \approx \frac{1}{N}\underline{1},
\end{eqnarray}
where N is the dimension of the Hilbert space.
(In practice $N$ is infinite) So the interpretation formula becomes
\begin{eqnarray}\label{equation}
&&
\mathrm{Prob}(\CO_1\in\tilde M_1,
\CO_2\in\tilde M_2,\cdots,
\CO_n\in\tilde M_n
)
\nonumber\\&&
=\mathrm{Tr}(
P_{\CO_n\in\tilde M_n}
\CU_{t_{n-1}\,\mathrm{to}\,t_n}
P_{\CO_{n-1}\in\tilde M_{n-1}}
\cdots
P_{\CO_{1}\in\tilde M_{1}}
\CU_{-\infty\,\mathrm{to}\,t_1}
\nonumber\\&&~~~~~~~~
\CU^\dag_{-\infty\,\mathrm{to}\,t_1}
P_{\CO_1\in\tilde M_1}
\cdots
P_{\CO_{n-1}\in\tilde M_{n-1}}
\CU^\dag_{t_{n-1}\,\mathrm{to}\,t_n}
)
\nonumber\\&&~~~
/\mathrm{Tr}
(
\CU_{t_n\,\mathrm{to}\,\infty}
\cdots
\CU^\dag_{-\infty\,\mathrm{to}\,t_1}
\cdots
\CU^\dag_{t_{n-1}\,\mathrm{to}\,t_n}
)~.~~~
\end{eqnarray}
where we used that 
\begin{eqnarray}
P_{\CO_n\in\tilde M_n}
=P_{\CO_n\in\tilde M_n}^2~.
\end{eqnarray}

\subsection{Conditional Probability}
With formulas like (\ref{2.13}) or (\ref{equation}) we can easily form also 
conditional probabilities such as 
\begin{eqnarray}
&&
\mathrm{Prob}(\CO_{p+1}\in\tilde M_{p+1},
\cdots, \CO_n\in\tilde M_n
| \CO_1\in\tilde M_1 
,\cdots,
\CO_p\in\tilde M_p )
\nonumber\\&&
= \mathrm{Prob}(
\CO_1\in\tilde M_1
,\cdots,
\CO_p\in\tilde M_p
,\cdots,
\CO_n\in\tilde M_n)
\nonumber\\&&~~~
/\mathrm{Prob}(\CO_1\in\tilde M_1
,\cdots,
\CO_p\in\tilde M_p)
~.~~~
\end{eqnarray}
In order to determine what happens if we know the wave function in some 
moment.
Let us as an example consider the idealized situation of a case in which 
we know -- by preparation set up -- the whole state of the universe of
one moment of time.
This we could imagine being described by taking a series of 
$P_{\CO_i\in\tilde M_i}$ projection of the same moment of time,
the moment in which we suppose that we know the wave function.
For consistency and for being able to take the limit of them being at
same time -- and therefore with an ill-determined algebraic order in
(\ref{2.13}) we must assume these same time projectors to commute.
If we consider the situation that we already know that the system has
all these $\CO_i$ in the small regions.
$\tilde M_i$ because we know the wave function at their common time
$t_{\mathrm{com}}$ then we are after that discussing only the conditional
probabilities with the set of relations $\CO_i\in\tilde M_i$, 
$i=1, \cdots, p$, taken as fixed.

For simplicity let us consider the simple case that we just ask for 
if a variable $\CO_n$ at a later time being in $\tilde M_n$
or not then the conditional probability is in (\ref{2.13}) form
\begin{eqnarray}
&&
\mathrm{Prob}(\CO_n\in\tilde M_n
| \psi )
\nonumber\\&&
=
\mathrm{Tr}
( \CU^\dag_{t_n\,\mathrm{to}\,\infty}
P_{\CO_n\in\tilde M_n}
\CU_{t_\mathrm{com}\,\mathrm{to}\,t_n}
P_{\CO_1\in\tilde M_1}
P_{\CO_2\in\tilde M_2}
\cdots
P_{\CO_p\in\tilde M_p}
\CU_{-\infty\,\mathrm{to}\,t_\mathrm{com}}
\nonumber\\&&~~~~
\CU^\dag_{-\infty\,\mathrm{to}\,t_\mathrm{com}}
P_{\CO_p\in\tilde M_p}
\cdots
P_{\CO_2\in\tilde M_2}
P_{\CO_1\in\tilde M_1}
\CU^\dag_{t_\mathrm{com}\,\mathrm{to}\,t_n}
P_{\CO_n\in\tilde M_n}
\CU^\dag_{t_n\,\mathrm{to}\,\infty})
\\&&~~
\biggm/\mathrm{Tr}(
\CU_{t_\mathrm{com}\,\mathrm{to}\,t_n}
P_{\CO_1\in\tilde M_1}
\cdots
P_{\CO_p\in\tilde M_p}
\CU_{-\infty\,\mathrm{to}\,t_\mathrm{com}}
\CU^\dag_{-\infty\,\mathrm{to}\,t_\mathrm{com}}\nonumber\\&&
\qquad\qquad
P_{\CO_p\in\tilde M_p}
\cdots
P_{\CO_1\in\tilde M_1}
\CU^\dag_{t_\mathrm{com}\,\mathrm{to}\,t_n}~.
\nonumber
\end{eqnarray}
Herein we can substitute
\begin{eqnarray}
|\psi \rangle \langle \psi|= 
P_{\CO_1\in\tilde M_1}
P_{\CO_2\in\tilde M_2}
\cdots
P_{\CO_p\in\tilde M_p}
\end{eqnarray}
and obtain using as usual for traces Tr(AB)=Tr(BA) 
\begin{eqnarray}
&&
\mathrm{Prob}(\CO_n\in\tilde M_n
| \psi )
\nonumber\\&&
=
\langle \psi | 
\CU^\dag_{t_\mathrm{com}\,\mathrm{to}\,t_n}
P_{\CO_n\in\tilde M_n}
\CU^\dag_{t_n\,\mathrm{to}\,\infty}
\CU_{t_n\,\mathrm{to}\,\infty}
P_{\CO_n\in\tilde M_n}
\CU_{t_\mathrm{com}\,\mathrm{to}\,t_n}
| \psi \rangle\nonumber\\&&
\langle \psi |
\CU_{-\infty\,\mathrm{to}\,t_\mathrm{com}}
\CU^\dag_{-\infty\,\mathrm{to}\,t_\mathrm{com}}
| \psi \rangle
\nonumber\\&&~~
\biggm/( \langle \psi |
\CU^\dag_{t_\mathrm{com}\,\mathrm{to}\,\infty}
\CU_{t_\mathrm{com}\,\mathrm{to}\,\infty}
| \psi \rangle 
\langle \psi |
\CU_{-\infty\,\mathrm{to}\,t_\mathrm{com}}
\CU^\dag_{-\infty\,\mathrm{to}\,t_\mathrm{com}}
| \psi \rangle )
\nonumber\\&&
\equiv  \langle \psi |
\CU_{t_\mathrm{com}\,\mathrm{to}\,t_n}
P_{\CO_n\in\tilde M_n}
\CU^\dag_{t_n\,\mathrm{to}\,\infty}
\CU_{t_n\,\mathrm{to}\,\infty}
P_{\CO_n\in\tilde M_n}
\CU_{t_\mathrm{com}\,\mathrm{to}\,t_n}
| \psi \rangle\nonumber\\&&
\qquad\qquad\bigm/ \langle \psi |
\CU^\dag_{t_\mathrm{com}\,\mathrm{to}\,\infty}
\CU_{t_\mathrm{com}\,\mathrm{to}\,\infty}
| \psi \rangle~.
\nonumber\\
\end{eqnarray}
We may rewrite this expression in a suggestive way of the how it is modified
relative to usual quantum mechanics by defining the final state density
matrix from time $t_n$
\begin{eqnarray}
\rho_{f\,\mathrm{from}\,t_n}
\equiv 
\CU^\dag_{t_n\,\mathrm{to}\,\infty}
\CU_{t_n\,\mathrm{to}\,\infty}
\end{eqnarray}
in the following way 
\begin{eqnarray}
\mathrm{Prob}(\CO_n\in\tilde M_n
| \psi )=
\frac{ \langle \psi |
\CU_{t_{\mathrm{com}}\,\mathrm{to}\,t_{n}}
P_{\CO_n\in\tilde M_n}
\, \rho_{f\,\mathrm{from}\,t_{n}}\,
P_{\CO_n\in\tilde M_n}
\CU^\dag_{t_{\mathrm{com}}\,\mathrm{to}\,t_n }|\psi\rangle
}{
\langle \psi |
\CU_{t_\mathrm{com}\,\mathrm{to}\,t_n}
\, \rho_{f\,\mathrm{from}\,t_n}\,
\CU^\dag_{t_\mathrm{com}\,\mathrm{to}\,t_n}|\psi\rangle
}\nonumber\\
{   }
\end{eqnarray}
Here $\CU_{t_\mathrm{com}\,\mathrm{to}\,t_n}$ is really a non unitary
$S$-matrix or development matrix for the time interval $t_\mathrm{com}$
at which we have $\psi$ to $t_n$ at which we look for $\CO_n$.
It is easy to see that we could have replaced $P_{\CO_n\in\tilde M_n}$
by a large set of commuting projections at a second common time $t_f$
destined to the single wave function $| \psi_f \rangle$ and thus
allowing to replace the series of projections put in place of
$P_{\CO_n\in\tilde M_n}$ by $| \psi_f \rangle \langle \psi_f |$.
Then we get for the probability of the transition from
$|\psi\rangle$ to $|\psi_f\rangle$
\begin{eqnarray}\label{pm99}
\mathrm{Prob}\left(|\psi_f\rangle\big||\psi\rangle\right)&=&
\frac{\langle\psi|\CU_{t_\mathrm{com}\,\mathrm{to}\,t_n}|\psi_f\rangle
\langle\psi_f|\rho_{f\,\mathrm{from}\,t_n}|\psi_f\rangle
\langle\psi_f|\CU^\dag_{t_\mathrm{com}\,\mathrm{to}\,t_n}|\psi\rangle
}{\langle\psi|\CU_{t_\mathrm{com}\,\mathrm{to}\,t_n}\rho_{f\,\mathrm{from}\,t_n}
\CU^\dag_{t_\mathrm{com}\,\mathrm{to}\,t_n}
|\psi\rangle}
\nonumber\\
&=&\left|
\langle\psi|\CU_{t_\mathrm{com}\,\mathrm{to}\,t_n}|\psi_f\rangle
\right|^2
\frac{\langle\psi_f|
\rho_{f\,\mathrm{from}\,t_n}|\psi_f\rangle}
{\langle\psi|\CU_{t_\mathrm{com}\,\mathrm{to}\,t_n}
\rho_{f\,\mathrm{from}\,t_n}
\CU^\dag_{t_\mathrm{com}\,\mathrm{to}\,t_n}|\psi\rangle
}\nonumber\\
{   }
\end{eqnarray}
Now we can compare this expression with the usual transition probability
expression when $S$ is only real $=S_R$,
\begin{eqnarray}
\mathrm{Prob_\mathrm{usual}}
\left(|\psi_f\rangle\big||\psi\rangle\right)
&=&
\langle\psi|\CU_{t_\mathrm{com}\,\mathrm{to}\,t_n}|\psi_f\rangle
\cdot
\langle\psi_f|\CU^\dag_{t_\mathrm{com}\,\mathrm{to}\,t_n}|\psi\rangle
\nonumber\\&=&
\left|\langle\psi|\CU_{t_\mathrm{com}\,\mathrm{to}\,t_n}|\psi_f\rangle\right|^2
\end{eqnarray}
Denoting the transition operator
\begin{eqnarray}
S&\equiv&
\CU^\dag_{t_\mathrm{com}\,\mathrm{to}\,t_n}
\end{eqnarray}
this means we have
\begin{eqnarray}
\mathrm{Prob}
\left(|\psi_f\rangle\big||\psi\rangle\right)&=&
\left|\langle\psi_f|S|\psi\rangle\right|^2
\cdot
\frac{\langle\psi_f|\rho_{f\,\mathrm{from}\,t_n}|\psi_f\rangle}
{\langle\psi|\rho_{f\,\mathrm{from}\,t_n}|\psi\rangle}
\end{eqnarray}
compared to the usual expression
\begin{eqnarray}
\mathrm{Prob}_\mathrm{usual}
\left(|\psi_f\rangle\big||\psi\rangle\right)&=&
\left|\langle\psi_f|S|\psi\rangle\right|^2 .
\end{eqnarray}
The deviations are thus the following:
\begin{enumerate}
  \item With our imaginary part in $S$ there is no longer unitality, i.e.
\begin{eqnarray}
S^\dag \neq S^{-1}.
\end{eqnarray}
The transition $S$ is calculated by the Feynman path integral with the
full
\begin{eqnarray}
S = S_R + iS_I .
\end{eqnarray}
  \item There is the extra wright factor
 $\langle \psi_f | \rho_{f\,\mathrm{from}\,t_n}$ describing the effect of
 the happenings and the $S_I$ in the future of the ``final measurement"
 $| \psi_f \rangle$.
  \item There is the only on the initial state $| \psi \rangle$ dependent
``normalization factor" in the denominator
\begin{eqnarray}
\langle\psi|S^\dag\rho_{f\,\mathrm{from}\,t_n}S|\psi\rangle
\end{eqnarray}
This denominator is indeed a normalization factor normalizing the total
probability for reaching a complete set -- an orthonormal basis --
of final states  $| \psi_{fk} \rangle$, $k=1, 2, \cdots$
which we for simplicity choose as eigenstate of 
$\rho_{f\,\mathrm{from}\,t_n}$
so that $\langle | \psi_{fk} \rho_{f\,\mathrm{from}\,t_n} | \psi_{fk} \rangle$
gets a diagonal matrix. 
Then namely
\begin{eqnarray}
&&
\sum_{k=1,2,\cdots}\mathrm{Prob}\left(|\psi_{f,k}\rangle\big||\psi\rangle\right)
\nonumber\\&&=
\frac{1}{\langle\psi|S^\dag \rho_{f\,\mathrm{from}\,t_n}S|\psi\rangle}\cdot
\sum_k\left|\langle\psi_{f,k}|S|\psi\rangle\right|^2
\langle\psi_{f,k}|\rho_{f\,\mathrm{from}\,t_n}|\psi_{f,k}\rangle
\nonumber\\&&=
\frac{1}{\langle\psi|S^\dag \rho_{f\,\mathrm{from}\,t_n}S|\psi\rangle}
\sum_k
\langle\psi|S^\dag|\psi_{f,k}\rangle
\langle\psi_{f,k}|\rho_{f\,\mathrm{from}\,t_n}|\psi_{f,k}\rangle
\langle\psi_{f,k}|S|\psi\rangle\nonumber\\
{   }
\end{eqnarray}
which by using that the off diagonal elements of 
$\rho_{f\,\mathrm{from}\,t_n}$ were chosen to be zero can be rewritten as a 
double sum -- i.e. over both $k$ and $k'$ --
\begin{eqnarray}
&&
\sum_{k}\mathrm{Prob}\left(|\psi_{f,k}\rangle\big||\psi\rangle\right)
\langle\psi|S^\dag\rho_{f\,\mathrm{from}\,t_n}S|\psi\rangle
\nonumber\\&&=
\sum_{k,k'}
\langle\psi|S^\dag
|\psi_{f,k}\rangle\langle\psi_{f,k}|\rho_{f\,\mathrm{from}\,t_n}
|\psi_{f,k'}\rangle\langle\psi_{f,k'}|
S|\psi\rangle
\nonumber\\&&=
\langle\psi|S^\dag
\rho_{f\,\mathrm{from}\,t_n}
S|\psi\rangle
\end{eqnarray}
\end{enumerate}

Thus we see that this denominator just ensures that the total probability 
for all that can happen at time $t_n$ starting from $| \psi \rangle$
at $t_\mathrm{com}$ becomes just one.

\subsection{Simplifying formula for conditional probability by approximating
future}
We have already suggested that we should approximate
\begin{eqnarray}
\rho_{f\,\mathrm{from}\,t_n}\approx \frac{1}{N}\underline{1}
\end{eqnarray}
provided the future after $t_n$ is so that we can practically consider the
$S_I$-effects small or so much delayed into the extremely far future that
our above ergodicity argument can be used.
By such an approximation we remove the deviation number 2 above given by
$\langle  \psi_{f}| \rho_{f\,\mathrm{from}\,t_n} | \psi_{f} \rangle$
because we approximate this matrix element
$\langle  \psi_{f}| \rho_{f\,\mathrm{from}\,t_n} | \psi_{f} \rangle$
by a constant as a function of $|\psi_f \rangle$.
After this approximation we get
\begin{eqnarray}
\mathrm{Prob}\left(|\psi_{f}\rangle\big||\psi\rangle\right)
=
\frac{\left|\langle\psi_f|S|\psi\rangle\right|^2}
{\langle\psi|S^\dag S|\psi\rangle}
\end{eqnarray}
We should have in mind that $S^+S \neq 1$ in general since with
the imaginary part of the action the Hamiltonian will be non-hermitean
and $S$ non unitary.
Thus the usual $\left|\langle\psi_f|S|\psi\rangle\right|^2$
would by itself not deliver total probability for what comes out of 
$|\psi\rangle$ to be unity.
Only after the division by the normalization
$\langle\psi|S^\dag S|\psi\rangle$
would it become normalized to unity.

\section{Can we make an unsquared form?}
\label{sec:10}
The formulas for the extraction of probabilities from our Feynman path 
integral with imaginary part of action $S_I$ also were derived by 
considerations of statistical addition with essentially random phases
of various classical path.
But our crucial formula, say (\ref{2.13}), for probabilities is
seemingly surprisingly complicated in as far as each projection operator
occurs twice in the trace in the numerator.
Even the  simplest example of asking if some variable $\CO$ at time
$t$ falls into the range $\bar M$ gets the expression
\begin{eqnarray}\label{*}
&&
\mathrm{Prob}(\CO\in\bar M)
\nonumber\\&&
=\mathrm{Tr}(\CU_{t\,\mathrm{to}\,\infty}P_{\CO\in\bar M}\CU_{-\infty\,\mathrm{to}\,t}
\CU^\dag_{-\infty\,\mathrm{to}\,t}P_{\CO\in\bar M}\CU^\dag_{t\,\mathrm{to}\,\infty})
/
\mathrm{Tr}(\CU_{-\infty\,\mathrm{to}\,\infty}\CU^\dag_{-\infty\,\mathrm{to}\,\infty})
\nonumber\\
\end{eqnarray}
containing the projection operator 
$P_{\CO\in\bar M}$
twice as factor in the expression.
If we make the approximation of no 
$S_I$-effects
in the 
$t$ to $\infty$
time range by taking
\begin{eqnarray}
\rho_{f\,\mathrm{from}\,t_n}=
\CU^\dag_{t\,\mathrm{to}\,\infty}\CU_{t\,\mathrm{to}\,\infty}
\end{eqnarray}
and approximating it by being proportional to the unit matrix then,
however, the two projection operators come together and we could formally
replace their product by just one of them.
So in the case of the in this way approximated future we could write
\begin{eqnarray}
\mathrm{Prob}(\CO\in\bar M)&=&
\mathrm{Tr}(P_{\CO\in\bar M}
\CU_{-\infty\,\mathrm{to}\,t}
\CU^\dag_{-\infty\,\mathrm{to}\,t})/
\mathrm{Tr}(\CU_{-\infty\,\mathrm{to}\,t}
\CU^\dag_{-\infty\,\mathrm{to}\,t}).
\end{eqnarray}
If $\CO$ were a variable among the variables used as the path-description
in the Feynman path integral the formula (\ref{*}) would by
functional integral be written
\begin{eqnarray}
\mathrm{Prob}(\CO\in\bar M)&=&
\frac{\left|
\int P_{\CO\in\bar M}\e^{iS}\CD\phi
\right|^2}{\left|
\int \e^{iS}\CD\phi
\right|^2}.
\end{eqnarray}
Strictly speaking these Feynman integrals should be summed over all
end of time configurations, but with significant
$S_I$ presumably this summation would be dominated by a  few ``true"
initial and states at 
$\pm\infty$ and the summation would not be so important.

So strictly speaking we have
\begin{eqnarray}
\mathrm{Prob}(\CO\in\bar M)&=&
\frac{\sum_\mathrm{init,final}\left|
\int^\mathrm{final}_\mathrm{initial} P_{\CO\in\bar M}\e^{iS}\CD\phi
\right|^2}{\sum_\mathrm{init,final}\left|
\int \e^{iS}\CD\phi
\right|^2}.
\label{pm115}
\end{eqnarray}

\section{The Higgs width broadening}
\label{sec:11}
As an example of application of the 
$S_I$-caused modification of the usual transition matrices we may consider
the decay of a particle 
-- which we for reasons to be explained below take to be the Weinberg-Salam
Higgs particle --
which has from $S_I$ induced an imaginary term in the mass (or energy).
Let us say take this term to have the effect of delivering a term being
a positive constant number multiplied by $-i$ in the Hamiltonian.
In the Schrodinger equation
\begin{eqnarray}
i\frac{d\psi}{dt}=H\psi
\end{eqnarray}
such a term will cause the wave function $\psi$ to decrease with time
so that it will decay exponentially with time $t$.
If the particle in addition decays ``normally" into decay products,
say $b\bar b$ as the Higgs particles
do the exponential decay rate will be the sum 
$\Gamma_{\mathrm{normal}}+\Gamma_{S_I}$
of the $S_I$-induced width
$\Gamma_{S_I}$ and the ``normal" decay width
$\Gamma_\mathrm{normal}$.
Let us for simplicity take as an approximation that the real part of
the mass is very large compared to both the ``normal" and the
$S_I$-induced widths so that we can work effectively non relativistically
with a resting Higgs particle.
We can let it be produced in a short moment of time which is short
compared to the inverse widths
$\frac{1}{\Gamma_{S_I}}$ and $\frac{1}{\Gamma_\mathrm{normal}}$
while still allowing the particle may be considered at rest approximately.

If we at first used the ``usual" formula
$|\langle\psi_f|S|\psi\rangle|^2$ for the decay process and calculate the
total probability for the particle to decay into anything one will find 
that this probability is only
$\frac{\Gamma_\mathrm{normal}}{\Gamma_\mathrm{normal}+\Gamma_{S_I}}$
because the average lifetime has been reduced by this factor, namely
from $\frac{1}{\Gamma_\mathrm{normal}}$ to 
$\frac{1}{\Gamma_\mathrm{normal}+\Gamma_{S_I}}$.
Since of course the usual particle with 
$\Gamma_{S_I}=0$ will decay into something with just probability unity, 
we thus need a normalization factor
$\langle\psi|S^\dag S|\psi\rangle$ to rescale the total probability to
be (again) unity in our imaginary action theory.

By Fourier transforming from time $t$ to energy the Higgs decay time
distribution we obtain in our model again a Breit-Wigner energy
distribution
\begin{eqnarray}
P(E)&=&
\frac{\Gamma_\mathrm{normal}+\Gamma_{S_I}}
{2\pi\left[(E-m_\mathrm{Higgs})^2
+\left(\frac{\Gamma_\mathrm{normal}+\Gamma_{S_I}}{2}\right)^2\right]}
\end{eqnarray}

If indeed we effectively should have such an $S_I$-induced imaginary part
in the mass of the Higgs, then the Higgs-width could be made bigger than
calculated in the usual width $\Gamma_\mathrm{normal}$.
This is an effect that might have been already seen in the LEP-collider
provided one has indeed seen some Higgses in this accelerator.
Indeed there has been found an excess of Higgs-like events with masses
slightly below the established lower bound for the Higgs mass of
114 GeV/c. 

\subsection{The effect of the 
$\langle\psi_f|\rho_{f\,\mathrm{from}\,t_n}|\psi_f\rangle$
suppression factor}
As an example of a (perhaps realistic) case of an effect of the factor
$\langle\psi_f|\rho_{f\,\mathrm{from}\,t_n}|\psi_f\rangle$
we could imagine that two particles are coming together organized to
hit head on
-- say in a relative $s$-wave -
able to potentially form two different resonances of which say one is a
Higgs which as above is assumed to have an imaginary term in its mass.
Now it is easy to see that the 
$\rho_{f\,\mathrm{from}\,t_n}$
(here $t_n$ is the moment the collision just formed one of the two
resonances thought upon as physical objects) will have
\begin{eqnarray}
\langle\psi_{f\,\mathrm{Higgs}}|\rho_{f\,\mathrm{from}\,t_n}|\psi_{f\,\mathrm{Higgs}}\rangle
<
\langle\psi_{f\,\mathrm{all}}|\rho_{f\,\mathrm{from}\,t_n}|\psi_{f\,\mathrm{all}}\rangle
\end{eqnarray}
where
$|\psi_{f\,\mathrm{Higgs}}\rangle$ and $|\psi_{f\,\mathrm{all}}\rangle$
represent respectively the two possible resonances the Higgs and the
alternative resonance.
Compared to the usual calculation of the transition to one of the resonances
-- essentially the square of the coupling constant --
the Higgs-resonance will occur with suppressed probability because of the
$\langle\psi|\rho_{f\,\mathrm{from}\,t_n}|\psi\rangle$-factor in the formula
(\ref{pm99}).
Really if the collision were safely organized that the collision occurs
because of $s$-wave impact preensured the total probability for one or the
other of the two resonances to be formed would be with properly normalized
probability 1 because of the
$\langle\psi|S^\dag S|\psi\rangle$
normalization factor.
However, the effect of
$\langle\psi_f|\rho_{f\,\mathrm{from}\,t_n}|\psi_f\rangle$
would be to increase the probability to form the alternative resonance while
decreasing the formation of the Higgs.

\section{Approaching a more beautiful formulation}
\label{sec:12}
Taking the regions in which $\CO$ may lie or not $\bar M$ as infinitesimally
extended we would the formula for the probability density in the form
\begin{eqnarray}
\mathrm{Prob}(\CO=\CO_0)&=&
\frac{\sum_{i,f}\left|\int_{\mathrm{BOUNDARY:}i,f}\e^{iS[\phi]}P_{\CO\in\bar M}\CD\phi\right|^2}
{\sum_{i,f}\left|\int_{\mathrm{BOUNDARY:}i',f'}\e^{iS[\phi]\CD\phi}\right|^2}
\nonumber\\
&\propto &
\frac{\sum_{i,f}\left|\int_{\mathrm{BOUNDARY:}i,f}\e^{iS[\phi]}\delta(\CO-\CO_0)\CD\phi\right|^2}
{\sum_{i,f}\left|\int_{\mathrm{BOUNDARY:}i',f'}\e^{iS[\phi]}\CD\phi\right|^2}
\label{mpns1}
\end{eqnarray}

We may claim that this kind of formula the probability density for finding
$\CO$ taking a value infinitesimally close to $\CO_0$ is a bit unaestetic
because of having the projection operator $P_{\CO\in\bar M}$ or the 
equivalent $\delta(\CO-\CO_0)$ occurring \underline{twice}
while one might have said it would be simpler to have just one
$\delta(\CO-\CO_0)$ or $P_{\CO\in\bar M}$ factor in the expression.

We should now seek to reformulate our expression with these factors
occurring twice into a simpler one with only single occurrence of 
$P_{\CO\in\bar M}$ or $\delta(\CO-\CO_0)$.
To perform this hoped for derivation we first argue that for nonoverlapping
$\CO$-value regions $\bar M_1$ and $\bar M_2$
\begin{eqnarray}
&&
\sum_{i,f}\left(
\int_{\mathrm{BOUNDARY}:i,f}P_{\CO\in\bar M_1}\e^{iS[\phi]}\CD\phi
\right)^*
\cdot
\int_{\mathrm{BOUNDARY}:i,f}P_{\CO\in\bar M_2}\e^{iS[\phi]}\CD\phi
\nonumber\\&&
\approx 0~~~
\mbox{for}~~\bar M_1\cap\bar M_2=\emptyset ~.
\label{mpns4}
\end{eqnarray}
This is argued for by maintaining that giving $\CO$ a different value at
time $t$ very typically by ``butterfly effect"
-- Lyapunov exponent -- will cause very different states $f$ and $i$ at
$\mp\infty$ respectively.
If the two factors in (\ref{mpns4}) have very different final $f$ and
initial $i$ states dominate at the boundaries and even random phases
the total sum is indeed much smaller than what one would obtain if
$\bar M_1$ and $\bar M_2$ were taken to be the same region
$\bar M_1=\bar M_2=\bar M$.
If we now use the zero expression (\ref{mpns4}) by adding such terms
into the numerator and analogously in the denominator of (\ref{mpns1})
we can formulate this expression for the probability of $\CO$ being in
$\bar M$ into an expression involving a summation over the value or region
for $\CO$ in one of the occurrences
\begin{eqnarray}
&&\mathrm{Prob}(\CO=\CO_0^{(2)})
=\nonumber\\&&
\Bigg(
\sum_{i,f}\int d\CO_0^{(1)}
\left(
\int_{\mathrm{BOUNDARY}: i,f}
\delta(\CO-\CO_0^{(1)})\e^{iS[\phi]}
\CD\phi
\right)^*
\nonumber\\&&~~~\cdot
\int_{\mathrm{BOUNDARY}: i,f}
\delta(\CO-\CO_0^{(2)})
\e^{iS[\phi]} \CD\phi 
\Bigg)
\\&&
\Biggm/
\left(\sum
_{i',f'}\left(\int_{\mathrm{BOUNDARY}: i',f'}
\e^{iS[\phi]}\CD\phi\right)^*\int_{\mathrm{BOUNDARY}: i',f'}
\e^{ iS[\phi]} \CD\phi
\right)~.
\nonumber
\end{eqnarray}

But now obviously we have
\begin{eqnarray}
\int d\CO_0^{(1)}\delta(\CO-\CO_0)=1
\end{eqnarray}
and thus we get
\begin{eqnarray}
&&\mathrm{Prob}(\CO-\CO_0^{(2)})=\nonumber\\&&
\frac{\sum
_{i,f}\left(\int_{\mathrm{BOUNDARY}: i,f}
\e^{iS[\phi]}\CD\phi\right)^*
  \int_{\mathrm{BOUNDARY}: i,f}
\delta(\CO-\CO_0^{(2)})\e^{iS[\phi]}\CD\phi          }
{\sum
_{i',f'}\left(\int_{\mathrm{BOUNDARY}: i',f'}
\e^{iS[\phi]}\CD\phi\right)^*  \int_{\mathrm{BOUNDARY}: i',f'}
\e^{iS[\phi]}\CD\phi  }
\nonumber\\
\end{eqnarray}
In this expression we have achieved to have
$\delta(\CO-\CO_0^{(2)})$ only occurring once as factor.
We could therefore trivially extract from it an expression for the
average of the $\CO$-variable
\begin{eqnarray}
\langle\CO\rangle&=&
\int\CO_0^{(2)}\mathrm{Prob}(\CO-\CO_o^{(2)})d\CO_0^{(2)}
\nonumber\\&=&
\frac{\sum
_{i,f}\left(\int_{\mathrm{BOUNDARY}: i,f}
\e^{iS[\phi]}\CD\phi\right)^*
}
{\sum
_{i',f'}\left(\int_{\mathrm{BOUNDARY}: i',f'}
\e^{iS[\phi]}\CD\phi\right)^*}
\nonumber\\&&
\cdot
\frac{\int_{\mathrm{BOUNDARY}: i,f}
\CO\e^{iS[\phi]}\CD\phi}{
\int_{\mathrm{BOUNDARY}: i',f'}
\e^{iS[\phi]}\CD\phi}
\end{eqnarray}
If we could somehow remove the after all identical complex conjugate functional
integrals
\begin{eqnarray}
\left(\int_{\mathrm{BOUNDARY}: i, f}\e^{iS[\phi]}\CD\phi\right)^*
\label{mp138}
\end{eqnarray}
and
\begin{eqnarray}
\left(\int_{\mathrm{BOUNDARY}: i', f'}\e^{iS[\phi]}\CD\phi\right)^*
\label{mp138 prime}
\end{eqnarray}
only deviating by the dummy initial and final state designations respectively
$(i,f)$ and $(i',f')$, then we could achieve the simple expression (\ref{1.5}).
But in order to argue for such removal being possible we would have to 
speculate say that some -- we could say the true --
boundary condition combination for the functional integrals 
(\ref{mp138}, \ref{mp138 prime}) completely dominates.
This is actually not at all unrealistic since indeed the $S_I$ will tend to
very few paths dominate.
In such a case of dominance we would have a set of dominant
$(f,i)$.
Presumably to make the chance that there should be such dominance we
should allow ourselves to be satisfied with a linear combinations of
$i$-state and of $f$-states to dominate.
But now if indeed we could do that and call these linear combinations
$(f_\mathrm{dom},i_\mathrm{dom})$, then we could approximate
\begin{eqnarray}
\langle\CO\rangle\approx 
\frac{\int_{\mathrm{BOUNDARY}\,f_\mathrm{dom},i_\mathrm{dom}}\CO\e^{iS[\phi]}\CD\phi}
{\int_{\mathrm{BOUNDARY}\,f_\mathrm{dom},i_\mathrm{dom}}\e^{iS[\phi]}\CD\phi}.
\label{mp141}
\end{eqnarray}
Now we would like not to have the occurrence in this expression of the rather 
special states $(f_\mathrm{dom},i_\mathrm{dom})$.
However, these dominant boundary conditions are precisely the dominant boundary
conditions for the denominator integral, because it were really just the 
complex conjugate for the latter for which we looked for the dominant 
boundaries.

So if we let the boundaries free then at least the denominator should become
dominantly just as if we had used the boundaries
$(f_\mathrm{dom},i_\mathrm{dom})$.
It even seems that because of the smoothness and boundedness of the variable
$\CO$ as functional of $\phi$ the dominant boundaries $(i, f)$ should not
be much changed by inserting an extra factor $\CO$ so that also by letting
the boundaries free in the numerator functional integral
$\int\CO[\phi]\e^{iS[\phi]}\CD\phi$ would
not change much the dominant boundaries from those of the same integral with
the $\CO$-factor removed.
But the removal of this $\CO$ leads to the denominator functional integral,
for which we already saw that the dominating boundary behavior were
$(f_\mathrm{dom},i_\mathrm{dom})$.
Thus we have argued that we can rewrite (\ref{mp141}) into
\begin{eqnarray}
\langle\CO\rangle=
\frac{\int\CO\e^{iS[\phi]}\CD\phi}{\int\e^{iS[\phi]}\CD\phi}
\label{mp145}
\end{eqnarray}
where it is understood that the boundaries for
$t\to\pm\infty$ are ``free".
Then we suggested they would automatically go to be dominated by
$(f_\mathrm{dom},i_\mathrm{dom})$ thus fitting on to the formulas with
double occurrence of $\delta(\CO-\CO_0)$'s.

The argumentation that the factor $\CO$ does not matter for the dominant
behavior at $\pm \infty$ may sound almost contradictory to our assumption
using the ``butterfly effect" to derive the rapid variation of (\ref{4}) 
which meant that
an insertion of $\delta(\CO-\CO_0)$ would drastically change 
behavior, including that of the boundary.

It is, however, not totally unreasonable that a sharp function
$\delta(\CO-\CO_0)$ which is zero in most places could modify the 
boundary conditions, while a smooth one $\CO$, almost never zero would
not modify them.
Basically we hope indeed for that the $S_I$-caused weighting is so
severely restricting the set of significant paths, that it practically
means that a single path, ``the realized path" is selected.
In this case the insertion of the factor $\CO$ into the functional
integral would just multiply it by the value of $\CO$ on ``realized path".
If you however insert $\delta(\CO-\CO_0)$ and it as most likely the case
$\CO_0$ is not the value of $\CO$ on the realized path then we kill by the 
zero-value of $\delta(\CO-\CO_0)$ at the realized path would totally kill
the dominant contribution.
Then of course the possibility for a completely different path is opened
and the orthogonality used in (\ref{mpns4}) gets realistic.

As conclusion of the just delivered derivation of (\ref{mp145}) 
we see that the starting point in the beginning the articles is indeed
consistent under the suggested approximations with the forms derived 
from the semiclassical start.

\section{The monopole argument for suppressing the $S_I$ in the
Standard Model}
\label{sec:13}
We have already above in section~\ref{sec:3}
argued that due to the material in the present era, and the future too,
being either massless or protected from decay by in practice conserved
quantum numbers and due to weakness of the interactions the contribution
to $S_I$ from these eras must be rather trivial.

It were also for the above argument important that the non-zero mass 
particles were non-relativistic in these eras.
That above argument may, however, not be sufficient for explaining that
no effect of our $L_I$ or $S_I$ would have been seen so far.
We have indeed had several high energy accelerators such as 
ISR (=intersecting storage ring at CERN) in which massive particles
-- such as protons --
have been brought to run for days with relativistic speeds.
That means that they would during this running in the storage rings say 
have had eigentime contributions significantly lower than the coordinate
 time or rather the time on earth.
This would presumably easily have given significant contributions to 
$S_I$ which going to the exponent would suppress
-- or priori perhaps enhance --
the probability of developments, solutions, to equations of motion,
leading to the running of such storage rings.
Since the protons have not already been made to run around dominantly
relativistically we should deduce that most likely the storage rings
would lead to increasing $S_I$ and thus lowering of the probability
weight.
Thus one would expect that the initial conditions should have been so 
adjusted as to prevent funding for this kind of accelerators, at least
for them running long time.
Contrary to Higgs producing accelerators which have so far not been able
to work on big scale (may be L.E.P. produced a few Higgses for a short time)
the accelerators with relativistic massive particles have seemingly
worked without especially bad luck.
In order to rescue our model it seems therefore needed to invent a crutch
for it of the type that there are actually no $L_I$-contributions
involving the particles for which the massive relativistically running
accelerators have been realized.
We have actually two mechanisms to offer which at the end can argue away
our $L_I$ or $S_I$ effects for all the hitherto humanly produced or found
particles, leaving the hopes for finding observable effects
-- bad luck for accelerators, mysterious broadening of resonance peaks --
to experiments involving the Higgs particle or particles outside the
Standard Model.
The point is indeed that we shall argue away the effects of $S_I$ for
gauge particles and for Fermions (coupled to them).

The suppression rules to be argue for are:
\begin{enumerate}
  \item[1)] Supposing the existence of monopoles we deduce that the
corresponding full gauge coupling constants must be real, basically as
a consequence of the Dirac relation.
  \item[2)] For fields which like the Fermion fields in renormalizable
theories occur homogeneously in the Lagrangian density
$\CL_R+i\CL_I$ this Lagrangian
density can be shown to be zero by inserting the equations of motion.
\end{enumerate}

\subsection{Spelling out the suppression rules}
Spelling out a bit the suppression rules let us for the monopole based 
argument remind the reader that although we consider a complex Lagrangian
density $\CL_R+i\CL_I$ for instance 
the electric and magnetic fields and the four potential $A_\mu(x)$
for electrodynamics are still real as usual.
Now if we have fundamental monopoles there must exist corresponding Dirac
strings which, however, as is well known must be unphysical.
The explicit flux in the Dirac string must
to have the Dirac string unobservable 
-- to be unphysical --
be compensated by an at the string  singular behavior of the four 
potential $A_\mu$
around the Dirac string.
The singular flux to compensate the extra flux in the Dirac string can,
however, only be real since the $A_\mu$-field is real and it is
given by a curve integral
$\oint A_\mu dx^\mu$
around the Dirac string.
Now as is well known the fluxes mentioned equal the monopole charges.
Thus the monopole charge $g$ must be real.
But then the Dirac relation
\begin{eqnarray}
eg=2\pi n,~~~n\in \bZ
\end{eqnarray}
tells that also the electric charge $e$ must be real.
Now, however, in the formalism with the electric charge absorbed into
the four potential
$\hat A_\mu=eA_\mu$
the coefficient on the
$F_{\mu\nu}^2$-term in the Lagrangian density is
$-\frac{1}{4e^2}$
so that the pure electromagnetic, kinetic, Lagrangian density
\begin{eqnarray}
\left(\CL_R+i\CL_L\right)\Big|_\mathrm{pure\,e.m.}=
-\frac{1}{4e^2}F^2_{\mu\nu}
\end{eqnarray}
becomes totally real.
I.e.
\begin{eqnarray}
\CL_L|_\mathrm{pure\,e.m.}=0.
\end{eqnarray}
We may skip or postpone a similar argument for non-abelian, Yang Mills,
theories to another paper, but really you may just think of some abelian
subgroup and make use of gauge invariance.

Concerning the rule 2) for homogeneously occurring fields, such as the
Fer\-mion
fields in renormalizable theories the trick is to use the equations of motion.
For example the part of the Lagrangian density $\CL_R+i\CL_I$ involving
a Fermion field $\psi$ is of the form
$\CL_F=Z\cdot\bar\psi(i\Ds -m)\psi$
where $Z$ is a constant and 
$D_\mu$ the covariant derivative and of course
$\Ds=\gamma^\mu D_\mu$.
This Fermionic Lagrangian density is homogeneous of rank two in the Fermion 
field $\psi$.
The Euler-Lagrange equations, the equations of motion for the Fermion
fields are derived from functional differentiation w.r.t. the field $\psi$
\begin{eqnarray}
\frac{\delta S}{\delta\psi(x)}=0
\end{eqnarray}
and end up giving equations of motion of the form
\begin{eqnarray}
\frac{\partial\CL_F}{\partial\psi}=0
\end{eqnarray}
or
\begin{eqnarray}
\frac{\partial\CL_F}{\partial\bar\psi}=0
\end{eqnarray}
(really these forms are only trustable modulo total divergences but that
is enough) leading as is well known to
\begin{eqnarray}
\bar\psi(i\Ds-m)=0
\end{eqnarray}
or
\begin{eqnarray}
(i\Ds-m)\psi=0.
\end{eqnarray}
But now it is a general rule that a homogeneous expression,
$\CL_F$ say, can be recovered from its partial derivatives
\begin{eqnarray}
\sum\frac{\partial\CL_F}{\partial\psi}\psi+
\sum\bar\psi\frac{\partial\CL_F}{\partial\bar\psi}
=\mathrm{rank}\cdot \CL_F
\end{eqnarray}
where rank is in the present case 
rank$=2$.
Such a recovering for homogeneous Lagrangian densities, however,
means that the Lagrangian density
-- at least modulo total divergences --
can be expressed by the equation of motions, which are zero,
if obeyed.
But then at least in the classical approximation the Lagrangian density
is zero at least modulo total divergences.
This means that the total 
$S_R+iS_I$ contribution from the just discussed homogeneous terms 
end up zero.
Especially the imaginary part also ends up zero, although
its form does not have to be zero.
It is only insertion of equations of motion that makes it zero.

\section{Conclusion}
\label{sec:14}
We have put up a formalism for a non-unitary model based on extending the
Lagrangian and thereby the action
to be complex by allowing complex coefficients in the Lagrangian density
$\CL_R+i\CL_I$.

We used two starting points for how to extract probabilities and expectation
values from the Feynman path way integral in our ambitious model that 
shall even be able to tell what really happens rather than just the 
equations of motion.
The first were the interpretation that an operator $\CO(t)$ should have the
expectation value
\begin{eqnarray}
\langle\CO\rangle=
\frac{\int\CO(t)\e^{iS[\phi]}\CD\phi}
{\int\e^{iS[\phi]}\CD\phi}
\end{eqnarray}
but this expression is a bit dangerous in as far as it is a priori not
guaranteed to be real even though the quantity 
$\CO(t)$ is real.
The second approach would rather have a series of projections onto small
regions $\bar M_i$
for operator $\CO_i(t_i)$ denoted
$P_{\CO_i\in\bar M_i}$
inserted into the functional integral but then this integral
is numerically squared for any combination $(i, f)$ of boundary
behaviors at respectively $-\infty$ and $+\infty$ times.
That is to say that the insertions are to be performed into the integral
$\int\e^{iS[\phi]}\CD\phi$
so as to replace the latter by
$\int\prod_iP_{\CO_i\in\bar M_i} \e^{iS[\phi]}\CD\phi$
just as in the first approach, but then one forms the numerical square summed
over the initial $i$ and final $f$ behaviors
\begin{eqnarray}
\sum_{i,f}\left(
\int_{\mathrm{BOUNDARY}: i,f}\e^{iS[\phi]}\CD\phi
\right)^*
\int_{\mathrm{BOUNDARY}: i,f}\e^{iS[\phi]}\CD\phi
\label{product}.
\end{eqnarray}

The probability distribution is then obtained by inserting the projection
operators into $both$ factors in (\ref{product}) and then as normalization
divide the (\ref{product})  having these insertion with (\ref{product})
not having the insertions.

Under some suggestive assumptions we argued that the two approaches
approximately will agree with each other.
The most important formula derived is presumably the formula to
replace usual unitary $S$-matrix or $\CU$-matrix transition between
two moments in time in our model.
This formula turns out for transition an initial state
$|\psi\rangle$
to a final
$|\psi_f\rangle$
to be
\begin{eqnarray}
\mathrm{Prod}(|\psi_f\rangle,|\psi\rangle)=
\frac{|\langle\psi_f|S|\psi\rangle|^2
\langle\psi_f|\rho_{f\,\mathrm{from}\,t_f}|\psi_f\rangle}
{\langle\psi|S^\dag S|\psi\rangle}
\end{eqnarray}

We used that to derive the broadening in our model of the Higgs-width.

As an outlook we may mention some of the expectations of our model 
used in a more classical language in our earlier publications: If the Higgs 
-- especially freely running Higgses --- decrease significantly the probabilty
(7.21) then the initial state should be organized so that Higgs production be 
largely avoided. This would actually make the prediction that some how or the 
other an accident will happen and the LHC-accelerator will be prevented from 
comming to full energy and luminosoty.

\section*{Acknowledgments}

The authors acknowledge the Niels Bohr Institute (Copenhagen) and 
Yukawa Institute for Theoretical Physics for their hospitality 
extended to one of them each.
The work is supported by 
Grand-in-Aids for Scientific Research on Priority Areas, 
Number of Area 763 ``Dynamics of Strings and Fields", 
from the Ministry of Education of Culture, Sports, Science 
and Technology, Japan.

\title{Coupling Self-tuning to Critical Lines From Highly Compact %
Extra Dimensions}
\author{K. Petrov}
\institute{%
Niels Bohr Institute\\
Blegdamsvej 17 \\
2100 Copenhagen \O \\
Denmark\\
\textrm{e-mail:} \texttt{kpetrov@nbi.dk}
}

\titlerunning{Coupling Self-tuning to Critical Lines \ldots}
\authorrunning{K. Petrov}
\maketitle

\begin{abstract}
We discuss possible origins of Multiple Point Principle. Inspired by
  results from finite temperature lattice gauge theory we conjecture possible
  physical reasons which could lead to the Principle
\end{abstract}

\section{Introduction}
One of many plagues of the Standard Model as we know it is the excessive
number of parameters. It is usually understood that it is a manifestation of
Standard Model being some effective theory for the ``true'' underlying model,
such as GUTs. One very promising way of explaining how parameters get
fine-tuned is so-called Mulptiple Point Principle\cite{mpp} introduced and
advocated by Nielsen et al. As it is widely discussed in the very same
proceeding we will not focus on its details and implications but only mention
the essential for our conjecture property. The core statement of MPP is the
assumption that many degenerate vacua exist and that they are separated by
first order phase transition. Couplings of the model do self-tune themself to
the critical lines, or, in existence of many vacua - to multi-critical points.
Here we suggest how this situation may indeed be realized in nature and
discuss the implications.

\section{Dimensional Reduction - a detour}
Before we turn to the main subject it is important to discuss certain results
which come from finite temperature gauge theory. We need to underline that
while these results were largely obtained using Lattice Gauge Theory the
mechanism is relevant to continuum and lattice here is just a way to
non-perturbatively study phase diagram.  For simplicity we will consider
2+1-dimensional spacetime, though the results are similar in 3+1 dimensions
(nobody ever analyzed 4+1 or more). Our model, which we will regard as
``physics'' would be pure Yang-Mills SU(3). It can be formulated in terms of
closed loops on the (hyper)cubic lattice. We will denote its extentions in
space and in time directions by $L_{s}$ and $L_{0}$ correspondingly. The
lattice is periodic in all directions, however differently. Spacial extent is
finite only for practical purposes and generally is sent to infinity. Time
extent behaves as usually in finite temperature field theory and is compact,
with its length being equal the inverse temperature%
\begin{equation}
L_{0}a=1/T\label{invT}%
\end{equation}
On each link of it we put a matrix $U(x;\mu)$ which belongs to the $SU(3)$
group and is linked to the gauge fields $A$ by%
\begin{equation}
U(x;\mu)=\exp ig_{0}A(x)
\end{equation}
where $g_{0}$ is the bare strong coupling constant.
\begin{equation}
S_W^{3}=-\frac{6}{g_{0}^{2}}\sum_{x\in\Lambda,\mu,\nu}Tr\left(U(x;\mu)U(x+\mu;\nu)U(x+\nu;\mu)^{-1}U(x;\nu)^{-1}+H.c.\right) 
\end{equation}
yields usual square of the field strength tensor in naive continuum limit.
Phenomena, symmetries and phases which occur in this model we will consider to
be ``physical''.

It has a deconfinement phase transition separating confined and deconfined
phase with $Z(3)$ being relevant symmetry. If we go to high temperatures,
about twice the critical one we can do a following trick, known as dimensional
reduction.  As you can see from Eq.\ref{invT} the time extent shrinks
drastically. The theory becomes perturbative and time-like degrees of freedom
may be integrated out. This has no relation to lattice and integrating is
often done in continuum. First we set time-like gauge field to be independent
of the imaginary time coordinate $x_{0}$%
\begin{equation}
A_{0}(x_{0},\vec{x})=A_{0}(\vec{x})\label{static}%
\end{equation}
and then eliminate the remnants of the
gauge freedom by adding the so-called Landau constraint
\begin{equation}
\sum_{x_{0}}\sum_{\hat{\imath}=1}^{2}\left[  A_{i}\left(  x\right)
-A_{i}\left(  x-a\hat{\imath}\right)  \right]  =0\label{stalg}%
\end{equation}
Together with (\ref{static}) this condition is called Static Time-Averaged
Landau Gauge. It is essential that all gauge degrees of freedom are fixed and
what is left is, or should be, physical.

Few Feynman diagrams later (for details reader may consult \cite{phd}) we get
an effective theory, formulated in two dimensions with an extra field which
lives on sites. The resulting action depends on the parameters of the original
theory $L_{0}$ and $g_0$. We make a change of variables
\begin{equation}
A_{0}\left( x\right)  =\phi(x)\sqrt{\frac{g_0^2}{L_0}}\label{phi}%
\end{equation}
to make effective model look in a more familiar way.

Now we can express $S_{eff}^{2}$ as the function of this
field and the two-dimensional gauge field $U.$ It consists of the following
three parts:

\begin{itemize}
\item pure gauge part $S_{W}^{2},$ which we had already in "naive" reduction%

\begin{equation}
S_{W}^{2}=\frac{6}{g_0^2}L_{0}\sum_{P}Tr\left(  1-\frac{1}{3}ReU(x;\hat
{1})U(x+\hat{1};\hat{2})U(x+\hat{2};\hat{1})^{-1}U(x;\hat{2})^{-1}\right)
\end{equation}

\end{itemize}

\begin{itemize}
\item gauge covariant kinetic term, depending both on $U$ and $\phi,$ obtained
by expanding the $S_{W}^{3}$ to the second order in $A_{0}$ 

\begin{equation}
S_{U,\phi}=\sum_{x}\sum_{i=1,2}Tr\left(  D_{i}\left(  U\right)  \phi
(x)\right)
\end{equation}
with $D_{i}$ being covariant derivative
\begin{equation}
D_{i}\left(  U\right)  \phi(x)=U(x;\hat{\imath})\phi(x+a\hat{\imath}%
)U(x;\hat{\imath})^{-1}-\phi(x)
\end{equation}

\item self-interaction term $S_{\phi}$, often called
\textquotedblright Higgs potential\textquotedblright\ 
\end{itemize}

\begin{equation}
S_{\phi}=h_{2}\sum_{x}Tr\phi^{2}(x)+h_{4}\sum_{x}\left(  Tr\phi^{2}(x)\right)
^{2}%
\end{equation}
where couplings are fixed by the parameters of the original model.

Now we have an effective action for our original theory, which lives in
lower-dimen\-sional space and reproduced very accurately results of the
higher-dimensional model. To understand it better one can make the couplings
free and study the model per se. Then, if you do a non-perturbative analysis
of the effective theorytwo unexpected results appear. The theory has new phase
transition, which was not there in the original model with relevant symmetry
being time-reflection, here manifested as $\phi\longrightarrow-\phi$. It can
be spontaneusly broken and Higgs field gets vacuum expectation value.  The
transition is of a strong first order.  Further important property of this
transition is that original coupling s are pretty much on the transition line.
Moreover, they turn out to be in a wrong phase. This suggests that up to
numerical ambiguity and errors from the perturbation theory in the course of
dimensional reduction - they may be {\em exactly} on the transition line.
This transition is of course not a physical one, as it is not present in the
underlying theory which we define as ``physics''. Now lets us go to the
conjecture itself.
\section{The Conjecture}

While the above described scenario has technically nothing to do with nothying
but pure gluodynamics we can reformulate it. First we note that there is
nothing special in the definition of temperature and instead of talking about
high-temperatures effective theory we will view the model as zero-temperature
one with one highly compact dimension. This way we can translate forementioned
results into something more generic.
\begin{itemize}
\item Gauge theory in D dimensions of which one is highly compactified may be
  rewritten as effective theory in $D-1$ dimensions.
\item Effective theory will be formulated via original variables and an
  additional field in the algebra of the gauge group with new couplings.
\item There will be a first order phase transition with one of the phases
  being unphysical.
\item For an observer which lives in $D-1$ dimensional world and somehow
  guessed the effective theory it will be a ``miracle'' that couplings
  self-tune themself to the critical line.
\end{itemize}
This scenario has been extensively tested in 3 dimensions with $SU(3)$ being a
gauge group and in 4 dimensions with $SU(2)$\cite{phd}. It is a bold move to
generalize it to arbitrary number of dimensions as renormalizability, which
differs from dimension to dimension, may change the situation. However
forementioned cases are also very different so one may hope it is a universal
behaviour.

If so, and the number of compact dimensions is more than one it may give rise
to a number of phases with multi-critical points then working as attractors
for effective couplings.

Then it can be the physical reason behind Multiple Point Principle. The
difference between our conjecture and the MPP is of course that only one phase
is actually physical and the remaining ones are the artefacts of the reduction
of the compactified space.  Once one has a candidate theory and mapped its
phase diagram one can get extra information about how the couplings are fixed.
However one should keep in mind that the dynamics in these phases is
completely unphysical and cannot be used for phenomenological purposes.


\backmatter

\thispagestyle{empty}
\parindent=0pt
\begin{flushleft}
\mbox{}
\vfill
\vrule height 1pt width \textwidth depth 0pt
{\parskip 6pt

{\sc Blejske Delavnice Iz Fizike, \ \ Letnik~7, \v{s}t. 2,} 
\ \ \ \ ISSN 1580--4992

{\sc Bled Workshops in Physics, \ \  Vol.~7, No.~2}

\bigskip

Zbornik 9. delavnice `What Comes Beyond the Standard Models', 
Bled, 16.~-- 26.~september 2006

Proceedings to the 9th workshop 'What Comes Beyond the Standard Models', 
Bled, September 16.--26.,  2006

\bigskip

Uredili Norma Manko\v c Bor\v stnik, Holger Bech Nielsen, 
Colin D. Froggatt in Dragan Lukman 

Publikacijo sofinancira Javna agencija za raziskovalno dejavnost Republike Slovenije 

Brezpla\v cni izvod za udele\v zence 

Tehni\v{c}ni urednik Vladimir Bensa

\bigskip

Zalo\v{z}ilo: DMFA -- zalo\v{z}ni\v{s}tvo, Jadranska 19,
1000 Ljubljana, Slovenija

Natisnila Tiskarna MIGRAF v nakladi 100 izvodov

\bigskip

Publikacija DMFA \v{s}tevilka 1652

\vrule height 1pt width \textwidth depth 0pt}
\end{flushleft}


\end{document}